\documentclass[a4paper,12pt]{article}
\pdfoutput=1
\def\parn              {  \par\noindent }

\def\parsmallskipn      {  \par\smallskip\noindent  }
\def\al{\alpha}
\def\be{\beta}
\def\ga{\gamma} \def\Ga{\Gamma}
\def\ep{\epsilon}
\def\lam{\lambda}
\def\Lam{\Lambda}
\def\sig{\sigma}
\def\Sig{\Sigma}

\def\calA{{\cal A}}  \def\calC{{\cal C}} 
 \def\calE{{\cal E}} \def\calF{{\cal F}}
\def\calG{{\cal G}}  
\def\calJ{{\cal J}} \def\calK{{\cal K}} \def\calL{{\cal L}}
\def\calM{{\cal M}}  \def\calO{{\cal O}}
  \def\calR{{\cal R}}
\def\calS{{\cal S}}  
\def\calV{{\cal V}}  
  
\def\jtil{\tilde{j}}

\def\ntil{\tilde{n}}

\def\Vtil{\widetilde{V}}

\def\Omtil{\widetilde{\Omega}}

\def\psitil{\tilde{\psi}}

\def\aldot{{\dot{\alpha}}}

\def\Sighat{\hat{\Sigma}}

\def\sighat{\hat{\sigma}}
\def\psihat{\hat{\psi}}

\def\Tbar{\bar{T}}
\def\ubar{\bar{u}}

\def\zbar{{\bar{z}}}
\def\Zbar{\bar{Z}}

\def\delbar{\bar{\del}}

\def\del        {  \partial }
\def\half       {  {1\over 2}  }
\def\rootof#1   {  \left( #1 \right)^{1/2}  } 
\def\trace      {  \mbox{Tr}\,  }

\def\abs#1      {  \vert #1 \vert  }
\def\ie         {{\it i.e.}\,\,}
\def\evalat#1   {  \left\vert_{#1} \right. } 
\def\e          { {\rm e}  }

\def\comma          {\, ,}
\def\period         {\, .}
\def\lsim      {\lower .65ex \hbox{\ $\stackrel{<}{\sim}$\ } }
\def\gsim      {\lower .65ex \hbox{\ $\stackrel{>}{\sim}$\ } }
\def\det       {{\rm det}\, }

\def\diag{{\rm diag}\,}
\def\bra#1{{\langle #1 | } }
\def\ket#1{{| #1 \rangle } }

\def\acom#1#2{{ \left\{ #1,#2\right\} } }
\def\matel#1#2#3  {{\langle #1 | #2 | #3 \rangle } }

\def\lrvec#1    {\hbox{$\stackrel{\leftrightarrow}{#1}$}}
\def\lvec#1     {\hbox{$\stackrel{\leftarrow}{#1}$}}

\def\vecii#1#2      {  \left(\begin{array}{c}#1\\#2\end{array}\right)  }
\def\veciii#1#2#3   {  \left(\begin{array}{c}#1\\#2\\#3\end{array}
                     \right)  }
\def\veciv#1#2#3#4  {  \left(\begin{array}{c}#1\\#2\\#3\\#4
                                 \end{array}\right)  }
\def\vecfv#1#2#3#4#5 {  \left(\begin{array}{c}#1\\#2\\#3\\#4\\#5
                                 \end{array}\right)  }

\def\matrixii#1#2#3#4            {  \left(\begin{array}{cc}#1&#2\\#3&#4
                                       \end{array}\right) }
\def\matrixiii#1#2#3#4#5#6#7#8#9 {  \left(\begin{array}{ccc}#1&#2&#3\\
                                     #4&#5&#6\\#7&#8&#9\end{array}
                               \right)  }
\def\mativ#1#2#3#4               {  \left(\begin{array}{cccc}
                                       #1\\#2\\#3\\#4\end{array}\right) }
\def\matv#1#2#3#4#5              {  \left(\begin{array}{ccccc}
                                     #1\\#2\\#3\\#4\\#5\end{array}
                              \right)  }
\def\eqabegin         {  \begin{eqnarray}  }
\def\eqaend           {  \end{eqnarray}  }
\def\nn               {  \nonumber  }
\def\bracetwo#1#2     {  \left\{ \begin{array}{l} #1 \\ #2 \end{array}
                         \right.  }
\def\bracetwocases#1#2#3#4  {   \left\{ \begin{array}{ll} #1 &
                                 \qquad #2 \\
                                 #3 & \qquad #4 \end{array} \right.  }
\def\bracebegin#1     {  \left\{ \begin{array}{#1}   }
\def\braceend         {  \end{array}\right.   }
\def\boxit#1#2      {  \vbox{\hrule\hbox{ \hskip -4.1pt \vrule\kern3pt 
                     \vbox
                    {  \hsize #1 \strut\kern3pt #2 \kern3pt\strut  }
                       \kern3pt  \vrule} \hrule  } }
\def\centerbox#1#2  {  \mbox{  }\par\bigskip  \hfil \boxit{#1}{#2} \hfil
                       \par\bigskip\noindent }
\def\rightbox#1#2   {  \hfill\boxit{#1}{#2}  }
\def\leftbox#1#2    {  \boxit{#1}{#2}  }
\def\fullbox#1      {  \boxit{\textwidth}{#1}  }
\newcommand{\nullify}[1]{}
\def\mpg#1#2{\begin{minipage}[t]{#1} #2  \end{minipage} }

\def\bfall{\boldmath\bf }
\def\nxt{\parsmallskipn}

\def\epsfig#1#2#3{
{\lower #3 \hbox{
 \mpg{#1}{\begin{center} \includegraphics[width=#1,clip]{#2.eps} \\
 Fig. #2\end{center} }}}}
\def\papertitlepage{\baselineskip 3.5ex \thispagestyle{empty}}
\def\Title#1{\baselineskip 1cm \vspace{1.5cm}\begin{center}
 {\Large\bf #1} \end{center} 
\vspace{0.5cm}}
\def\Authors#1{\begin{center} {\it #1} \end{center}}
\def\Abstract{\vspace{1.0cm}\begin{center} {\large\bf Abstract} 
           \end{center} \par\bigskip}
\def\Komabanumber#1#2#3{\hfill \begin{minipage}{4.2cm} UT-Komaba #1
              \parn #2 
              \parn #3 \end{minipage}}
\renewcommand{\thefootnote}{\fnsymbol{footnote}}
\def\slprod#1#2{\langle #1, #2 \rangle}

\def\bfall{\boldmath\bf}
\usepackage{ifpdf}
\ifpdf
\pdfoutput=1
\usepackage{graphicx} 
\usepackage[setpagesize=false,pagebackref=false, linktocpage, bookmarksopen=true, colorlinks=true, linkcolor=darkblue,citecolor=darkblue,urlcolor=darkblue]{hyperref}
\usepackage{hyperref}
\newcommand{\arXiv}[2]{\href{http://arxiv.org/abs/#1}{{\tt arXiv:#2}}}
\newcommand{\hep}[2]{\href{http://arxiv.org/abs/#1}{{\tt #2}}}
\def\picture#1#2{\includegraphics[#1]{#2.pdf}}
\else
\usepackage{graphicx}
\newcommand{\arXiv}[2]{[{\tt arXiv:#2}]}
\newcommand{\hep}[2]{[{\tt #2}]}
\def\picture#1#2{\includegraphics[#1]{#2.eps}}
\fi
\usepackage{graphicx}
\usepackage{latexsym}
\usepackage{amsmath}
\usepackage{amssymb}
\usepackage[usenames]{color}
\usepackage{ascmac}
\usepackage{mathrsfs}
\definecolor{darkgreen}{rgb}{0,0.5,0}
\definecolor{darkblue}{cmyk}{0.9,0.9,0,0}
\definecolor{darkred}{rgb}{0.6,0,0.3}
\usepackage{tikz}
\usetikzlibrary{decorations.markings}
\usetikzlibrary{snakes} 
\newcommand{\Cross}{$\mathbin{\tikz [x=1.4ex,y=1.4ex,line width=.2ex, red] \draw (0,0) -- (0.8,0.8) (0,0.8) -- (0.8,0);}$}
\usepackage{cite}
\definecolor{MyRed}{cmyk}{0,1,1,0.15}
\definecolor{MyBlue}{cmyk}{1,1,0,0.25}

\def\Xbar{\bar{X}}
\def\xbar{\bar{x}}

\def\barz{\bar{z}}
\def\delbar{\bar{\partial}}
\def\fn#1{\footnote{#1}}

\def\beq#1{\begin{align}#1\end{align}}
\def\pmatrix#1#2{\left( 
\begin{array}{#1}
#2\end{array} 
\right)}

\def\ads3{Euclidean $AdS_3$}

\def\figref#1{figure \ref{#1}}

\def\iDelta{\scalebox{0.7}[1]{\text{$\mathit{\Delta}$}}}

\newcommand{\beqa}{\begin{eqnarray}}
\newcommand{\eeqa}{\end{eqnarray}}
\newcommand{\bea}{\begin{array}}
\newcommand{\eea}{\end{array}}
\newcommand{\beqn}{\begin{equation}}
\newcommand{\eeqn}{\end{equation}}
\def\Sigtil{{\tilde{\Sigma}}}
\def\bbY{\mathbb{Y}}
\def\sl#1{\langle #1 \rangle}

\def\iDelta{\scalebox{0.7}[1]{\text{$\mathit{\Delta}$}}}

\def\costh2{\cos{\theta_0 \over 2}}
\def\sinth2{\sin{\theta_0 \over 2}}
\def\bbYref{\mathbb{Y}^{\rm ref}}
\def\bbYdiag{\mathbb{Y}^{\rm diag}}
\def\SLCR{{\rm SL(2,C)}_R} 
\def\SLCL{{\rm SL(2,C)}_L}
\def\SL2C{{\rm SL(2,C)}}
\def\SUR{{\rm SU(2)}_R} 
\def\SUL{{\rm SU(2)}_L} 
\def\psiref{\psi^{\rm ref}}
\def\wrons#1#2{\langle #1, #2 \rangle}
\def\AdS3{$AdS_3$}
\def\EAdS3{$E\!AdS_3$}
\def\iDelta{\scalebox{0.9}{\text{$\mathit{\Delta}$}}}
\def\psitilref{\tilde{\psi}^{\rm ref}}
\def\tr{{\rm tr}\,} 
\def\bbP{\mathbb{P}}
\def\Zbar{{\bar{Z}}}
\def\Xbar{{\bar{X}}}
\def\ndiag{n^{\rm diag}}
\def\ntildiag{\tilde{n}^{\rm diag}}

\def\nref{n^{\rm ref}}

\def\psidiag{\psi^{\rm diag}}

\def\itil{\tilde{i}}
\def\vecx{\vec{x}}
\def\bbX{\mathbb{X}}
\def\jhat{\hat{j}}
\def\sl#1{\langle #1 \rangle}
\def\no#1{\textbf{\large :}#1\textbf{\large :}}
\def\easy{\mathcal{A}_{\varpi}}
\def\hard{\mathcal{A}_{\eta}}
\def\kin{\mathcal{V}_{\textrm{kin}}}
\def\dyn{\mathcal{V}_{\textrm{dyn}}}
\def\sym#1{\Big<\!\!\!\Big< #1 \Big>\!\!\!\Big> }
\def\ev#1{\left.#1\right|}
\def\Global{{\sf Global }}
\def\Local{{\sf Local }}
\def\Double{{\sf Double }}
\def\Extra{{\sf Extra }}
\def\bGlobal{\overline{\text{\sf Global}} }
\def\bLocal{\overline{\text{\sf Local}} }
\def\bDouble{\overline{\text{\sf Double}} }
\def\bExtra{\overline{\text{\sf Extra}} }
\def\swkb{{\text{\tiny WKB}}}
\def\id{1}
\def\Re{{\rm Re}\,} 
\def\Im{{\rm Im}\,}
\begin{document}
%\nullify{
%%%%%%%%%%%%%%%%%%%%%%%%%%%%
\papertitlepage
\vspace*{-0.9cm}
\Komabanumber{13-16}{December, 2013}{}
%%%%%%%%%%%%%%%%%%%%%%%%%%%%
\Title{Three-point functions in the SU(2) sector \\
at strong coupling } 
%\vspace{1cm}
\Authors{{\sc Yoichi Kazama\footnote[2]{\textcolor{darkblue}{kazama@hep1.c.u-tokyo.ac.jp}}
 and Shota Komatsu\footnote[3]{\textcolor{darkblue}{skomatsu@hep1.c.u-tokyo.ac.jp}}
\\ }
\vskip 3ex
 Institute of Physics, University of Tokyo, \\
 Komaba, Meguro-ku, Tokyo 153-8902 Japan \\
  }
%\vspace{1.3cm}

%%%%%%%%%%%%%%%%%%%%%%%%%%%%%%%%%%%%%%%%%(*.*)
\numberwithin{equation}{section}
\numberwithin{figure}{section}
\numberwithin{table}{section}
%%%%%%%%%%%%%%%%%%%%%%%%%%%%%%%%%%%%%%%%%
%%%%%%%%%%%%%%%%%%%%%%%%%%%%%%%%%%%%%%%
\parskip=0.9ex
%%%%%%%%%%%%%%%%%%%%%%%%%%%%%%%%%%%%%%%
\baselineskip 3.5ex
%%%%%%%%%%%%%%%%%%%%%%%%%%%%%%%%%%%%%%%
\Abstract
Extending  the methods developed in our previous works (\href{http://arxiv.org/abs/1110.3949}{{\tt 1110.3949}}, 
 \href{http://arxiv.org/abs/1205.6060}{{\tt 1205.6060}}),  we compute the three-point functions at strong
 coupling of the non-BPS states with large quantum numbers corresponding to the composite operators belonging to  the so-called SU(2) sector in the $\mathcal{N}=4$ super-Yang-Mills theory in four dimensions. This is achieved by the semi-classical evaluation of the three-point functions  in the dual string theory in the $AdS_3 \times S^3$ spacetime, using the general  one-cut  finite gap solutions  as the external states. In spite of the complexity of the contributions from various parts in the intermediate stages, the final answer for the three-point function takes a remarkably simple form,  exhibiting the structure reminiscent of the one obtained at weak coupling.  In particular, in the Frolov-Tseytlin limit the result is expressed in terms of markedly  similar integrals,  however with different  contours of  integration.
We discuss a natural mechanism for introducing additional singularities on the worldsheet without affecting the infinite number of conserved charges,  which can  modify the contours of integration. 
\newpage
%%%%%%%%%%%%%%%%%%%%%%
\baselineskip 3.5ex
%%%%%%%%%%%
\renewcommand{\thefootnote}{\arabic{footnote}}
%%%%%%%%table-of-contents%%%%%%%%
\thispagestyle{empty}
\enlargethispage{2\baselineskip}
\renewcommand{\contentsname}{\hrule {\small \flushleft{Contents}}}
{\footnotesize \tableofcontents}
\nxt\hrule
\newpage
%%%%%%%%%%%%%%%%%%%%%%%%%
%\setcounter{page}{1}
%%%%%%%%%%%
\section{Introduction\label{sec:intro}}
%%%%%%%%%%%%%%%
\subsection{Introduction and motivation\label{subsec:intro-mot}}
%%%%%%%%%%%%%%%
After more than fifteen years since the advent of the AdS/CFT correspondence\cite{AdSCFT1,AdSCFT2,AdSCFT3}, 
 we now have a large number of  examples of this type of duality in various dimensions. In  the majority of  these examples,  the bulk and the boundary theories  share  the same (super)conformal symmetry, 
 showing the obvious importance of such a  symmetry.
 On the other hand,  the behaviors of the theories in each specific 
correspondence  are actually rather different,  especially in different dimensions. 
This evidently is due to different dynamics and it in turn 
urges us to understand the common dynamical  structure at the root of the 
 duality which 
 is represented in different  fashions in the bulk and on the boundary
 in various examples. 

With such a purpose in mind, in this article we shall  study the three 
point functions of certain semiclassical non-BPS states in the strong coupling 
regime in the context of the  duality between the string theory in 
$AdS_5 \times S^5$  and the $\mathcal{N}=4$ super Yang-Mills theory in four dimensions. 
More specifically,  we will deal with the string theory 
 in the $E\!AdS_3 \times S^3$ subspace, where $E\!AdS_3$ stands for the Euclidean $AdS_3$. It should be dual to the so-called 
 SU(2) sector\footnote{Actually the global symmetry of this sector is SO(4) $=$ SU(2) $\times$ SU(2), as we will emphasize later.} of the super Yang-Mills theory.  
This should certainly be of great interest 
in view of the fact that the results for the corresponding 
 quantities at weak coupling 
have recently become available\cite{Tailoring3, Kostov1,Kostov2, Serban}.  Detailed comparison 
of the results in two regimes may allow us to identify the common non-trivial 
structure beyond kinematics. 
 
As it will be evident,  the computation of the three-point functions 
 of non-BPS states in string theory in a curved spacetime is  quite non-trivial even at the leading semi-classical level. In the first of such attempts\cite{JW}, 
the contribution from the $AdS_2$ part was evaluated for the string in 
 $AdS_2\times S^k$, where the string is assumed to be rotating 
 only in $S^k$. Since the contribution from the sphere part was not 
computed in \cite{JW},  the complete  answer for the three-point function was not given. 
In this context, our present work can be regarded as (the extended version 
 of)  the completion of  the work initiated by \cite{JW}. 

At about the same time, computation of the three-point functions 
 for different type of heavy external states was attempted by the present authors 
\cite{KK1}. We took as the external states the so-called  Gubser-Klebanov-Polyakov (GKP)  strings\cite{GKPstring}   spinning within $AdS_3$  with large spins.  In this work, the contribution to the three-point function from the action  evaluated on the saddle point configuration  was computed by a method similar to 
 the one in \cite{JW}. However, unlike the case of \cite{JW}, 
 the GKP string is not point-like on the boundary, 
and hence the contributions from the non-trivial vertex operators were 
needed to give the complete answer. Since the precise form of such 
 vertex operators were not known, again the computation had to be 
 left unfinished. This difficulty was later overcome by the development of 
 a new integrability-based method built on the state-operator correspondence  and the contribution of the  non-trivial wave functions of the external  states
 was obtained\cite{KK2}. Combined  with the 
 contribution from the action evaluated previously,  this gave the full answer for the three-point function  of  the GKP strings in the large spin limit\cite{KK2}. 

These  works paved the way  for the present investigation for 
 more general external states  in the product space $E\!AdS_3 \times S^3$.
However, the applications  of our general methods developed in \cite{KK1, KK2} to the present case are not quite  straightforward. One difficulty is that the 
 external states in the $S^3$ sector, which are taken to be general 
 one-cut finite gap solutions \cite{KMMZ, DV1,DV2,Vicedo} for the purpose of making comparison with the weak coupling  result, are much more structured than the large  spin limit of the GKP solutions.  In particular, this makes the analysis of the analyticity property of the basic quantities on the spectral curve  considerably  more  complex.  Another 
 new ingredient concerns the logic of the determination of the internal wave functions  for the $S^3$ sector, which look rather different from those for the AdS
 part. For the AdS part, the wave functions explicitly depend on 
 the positions on the boundary at which  the external string states  land and 
 the form of the dependence is well-known from the conformal symmetry, namely 
 the appropriate power of the difference of these positions. 
For the $S^3$ part, such landing positions do not exist and one must 
 reconsider how to determine the proper wave functions.  
We shall develop a unified method with which one can construct 
 the wave function for a string in a general spacetime. 
Moreover, our method cleanly factorizes the kinematical and the dynamical contributions to the wave functions. This feature is important both conceptually and  practically.  
Also, it should be
 mentioned that this is the first time where we have to combine 
 the contributions from the two different sectors, $S^3$ and \EAdS3, 
 which nevertheless are interconnected through the Virasoro constraints. 
We shall see that when these contributions are put together, considerable simplifications occur, showing the intimate interrelation between them, as expected. 

The end result of our rather involved  computation is a remarkably simple 
 formula for the three-point function, which exhibits intriguing features.  First, one recognizes  the expressions to be quite analogous to those that appear in the weak coupling result,  even before taking any special limits. 
A priori it is not obvious why the result in the strong coupling limit should resemble the weak coupling answer so closely. This  resemblance  becomes 
more conspicuous upon taking the so-called Frolov-Tseytlin limit, 
where the angular momentum $J$ for the $S^3$ rotation is quite large so that the ratio $\sqrt{\lam} /J$, where $\lam$ is the 't Hooft coupling,  is small.
In this limit the integrands of the integrals expressing the answer 
become almost identical. 
However, the integration contours do not quite match. This is not 
immediately a contradiction since  there is no rigorous argument 
why three-point functions should agree exactly in that  limit. 
Nevertheless, it is of interest to look for a possible mechanism  to modify   the contours. One important fact to be noted in this regard is that,   in addition to the ordinary one-cut solutions we used for our external states,  there exist different 
types of one-cut solutions which can be obtained by taking certain 
 degeneration limits of multi-cut solutions. Since the values of the infinite number of conserved charges  do not change in this limiting procedure, these solutions should be considered on equal footing with the corresponding 
ordinary one-cut solutions. 
The important difference, however,  is that such a  ``degenerate solution"  has one or more additional singularities on the worldsheet.  
Since the determination of the contours of integration depends crucially on 
 the analytic structure of the saddle point configuration,  this phenomenon 
provides an example of a natural mechanism by which the contour of integration in the formula for the three-point function can be modified. 
 This issue, however, should be studied  further in future investigations. 

Now as this article has become rather lengthy  due to various steps of somewhat  involved analyses,  it should be helpful  to give a brief preview of  the basic procedures and exhibit the main result. The next subsection 
 will be devoted to this purpose. 
%%%%%%%%%%%%%%%
\subsection{Preview and the main result\label{subsec:intro-pre}}
%%%%%%%%%%%%%%
\subsubsection{The set-up}
%%%%%%%%%%%%%%
The three-point function we wish  to compute in the semi-classical 
 approximation has the following structure:
\begin{align}
G(x_1, x_2, x_3) &= 
 e^{-S[X_\ast]_\ep }  \prod_{i=1}^3 \calV_i[X_\ast; x_i, Q_i ]_\ep  \period \label{corfundecomp}
\end{align}
It consists of the contribution of the action and that of the vertex operators, 
 evaluated on the saddle point configuration denoted  by $X_\ast$. 
The subscript  $\ep$ signifies a small cut-off which regulates the 
 divergences contained in  $S$ and $\calV_i$.  As we shall show, these divergences cancel against each other and the total three-point function is completely finite. The vertex operator  $\mathcal{V}_i[X_\ast; x_i, Q_i ]_\ep$ is assumed to carry 
  a large charge $Q_i$ of order  $O(\sqrt{\lam})$   and is  located at $x_i$ on the boundary of the AdS space. 

In the case of a string in $E\!AdS_3 \times S^3$,
 the action and the  vertex operators are split into the $E\!AdS_3$ part and the $S^3$ part. Their contributions are connected solely through the Virasoro 
 constraint  $T(z)_{E\!AdS_3} + T(z)_{S^3} =0$ (and its anti-holomorphic 
 counterpart).  In the semi-classical approximation, an external state 
 is  characterized by  the asymptotic behavior of a classical solution, which 
should be the saddle point configuration for its two-point function. 
However, conformally invariant vertex operator which creates such a state  is practically impossible to construct at present.  Moreover, even if one had the 
vertex operator, it is of no use since the explicit saddle point solution $X_\ast$ 
 on which to evaluate the vertex operator (and the action) cannot be obtained 
 by existing technology.  

Such difficulties, although seemingly insurmountable, can be overcome 
 with the aid  of the integrable and analytic structure of the system. 
For this purpose, it is convenient to formulate the string theory in question 
 as a non-linear sigma model. Since the treatment of the $S^3$ part and the 
$E\!AdS_3$ part are essentially the same in this regard, we shall focus primarily on 
 the $S^3$ part in this summary.  The basic information is then contained  in the right-current 
 $j\equiv \bbY^{-1} d\bbY$ and the left-current $l \equiv 
 d\bbY \bbY^{-1}$, where $\bbY$ is the $2\times 2$ matrix with unit 
 determinant composed 
 of the embedding coordinates $Y^I\,  (I=1,2,3,4)$ of $S^3$ in the manner
\begin{align}
\bbY = \matrixii{Z_1}{Z_2}{-\bar{Z}_2}{\bar{Z}_1}
\comma \qquad Z_1=Y^1+iY^2\comma \quad Z_2=Y^3+iY^4
\period
\end{align}
$\bbY$ transforms under the global symmetry group SO(4)$=$ SU(2)$_{\rm L}$   $\times$ SU(2)$_{\rm R}$  as  $\bbY \rightarrow V_L \bbY V_R$, with 
$V_L \in $SU(2)$_{\rm L}$, $V_R \in$ SU(2)$_{\rm R}$. 
The equation of motion for $Y^I$ can  then be expressed in the Lax form, 
 with the complex spectral parameter $x$, as 
\begin{align}
\begin{aligned}
&\left[ \del +\calJ_z^r, \delbar +\calJ_\zbar^r  \right] =0\comma \\
& \calJ = \calJ_z^r dz + \calJ_\zbar^r d\zbar\comma 
\quad \calJ_z^r =  {j_z \over 1-x} \comma \quad 
\calJ_\zbar^r ={j_\zbar \over 1+x} \comma 
\end{aligned}
\end{align}
which makes the classical integrability of the system manifest. The information 
 of the infinite number  of conserved charges is encoded in the 
monodromy matrix $\Omega(x) ={\rm P} \exp (-\oint \calJ(x) )$, the 
 eigenvalues of which are given by $\Omega(x) \sim {\rm diag} (e^{ip(x)}, e^{-ip(x)})$,  where   $p(x)$ is the quasi-momentum.  
One can then  define 
 the spectral curve $\Ga$  by $\det (y{\bf 1} -\Omega(x) )=0$, which describes 
 a two-sheeted  Riemann surface in the variable $x$ with a number of cuts with additional singularities. To each such curve corresponds a classical  ``finite gap solution"\cite{KMMZ,DV1,DV2,Vicedo}, which can be constructed  in terms of the solutions
 of the so-called (right and left) auxiliary linear problems, to be abbreviated 
 as ALP throughout,  given by 
\begin{align}
&(i)\quad \left(\del + {j_z \over 1-x}\right) \psi =0\comma \quad 
\left(\delbar + {j_\zbar \over 1+x}\right) \psi =0\comma \label{RALP} \\
&(ii)\quad \left(\del + {x l_z \over 1-x}\right) \psitil =0\comma \quad 
\left(\delbar - {x l_\zbar \over 1+x}\right) \psitil =0  \period \label{LALP}
\end{align}
The solutions $\psi$ and $\psitil$ are expressed  in terms of the Riemann theta functions 
 and the exponential functions,  which depend on  the  data of the curve such as the location of the branch points and other singularities. 
We will be interested in the ``one-cut solution",  the curve for which has a single  square root branch cut of finite size\footnote{As already mentioned in the introduction, this class can contain solutions which are obtained from 
  $m$-cut solutions by shrinking $m-1$ of them to infinitesimal size. 
They may play important roles in obtaining all possible three-point functions 
 of this category.},  since the vertex operators producing such solutions should correspond to the composite operators in the SU(2) sector in $\mathcal{N}=4$  super Yang-Mills theory. 

Now with this setup, let us sketch  how one can compute the three 
 point functions  with the above one-cut solutions\footnote{When it is not
  confusing,  we use one-cut solution to refer either to the one-cut solution of ALP 
 or the solution of the original equation of motion reconstructed in terms of 
 such solutions.} 
 as external legs. 
%%%%%%%%%%%%%
\subsubsection{Evaluation of the contribution of the action }
%%%%%%%%%%%%%%
First consider the evaluation of the action part. 
As  described  in \cite{KK1} for the GKP string
 and will be detailed for the case of our interest, the action integral 
 can be written in the form $S \sim \int_\Sigtil \varpi \wedge \eta$, where 
 $\varpi$ and $\eta$ are, respectively  a holomorphic 1-form and  a closed 1-form defined on the double cover $\Sigtil$ of the worldsheet. By using the Stokes theorem, this can be rewritten as a contour integral 
 $S \sim \int_{\del \Sigtil} \Pi \eta $, where $\del \Sigtil$ is the boundary of $\Sigtil$ and the function $\Pi = \int^z \varpi$ is single-valued on $\Sigtil$. 
This expression for the action can be further rewritten, using a generalization of  the Riemann bilinear identity developed in \cite{KK1},  into a sum of 
products of certain contour integrals. The important point is that the 
 contours of these integrals interconnect the vertex insertion points $z_1, z_2, z_3$, thereby  correlating  the behaviors  around these points. Therefore, to compute 
the integral it is natural to study the behavior of the eigenfunctions of the 
 ALP around $z_i$ and more importantly along the paths connecting $z_i$ and 
$z_j$. 

Although we do not know the exact saddle point solution for the three-point 
 function, we do know the behavior in the vicinity of each $z_i$ since 
 it should be the same as  the one-cut solution discussed  above. 
This provides  the form of the currents  needed to analyze 
the ALP around $z_i$. 
Clearly there are two independent solutions around each $z_i$ and one 
 can compute the local monodromy matrix $\Omega_i$  belonging 
 to SL(2,C),  which mixes  these solutions upon going around $z_i$.  Then one can take the basis of the solutions of ALP  at $z_i$ to be
 the  eigenvectors of $\Omega_i$, denoted by $i_\pm$, belonging 
 to the eigenvalues  $e^{\pm ip_i(x)}$ of $\Omega_i$. These eigenvectors
 are normalized\fn{For the normalization of each eigenvector, see section \ref{subsec:gen-three}.} with respect to the SL(2,C) invariant product 
$\slprod{\psi}{\chi} \equiv \det (\psi, \chi)$, to be refereed to as Wronskian
 throughout this article,  as
$\slprod{i_+}{i_-} =1$.

To gain information about the solution of ALP valid in the entire worldsheet, 
one can make the ``WKB expansion"  with $\zeta =(1-x)/(1+x)$ as the small parameter corresponding to $\hbar$. One then finds that the same contour integrals with which the action is expressed appear in the WKB expansion of the Wronskians 
$ \sl{i_\pm, j_\pm}$.  Therefore our task is reduced to their computation. 
%%%%%%%%%%%%%%%

The crucial information about such Wronskians is  contained in the global consistency condition of the monodromy  matrices  given by 
\begin{align}
\Omega_1 \Omega_2 \Omega_3 =1 \period
\end{align}
Since $\Omega_i$'s cannot in general be diagonalized simultaneously, this 
 serves as  a highly non-trivial constraint. In fact this condition allows one 
 to express  certain products 
 of two Wronskians in terms of the local quasi-momenta 
$p_i(x)$'s, an example of which is given by 
\begin{align}
&\sl{1_+\comma 2_+}\sl{1_-\comma 2_-} = \frac{\sin \frac{p_1(x)
+p_2(x)+p_3(x)}{2}\sin \frac{-p_1(x)-p_2(x)+p_3(x)}{2}}{\sin p_1(x) \sin p_2(x)} \period \label{sl12}
\end{align}
It turns out that the knowledge of the Wronskians, such as $\sl{1_+\comma 2_+}$, between the eigenfunctions at different insertion points is of utmost 
importance.  All the basic quantities, namely the contour integrals giving  the 
contribution of the action and  the wave functions, to be discussed shortly, 
 can be expressed in terms of the Wronskians. 

Therefore the crucial task is 
 to separate out, from the relations such as (\ref{sl12}), the individual 
 Wronskian $\sl{i_\pm, j_\pm}$. This can be achieved if we know 
which of the two factors  is responsible for each zero and the pole  on the spectral curve,  produced by the expression on the right hand side. 
This information dictates the analyticity property of the individual 
Wronskian in $x$ and by solving the appropriate Riemann-Hilbert problem
 we can obtain the Wronskians. 

 As an example, consider the poles produced 
 by the zeros of $\sin p_1(x)$ on the right hand side of (\ref{sl12}), namely 
 at $p_1(x_{\rm pole}) = n\pi$. These are the singular points of the spectral curve
 where the monodromy matrix $\Omega_1(x_{\rm pole})$ takes  the form of
 a Jordan block and the bigger of the two eigenvectors  $1_\pm$ diverges. 
This means that the Wronskian on the left hand side of (\ref{sl12}) involving  such a ``big solution"  must be responsible for these poles. Now which eigenvector is  big  and which
 is small near $z_i$  depends on the value of $x$. 
In the case of the ordinary one-cut solution its explicit form tells us that 
it is dictated by the sign of ${\rm Re}\, q_i(x)$. This means that 
across the line ${\rm Re}\, q_i(x)=0$, the analytic property of the 
 eigenfunctions $i_\pm$ changes. We can then extract the 
 regular part of the Wronskian between ``small solutions" by using the well-known 
technique of Wiener-Hopf decomposition,  which takes the form of 
 a convolution integral  with the contour along the line ${\rm Re} q_i(x)=0$. 
Due to the two-sheeted nature of the spectral curve, the kernel of the decomposition formula must be appropriately generalized. 

Now the remaining analysis, namely that  of the zeros of the right hand side 
 of (\ref{sl12}),  is similar in spirit but is much more complicated because 
 it involves the interplay between the three local quasi-momenta $p_1(x), p_2(x), p_3(x)$ and requires a certain knowledge of the global properties of the 
 solutions of the ALP on the spectral parameter plane. To properly deal with 
 this problem,  we will introduce a notion of the ``exact WKB curve". 
Also,  since  each $p_i(x)$ is double valued, the convolution kernel will  be defined on an eight-sheeted Riemann surface.  Moreover it turns out that the contour  of integration must be determined not just by ${\rm Re}\, q_i(x)=0$ for  each $i$ but also by certain global ``connectivity conditions" expressed in terms  of the quantity $N_i \equiv |{\rm Re}\, p_i(x)|$.  Despite such technical complexities, we will be able to compute the desired Wronskians in terms of the quasi-momenta $p_i(x)$. 

With the procedures described above, one obtains the contribution from 
 the action for the $S^3$ part.  Further,  in an analogous manner, the corresponding contribution from  the $E\!AdS_3$ part can be computed. 
%%%%%%%%%%%%%%%
\subsubsection{Evaluation of the contribution of the vertex operators}
%%%%%%%%%%%%%%%
Let us now turn to the computation of  the contribution of the vertex operators.  To this end,  we extend the powerful method developed in our previous work\cite{KK2} for
 the GKP string to more general string.  It is based on the state-operator correspondence and  the construction of the corresponding wave function in terms of the  action-angle variables.  If one can construct the action-angle variables 
$(S_i, \phi_i)$, the wave function can be constructed simply as 
\begin{align}
\Psi[\phi] &= \exp \left( i \sum_i S_i \phi_i -\calE(\{S_i\}) \tau 
\right)\comma
\end{align}
where $\calE(\{S_i\})$ is the worldsheet energy\footnote{The sum of such energies of course vanishes for the total system due to the Virasoro 
constraint.}. 
 Although the construction of such variables for a non-linear system
 is prohibitively hard in general,  for  integrable systems of the present type 
 there exists a beautiful method \cite{DV1, DV2, Vicedo},  based 
 on the Sklyanin's separation of variables\cite{Sklyanin},  which allows us 
 to construct them from the 
  Baker-Akhiezer eigenvector $\psi$,  which is the solution of ALP satisfying  the monodromy
 equation of the form 
$\Omega(x;\tau, \sigma)  \psi(x;\tau, \sigma)
= e^{i p(x)} \psi(x;\tau, \sigma) $. 
 More precisely,  the dynamical information is  encoded in the 
function $n \cdot \psi(x,\tau)$, where $n =(n_1, n_2)$, to be specified  later,  is referred to as 
the ``normalization vector".  It is known that 
for an $m$-cut solution 
$n \cdot  \psi(x,\tau)$ as a function of $x$  has 
 $m$ zeros at certain  positions $x=\{\ga_1, \ga_2, \ldots, \ga_{m}\}$
 and the dynamical variables $z(\ga_i)$ and $p(\ga_i)$, where 
 $z=\sqrt{\lambda}\left( x+x^{-1}\right)/(4\pi)$, can be shown to form canonical conjugate pairs. Then by making a suitable canonical 
transformation, one can construct  the action-angle variables  $(S_i, \phi_i)$, 
 where, in particular,  the angle variables are  given by the  generalized Abel map 
\begin{align}
\phi_i &= 2\pi \sum_{j=1}^m \int_{x_0}^{\ga_j} \omega_i 
\comma \qquad i=1,2,\ldots, m \period 
\end{align}
Here, $\omega_i$ are  suitably normalized holomorphic differentials (with certain singularities 
 depending on the specific problem) and $x_0$ is an arbitrary base point. 
%%%%%%%%%%%%%%%%%%%%%%%%%%%%%%%%
In the case of the one-cut solution of our interest, we have one angle 
 variable $\phi_R$ associated with the right ALP shown in (\ref{RALP}) 
and one left angle variable $\phi_L$ associated with the left ALP 
described in (\ref{LALP}).  Hereafter we will only refer to the ``right sector"
 for brevity of explanation. 

 Now as we shall describe in section \ref{subsec:one-cut},  we can write down a simple formula which reconstructs the classical string solution from the Baker-Akhiezer 
vector\footnote{Precisely speaking,  we need the Baker-Akhiezer vectors for both  the left and the right ALP, but we ignore such a detail here.} $\psi$.  Therefore,  with a choice of the normalization vector $n$,   one can  associate the angle variable $\phi_R(n)$  to a classical solution, through the (zeros of the) quantity $n \cdot \psi$. 

Let $\bbY$ denote the form of
 the three-point saddle solution near the vertex insertion point $z_i$. We will  call this  part of the solution the $i$th leg. As  we have to normalize the three-point function by the two-point function for each leg, what we wish to compute  is  the angle variable $\phi_R(n)$ associated to $\bbY$ {\it relative to} the one  $\phi^{\rm ref}_R(n)$  associated to the ``reference  two-point solution"  $\bbYref$ which is created  by the  {\it same} vertex  operator\footnote{The same vertex 
 operator can produce  slightly different semi-classical behavior around it, 
 depending on whether it resides in the two-point function or in the three
 point function.} at  $z_i$. 

Now the vertex operators of our 
 interest  are those which correspond to the gauge-invariant composite operators  in the ${\rm SU}(2)$ sector of the super Yang-Mills theory. 
As we discuss  in detail in section \ref{sec:vertex},  the basic operators of that category 
are the charge-diagonal  operators which are ``highest weight"  with respect  to the global symmetry group $\SUR \times \SUL \simeq {\rm SO}(4)$.  Focusing just
 on the $\SUR$ property, one can characterize such an operator by  what we call 
a ``polarization spinor",   in this case  $n^{\rm diag} = (1,0)^t$,  which is annihilated by the raising operator of $\SUR$.  More general operator of our interest can then be obtained from  such a diagonal  operator by an $\SUR$ ($\times \SUL$) rotation and is characterized by the polarization spinor $n$,  obtained from $n^{\rm diag}$ 
by the  corresponding  $\SUR$ rotation.  In this way,  each vertex operator is associated  with such a spinor $n$. 

What is important is that this ``polarization spinor" $n$ can be  shown to be 
identical to  the ``normalization vector" $n$ which determines the 
angle variable $\phi_R(n)$ through the quantity $n \cdot \psi$. 
As was elaborated in our previous work \cite{KK2}, 
once the normalization vector $n$  is specified, the relative shift  $\Delta \phi_R
= \phi_R -\phi^{{\rm ref}}_R$ of  the  angle variable for  the three-point solution $\bbY$ around $z_i$  from  that for the two-point reference solution $\bbYref$ 
can be computed from the knowledge of  the transformation matrix $V \in 
\text{SL(2,C)} $ which 
 connects  $\bbY$ and $\bbYref$  in the manner $\bbY = \bbYref V$ 
in the vicinity of  $z_i$. As it will be shown in section \ref{sec:vertex}, the allowed form of   $V$ can be deduced  from the property  that both $\bbY$ 
 and $\bbYref$ are produced from the same vertex operator characterized by 
 the polarization spinor $n$. 

Then, by using the master formula developed in \cite{KK2}, we can 
 express $\Delta \phi_R$ in terms of $n$,  the solutions 
 of the Baker-Akhiezer functions corresponding to $\bbYref$,  and the 
 parameters describing $V$.  Applying this procedure to each leg of the 
 three-point function and  by making use of a relation between the 
normalization vector $n$ and the value of $\psi$ at $x=\infty$, 
we can express  the wave function (for the right sector)  in terms of 
 the Wronskians as 
\begin{align}
\Psi_R^{S^3} &=\prod_{\{i,j,k\}} \left( {\wrons{j_-}{k_-} \over \wrons{i_-}{j_-} \wrons{k_-}{i_-}  }\bigg|_\infty  {
\wrons{n_i}{n_j}  \wrons{n_k}{n_i} \over \wrons{n_j}{n_k} }\right)
^{ R_i+R_j -R_k}   \period 
\end{align}
Here $n_i$ is the polarization spinor associated with the vertex operator $\calV_i$ at $z_i$ and $R_i$ is the absolute value of the $\SUR$ charge carried by $\calV_i$.  Note that the kinematical part expressed in terms 
 of $\wrons{n_i}{n_j}$ is clearly separated from the dynamical part, 
which again is composed of the Wronskians of the solutions of the ALP. 
The wave function for the $E\!AdS_3$ part can be  obtained in a similar fashion. In that case, the Wronskians $\wrons{n_i}{n_j}$ can be expressed in terms  of the difference of the landing positions  of the three legs on the 
boundary of $E\!AdS_3$ and yield the familiar coordinate  dependence of the 
 three-point functions. 
%%%%%%%%%%%%%%
\subsubsection{Final result for the three-point function}
%%%%%%%%%%%%%%
We now have all the ingredients for the evaluation of  three-point functions. 
Substituting the explicit expressions of the Wronskians $\wrons{i_\pm}{j_\pm}$ into the action and  the wave function and assembling  the contributions 
 from the $S^3$ part and the $E\!AdS_3$ part together, we find that  remarkable simplifications take place in the sum. The final result for the general 
 one-cut external states is thus found to be 
\beq{
\begin{aligned}
\langle \calV_1 \calV_2 \calV_3  \rangle = &\frac{1}{N}\frac{C_{123}}{|x_1-x_2|^{\Delta_1 +\Delta_2-\Delta_3}|x_2-x_3|^{\Delta_2 +\Delta_3-\Delta_1}|x_3-x_1|^{\Delta_3 +\Delta_1-\Delta_2}}\\
&\times \sl{n_1\comma n_2}^{R_1 +R_2-R_3}\sl{n_2\comma n_3}^{R_2 +R_3-R_1}\sl{n_3\comma n_1}^{R_3 +R_1-R_2}\\
&\times\sl{\tilde{n}_1\comma \tilde{n}_2}^{L_1 +L_2-L_3}\sl{\tilde{n}_2\comma \tilde{n}_3}^{L_2 +L_3-L_1}\sl{\tilde{n}_3\comma \tilde{n}_1}^{L_3 +L_1-L_2}\comma
\end{aligned}
\label{Finalkin}
}
where the prefactor $1/N$ comes from the string coupling constant $g_s$ and 
the logarithm of the structure constant $C_{123}$ is given by 
 
{\footnotesize
\beq{
&\ln C_{123}=\nn\\
&\int_{\mathcal{M}^{uuu}_{---}}\hspace{-20pt}\frac{z(x)\left( dp_1 +dp_2 +dp_3\right)}{2\pi i} \ln \sin \left(\frac{p_1+p_2+p_3}{2} \right)
+\int_{\mathcal{M}^{uuu}_{--+}}\hspace{-20pt}\frac{z(x)\left( dp_1 +dp_2 -dp_3\right)}{2\pi i} \ln \sin \left(\frac{p_1+p_2-p_3}{2} \right)\nn\\
&+\int_{\mathcal{M}^{uuu}_{-+-}}\hspace{-20pt}\frac{z(x)\left( dp_1 -dp_2 +dp_3\right)}{2\pi i} \ln \sin \left(\frac{p_1-p_2+p_3}{2} \right)+\int_{\mathcal{M}^{uuu}_{+--}}\hspace{-20pt}\frac{z(x)\left( -dp_1 +dp_2 +dp_3\right)}{2\pi i} \ln \sin \left(\frac{-p_1+p_2+p_3}{2} \right)\nn\\
&-\int_{\hat{\mathcal{M}}^{uuu}_{---}}\hspace{-20pt}\frac{z(x)\left( d\hat{p}_1 +d\hat{p}_2 +d\hat{p}_3\right)}{2\pi i} \ln \sin \left(\frac{\hat{p}_1+\hat{p}_2+\hat{p}_3}{2} \right)
-\int_{\hat{\mathcal{M}}^{uuu}_{--+}}\hspace{-20pt}\frac{z(x)\left( d\hat{p}_1 +d\hat{p}_2 -d\hat{p}_3\right)}{2\pi i} \ln \sin \left(\frac{\hat{p}_1+\hat{p}_2-\hat{p}_3}{2} \right)\nn\\
&-\int_{\hat{\mathcal{M}}^{uuu}_{-+-}}\hspace{-20pt}\frac{z(x)\left( d\hat{p}_1 -d\hat{p}_2 +d\hat{p}_3\right)}{2\pi i} \ln \sin \left(\frac{\hat{p}_1-\hat{p}_2+\hat{p}_3}{2} \right)-\int_{\hat{\mathcal{M}}^{uuu}_{+--}}\hspace{-20pt}\frac{z(x)\left( -d\hat{p}_1 +d\hat{p}_2 +d\hat{p}_3\right)}{2\pi i} \ln \sin \left(\frac{-\hat{p}_1+\hat{p}_2+\hat{p}_3}{2} \right)\nn\\
& -2\sum_{j=1}^{3}\int_{\Gamma_{j_-}^u} \frac{z(x) \,dp_j}{2\pi i} \ln \sin p_j+2\sum_{j=1}^{3}\int_{\hat{\Gamma}_{j_-}^u} \frac{z(x) \,d\hat{p}_j}{2\pi i} \ln \sin \hat{p}_j + \textsf{Contact}\period
}}
The notations used in the above expressions are as follows. 
In the equation (\ref{Finalkin}), $\Delta_i$ 
 is the conformal dimension of the $i$-th vertex operator $\calV_i$ and 
$n_i$ and 
$\ntil_i$ are  the polarization spinors  for $\calV_i$ with respect to $\SUR$ and  $\SUL$. In the expression for $\ln C_{123}$, $p_i$ and 
 $\hat{p}_i$ are the quasi-momenta  for the $i$-th leg for the 
 $S^3$ part and the $E\!AdS_3$ part respectively. $z(x)$ is the Zhukovsky variable given in \eqref{Zhukvar}. The symbols  $\mathcal{M}^{uuu}_{\pm \pm \pm}$and $\Gamma_{j_-}^{u}$ denote the contours of integration for the $S^3$ part
 and $\hat{\mathcal{M}}^{uuu}_{\pm \pm \pm}$ and $\hat{\Gamma}_{j_-}^{u}$ are the contours 
 for the $E\!AdS_3$ contribution. The last term $\textsf{Contact}$ stands 
 for some  special terms  which depend on the detail of the external states.  
It should be noted that the result above for the three-point function for the 
 operators  corresponding to general one-cut solutions is already reminiscent of the expression in the weak coupling regime.  In section \ref{sec:compare},  we demonstrate that 
our  formula  gives  the correct  result for the case of three BPS operators 
and that it  reduces to the two-point function in the limit when the charge of one 
 of the operators becomes negligibly small.  Further,  we  analyze  the Frolov-Tseytlin 
 limit for the case of one non-BPS and two BPS operators and find that the integrals giving the three-point coupling take extremely similar forms,  except for  different 
 contours of integration.  For this issue, we  point out the existence of a natural mechanism by which the contours can be modified. 

Now we briefly indicate the organization of the rest of this  article: 
In section \ref{sec:basic}, we begin with the description of the string in 
$E\!AdS_3 \times S^3$ spacetime and discuss the one-cut solutions 
 we will consider in this work. In section \ref{sec:action}, we will 
study the contribution of the action for the 
  $S^3$ part to the  three-point function and show that the 
 action can be re-expressed in terms of certain contour integrals. 
 In section \ref{sec:vertex}, we describe the evaluation of the wave functions 
for the $S^3$ part. Characterizing the vertex operator by 
a polarization spinor and identifying it with the normalization vector 
 determining the angle variable, we apply the master formula for the shift 
 of the angle variables developed  in our previous work to construct 
 the wave functions. 
Section \ref{sec:wron} will be devoted to the explicit evaluation  of the Wronskians. 
The main task is to find the analyticity property of the 
 Wronskian from the improved WKB analysis of the ALP. Using this 
 information, we can project out the individual Wronskian from the  
expression of the product of Wronskians in terms of
 the quasi-momenta $p_i(x)$  by the use of  the Wiener-Hopf decomposition. 
In section \ref{sec:3pt},  all the results obtained up to this point are put together to 
 produce the final result for the three-point functions of the general one-cut external states.  In section \ref{sec:compare}, in addition to some basic checks of our result, 
 we present the analysis of the Frolov-Tseytlin limit and discuss its outcome.  
Finally, in section \ref{sec:discussion} we make some important comments  on  our present work and indicate possible future directions. 
Several appendices are provided to supply  some additional details. 
%%%%%%%%%%%%%%%%%%%%%%%%%%%%%
%%%%%%%%%%%%%%%%%
\section{{\bfall String in $E\!AdS_3 \times S^3$ and classical 
 solutions}\label{sec:basic}}  
%%%%%%%%%%%%%%%%%%%
We begin by setting up the formalism to deal with the strings 
 in  $E\!AdS_3 \times S^3$ in subsection \ref{subsec:prelim-string} and describe the classical solutions 
 we will use as the external states of the three-point functions in subsection \ref{subsec:one-cut}. We then give a brief account on the basic set-up of the three-point function in subsection \ref{subsec:gen-three}. 
%%%%%%%%%%%%%%%%%%%
\subsection{String in $E\! AdS_3 \times S^3$  spacetime\label{subsec:prelim-string}}
%%%%%%%%%%%%%%%%%%%%%%%%%%
\subsubsection{Preliminaries}
%%%%%%%%%%%%%%%
In this article, we will exclusively deal with the string propagating 
 in the product space of the Euclidean $AdS_3$ subspace of $AdS_5$ (to be denoted by $E\!AdS_3$)  and the sphere $S^3$. 
 If we describe the $AdS_5$ in terms of the embedding coordinates  
by $X^M \eta_{MN} X^N = -1$, where the superscript $M$ is taken 
 to run as  $M=-1, 0, 1,2,3,4$ and the metric 
is given by $\eta_{MN} = \diag (-1, -1, 1,1,1,1)$, then the $E\!AdS_3$ subspace is defined by setting $X^0=X^3=0$. Therefore we will parametrize the $E\!AdS_3 \times S^3$ space in the following way\footnote{In order to conform 
 to the standard convention, we have chosen the range 
 of the $S^3$ embedding coordinates $Y^I$ to be $I=1,2,3,4$, some of which  coincide in name to the range of the $E\!AdS_3$ coordinates. We believe this  will not cause  confusion.}
:
\begin{align}
E\! AdS_3:&\quad X^\mu X_\mu  \equiv X^\mu \eta_{\mu \nu} X^\nu 
 = -1 \comma \quad \mu, \nu=-1,1,2,4  \comma \\
& \quad \eta_{\mu\nu} = \diag (-1, 1, 1,1) \comma \nn\\
S^3:&\quad Y^I Y_I \equiv Y^I \delta_{IJ} Y^J = 1 \comma 
\quad I,J=1,2,3,4  \period
\end{align}
The Poincar\'e coordinates $(x^r, z)= (x^0, x^1, x^2, x^3, z) $
 of $AdS_5$ are defined in the usual way as 
\begin{align}
 {X^{-1} +X^4} & ={1\over z}   \comma \qquad 
X^{-1} -X^4 = z+ {x^r x_r \over z}
\comma \qquad  X^r = {x^r \over z} \comma  \label{Poincare}
\end{align}
where $z=0$ corresponds to the boundary of $AdS_5$. 
When restricted to the $EAdS_3$ subspace\footnote{If desired, one can also deal 
 with  the $AdS_3$ subspace given by $X^2=X^3=0$, with the 
Minkowski boundary plane  $(x^0, x^1)$.},   its boundary 
 is the Euclidean plane parametrized by $(x^1, x^2)$. 

The action of a  string in this space  is given by 
\begin{align}
S &= {\sqrt{\lam} \over \pi} \int d^2 z 
\left( \del X^\mu \delbar X_\mu + \Lam (X^\mu X_\mu +1) 
 + \del Y^I \delbar Y_I + \tilde{\Lam} (Y^I Y_I -1) \right) \comma 
\label{action}
\end{align}
where $\Lam$ and $\tilde{\Lam}$ are Lagrange  multiplier fields. 
Upon  eliminating them  the equations of motion 
become 
\begin{align}
&\del \delbar X^\mu - (\del X^\nu \delbar X_\nu) X^\mu =0 \comma \\
&\del \delbar Y^I + (\del Y^J \delbar Y_J) Y^I =0  \period
\end{align}
For physical configurations,   we must  in addition  impose the Virasoro constraints, which require  that the sum of the stress-energy tensors for the AdS part 
 and the sphere part must vanish. Namely, 
\begin{align}
T_{AdS}(z)  + T_{S}(z) &= 0 \comma \qquad \Tbar_{AdS}(\zbar)
+\Tbar_{S}(\zbar) =0 \comma \\
T_{AdS}(z) &= \del X^\mu \del X_\mu \comma \qquad 
T_{S} (z) = \del Y^I \del Y_I \comma \\
\Tbar_{AdS}(\zbar) &= \delbar X^\mu \delbar X_\mu 
\comma \qquad \Tbar_{S}(\zbar) = \delbar Y^I \delbar Y_I \period
\end{align}
For the AdS part, we shall take the external states to be those without 
 the two-dimensional spins. Then near the vertex insertion point the saddle point solution should approach the two-point solution, which is known to be point-like. 
The forms of  $T_{AdS}(z)$  and $\Tbar_{AdS}(\zbar)$ for such a two-point solution are uniquely determined by their  transformation properties 
as a $(2,0)$ and a $(0,2)$ tensor respectively and are given  in terms 
 of the conformal dimension $\Delta$ of the vertex operator as 
\begin{align}
T_{AdS, {\rm 2pt}}(z) &={\kappa^2 \over z^2} \comma 
\qquad \Tbar_{AdS,{\rm 2pt}} = {\kappa^2 \over \zbar^2} \comma  
\qquad \kappa = \frac{\Delta}{2\sqrt{\lam}} \period
\end{align}
Therefore, taking into account the Virasoro condition,  near each vertex insertion point $z_i$  we must have 
\begin{align}
T_{AdS}(z) \sim {\kappa_i^2 \over (z-z_i)^2} 
\comma \qquad T_{S}(z) \sim {-\kappa_i^2 \over (z-z_i)^2}
\qquad \mbox{as}\ z\rightarrow z_i 
\comma \label{stressbehavior}
\end{align}
and similarly for the anti-holomorphic parts. 
In the case of  three-point functions,  the information of such  asymptotic behaviors suffices
 to determine the form of the energy-momentum tensor exactly everywhere. 
For the $E\! AdS_3$, the holomorphic part  takes the form 
\begin{align}
T(z) &=  \left( {\kappa^2_1 z_{12}z_{13} \over z-z_1} 
+ {\kappa^2_2 z_{21}z_{23} \over z-z_2} +{\kappa^2_3 z_{31}z_{32} \over z-z_3}  \right) {1\over (z-z_1)(z-z_2)(z-z_3) } \comma \\
z_{ij} &\equiv z_i-z_j \period
\end{align}
Here and hereafter, we shall omit the subscript $AdS$ for the stress tensor 
for the AdS part and simply write $T(z)$ and $\Tbar(\zbar)$ for $T_{AdS}(z)$ and $\Tbar_{AdS}(\zbar)$. 

We  now discuss the methods  for constructing  the solutions of the 
 equations of motion with the use of the classical integrability of the system. 
There exist  two apparently different formalisms. One is the sigma model 
formulation\cite{BPR,KMMZ}  and the other is the so-called Pohlmeyer reduction\cite{Pohlmeyer,DS}. 
The former deals with variables which transform covariantly under the 
 global symmetry transformations, whereas the latter employs  invariant 
 variables. Because of this feature they have advantages and disadvantages 
 depending on the problem one would like to solve. We shall employ both. 
It should be remarked however that they are actually connected  by a ``gauge transformation",  as shown in Appendix \ref{apsubsec:con-eigen}.
%%%%%%%%%%%%%
\subsubsection{Sigma model formulation}
%%%%%%%%%%%%%
Consider first the sigma model formulation. We will focus on the $S^3$ part, as  the $E\! AdS_3$ part can be treated similarly. The embedding coordinates $\{Y_I\}$ are conveniently  assembled into a  $2\times 2$ matrix 
with unit determinant given by 
\begin{align}
\bbY &= \matrixii{Z_1}{Z_2}{-\Zbar_2}{\Zbar_1} \comma \label{bbY}\\
Z_1 &= Y_1+iY_2 \comma \qquad Z_2 = Y_3 +iY_4 \comma 
\end{align}
which transforms under  the global symmetry group 
$SO(4) =\SUL \times \SUR$ as 
\begin{align}
\bbY' &= U_L \bbY U_R \comma \qquad U_R \in \SUR\comma \quad 
U_L\in \SUL \period \label{SUtransf}
\end{align}
The quantities of central importance are the ``right" and the ``left" currents  (or connections) $j$ and $l$ respectively, defined by 
\begin{align}
j &\equiv \bbY^{-1} d\bbY \comma \qquad l \equiv d\bbY \bbY^{-1} 
 \period \label{defcurrent}
\end{align}
Evidently, $j$ and $l$ are related by $l=\bbY j \bbY^{-1}$. 
Under the transformation (\ref{SUtransf})  they  transform covariantly as  
 $j \rightarrow U_R^{-1} j U_R$ and $l \rightarrow U_L l U_L^{-1}$. 
 Now reflecting the classical integrability of the system these equations 
 can be extended to  one parameter family of equations called Lax equations given by  
\begin{align}
&\begin{aligned}
& [ \del + \mathcal{J}_z^r(x) , \delbar + \mathcal{J}^r_{\zbar}(x) ] =0 \comma \\
& \mathcal{J}^r_z(x)  \equiv {j_z \over 1-x} \comma \quad \mathcal{J}^r_\zbar(x) \equiv 
{j_\zbar \over 1+x} \comma 
\end{aligned}\\
&\begin{aligned}
& [ \del + \mathcal{J}_l^l(x) , \delbar + \mathcal{J}^l_{\zbar}(x) ] =0 \comma \\
& \mathcal{J}^l_z(x)  \equiv {x l_z \over 1-x} \comma \quad \mathcal{J}^l_\zbar(x) \equiv 
-{x l_\zbar \over 1+x} \comma
\end{aligned}
\end{align}
where $x$ is the complex  spectral parameter. The two connections 
 $\mathcal{J}^r=\mathcal{J}_z^r dz + \mathcal{J}_\zbar^r d\zbar$ and $\mathcal{J}^l=\mathcal{J}_z^l dz + \mathcal{J}_\zbar^l d\zbar$ are related by the gauge transformation of the form 
 $\bbY (d+\mathcal{J}^r) \bbY^{-1} = d+\mathcal{J}^l$. 
It is useful to note that the energy-momentum tensors and hence 
 the Virasoro conditions can be expressed  in terms of the currents in a concise way. We have, in the cylinder coordinate, 
\begin{align}
T_S(z) &= -\half \trace (j_z j_z) = -\kappa^2 
\comma \qquad \Tbar_S(\zbar) = -\half \trace (j_\zbar j_\zbar) = -\kappa^2 \period \label{Tj}
\end{align}

Central to the construction and the analysis of the solutions of the 
 equations of motion are  the right and the left auxiliary linear problems, to be abbreviated as ALP, which are  coupled linear differential equations for vector functions:
\begin{align}
\mbox{right ALP}:\quad  &(\del + \mathcal{J}_z^r(x)) \psi =0 \comma \qquad 
(\delbar + \mathcal{J}^r_{\zbar}(x))\psi =0 \comma \label{rightALP}\\
\mbox{left  ALP}:\quad  &(\del + \mathcal{J}_z^l(x)) \psitil =0 \comma \qquad 
(\delbar + \mathcal{J}^l_{\zbar}(x))\psitil =0 \period \label{leftALP}
\end{align}
Compatibility of the system of ALP  implies  the original equations of motion. 
Upon developing $\psi$ and $\psitil$  from a point $z_0$ along  a closed spacelike curve, we  obtain the right and the left  monodromy matrices $\Omega(x)$ and $\Omtil(x)$
 respectively as
\begin{align}
\Omega(x;z_0) &= {\rm P} \exp \left( -\oint  J^r \right)
 = {\rm P} \exp \left( -\oint \left( {j_z dz \over 1-x} + {j_\zbar d\zbar \over 1+x} \right) \right)
\comma  \label{ROmega}
\\
\Omtil(x;z_0)&= {\rm P} \exp \left( -\oint J^l 
\right)=  {\rm P} \exp \left( -\oint \left( {x l_z dz \over 1-x} - {x l_\zbar d\zbar \over 1+x} \right) \right)
 =\bbY \Omega(x;z_0) \bbY^{-1}   \period \label{LOmega}
\end{align}
By virtue of the flatness of the connection, expansion of $\Omega(x)$ as a function of $x$ around any point yields an infinite number of conserved charges  as coefficients. In particular, expansions around $x=\infty$ and $x=0$ yield, in the leading behavior, the Noether charges for the global  $\SUR$ and $\SUL$,  respectively,  defined by  
\begin{align}
Q_R &\equiv {\sqrt{\lam} \over 4\pi} \oint \ast j = {\sqrt{\lam} \over 4\pi} \int_0^{2\pi} d\sig j_{\tau} \comma  \\
Q_L &\equiv {\sqrt{\lam} \over 4\pi} \oint \ast l = {\sqrt{\lam} \over 4\pi} \int_0^{2\pi} d\sig l_{\tau} \period
\end{align}
Indeed, expanding  $\Omega(x;z_0)$  around 
$x=\infty$ and $x=0$ and using the definitions above,  we get
\begin{align}
\Omega(x;z_0) &= 1 - {1\over x} {4\pi \over \sqrt{\lam}} Q_R + O(x^{-2}) \comma \qquad (x \rightarrow \infty) \comma \label{Omegainfty}\\
\bbY \Omega(x;z_0) \bbY^{-1}&= 1 + x {4\pi \over \sqrt{\lam}} Q_L + O(x^2)
\comma \qquad (x \rightarrow 0) \period \label{Omegazero}
\end{align}
By diagonalizing $\Omega(x;z_0)$, we can obtain a quantity independent of 
 $z_0$. Since $\det \Omega(x;z_0)=1$, 
 its eigenvalues must be of  the structure 
\begin{align}
u(x;z_0) \Omega(x;z_0)u(x;z_0)^{-1} = \matrixii{e^{ip(x)}}{0}{0}{e^{-ip(x)}} \comma  
\end{align}
where $p(x)$ is called the quasi-momentum. 
Comparing with the diagonalized 
form of the expressions (\ref{Omegainfty}) and (\ref{Omegazero}), 
the behaviors of $p(x)$ around $x=\infty$ and $x=0$ are of the form 
\begin{align}
p(x)-p(\infty) &= -{1\over x} {4\pi \over \sqrt{\lam}} R +  O(x^{-2}) \comma \qquad (x \rightarrow \infty) \comma  \label{pxinf}\\
p(x) -p(\infty) &= 2\pi m + x {4\pi \over \sqrt{\lam}} L + O(x^2)
\comma \qquad (x \rightarrow 0)  \comma  \label{pxzero}
\end{align}
where $m$ is an integer and  the  right and the left  charges  $R$ and $L$  are
 the (positive) eigenvalues of $Q_R$ and $Q_L$ respectively.

For the study of the ALP and construction of the finite gap solutions 
of our interest, the analytic property of the quasi-momentum 
is of critical  importance. Such a structure is  encoded 
 in the spectral curve defined by 
\begin{align}
\Ga:\quad \Ga(x,y) &= \det (y {\bf 1} -\Omega(x;z_0) )=0 \comma
\end{align}
 which is equivalent to $(y-e^{ip(x)})(y-e^{-ip(x)}) =0$. In the present 
case, it can be regarded as  a two-sheeted Riemann surface with various  singularities. From the definition of the monodromy matrix (\ref{ROmega}) and (\ref{LOmega}) and the constraints  (\ref{Tj}), it is clear that $p(x)$ has poles at $x=\pm 1$ with the magnitude of the residue equaling  $-2\pi \kappa$. 
 Since $p(x)$ lives on a two-sheeted surface,  we specify its  branch by defining  the signs at these singularities. 
 We shall employ  the definition
\begin{align}
p(x) & \sim {-2\pi \kappa \over x-1} + O ((x-1)^0) \comma 
\qquad (x \rightarrow 1^+) \comma \\
p(x) & \sim {-2\pi \kappa \over x+1} + O ((x+1)^0) \comma 
\qquad (x \rightarrow -1^+) \period 
\end{align}
where the $+$ superscript  on $ 1^+$ signifies  that  the point is on the first sheet. Similarly,  we shall use $-$ superscript for points on the second sheet. 
We will give a more detailed discussion of the structure of $p(x)$ for the one-cut solutions  of our interest in subsection \ref{subsec:one-cut}. 

From the structure of the spectral curve and the quasi-momentum $p(x)$ defined upon it, one can extract  important information.  For this purpose, we first  define the $a$- and $b$-cycles in the usual way. For the hyperelliptic curve of our interest, an $a$-cycle is defined as a cycle which goes around the cut on the same sheet. On the other hand, a $b$-cycle is defined as the one which starts from  a point on the first sheet, goes into the second sheet through the cut and eventually comes back to the same point on th first sheet. 
Clearly, around an $a$-cycle, we have $\oint_{a_i}  dp =0$. In contrast, the integral along  the $b$-cycle does not vanish in general and gives 
$\oint_{b_i} dp = 2\pi n_i$, where $n_i$ is an integer called the mode number. Now using the $a$-type cycles, one can define a set of conserved 
charges called the {\it filling fractions} as 
\begin{align}
S_i \equiv \frac{i}{2\pi} \oint_{a_i} pdz\left( =\oint_{a_i} \frac{z dp}{2\pi i}\right)  \comma 
\label{fillfrac}
\end{align}
where 
\begin{align}
z &\equiv \frac{\sqrt{\lambda}}{4\pi}\left( x + {1\over x}\right) \label{Zhukvar}
\end{align}
is the Zhukovsky variable.  In particular, the filling fractions $S_\infty$ 
 and $S_0$ defined with the contours $a_\infty$ and $a_0$, which 
 encircle the point at $\infty$ and $0$ respectively, are of  special importance 
since they are related to the global $\SUR$ and $\SUL$ charges in the following 
 way, as can be checked using \eqref{pxinf} and \eqref{pxzero}:
\begin{align}
S_\infty &= -R \comma \qquad S_0 = L \period
\label{SinfS0}
\end{align}
%
%%%%%%%%%%%%%%%%%
\subsubsection{Pohlmeyer reduction for a string in $S^3$} 
%%%%%%%%%%%%%%%%%
The sigma model formulation we have sketched above is convenient 
 for analyzing the property of the system  under the global symmetry 
 transformations. Hence it will be used as the basis of the construction 
 of the wave function corresponding to the vertex operators in section \ref{sec:vertex}. 
On the other hand, for the analysis of the contribution of the action, which 
 is invariant under the global transformation, the formalism of the {\it Pohlmeyer reduction} will be more convenient. 

The essential idea of the Pohlmeyer reduction is to describe the motion of the 
  string  in a suitably defined moving frame. This then leads to the Lax equations in terms of the connections which are invariant under the global symmetry 
 transformations. Below we shall only sketch the procedures and then summarize  the basic equations we will need later. Further details will be given in Appendix \ref{sec:pohl}. 

In what follows we shall denote a 4-component field $A^I$ simply as $A$
 and use the notations $A\cdot B=A^I B_I,  A^2=A^I A_I$. 
The  basic moving  frame of  4-component fields, 
to be called $q_i,( i=1,2,3,4)$, are taken as  
$q_1 \equiv Y, q_2 \equiv a \del Y + b \delbar Y, 
 q_3 \equiv c \del Y + d \delbar Y$ and $q_4  \equiv N$, where 
$N$ is the  unit vector orthogonal to $Y, \del Y$ and $\delbar Y$, 
and the (field-dependent) coefficients $a,b,c,d$ are chosen so  that the simple conditions
$q_2 \cdot q_3  = -2, q_2^2=q_3^2=0$  are satisfied. (Note that since $
Y^2 =1$, we automatically  have $q_1^2=1, q_1 \cdot q_2 = q_1\cdot q_3=0$.) Let us define an SO(4)-invariant field $\ga$ by the relation 
\begin{align}
\del Y \cdot \delbar Y &= \sqrt{T\Tbar} \cos 2\ga \period \label{defgamma}
\end{align}
Then, the coefficients $a,b,c,d$ can be expressed in terms of $T, \Tbar$ and $\ga$,  giving  $q_2$ and $q_3$ of the form 
\begin{align}
q_2 &= -\frac{i}{\sin 2\gamma}\left[\frac{e^{i\gamma }}{\sqrt{T}}\del Y+ \frac{e^{-i\gamma }}{\sqrt{\bar{T}}}\delbar Y\right]\comma \\
q_3 &= \frac{i}{\sin 2\gamma}\left[\frac{e^{i\gamma }}{\sqrt{\bar{T}}}\delbar Y+ \frac{e^{-i\gamma }}{\sqrt{T}}\del Y\right] \period
\end{align}
Once the moving frame is prepared, one can compute the derivatives of 
 $q_i$ and express them in terms of $q_i $ again. The result can be 
assembled into the following equations 
\begin{align}
\del W + B_z^L W + W B_z^R =0 \comma \qquad 
\delbar W + B_\zbar^L W + W B_\zbar^R =0 \comma  \label{Weq}
\end{align}
where $W$ is given by 
\begin{align}
W &= \half \matrixii{q_1+iq_4}{q_2}{q_3}{q_1-iq_4} \comma 
\end{align}
and  $B_{z, \zbar}^{L, R}$ are matrices whose components  are 
 expressed in terms of $T, \Tbar$ and $\ga$. (Explicit forms are given in Appendix \ref{sec:pohl}.)  From the equations (\ref{Weq}) one deduces  that the left and the right connections $B^L$ and $B^R$, given in \eqref{BLz}--\eqref{BRzbar},   are flat, namely 
\begin{align}
[\del +B^L_z, \delbar +B_\zbar^L] =0 \comma \qquad 
[\del +B^R_z, \delbar +B_\zbar^R] =0 \period
\end{align}
These relations give the equations of motion for the  invariant fields in the form 
\beq{
\begin{aligned}
&\del \delbar \gamma + \frac{\sqrt{T\bar{T}}}{2}\sin 2\gamma + \frac{2\rho \tilde{\rho}}{\sqrt{T\bar{T}}}\frac{1}{\sin 2\gamma}=0\comma\\
&\del \tilde{\rho}+ \frac{2\delbar \gamma}{\sin 2\gamma}\rho =0\comma \qquad \delbar \rho+ \frac{2\del \gamma}{\sin 2\gamma}\tilde{\rho} =0\comma
\end{aligned}
\label{singordon}
}
where $\rho$ and $\tilde{\rho}$ are defined by 
\begin{align}
\rho\equiv \frac{1}{2}N \cdot \del^2 Y\comma \quad \tilde{\rho}\equiv \frac{1}{2}N \cdot \delbar^2 Y\period\label{rho}
\end{align}

Just as in the case of the sigma model formulation, the integrability of the 
 system allows one to introduce a spectral parameter $\zeta$, 
 related to $x$ by 
\beq{
\zeta = \frac{1-x}{1+x}\comma\label{relzetax}
}
without spoiling the flatness 
 conditions. The Lax equation so obtained is given by 
\begin{align}
[ \del +B_z(\zeta), \delbar +B_\zbar(\zeta)] =0 \comma \label{pohlconnect}
\end{align}
where 
 \beq{\begin{aligned}
&B_z (\zeta)\equiv \frac{\Phi_z}{\zeta} +A_z \comma \quad B_{\barz}(\zeta) \equiv \zeta \Phi_{\barz} + A_{\barz}\comma\\
&\Phi_z \equiv \pmatrix{cc}{0&-\frac{\sqrt{T}}{2}e^{-i\gamma}\\-\frac{\sqrt{T}}{2}e^{i\gamma}&0}\comma\quad \Phi_{\barz} \equiv \pmatrix{cc}{0&\frac{\sqrt{\bar{T}}}{2}e^{i\gamma}\\\frac{\sqrt{\bar{T}}}{2}e^{-i\gamma}&0}\comma\\
&A_z \equiv \pmatrix{cc}{-\frac{i\del \gamma}{2}&\frac{\rho e^{i\gamma}}{\sqrt{T}\sin 2\gamma}\\\frac{\rho e^{-i\gamma}}{\sqrt{T}\sin 2\gamma}&\frac{i\del \gamma}{2}}\comma \quad A_{\barz}\equiv \pmatrix{cc}{\frac{i\delbar \gamma}{2}&\frac{\tilde{\rho} e^{-i\gamma}}{\sqrt{\bar{T}}\sin 2\gamma}\\\frac{\tilde{\rho} e^{i\gamma}}{\sqrt{\bar{T}}\sin 2\gamma}&-\frac{i\delbar \gamma}{2}}\period 
\end{aligned}
}

One can consider the auxiliary linear problem also for the Pohlmeyer connections \eqref{pohlconnect},
\beq{
\left( \del + B_z (\zeta )\right)\hat{\psi}=0\comma \quad \left( \delbar + B_{\barz} (\zeta )\right)\hat{\psi}=0\comma \label{PMALP}
}
where $\hat{\psi}$ denotes the solution in this formulation. 
As shown in Appendix \ref{apsubsec:con-eigen}, the Pohlmeyer connections \eqref{pohlconnect} are actually related to the connections in the sigma model formulation, \eqref{rightALP} and \eqref{leftALP}, by   gauge transformations. Correspondingly,  the solutions to the ALP are also related by gauge  transformations  as
\beq{
\psi = \mathcal{G}^{-1}\hat{\psi}\comma \quad \tilde{\psi}=\tilde{\mathcal{G}}^{-1}\hat{\psi}\comma\label{psipsitil}
}
where $\psi$ and $\tilde{\psi}$ are the solutions to the right  and the left ALP respectively and $\mathcal{G}$ and $\tilde{\mathcal{G}}$ 
are the gauge transformations,  the explicit form of which are given in Appendix \ref{apsubsec:con-eigen}. 
 Here and hereafter,  we shall often refer to the use of the Pohlmeyer formulation as 
choosing  the {\it Pohlmeyer gauge}. 
%%%%%%%%%%%%%%%%%%%
\subsection{One-cut finite gap solutions  in $S^3$\label{subsec:one-cut}}
%%%%%%%%%%%%%%%%
We now describe a particular class of solutions to the equations of motion 
and the Virasoro constraints, which can be constructed  by the so-called  
 finite gap integration method\cite{DV1,DV2,Vicedo}. These
 solutions describe  the local behaviors of the saddle point
 solution for the three-point function in the vicinity of the vertex insertion point. The class of our interest is characterized by the associated spectral curve having one square-root branch cut of finte size and will be referred to 
 as a {\it one-cut solution}.  We will first consider  the ``basic"  one-cut solutions,  which are customarily referred to as  genus $0$ solutions, and study their  properties in detail.  Then,  we describe  another class of one-cut solutions which are obtained from multi-cut solutions by certain degeneration procedure. We show that they contain additional singularities on the worldsheet, 
which may play an important role when we compare the three point functions 
 at strong and weak couplings in section \ref{sec:compare}.  
%%%%%%%%%%%%%%%%
\subsubsection{Basic one-cut solution and ``reconstruction" formula}
%%%%%%%%%%%%%%%%%
A powerful method for constructing a large class of classical solutions  in the sigma model formulation is the so-called finite gap integration method. (For a comprehensive review, see \cite{Vicedo}.)  The method consists of two steps. 
As the first step, the solutions to the left and the right ALP, called 
 the Baker-Akhiezer  functions,  are constructed 
 by treating the problems as Riemann-Hilbert problems on a  finite genus
 Riemann surface.  Namely, by 
proving that the function satisfying all the required analytic properties 
is unique, one constructs such a function in terms of the Riemann theta 
 functions and the exponential functions.  Then, as the second step, 
one develops  the  ``reconstruction" formula\footnote{Although it is  usually 
 referred to as the ``reconstruction" formula, in practice it is used as a 
solution-generating formula. This is because the Baker-Akhiezer functions are 
 constructed not by solving ALP with specific known connections but by more generic methods.}
,  which constructs the solutions  to the original equations of motion from the knowledge of the Baker-Akhiezer functions. In this subsection, we will describe the simplest class of solutions 
 corresponding to the case of genus zero Riemann surface, or a two-sheeted 
 surface with one square-root branch cut. Such solutions will be referred to as  the basic   one-cut solutions. 

Consider first the right ALP given in (\ref{RALP}) and 
let  $\psi_\pm(x,z,\zbar)$ be the Baker-Akhiezer vector which 
 are at the same time the eigenvectors of the monodromy matrix 
$\Omega(x)$  corresponding to  the eigenvalues $e^{\pm ip(x)}$ respectively. 
According to the general theory of finite gap integration, $\psi_\pm$ 
 corresponding to the one-cut solution are given  by simple exponential functions as 
\begin{align}
\psi_+(x;\tau,\sig) &= \vecii{ c_1^+ \exp\left( {i \sig \over 2\pi} 
\int_{\infty^+}^x dp + {\tau \over 2\pi} 
\int_{\infty^+}^x dq \right) }{ c_2^+ \exp\left( {i \sig \over 2\pi} 
\int_{\infty^-}^x dp + {\tau \over 2\pi} 
\int_{\infty^-}^x dq \right) } \comma \label{solpsip} \\
\psi_-(x;\tau,\sig) &= \psi_+(\sighat x;\tau,\sig)  \period
\end{align}
where $c_i^+$ are constants, $\sighat x$ denotes the point $x$ on the 
 opposite sheet, and $\infty^+ (\infty^-)$ is the point at infinity on the first (resp. second) sheet.  The quantity 
 $dp$ is the differential of the quasi-momentum $p(x)$, 
 while $dq$ is the differential of the quasi-energy $q(x)$. Just like $p(x)$, 
 the quasi-energy $q(x)$ is defined  by the pole behavior at $x=\pm 1^+$ 
 of the form 
\begin{align}
q(x) & \sim {-2\pi \kappa \over x-1} + O ((x-1)^0) \comma 
\qquad (x \rightarrow 1^+) \comma \\
q(x) & \sim {+2\pi \kappa \over x+1} + O ((x+1)^0) \comma 
\qquad (x \rightarrow -1^+) \period
\end{align}
The structure and the signs of  the residue at $x=\pm 1$ for $q(x)$ 
 are determined so that the holomorphicity of the solution (\ref{solpsip}) 
at  $x\simeq \pm 1$  is as dictated  by the ALP. For example at $x=1$ 
the holomorphic part of the ALP is dominating  and hence the 
Baker-Akhiezer vector should be holomorphic. This is in fact realized 
 since $p(x) = q(x)$ near $x=1$ and hence the exponent of $\psi_\pm$ 
is a function of the combination $z=\tau+ i\sig$. In the same way, 
 at $x=-1$ the exponent of $\psi_\pm$ becomes anti-holomorphic  as desired.

Now for the left ALP, the Baker-Akhiezer eigenvectors, denoted by 
  $\psitil_\pm(x,z,\zbar)$, are given by 
\begin{align}
\psitil_+(x;\tau, \sig) &=  \vecii{ c_1^- \exp\left( {i \sig \over 2\pi} 
\int_{0^+}^x dp + {\tau \over 2\pi} 
\int_{0^+}^x dq \right) }{ c_2^- \exp\left( {i \sig \over 2\pi} 
\int_{0^-}^x dp + {\tau \over 2\pi} 
\int_{0^-}^x dq \right) } \comma \\
\psitil_-(x;\tau,\sig) &= \psitil_+(\sighat x;\tau,\sig) \comma 
\end{align}
where the notations are similar and should be self-explanatory. 

We will be interested in the case where the branch cut runs between 
  $u$ and its complex conjugate $\ubar$ on the spectral curve. 
Such a cut is described by a factor of the form 
\begin{align}
y(x) &\equiv \sqrt{(x-u)(x-\ubar)} \period
\end{align}
We define the branch of $y(x)$ to be such that the  sign of $y(x)$ is $+1$  at  $x=1^+$. Then $p(x)$ and $q(x)$ satisfying the prescribed analyticity properties are fixed to be 
\begin{align}
p(x) &= -2\pi \kappa y(x) \left( {1\over |1-u|} {1\over x-1} 
 + \ep {1\over |1+u|} {1\over x+1} \right) \comma  \label{onecutpx}\\
q(x) &= -2\pi \kappa y(x) \left( {1\over |1-u|} {1\over x-1} 
 - \ep {1\over |1+u|} {1\over x+1} \right) \comma \label{onecutqx}\\
\ep &= \bracetwo{+1\quad \mbox{for}\quad |\Re u|>1}{-1\quad \mbox{for}\quad |\Re u|<1} \period \label{defep}
\end{align}
Here we fixed $p(x)$ and $q(x)$ such that they vanish at the branch points although the analyticity properties only determine the differential $dp$ and $dq$. This choice is suitable for the purpose of this paper since the solutions to the ALP in the Pohlmeyer gauge.  
The forms of $p(x)$ and $q(x)$ depend on whether the cut is placed to the right or to the left of  $x=1$. Substituting these forms into the formulas for $\psi_\pm$ and $\psitil_\pm$ we get the one-cut solutions for the ALP. 

Let us now describe  the second step,  the (re)construction of the 
solutions of the  equations of motion from the Baker-Akhiezer vectors. 
Although this has been discussed in the literature\cite{DV1,DV2, Vicedo},  we present below a  more transparent formula. Let us form 
a  $2\times 2$ matrix $\Psi$ in terms of the two independent Baker-Akhiezer column vectors $\psi_\pm$  satisfying the right ALP as 
$\Psi = ( \psi_+\, \psi_-) $
and consider the quantity 
\begin{align}
\tilde{\Psi} \equiv \bbY \Psi \period
\end{align}
 Then, by 
 using the definitions $l_{z} =\del \bbY \bbY^{-1}$ and 
 $j_{z} =\bbY^{-1} \del \bbY$,  we can easily show that 
\begin{align}
&\left( \del + {x l_z \over 1-x} \right) \tilde{\Psi} 
 = \bbY \left( \del + {j_z \over 1-x} \right) \Psi =0 \comma  
\label{PsitileqR} \\
&\left(  \delbar -{x l_\zbar \over 1+x} \right) \tilde{\Psi} 
 =  \bbY  \left( \del + {j_\zbar \over 1+x} \right) \Psi =0 
\period \label{PsitileqL} 
\end{align}
If we express $\tilde{\Psi}$ in terms of  two column vectors  $\tilde{\psi}_\pm$ 
 as $\tilde{\Psi} = (\tilde{\psi}_+ \, \tilde{\psi}_-)$,  the above equations  show  that 
 $\tilde{\psi}_\pm$ are actually two independent solutions to  the left ALP. 
This means that there exist solutions $\psi_\pm$ and $\tilde{\psi}_\pm$ 
to the right and the left ALP respectively  so that $\bbY$ can be expressed 
 as 
\begin{align}
\bbY &= \tilde{\Psi} \Psi^{-1}  \period   \label{reconst}
\end{align}
This general relation by itself, however,  is not  useful since even if we provide a solution $\Psi$ explicitly, finding $\tilde{\Psi}$ which satisfies 
\eqref{reconst} tantamounts to finding  $\bbY$ itself.  
Now the formula \eqref{reconst} turns into a genuine reconstruction 
formula  when we consider the special values of the spectral parameter $x$. 
If we set $x=0$, it is evident from the form of ALP redisplayed above 
 in \eqref{PsitileqR} and \eqref{PsitileqL} that the left ALP equations for $\tilde{\Psi}$  reduce  to $\del \tilde{\Psi} = \delbar \tilde{\Psi}=0$, and 
hence $\tilde{\Psi}(x=0)$ becomes a constant matrix. Therefore the 
 solution $\bbY$ is reconstructed from the  right ALP solution $\Psi$ as  $\bbY(z,\zbar) = \tilde{\Psi}(x=0) \Psi^{-1}(z,\zbar;x=0)$, where the constant matrix $\tilde{\Psi}(x=0)$ represents  the freedom of making a global transformation from left.  Similarly, by setting $x=\infty$, we can make the right ALP 
 equations trivial, namely  $\del \Psi = \delbar \Psi =0$. Then $\Psi (x=\infty)$ becomes a constant matrix and $\bbY$ can be reconstructed from 
 the  left ALP solution $\tilde{\Psi}$ as $\bbY(z, \zbar) = \tilde{\Psi}(z,\zbar;x=\infty) \Psi^{-1}(x=\infty)$. Summarizing, we have two types of simple reconstruction  formulas 
\begin{align}
\bbY(z,\zbar) &=  \tilde{\Psi}(0) \Psi^{-1}(z,\zbar;0)  \comma 
\label{reconsteqR}\\
\bbY(z,\zbar)&=\tilde{\Psi}(z,\zbar;\infty) \Psi^{-1}(\infty) \period
\label{reconsteqL}
\end{align}

By using the reconstruction formula given above, one can write down 
 the general basic one-cut solution explicitly. It can be written  in the form\cite{KMMZ,Vicedo}
\begin{align}
\bbY &= \matrixii{\costh2 e^{\nu_1 \tau  + im_1\sig}}{ \sinth2 e^{\nu_2 \tau  + im_2 \sig}}{  - \sinth2 e^{-\nu_2 \tau  - im_2 \sig}}{
\costh2 e^{-\nu_1 \tau  - im_1\sig}}
\comma \label{Ysolexp}
\end{align}
where the parameters $\nu_i, m_i$ and $\theta_0$ must satisfy the 
 following conditions expressing the equations of motion and the Virasoro 
 conditions:
\begin{align}
 & \nu_1^2 -m_1^2 = \nu_2^2-m_2^2 \comma \label{eqmotion}\\
 & 4\kappa^2 = (\nu_1^2 +m_1^2) \cos^2 {\theta_0 \over 2} +  (\nu_2^2 +m_2^2) \sin^2 {\theta_0 \over 2} \comma  \label{Virone}\\
 & \nu_1m_1 \cos^2 {\theta_0 \over 2} 
+ \nu_2m_2 \sin^2 {\theta_0 \over 2} =0 \period \label{Virtwo}
\end{align}
Applying the reconstruction formula (\ref{reconsteqR}) 
 with the constant matrix $\tilde{\Psi}(0)$ taken to be the identity matrix and using the form of $\psi_+$  given in (\ref{solpsip}),  we easily find that the 
parameters  $m_i$ and $\nu_i$ can be  expressed in terms of $p(x)$ and $q(x)$ as 
\beq{
&m_1=\frac{1}{2\pi}\int_{0^{+}}^{\infty^{+}}dp\comma\qquad 
\nu_1=\frac{1}{2\pi}\int_{0^{+}}^{\infty^{+}}dq\comma\label{ppq1}\\
&m_2=\frac{1}{2\pi}\int_{0^{+}}^{\infty^{-}}dp\comma \qquad
\nu_2= \frac{1}{2\pi}\int_{0^{+}}^{\infty^{-}}dq\period \label{ppq2}
}
The right and the left Noether charges $R$ and $L$ can be computed directly from the solution \eqref{Ysolexp} and are given in terms of the parameters $\nu_i$, $m_i$ and $\theta_0$ in a universal manner as
\begin{align}
{R\over \sqrt{\lam}} &= \frac{1}{2}\left( -\nu_1 \cos^2{\theta_0 \over 2} +\nu_2\sin^2{\theta_0 \over 2}\right) \comma \\
{L\over \sqrt{\lam}} &= \frac{1}{2}\left(-\nu_1 \cos^2{\theta_0 \over 2} -\nu_2\sin^2{\theta_0 \over 2} \right)\period
\end{align}
Explicit expressions of $R$ and $L$ in terms of the position of the cut are given in Appendix \ref{apsubsec:cut}. As a result, we find that the charges $R$ and $L$ are positive irrespective of the position of the cut. This means that they should be regarded not as the charges themselves but as their {\it absolute magnitudes}. On the other hand, the relative magnitude of $R$ and $L$ depends on the position of the cut as
\begin{align}
&R<L \qquad \mbox{for}\quad |\Re u|>1 \comma \label{ug1}\\
& R>L \qquad \mbox{for}\quad |\Re u|<1 \period \label{ul1}
\end{align}
In section \ref{subsec:relation}, we will see that the difference in the relative magnitude corresponds 
 to the difference of the class of vertex operators for which the solution 
 is the saddle point of the two-point function. 
%%%%%%%%%%%%%%%%%
\subsubsection{One-cut solutions from multi-cut  solutions\label{subsubsec:multi}}
%%%%%%%%%%%%%%%%
We now discuss a more general  type of ``one-cut" solutions, namely 
 the ones with additional cuts of infinitesimal size besides a cut of finite 
size. As we shall discuss in section \ref{subsubsec:mech},  this type of solutions may play an important role in the comparison of the three-point functions at strong and weak couplings.  Besides such specific reason,  as these infinitesimal cuts do not contribute to any of the  (infinite number of) conserved charges, 
they should, on general grounds,  be considered on an equal footing with the corresponding solutions carrying the same charges.  As a matter of fact, it is much more natural to consider solutions with infinite number of infinitesimal 
 cuts, as they correspond to the infinite number of angle variables 
which must exist for a string theory even when their conjugate action variables have vanishing values\footnote{As already emphasized in \cite{KK2}, in order to construct  a three-point solution in the framework of the finite gap method, which is tailored for construction of  two-point solutions,   inclusion of infinite number of small infinitesimal cuts is necessary as one has to produce an additional singularity corresponding to the third vertex operator. }.    
Now adding an infinitesimal cut to the genus $g$ Riemann surface  is equivalent to shrinking  a cut in the genus $g+1$ surface\footnote{A similar discussion of this 
 process can be found in \cite{Vicedo2}.}. 
As we shall see, depending on the choice of the parameters  
we either get back an ordinary genus $g$ finite gap solution 
 or we obtain a new solution with additional singularities. 

In contrast to the one-cut solution corresponding to genus zero
we have been considering,  
for a genus $g$ finite gap solution with $g \ge 1$ the components 
 of the Baker-Akhiezer vector are given by the following expressions
 containing  ratios of Riemann theta functions  
$\Theta(\boldsymbol{z})$ in addition to the exponential part:
\begin{align}
\psi_1 &= h_+(x) {\Theta(\calA(x) +k\sig -i\omega \tau -\zeta_{\ga_{-}(0)}) \Theta (\calA(\infty^+) -\zeta_{\ga_{-}(0)} )
\over \Theta(\calA(x) -\zeta_{\ga_{-}(0)}) \Theta (\calA(\infty^+)+ k\sig -i\omega \tau-\zeta_{\ga_{-}(0)}) }\exp \left(  {i\sig\over 2\pi} \int^x_{\infty^+} dp 
+ {\tau \over 2\pi } \int^x_{\infty^+} dq \right)  \comma \label{psione} \\
\psi_2 &= h_-(x) {\Theta(\calA(x) +k\sig -i\omega \tau -\zeta_{\ga_{+}(0)}) \Theta (\calA(\infty^-) -\zeta_{\ga_{+}(0)} )
\over \Theta(\calA(x) -\zeta_{\ga_{+}(0)}) \Theta (\calA(\infty^-)+ k\sig -i\omega \tau-\zeta_{\ga_{+}(0)} )}\exp \left( {i\sig\over 2\pi} \int^x_{\infty^-} dp 
+ {\tau \over 2\pi } \int^x_{\infty^-} dq \right) \period \label{psitwo}
\end{align}
As it is not our purpose here to review the details of the finite gap construction, below we will only explain the minimum of the ingredients and refer the reader  to a review article such as \cite{Vicedo}.  Also, for simplicity and clarity, 
 we will focus on the case of the degeneration from $g=1$ to $g=0$. 
This suffices 
 to explain the essence of the construction and the generalization to 
 the case of higher genus is straightforward. 

For a  $g=1$ two-cut solution, the Riemann theta function $\Theta(\boldsymbol{z})$ reduces to the elliptic  theta function $\theta(z)$ defined by 
\begin{align}
\theta (z) &\equiv \sum_{m \in \mathbb{Z}}\exp \left( i m z + \pi i \Pi m^2 \right) \comma 
\end{align}
where $\Pi$ is the period given  by the integral of the holomorphic differential $w$ over the  $b$-cycle 
 of the torus 
\begin{align}
\Pi &= \oint_b w \period
\end{align}
As usual,   $w$ is  normalized by the   integral 
over the $a$-cycle as   
$\oint_a w =1$. $\calA(x)$ appearing in the argument of the $\Theta$-functions is the Abel map defined by 
\begin{align}
\calA(x) &= 2\pi \int_{\infty^+}^x w \period
\end{align}
$h_\pm(x)$ are normalization constants and 
$k$ and $\omega$ are the ``momentum" and the ``energy" defined 
 by the integrals 
\begin{align}
k &\equiv {1\over 2\pi} \oint_b dp \comma \qquad 
\omega \equiv {1\over 2\pi} \oint_b dq \period
\end{align}
A quantity of importance is  the constant $\zeta_{\ga_\pm(0)}$
 defined by 
\begin{align}
\zeta_{\ga_\pm(0)} &\equiv \calA(\ga_\pm(0)) + \calK \comma 
\end{align}
In this formula, 
$\calK$ is the ``vector of Riemann constants", which for a torus 
 is  simply  a number proportional to  the period $\Pi$ as\footnote{For its definition for a  general genus $g$ surface, see for example \cite{FK}.}
\begin{align}
\calK &= \pi \Pi \period  \label{defcalK}
\end{align}
Finally $\ga_\pm(0)$ are certain points\fn{Precisely speaking, $\ga_\pm(0)$ are certain divisors $\ga_\pm(t)$ depending 
 on the infinite set of higher times $t=(t_0, t_1, t_2,\ldots) $ evaluated 
 at $t=0$. For a detailed 
definition,  see \cite{Vicedo}.} on the Riemann surface, which determine the initial conditions for the solution.

Let us now study what happens when we pinch the $a$-cycle. In order to keep  the normalization condition $\oint_a w =1$ intact, $w$ must behave near the position  of the infinitesimal cut $x_c$ as 
\begin{align}
w &\sim \bracetwocases{{1\over 2\pi i} {1\over x-x_c}}{\mbox{for $x$
 on the first sheet}}{-{1\over 2\pi i} {1\over x-x_c}}{\mbox{for $x$
 on the second sheet}} \period
\end{align}
This  means that the imaginary part of the period $\Pi$ defined by 
 the integral over the $b$-cycle approaches positive infinity in the manner
\begin{align}
\Pi &= \oint_b w  \sim {1\over 2\pi i} \int^{x_c +\ep}_{x_c -\ep}  {dx \over x-x_c} 
 \sim -{i \over \pi } \ln \ep \rightarrow +i \infty \period
\end{align}
Now writing the $\theta$-function as 
\begin{align}
\theta(z) &= \sum_{m \in \mathbb{Z}}\exp \left( i m z + \pi i (\Re \Pi )m^2 \right) \cdot  \exp \left( -\pi \Im \Pi m^2 \right) \comma 
\end{align}
we see that the last factor vanishes as $\Im \Pi \rightarrow \infty$, 
 except for $m=0$.  Therefore in this limit we get  $\theta(z) \rightarrow 1$ 
and one gets the usual genus 0 solution with  only the exponential part. 

Now if we identify $z= k\sig -i\omega \tau$ in the  formulas for $\psi_i$ given in (\ref{psione}) and (\ref{psitwo}),  the arguments  of the $\theta$-functions 
containing $z$ are actually of the form  $z-a$,  with a  constant shift $a$ given by  $a= \zeta_{\ga^{\pm}(0)} +  \cdots$.  
What is important is that  $\zeta_{\ga^{\pm}(0)}$  diverges as we pinch the $a$-cycle. First, obviously $\Im \calK$ diverges as $\pi \Im \Pi$. 
Second, if $\ga_\pm(0)$ is at the position of the shrunk cut $x_c$, 
$\Im \calA(\ga_\pm(0))$ diverges just like $\pi \Im \Pi$:
\begin{align}
\calA(\ga_\pm(0)) &= 2\pi \int^{x_c + \ep}_{\infty^+} dw 
 \sim 2\pi {1\over 2\pi i} \ln \ep  \sim i \pi \Im \Pi \rightarrow i\infty \period
\end{align}
Since $\calA(\ga_\pm(0))$ is finite otherwise, we must distinguish 
 two cases: case (a)\, $\Im \zeta_{\ga^{\pm}(0)} 
\sim 2\pi \Im \Pi$ for $\ga_{\pm(0)} = x_c$
 and case (b)\, $\Im \zeta_{\ga^{\pm}(0)} 
\sim \pi \Im \Pi$ for $\ga_{\pm(0)} \ne x_c$. 
Therefore let us write  $a=l \Im \pi \Pi + c$, where $l=2$ or $l=1$ and  $c$ is a finite constant. Then the $\theta$-function with this shift can be 
 written as 
\begin{align}
\theta(z-a) &= \sum_{m \in \mathbb{Z}}\exp \left( i m (z -\pi \Re \Pi -c)
 + \pi i (\Re \Pi )m^2 \right) \cdot  \exp \left(-\pi \Im \Pi(m^2-l m) \right) \period 
\end{align}

First consider the case (a). It is easy to see that terms with negative $m$ all vanish in the limit 
 $\Im \Pi \rightarrow \infty$. On the other hand, the terms with $m=0$ and $m=2$ are finite and those with $m \ge 3$ vanish in the degeneration limit while the single term with $m=1$ diverges. In other words, 
\begin{align}
\theta (z-a) \rightarrow \hat{C} e^{iz} \comma \qquad \hat{C} \rightarrow \infty \period
\end{align}
 As the $\theta$-functions occur in pairs in the numerator and the 
 denominator  in $\psi_i$, their ratio goes to a $z$-independent finite constant in the degeneration limit and we get back the usual $g=0$ one-cut solution. 
In fact,  by repeating this type of process, one can  produce  a finite gap solution  from an infinite gap solution, which must be the generic situation for theories  with infinite degrees of freedom, such as  string theory. 

Next consider the case (b). For $l=1$, two terms in the series survive
in the limit $\Im \Pi \rightarrow \infty$, namely $m=0$ and $m=1$. Therefore we obtain a non-trivial function of the form 
\begin{align}
\theta (z-a) &\rightarrow 1+Ce^{iz} = 1+ C e^{ik\sig + \omega \tau} 
\comma 
\end{align}
where $C$ is a constant. In particular, this function can vanish at certain points, the number of which depend on the magnitude of $k$.  Such a  $\theta$-function in the denominator of the expressions for $\psi_i$  gives rise to additional  simple poles on the worldsheet. In distinction to the singularity due to 
 a vertex operator, these singularities do not carry any  charges (including 
 infinitely many higher charges) because the solution is   obtained 
 without changing the form of $p(x)$.  

Although we will not explicitly make use of the degenerate multi-cut solutions discussed above in the bulk of our investigation, they will be recognized in section \ref{subsubsec:mech} as providing an example of a concrete mechanism by which extra singularities can be naturally produced.  Existence of such singularities can modify the contours of  the integrals that express the three-point coupling and may play an important role in the interpretation of our final result. 
\subsection{Some properties of the eigenvectors of the monodromy matrix\label{subsec:gen-three}}
%%%%%%%%%%%%%%%%%%%%%
As described in the preceding subsections, the solutions of the ALP play the 
central role in the construction of the two-point solutions to the equations of motion. Now for the construction of the three-point functions, to be discussed starting from the next section,  what will be of vital  importance are the special linear combinations of the solutions of ALP, namely the eigenvectors of the local monodromy matrix $\Omega_i$, defined around each  vertex insertion point $z_i$. We will denote  such eigenvectors and  eigenvalues as 
$i_\pm$ and $e^{\pm ip_i(x)}$,  which satisfy the relations 
\beq{
\Omega_i i_{\pm}=e^{\pm p_i(x)} i_{\pm}\period\label{normcond1}
}
In what follows, we will describe some important properties of $i_\pm$ and 
 related states. 

Of crucial importance in the computation of three-point functions will be  the SL(2,C) invariant product for  $i_\pm$ and $j_\pm$ given by 
\beq{
\sl{i_\pm, j_\pm}\equiv \det \left( i_\pm,j_\pm\right)\period
}
 In the rest of the paper, we shall refer to this skew-product as {\it Wronskian}. Since the Wronskians are invariant under  gauge transformations,  we can use the results in various gauges interchangeably.  For example, from the 
 relation \eqref{psipsitil} between the 
eigenvectors in the sigma model formulation and the Pohlmeyer formulation, we have the equalities 
\beq{
\sl{i_{\pm},j_{\pm}}=\sl{\tilde{i}_{\pm},\tilde{j}_{\pm}}=\sl{\hat{i}_{\pm},\hat{j}_{\pm}}\period
}

For later convenience, let us  fix the normalization of the eigenvectors $i_{\pm}$. We will first impose the usual condition
\beq{
\sl{i_{+},i_{-}}=1\period\label{normcond2}
}
This, however, does not  fully fix the normalization of the individual eigenfunctions,  as we can rescale $i_{\pm}$ as  $i_+\to a i_+$ and $i_-\to a^{-1} i_-$, without violating the condition \eqref{normcond2}.  To determine the normalization completely, we will make use of the asymptotic behavior of $i_{\pm}$ around the puncture $z_i$. 
For this purpose, it is convenient to employ the Pohlmeyer gauge, as it  is invariant under the global symmetry transformation.  Now although the explicit form of the solution for the three-point function is not known, it can be approximated by the solution for the two-point function in the vicinity of the vertex operators. 
Therefore, we can determine the normalization of  $\hat{i}_\pm$ by 
demanding that they coincide with the corresponding two-point functions 
at the insertion point of the vertex operator:
\beq{
\hat{i}_{\pm} (x;\tau^{(i)},\sigma^{(i)}) \ \longrightarrow\   \hat{i}_{\pm}^{\rm 2pt}(x;\tau = \tau^{(i)},\sigma =\sigma^{(i)} )\period\label{normcond3}
}
In this formula,  $(\tau^{(i)},\sigma^{(i)})$ are the  local cylinder coordinates around $z_i$, defined by
\beq{
\tau^{(i)}+i\sigma^{(i)}=\ln \left( \frac{z-z_i}{\epsilon_i}\right)\period\label{localcoordinate}
}
Here we have chosen the origin of $\tau^{(i)}$ to be such that $\tau^{(i)}=0$  on the small circle $|z-z_i| = \ep_i$,  which will serve 
 to separate the contributions from the action and the wave function 
in subsequent sections. 
Using the results of Appendix \ref{apsec:1-cut}, the eigenvectors for the two-point function $\hat{i}_{\pm}^{\rm 2pt}$ can be computed  as
\beq{
&\hat{i}_{+}^{\rm 2pt}(x;\tau,\sigma) = \pmatrix{c}{\frac{e^{\pi i/8}}{\sqrt{2}}\left(\frac{x-\bar{u}_i}{x-u_i}\right)^{1/4}\left(\frac{\bar{u}_i^2-1}{u_i^2-1}\right)^{1/8}\\\frac{e^{\pi i/8}}{\sqrt{2}}\left(\frac{x-u_i}{x-\bar{u}_i}\right)^{1/4}\left(\frac{u_i^2-1}{\bar{u}_i^2-1}\right)^{1/8}}\exp \left( \frac{q_i(x)\tau + ip_i(x)\sigma}{2\pi}\right)\comma\label{psi1}\\
&\hat{i}_{-}^{\rm 2pt}(x;\tau,\sigma) = \pmatrix{c}{\frac{e^{-\pi i/8}}{\sqrt{2}}\left(\frac{x-\bar{u}_i}{x-u_i}\right)^{1/4}\left(\frac{\bar{u}_i^2-1}{u_i^2-1}\right)^{1/8}\\-\frac{e^{-\pi i/8}}{\sqrt{2}}\left(\frac{x-u_i}{x-\bar{u}_i}\right)^{1/4}\left(\frac{u_i^2-1}{\bar{u}_i^2-1}\right)^{1/8}}\exp \left( -\frac{(q_i(x)\tau + ip_i(x)\sigma)}{2\pi}\right)\comma\label{psi2}
}
where $u_i$ and $\bar{u}_i$ are the positions of the branch points of the quasi-momentum $p_i(x)$ for the $i$-th puncture. The conditions  \eqref{normcond3}, \eqref{psi1} and \eqref{psi2}  determine the normalization of $i_{\pm}$ completely. 
The important property of the eigenvectors so normalized is that 
 they transform in the following way when they cross the branch cut\fn{Note that the extra minus sign is necessary in the second equation of \eqref{change-sheet} in order to retain the condition \eqref{normcond2}.}:
\beq{
\left. \hat{i}_{+}(x)\right|_{\text{on 2nd sheet}}=\left. \hat{i}_{-}(x)\right|_{\text{on 1st sheet}}\comma \quad \left. \hat{i}_{-}(x)\right|_{\text{on 2nd sheet}}=-\left. \hat{i}_{+}(x)\right|_{\text{on 1st sheet}}\period\label{change-sheet}
}
This relation will be used in section \ref{subsec:sing} to determine the normalization of certain Wronskians.%%%%%%%%%%%%%%%%%%%%%%
\section{{\bfall Structures of the action for the $S^3$ part}\label{sec:action}}
%%%%%%%%%%%%%%%%%%%%%%%
Let us now start our study of the three-point functions. 
In what follows, we will denote  their structure as 
\begin{align}
\langle \calV_1 \calV_2 \calV_3\rangle &= \exp \left(F_{S^3} + F_{E\!AdS_3} \right) \comma 
\end{align}
where 
\begin{align}
F_{S^3}  &= \calF_{\rm action} + \calF_{\rm vertex} \comma \\
F_{E\!AdS_3} &=  \hat{\calF}_{\rm action} + \hat{\calF}_{\rm vertex} \period
\end{align}
 In this section, we focus on the contribution of the action for the $S^3$ part,  namely $\calF_{\rm action} $.  First, in subsection \ref{subsec:contour}, we rewrite the action as a boundary contour integral using the Stokes theorem and then apply  the generalized Riemann bilinear identity derived in \cite{KK1} to bring it to a more convenient form. 
Next we turn in subsection \ref{subsec:WKB} to the analysis of the WKB expansion of the auxiliary linear problem. We then find that  the same contour integrals we used to rewrite the action appear also in the WKB expansion of the Wronskians of the solutions to the ALP. Using this relation, we re-express the action in terms of the Wronskians in subsection \ref{subsec:action-wron}. The resultant expression will be used for the explicit  evaluation 
 of  the contribution of the action in section \ref{sec:3pt}.
%%%%%%%%%%%%%%%%%
\subsection{Contour integral representation of the action\label{subsec:contour}}
%%%%%%%%%%%%%%%%%%
For the three-point function of our interest, the (regularized) action for the $S^3$ part of the string 
is given by 
\beq{
S_{S^3}=\frac{\sqrt{\lambda}}{\pi}\int_{\Sigma\backslash\{\epsilon_i\}} d^2 z\del Y_I \delbar Y_I \comma \label{eq-3-1}
} 
where the symbol $\Sigma\backslash\{\epsilon_i\}$ 
denotes the worldsheet for the three-point function, which is a two-sphere with a small disk of radius $\epsilon_i$ cut out at each vertex operator insertion point $z_i$. Such a point will  often be referred to as a puncture also.
In \cite{JW} and \cite{CT}, such worldsheet cut-offs 
are related  to the spacetime cut-off in AdS in order to obtain the spacetime dependence of the correlation functions without introducing the vertex operators. In contrast, 
as we shall separately take into account the contribution of the vertex 
 operators,  $\epsilon_i$'s can be taken to be arbitrary  in our approach, as long as they are sufficiently small and the same  for the $S^3$ part and the $E\!AdS_3$ part. 

\begin{figure}[htbp]
\begin{minipage}{0.5\hsize}
  \begin{center}
   \picture{clip,height=5cm}{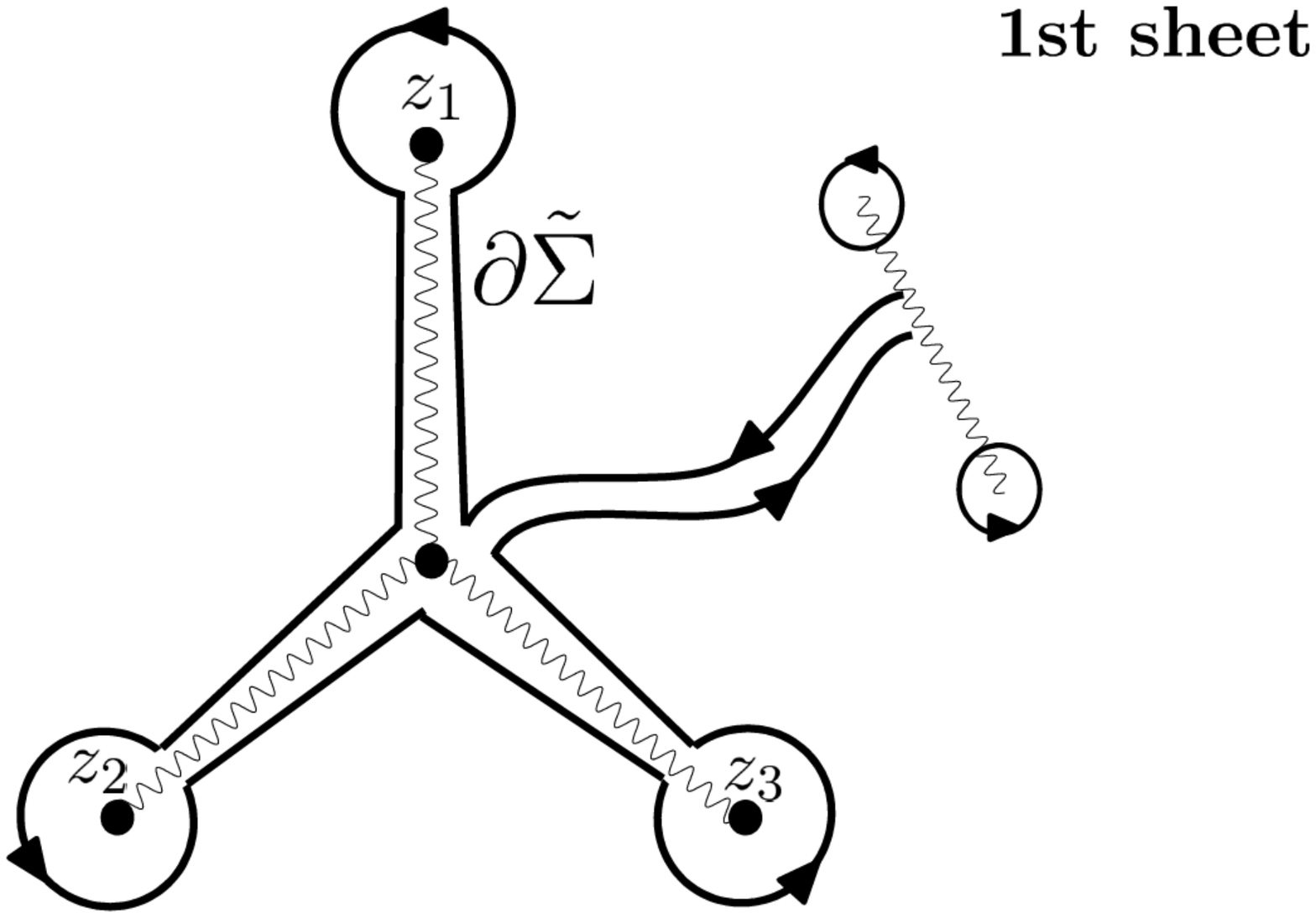}
  \end{center}
 \end{minipage}
 \begin{minipage}{0.5\hsize}
  \begin{center}
   \picture{height=5cm,clip}{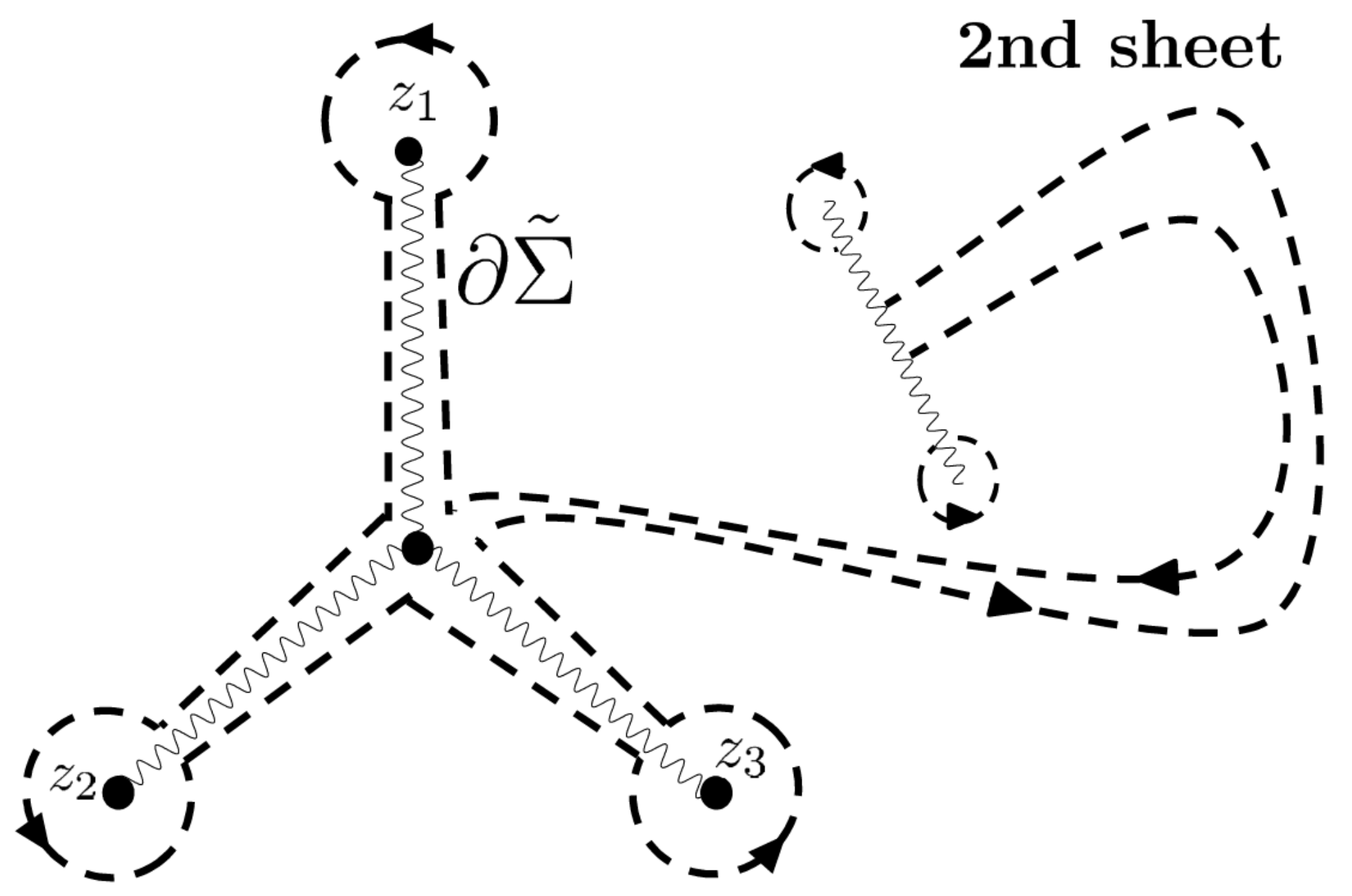}
  \end{center}
 \end{minipage}
\caption{The  coutour $\del \tilde{\Sigma}$ which forms the boundary of the double cover of the worldsheet $\tilde{\Sigma}$,  dipicted  as the union of the first and the second sheet. There are three logarithmic branch cuts attached to the puncture in the middle, in addition to one  square-root branch cut, shown as a wavy segment, through which the two sheets are connected.}
\label{fig-sigma}
\end{figure}

As the action is invariant under the global symmetry transformations, 
it is natural to express \eqref{eq-3-1} in terms of the quantities 
 used in  the Pohlmeyer reduction. From \eqref{defgamma}, we can indeed write 
\beq{
S_{S^3}= \frac{\sqrt{\lambda}}{\pi} \int_{\Sigma\backslash\{\epsilon_i\}} d^2 z\sqrt{T\bar{T}}\cos 2\gamma\period \label{eq-3-2}
}
We further rewrite \eqref{eq-3-2} by introducing the following 
 one-forms:
\beq{
&\varpi \equiv \sqrt{T}dz \comma\\
&\eta \equiv -\sqrt{\bar{T}}\cos 2\gamma d\barz +\frac{2}{\sqrt{T}}\left( -(\del \gamma)^2 + \frac{\rho^2}{T}\right) dz\period \label{defeta}
}
The second term on the right hand side of  (\ref{defeta}) is added to make $\eta$ closed, as one can  verify using the relation \eqref{singordon}. 
With these one-forms, we can re-express the action \eqref{eq-3-2} as a wedge product of the form 
\beq{
S_{S^3}=\frac{i\sqrt{\lambda}}{2\pi}\int_{\Sigma\backslash\{\epsilon_i\}} \varpi \wedge \eta\comma\label{eq-3-3}
}
where an extra prefactor $i/2$ comes from the definition of the volume form, $dz\wedge d\barz = -2i\, d^2 z$. Then denoting  the integral of $\varpi(z)$ as
\beq{
\Pi (z)= \int_{z_0}^{z}\varpi(z^{\prime}) dz^{\prime}\comma
}
the action can be rewritten, using  the Stokes theorem, as a contour integral along a boundary $\del \tilde{\Sigma}$ of a certain region $\tilde{\Sigma}$ (see \figref{fig-sigma}):
\beq{
S_{S^3}=\frac{i\sqrt{\lambda}}{4\pi}\int_{\tilde{\Sigma}} \varpi \wedge \eta =\frac{i\sqrt{\lambda}}{4\pi}\int_{\tilde{\Sigma}} d\left( \Pi \eta\right) =\frac{i\sqrt{\lambda}}{4\pi}\int_{\del \tilde{\Sigma}} \Pi \eta\period\label{convert}
}

\begin{figure}[htbp]
\begin{minipage}{0.5\hsize}
  \begin{center}
   \picture{clip,height=5cm}{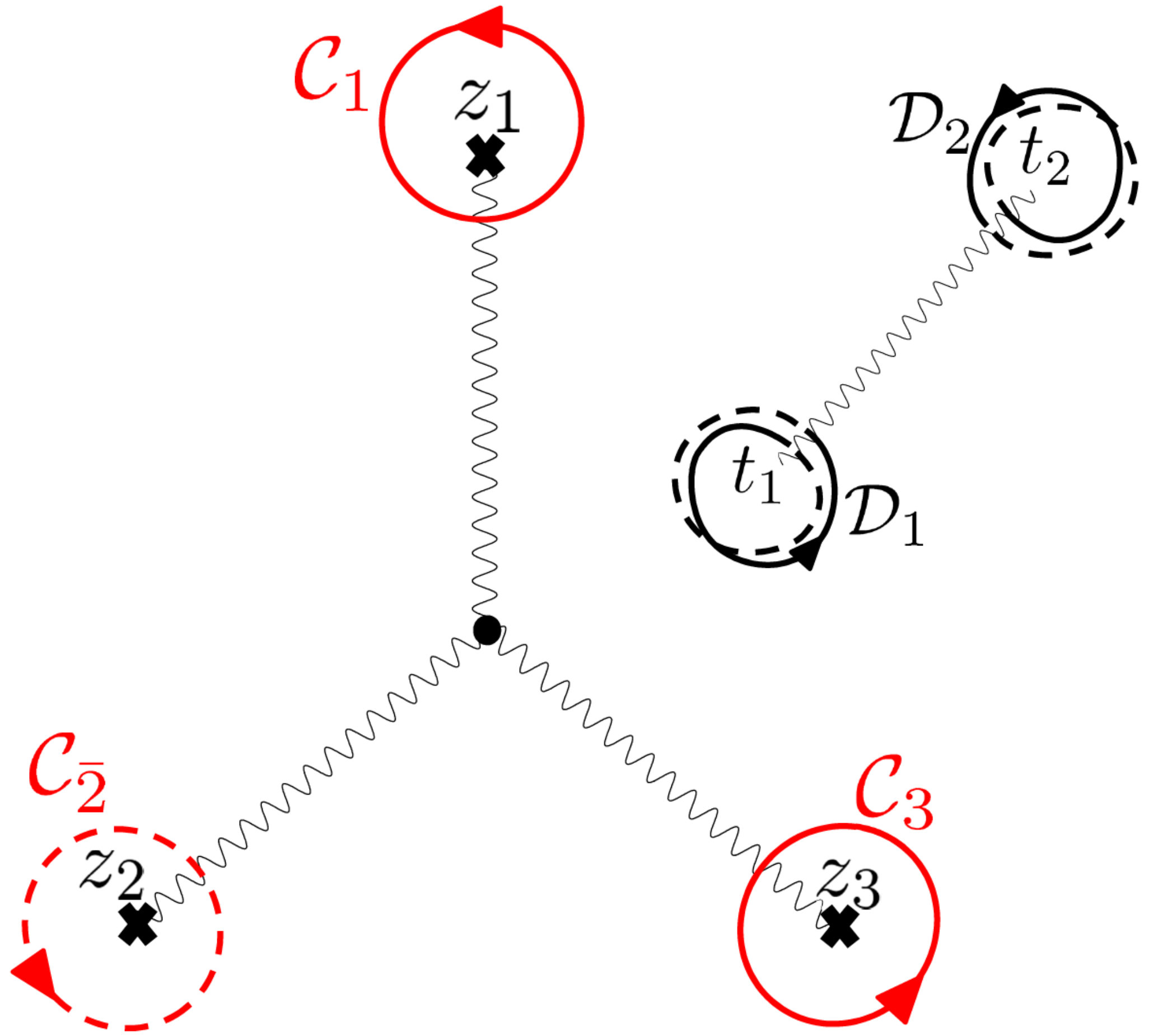}
  \end{center}
 \end{minipage}
 \begin{minipage}{0.5\hsize}
  \begin{center}
   \picture{height=5cm,clip}{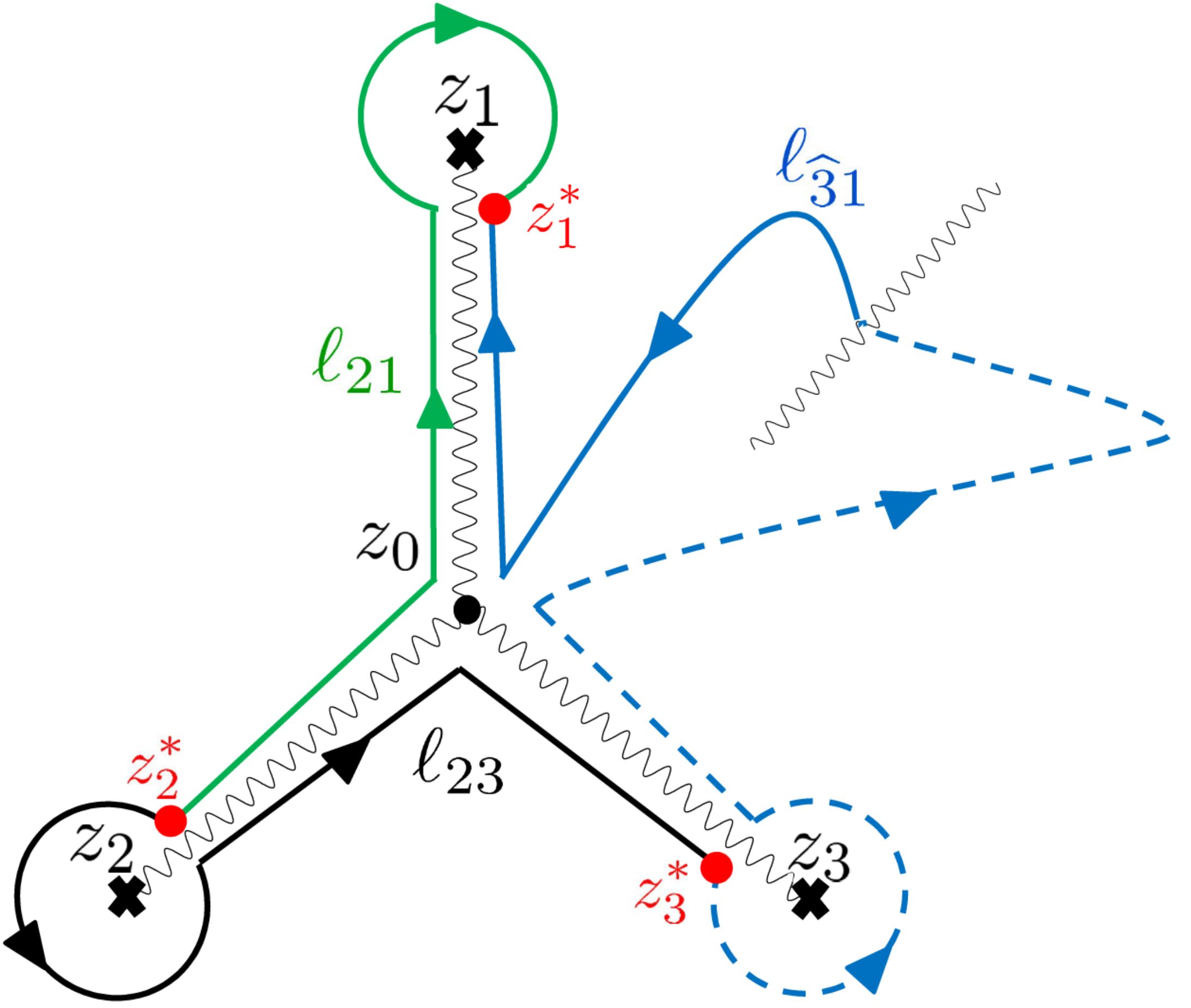}
  \end{center}
 \end{minipage}
\caption{ Definitions of the contours used  to rewrite the action: The contours which enclose the punctures ($\mathcal{C}_i$) are shown in the left figure and the ones which connect two punctures ($\ell_{ij}$) are shown in the right figure. In both figures, the portions of the contours on the second sheet are drawn  as dashed lines. Also depicted in the right figure are  the starting points and the end points of the contours, $z_i^{\ast}$'s.} 
\label{contours}
\end{figure}

To determine the proper region of integration $\tilde{\Sigma}$, we need to know the analytic structure of $\Pi(z)$, which in turn is dictated by that of $T(z)$.  As already explained in section \ref{sec:basic}, in the case of 
 three-point functions  the information of the asymptotic behavior of $T(z)$ at   each puncture  $z_i$ is sufficiently restrictive to 
determine $T(z)$ exactly to be of the form 
\beq{
&T(z)=\left(\frac{\kappa_1^2 z_{12}z_{13}}{z-z_1}+\frac{\kappa_2^2 z_{21}z_{23}}{z-z_2}+\frac{\kappa_3^2 z_{31}z_{32}}{z-z_3} \right) \frac{1}{(z-z_1)(z-z_2)(z-z_3)}\comma\\
&z_{ij}\equiv z_i-z_j\period\nn
}
From this, one can show that $\Pi(z)$ has three logarithmic branch cuts running from the punctures $z_i$,  and one square-root branch cut connecting two zeros of $T(z)$, to be denoted by $t_1$ and $t_2$. 
Therefore, we should  take $\tilde{\Sigma}$ to be the double cover ($y^2 = T(z)$) of the worldsheet $\Sigma$ with an appropriate boundary $\del \tilde{\Sigma}$, so that $\Pi(z)$ is single-valued on the whole integration region.
 In what follows, we will consider the case where the branch cut is located between $z_1$ and $z_3$ as depicted in \figref{contours}. 
 In such a case, the branch of the square-root of $T(z)$ can be chosen so that it behaves near the punctures on the first sheet as
\beq{
\begin{aligned}
\sqrt{T(z)}&\sim \frac{\kappa_{i}}{z-z_{i}} \qquad\text{as }z\to z_{i}\quad (i=1,3)\comma\\
&\sim \frac{-\kappa_2}{z-z_2}\qquad \text{as }z\to z_{2}\period
\end{aligned}
\label{branchT}
}
Although the discussion to follow is tailored for this particular case, the final result for the three-point function, to be obtained in section \ref{sec:3pt}, will turn out to be completely symmetric under the permutation of the punctures. 

At this point, we shall apply the generalized Riemann bilinear identity, derived in \cite{KK1}, to the integral \eqref{convert}.
As the derivation is lengthy, we refer the reader to \cite{KK1} for details and just present the result\fn{By decomposing the contours $\ell_{ij}$'s in \eqref{gRBI} into $d$ and $\ell_i$'s defined in \cite{KK1}, we arrive at the formula derived in \cite{KK1}.}. It can be written as 
\beq{
&\int _{\tilde{\Sigma}}\varpi\wedge \eta =\Local + \Double+\Global+\Extra\comma\label{gRBI}
}
where the definition of each term will be given successively below\fn{In \cite{JW} and \cite{CT}, the ordinary Riemann bilinear identity was applied to derive an expression similar to \eqref{gRBI} but without the terms \Local and \Double. In their cases, \Local and \Double vanish and the use of the ordinary Riemann bilinear identity is justified. On the other hand, these two terms do not vanish in our case and we must use the generalized Riemann bilinear identity.}. The first term, $\text{\sf Local}$, denotes the contribution from the product of  contour integrals, each of which is just around the puncture and hence 
called  ``local".  It is of the form 
\beq{
\text{\sf Local}=\sum_i \oint_{\mathcal{C}_i} \varpi\oint_{\mathcal{C}_i} \eta +\sum_{i<j}\left( \oint_{\mathcal{C}_i}\varpi\oint_{\mathcal{C}_j}\eta-(\varpi \leftrightarrow \eta)\right)\comma\label{Local}
}
where $\mathcal{C}_i$ is a contour encircling the puncture $z_i$ counterclockwise. 
Here and hereafter, the symbol  $(\varpi \leftrightarrow \eta)$ stands for  the contribution obtained by exchanging $\varpi$ and $\eta$ in the preceding term. The second term, $\text{\sf Double}$, denotes the double integrals around the punctures given by 
\beq{
\text{\sf Double}=-2\sum_{i}\oint _{\mathcal{C}_i}\eta \int^{z}_{z_{i}^{\ast}}\varpi\period\label{Double}
}
The third term, $\text{\sf Global}$, denotes the contribution from the product of contour integrals, one of which is along a contour connecting 
 two different punctures. It is given by 
\beq{
\text{\sf Global}=\left( \oint_{\mathcal{C}_1+\mathcal{C}_{\bar{2}}-\mathcal{C}_3}\hspace{-0.8cm}\varpi \hspace{0.5cm} \int_{\ell_{21}}\eta+ \oint_{\mathcal{C}_{\bar{2}}+\mathcal{C}_3-\mathcal{C}_1}\hspace{-0.8cm}\varpi \hspace{0.5cm}\int_{\ell_{23}}\eta+\oint_{\mathcal{C}_3+\mathcal{C}_1-\mathcal{C}_{\bar{2}}}\hspace{-0.8cm}\varpi \hspace{0.5cm}\int_{\ell_{\bar{3}1}}\eta\right)- \left(\varpi \leftrightarrow \eta\right)\period\label{Global}
}
More precisely, $\ell_{ij}$ denotes the contour connecting $z_i^{\ast}$ and $z_j^{\ast}$, where $z_i^{\ast}$ is the point near the puncture $z_i$ satisfying $z_i^{\ast}-z_i=\epsilon_i$. The barred indices indicate  the points on the second sheet of the double cover $y^2 = T(z)$.  For instance, $\mathcal{C}_{\bar{i}}$ is a contour encircling the point $z_{\bar{i}}$, which is on the second sheet right below $z_i$. 
Finally, the term  {\sf Extra} denotes  additional terms which come from the integrals around the zeros of $\sqrt{T}$, to be denoted by $t_k$, at which $\eta$
 becomes singular, and is given by 
\beq{
\text{\sf Extra}=\sum_k \oint _{\mathcal{D}_{k}} \Pi \eta\period\label{Extra}
}
Here $\mathcal{D}_k$ is the contour which encircles $t_k$ twice as depicted in \figref{contours}.

Among these four terms, {\sf Local} and {\sf Double} are expressed solely 
in terms of  the integrals around the punctures and are easy to compute. 
The explicit results, computed in Appendix \ref{apsubsec:landd}
are\fn{The one-forms $\varpi$ and $\eta$ flip the sign under the exchange of two sheets. Therefore \eqref{cint} is odd whereas \eqref{cint2} is even under such sheet-exchange. In \eqref{cint2}, $\kappa_i$ for $i=\bar{2}$ is set to be equal to $\kappa_2$.} 
\beq{
&\oint_{\mathcal{C}_i}\varpi = 2\pi i \kappa_i\comma \quad \oint_{\mathcal{C}_i}\eta = 2\pi i\kappa_i \Lam_i \comma \label{cint}\\
&\oint _{\mathcal{C}_i}\eta \int^{z}_{z_{i}^{\ast}}\varpi=-2\pi\kappa_i^2 \Lam_i   \comma\qquad \text{for }i=1,\bar{2}, 3 \period\label{cint2}
}
Here $\Lam_i$'s are given in terms of $\gamma_i$ and $\rho_i$,  defined in \eqref{defgami} and \eqref{defrhoi} respectively, as 
\beq{
\Lam_i = \cos 2\gamma_i +\frac{2\rho_i^2}{\kappa_i^4}\period\label{gi}}

It is important to note that \Local and \Double are real since 
 $\kappa_i$ and $g_i$ are all real.  Therefore they contribute exclusively  to the imaginary part of the action \eqref{convert} and hence only yield an overall phase of the three-point functions.  We shall neglect such quantities 
 in this paper. 

 Among the remaining two types of terms, {\sf Extra} can be explicitly evaluated as follows.  Since the worldsheet is assumed to be smooth except at the punctures, the quantity  $\sqrt{T\bar{T}}\cos 2\gamma$, which is the integrand  of the action integral   given in \eqref{eq-3-2},   should not vanish even at the zeros of $T(z)$. This in turn implies  that $\gamma$ is logarithmically divergent at such points in the manner
\beq{
\gamma \sim \pm \frac{i}{2}\ln |z-t_k| \quad \text{as }z\to t_k\period
}
Then, by approximating $T(z)$ as $T(z)\sim c(z-t_k)$ around $t_k$, we can write down the leading singular behavior of $\eta$ around $t_k$ as
\beq{
\eta \sim -\frac{2}{\sqrt{T}} (\del \gamma)^2 d\barz \sim  \frac{d\barz}{8\sqrt{c}(z-t_k)^{5/2}}\period
}
Thus the integral along $\mathcal{D}_k$ can be computed as
\beq{
\oint_{\mathcal{D}_k} \Pi \eta = \oint_{\mathcal{D}_k} \frac{2\sqrt{c}(z-t_k)^{3/2}}{3}\frac{d\barz}{8\sqrt{c}(z-t_k)^{5/2}}=-\frac{\pi i}{6}\period
}
Since  there exist  two zeros, \Extra is twice this integral and hence is  given by
\beq{
\Extra = -\frac{\pi i}{3}\period\label{exp-extra}
}

For later convenience, we shall  derive another expression for the action using a different set of one-forms  given by 
\beq{
&\bar{\varpi}=\sqrt{\bar{T}}d\barz\comma\\
&\tilde{\eta}=-\sqrt{T}\cos 2\gamma dz +\frac{2}{\sqrt{T}}\left( -(\delbar \gamma)^2 + \frac{\rho^2}{\bar{T}}\right) d\barz\comma
}
and then consider the average of the two expressions. Using  the forms above, the action can be written as
\beq{
S_{S^3}=-\frac{i\sqrt{\lambda}}{4\pi}\int_{\tilde{\Sigma}} \bar{\varpi} \wedge \tilde{\eta}\period\label{convert2}
}
As compared to \eqref{convert}, the expression \eqref{convert2} has an extra minus sign, which is due to the property  $dz\wedge d\barz=-d\barz \wedge dz$. Applying the generalized Riemann bilinear identity  to \eqref{convert2}, we get
\beq{
&-\int_{\tilde{\Sigma}}\bar{\varpi}\wedge \tilde{\eta}=-\left(\bLocal+\bDouble+\bGlobal+\bExtra\right)\comma\label{gRBI2}
}
where $\bLocal$, $\bDouble$ and $\bGlobal$ are given respectively by \eqref{Local}, \eqref{Double} and \eqref{Global} with $\varpi$ and $\eta$ replaced by  $\bar{\varpi}$ and $\tilde{\eta}$. 
The integrals of $\bar{\varpi}$ and $\tilde{\eta}$ around the punctures are given by\fn{\eqref{cint3} is odd and \eqref{cint4} is even under the exchange of the first and the second sheets, as in the case of the integrals of $\varpi$ and $\eta$ given in \eqref{cint} and \eqref{cint2}.}
\beq{
&\oint_{\mathcal{C}_i}\bar{\varpi} = -2\pi i \kappa_i\comma \quad \oint_{\mathcal{C}_i}\tilde{\eta} = -2\pi i\kappa_i \bar{\Lam}_i  \comma\label{cint3}\\
&\oint _{\mathcal{C}_i}\tilde{\eta} \int^{z}_{z_{i}^{\ast}}\bar{\varpi}=-2\pi \kappa_i^2 \bar{\Lam}_i\comma\qquad \text{for }i=1,\widehat{2}, 3\label{cint4}
}
where $\bar{\Lam}_i$'s are given in terms of $\gamma_i$ and $\tilde{\rho_i}$, defined  in Appendix \ref{apsubsec:pohl},   as  
\beq{
\bar{\Lam}_i=\cos 2\gamma_i + \frac{2\tilde{\rho_i}^2}{\kappa_i^4}\period\label{bargi}
}
Again $\bLocal$ and $\bDouble$ are real and they contribute only  to the overall phase.  
On the other hand,  $\bExtra$ can be evaluated  just like $\Extra$ and 
yields  $+\pi i/3$.  
Thus,  by averaging over the two expressions  \eqref{gRBI} and \eqref{gRBI2}  and neglecting  terms which contribute exclusively  to  the overall phase, we arrive at the following more symmetric expression:
\beq{
\frac{1}{2}\left(\int_{\tilde{\Sigma}}\varpi\wedge \eta  -\int_{\tilde{\Sigma}}\bar{\varpi}\wedge \tilde{\eta}\right)=-\frac{\pi i}{3}+\frac{1}{2}\left(\Global-\bGlobal \right)\period \label{gRBI3}
}
The quantity  \eqref{gRBI3} consists of various integrals along 
the contours $\mathcal{C}_i$ and $\ell_{ij}$. 
Among them, the ones  along $\mathcal{C}_i$ can be easily computed using \eqref{cint} and \eqref{cint3}. The integral of $\varpi$ along $\ell_{ij}$ can also be computed in principle as we know the explicit form of $\varpi$. Thus the major nontrivial task is the evaluation of $\int_{\ell_{ij}}\eta$ and $\int_{\ell_{ij}}\tilde{\eta}$. In the rest of this section, we will see how these integrals are related to the Wronskians of the form  $\sl{i_{\pm}\comma j_{\pm}}$, where $i_\pm$ are the Baker-Akhiezer eigenvectors at $z_i$ 
 of the ALP, corresponding to the eigenvalues $e^{\pm ip_i(x)}$. 
%%%%%%%%%%%%%%%%%%%
\subsection{WKB expansions of the auxiliary linear problem\label{subsec:WKB}}
%%%%%%%%%%%%%%%%%%%%
We now  perform the WKB expansion of the auxiliary linear problem and observe that the contour integrals of our interest, $\int_{\ell_{ij}}\eta$ and $\int_{\ell_{ij}}\tilde{\eta}$, appear in the expansion of the Wronskians between the eigenvectors of the monodromy matrices.

Let us first consider the WKB expansion of the solutions to the ALP.
For this purpose,  it is convenient to use the ALP of the Pohlmeyer reduction \eqref{PMALP}. The use of \eqref{PMALP} has two main virtues. First, as $\Phi$'s are given explicitly in terms of $T(z)$ and $\bar{T}(\barz)$, it is easier to perform the expansion around $\zeta=0$ or around $\zeta=\infty$. Second, since the connection \eqref{pohlconnect} is expressed solely in terms of  the quantities invariant under the global symmetry transformation, we can directly explore the dynamical aspect of the problem setting aside all the kinematical information.

We shall first perform the expansion around $\zeta=0$. To  facilitate this task, it is convenient to perform a further  gauge transformation and convert \eqref{PMALP} to the ``diagonal gauge", where the ALP take the form 
\beq{
\left( \del + \frac{1}{\zeta}\Phi^d_z +A^d_z\right) \hat{\psi}^d =0\comma \quad \left( \delbar + \zeta \Phi^d_{\barz} + A^d_{\barz}\right)\psihat^d = 0\period\label{diag-gauge}
}
In the above,  $\psihat^d$ in the diagonal gauge is defined by 
\beq{
\psihat^d \equiv \frac{1}{\sqrt{2}}\pmatrix{cc}{e^{i\gamma /2}&-e^{-i\gamma /2}\\e^{i\gamma /2}&e^{-i\gamma /2}}\hat{\psi} \comma 
}
and $\Phi^d$'s and $A^d$'s are given by
\beq{
\begin{aligned}
&\Phi^d_z=\frac{\sqrt{T}}{2}\pmatrix{cc}{1&0\\0&-1}\comma &\Phi^d_{\barz}=\frac{\sqrt{\bar{T}}}{2}\pmatrix{cc}{-\cos 2\gamma & i\sin 2\gamma\\-i\sin 2\gamma &\cos 2\gamma}\comma\\
&A^d_{z}=\pmatrix{cc}{-\frac{\rho}{\sqrt{T}}\cot 2\gamma &\frac{i\rho}{\sqrt{T}}-i\del \gamma\\ - \frac{i\rho}{\sqrt{T}}-i\del \gamma&\frac{\rho}{\sqrt{T}}\cot 2\gamma}\comma &A^d_{\barz}=\frac{-\tilde{\rho}}{\sqrt{\bar{T}}\sin 2\gamma}\pmatrix{cc}{1&0\\0&-1}\period
\end{aligned}
\label{diag-conn}
}
Note that the leading terms in the ALP equations as $\zeta \rightarrow 0$, 
 namely $\Phi^d_z$ for the first equation and $A^d_{\barz}$ for 
the second,  have been  diagonalized. Because of this feature,  the leading exponential behavior of the two linearly independent solutions  around $\zeta\sim 0$ can be readily determined as
\beq{
\hat{\psi}^d_1 \sim \pmatrix{c}{0\\1}\exp \left[ \frac{1}{2\zeta}\int^{z}_{z_0} \varpi\right]\comma \quad \hat{\psi}^d_2 \sim \pmatrix{c}{1\\0}\exp \left[\frac{-1}{2\zeta}\int^{z}_{z_0} \varpi\right]\comma \label{wkbsol}
}
By performing the WKB expansion around $\zeta\sim 0$ systematically, one can also determine the subleading terms of \eqref{wkbsol} in $\zeta$, as shown in Appendix \ref{apsub:direct}.

The quantities of prime interest in the subsequent discussions are the Wronskians of the eigenvectors of the monodromy matrices. 
To perform the WKB expansion of such Wronskians, we need to have a good control over the asymptotics of the Wronskians $\sl{i_{\pm}\comma j_{\pm}}$ around $\zeta=0$. For this purpose, both of the eigenvectors in the Wronskian need to be small solutions since big solutions can contain a multiple of small solutions and hence are ambiguous\cite{GMN,AM,AGM,AMSV}. When $\zeta$ is sufficiently close to zero, one can show that the plus solutions $i_{+}$ are the small solutions if $\Re \zeta$ is positive whereas it is the minus solutions $i_{-}$ which are small if $\Re \zeta$ is negative. 
Thus, the Wronskians that can be expanded consistently around $\zeta=0$ are $\sl{i_{+}\comma j_{+}}$'s for $\Re\zeta>0$ and $\sl{i_{-}\comma j_{-}}$'s for $\Re\zeta<0$. The detailed form of the expansion can be determined by employing the Born series expansion explained in Appendix \ref{apsub:born} and the results are given in the following simple form:
\beq{
&\text{For }\Re \,\zeta >0\comma\nn\\
&\quad \langle 2_{+} \comma 1_{+}\rangle = \exp  \left( - S_{2\to 1}\right)\comma \quad \langle 2_{+} \comma 3_{+}\rangle = \exp \left( -S_{2\to 3}\right)\comma \quad \langle 3_{+} \comma 1_{+}\rangle = \exp \left( -S_{\hat{3}\to 1}\right)\comma\label{++wkb}\\
&\text{For }\Re \,\zeta <0\comma\nn\\
&\quad \langle 2_{-} \comma 1_{-}\rangle = \exp \left( S_{2\to 1}\right)\comma \quad \langle 2_{-} \comma 3_{-}\rangle = \exp \left( S_{2\to 3}\right)\comma \quad \langle 1_{-} \comma 3_{-}\rangle = \exp \left( S_{\hat{3}\to 1}\right)\period\label{--wkb}
}
In these expressions,  $S_{i\to j}$ stands for the quantity 
\beq{
S_{i\to j}=\frac{1}{2\zeta}\int_{\ell_{ij}} \varpi + \int_{\ell_{ij}} \alpha + \frac{\zeta}{2}\int_{\ell_{ij}}\eta +\cdots \comma\label{wrexp}
}
where the one-form $\alpha$ is given in \eqref{defofal} in Appendix \ref{apsub:born}.
A remarkable feature of \eqref{wrexp} is that the integral of our interest $\int_{\ell_{ij}}\eta$ makes its appearance in the exponent $S_{i\rightarrow j}$.

Now to make use of the averaging procedure described in the previous 
subsection, we need the other type of integrals $\int_{\ell_{ij}}\tilde{\eta}$ which appear in  $\bGlobal$. To obtain them,  we need to expand the Wronskians this time around $\zeta=\infty$. Since the discussion is similar to the expansion around $\zeta=0$, we will not  elaborate on  the details and  simply give the results:
\beq{
&\text{For }\Re \,\zeta >0\comma\nn\\
&\quad \langle 2_{+} \comma 1_{+}\rangle = \exp  \left( -\tilde{S}_{2\to 1}\right)\comma \quad \langle 2_{+} \comma 3_{+}\rangle = \exp \left( -\tilde{S}_{2\to 3}\right)\comma \quad \langle 3_{+} \comma 1_{+}\rangle = \exp \left( -\tilde{S}_{\hat{3}\to 1}\right)\comma\label{++wkb2}\\
&\text{For }\Re \,\zeta <0\comma\nn\\
&\quad \langle 2_{-} \comma 1_{-}\rangle = \exp \left( \tilde{S}_{2\to 1}\right)\comma \quad \langle 2_{-} \comma 3_{-}\rangle = \exp \left( \tilde{S}_{2\to 3}\right)\comma \quad \langle 1_{-} \comma 3_{-}\rangle = \exp \left( \tilde{S}_{\hat{3}\to 1}\right)\comma\label{--wkb2}
}
Here  $\tilde{S}_{i\to j}$ is defined by
\beq{
\tilde{S}_{i\to j}=\frac{\zeta}{2}\int_{\ell_{ij}} \bar{\varpi} + \int_{\ell_{ij}} \tilde{\alpha} + \frac{1}{2\zeta}\int_{\ell_{ij}}\tilde{\eta} +\cdots \comma  \label{wrexp2}
}
where  $\tilde{\alpha}$ is a one-form given in \eqref{defofalbar} in Appendix \ref{apsub:born}. 
Making use of these two types of expansions, we will be able to rewrite the action in terms of the Wronskians,  as described  in the next subsection.
%%%%%%%%%%%%%%%%
\subsection{The action in terms of the Wronskians\label{subsec:action-wron}}
%%%%%%%%%%%%%%%%
We are now ready  to derive an explicit expression of the action in terms of the Wronskians. As shown in the previous subsection,  the integrals we used to rewrite the action, namely $\oint_{\ell_{ij}}\eta$ and $\oint_{\ell_{ij}}\tilde{\eta}$, can be extracted  from the Wronskians. For instance, consider the integral 
$\oint_{\ell_{21}}\eta$,  which  appears  in $\sl{2_{-},1_{-}}$. 
Differentiating  $\ln \sl{2_{-},1_{-}}$ 
 with respect to $\zeta$ using (\ref{--wkb}) and (\ref{wrexp}),  we get 
\begin{align}
\del_\zeta \ln \sl{2_{-},1_{-}} =-{1\over \zeta^2} \int_{\ell_{21}}\varpi  + {1\over 2} \int_{\ell_{21}}\eta + O(\zeta) \period
\end{align}
Therefore we can get the integral $\oint_{\ell_{21}}\eta$ by subtracting
 the first divergent term and then taking the limit $\zeta \rightarrow 0$. 
Similarly $\oint_{\ell_{21}}\tilde{\eta}$ can be obtained from 
$\sl{2_{-},1_{-}}$ in the $\zeta \rightarrow \infty$ limit. 
Such procedures can be compactly implemented  if we use the variable 
 $x$ instead of $\zeta$, which are related as in (\ref{relzetax}). 
Then, we can write 
\beq{
&\oint_{\ell_{21}}\eta = -4\no{\del_x \ln \sl{2_{-}\comma 1_{-}}}_{+}\comma \qquad \oint_{\ell_{21}}\tilde{\eta} = -4\no{\del_x \ln \sl{2_{-}\comma 1_{-}}}_{-}\comma
}  
where the ``normal ordering"  symbol  $\no{A(x)}_{\pm}$ is defined by 
\beq{
\no{A(x)}_{\pm}\equiv \lim_{x\to \pm 1} \left[ A(x)-(\text{double pole at $x=\pm1$})\right]\period \label{def-no}
}
This precisely  subtracts the divergent term mentioned above. 
Substituting such expressions to the definitions of \Global and $\bGlobal$, we can express them in terms of the Wronskians. Then, using \eqref{gRBI3}, we arrive at the following expression for the contribution from  the $S^3$ part of the action $\calF_{\textrm{action}}$:
\beq{
\calF_{\textrm{action}}=-S_{S^3} =\frac{\sqrt{\lambda}}{6}+\easy + \hard\period\label{faction}
}
The first term in \eqref{faction} expresses  the contributions of \Extra and $\bExtra$. The second term $\easy$ denotes the contribution of $\int_{\ell_{ij}}\varpi$ and $\int_{\ell_{ij}}\bar{\varpi}$ in \Global and $\bGlobal$ and is given by
\beq{
&\easy=\nn\\
&\frac{\sqrt{\lambda}}{4} \biggl( (\kappa_1 \Lam_1 + \kappa_2 \Lam_2 -\kappa_3 \Lam_3)\int_{\ell_{21}}\varpi +(\kappa_1 \Lam_1 - \kappa_2 \Lam_2 +\kappa_3 \Lam_3)\int_{\ell_{\hat{3}1}}\varpi \nn\\
&+(-\kappa_1 \Lam_1 + \kappa_2 \Lam_2 +\kappa_3 \Lam_3)\int_{\ell_{23}}\varpi \biggr)
 + \left( \Lam_i\rightarrow \bar{\Lam}_i \comma\, \varpi \rightarrow \bar{\varpi} \right)\comma\label{Api}
}
where $\Lam_i$ and $\bar{\Lam}_i$ are as given in (\ref{gi}) and (\ref{bargi}) and  $\left( \Lam_i\rightarrow \bar{\Lam}_i \comma\, \varpi \rightarrow \bar{\varpi} \right)$ in the last line denotes the terms obtained by replacing  $\Lam_i$ and $\varpi$ in the second line with $\bar{\Lam}_i$ and $\bar{\varpi}$ respectively.
The third term $\hard$ is the contribution of $\int_{\ell_{ij}}\eta$ and $\int_{\ell_{ij}}\tilde{\eta}$, which is expressed in terms of the Wronskians 
in the following way:
\beq{
\hard =\sqrt{\lambda}&\big[(\kappa_1 +\kappa_2 -\kappa_3)\left( \no{\del _x\! \ln \sl{2_{-}\comma 1_{-}}}_{+}-\no{\del _x\! \ln \sl{2_{+}\comma 1_{+}}}_{-}\right)\nn\\
&+(\kappa_1 -\kappa_2 +\kappa_3)\left(\no{\del _x \!\ln \sl{3_{-}\comma 1_{-}}}_{+}-\no{\del _x \!\ln \sl{3_{+}\comma 1_{+}}}_{-}\right)\nn\\
&+(-\kappa_1 +\kappa_2 +\kappa_3)\left( \no{\del _x\!\ln \sl{2_{-}\comma 3_{-}}}_{+}-\no{\del _x\!\ln \sl{2_{-}\comma 3_{-}}}_{-} \right)\big]\period\label{Aeta}
}
The general formula \eqref{faction} will later be used in section \ref{sec:3pt} to compute the three-point functions.
\section{Structure of the contribution from the vertex operators\label{sec:vertex}}
%%%%%%%%%%%%%
Having found the structure of the contribution of the action part, 
we shall now study  that of the vertex operators. 
%%%%%%%%%%%
\subsection{Basic idea and framework\label{subsec:idea}}
%%%%%%%%%%%
Before plunging into the details of the analysis, let us describe in this subsection the basic idea and the framework, which includes a brief review of
 the methods developed in our previous work \cite{KK2}. 

 As explained in detail in \cite{KK1}, the precise form of the 
 conformally invariant vertex operator corresponding to a string solution 
 in a curved spacetime, such as $AdS_3$ discussed there or \EAdS3 $\times S^3$ of our interest in this paper, is 
in general not known. In particular, for a non-BPS   solution with 
non-trivial  $\sigma$ dependence the corresponding vertex operator would contain infinite number of derivatives and is hard to construct. To overcome this 
 difficulty, we have developed in \cite{KK2} a powerful method of computing 
 the contribution of the vertex operators by using  the state-operator correspondence and the construction of the corresponding wave function in terms of the  action-angle variables. Although it was applied in \cite{KK2} 
to the case of the GKP  string in $AdS_3$, the basic idea of the method is 
applicable to more general situations, including the present one, albeit with 
appropriate modifications and refinements. 

Let us  briefly review the essential ingredients of the method. (For 
 details, see section 3 of \cite{KK2}.)
The state-operator correspondence, in the semi-classical approximation, 
 is expressed by the following equation:
\begin{align}
\mathcal{V}[q_\ast(z=0)] e^{-S_{q_\ast}(\tau <0)} = 
\Psi[q_\ast]\big|_{\tau=0} \period\label{stateoperatormap}
\end{align}
Here $q_\ast$ signifies the saddle point configuration, $\mathcal{V}[q_\ast(z=0)] $
is the value of the vertex operator inserted at the origin of the worldsheet 
 $z=e^{\tau+i\sig}=0$, corresponding to the cylinder time $\tau=-\infty$,  the factor $\exp[-S_{q_\ast}(\tau <0)]$ is the amplitude to 
 develop into the state on a unit circle and the 
$\Psi[q_\ast]\big|_{\tau=0}$ is the semi-classical wave function describing the state on that circle. In particular, if we can construct the action-angle variables 
$(S_i, \phi_i)$ of the system and use $\{\phi_i\}$ as $q$,  then the wave function evaluated at the cylinder time $\tau$ can be expressed simply as 
\begin{align}
\Psi[\phi] &= \exp \left( i \sum_i S_i \phi_i -\calE(\{S_i\}) \tau 
\right)\comma\label{involvingenergy} 
\end{align}
where the action variables $S_i$ and the worldsheet energy $\calE(\{S_i\})$ are constant.

In the case of the classical  string  in $R \times S^3$, 
 the  method for the construction of the action-angle variables 
 was developed in \cite{DV1,DV2,Vicedo}, employing 
 the so-called Sklyanin's separation of variables \cite{Sklyanin}. 
This method was adapted to the case of the GKP string in $AdS_3$
 in \cite{KK2} and, as we shall see, can be 
 applied to the present case of the string in \EAdS3 $\times S^3$ with 
 appropriate modifications. In this method, the essential dynamical 
 information is contained in the two-component Baker-Akhiezer vectors $\psi_\pm $, which satisfy the ALP for the right sector and are the eigenvectors of the monodromy matrix $\Omega$ 
\begin{align}
\Omega(x;\tau, \sigma)  \psi_\pm(x;\tau, \sigma)
&= e^{\pm i p(x)} \psi_\pm (x;\tau, \sigma) \period
\end{align}
 More precisely, 
 the dynamical information is  encoded in the normalized Baker-Akhiezer 
 vector $h(x;\tau)$, defined to be proportional to $\psi(x;\tau, \sigma=0)$
 (conventionally  taken to be  $\psi_+$) and  satisfying  the normalization condition
\begin{align}
n \cdot h &= n_1h_1 + n_2 h_2 =1  \comma \qquad 
 h ={1\over n\cdot  \psi} \psi \comma \label{defh}
\end{align}
where $n$ is called the {\it normalization vector}. 
For a finite gap solution associated to a genus $g$ algebraic curve, 
$h(x;\tau)$ as a function of $x$  is known to have 
 $g+1$ poles at the positions $x=\{\ga_1, \ga_2, \ldots, \ga_{g+1}\}$
 and the dynamical variables $z(\ga_i)$ and $p(\ga_i)$, where 
 $z$ is the Zhukovsky variable defined in \eqref{Zhukvar}, can be shown to form canonical conjugate pairs. Then by making a suitable canonical 
transformation, one can go to the action-angle pairs $(S_i, \phi_i)$, 
 where, in particular,  the angle variable is given by the 
 generalized Abel map 
\begin{align}
\phi_i &= 2\pi \sum_j \int_{x_0}^{\ga_j} \omega_i \period 
\end{align}
Here, $\omega_i$ are  suitably normalized holomorphic differentials (with certain singularities 
 depending on the specific problem) and $x_0$ is an arbitrary base point. 
Now since one can reconstruct the classical string solution from the Baker-Akhiezer vector $\psi$ (and $\psitil$, which is  the solution of the left ALP)  as 
 shown  in \eqref{reconsteqR} and \eqref{reconsteqL},   with a choice of the normalization vector $n$  one can 
 associate a set of angle variables $\phi_i$ to  a classical solution. 
In fact, the angle variables can be thought to be determined by the quantity $n \cdot \psi$, since the poles of the normalized vector $h$ occur at the 
 zeros of $n \cdot \psi$, as is clear from (\ref{defh}). As we are actually dealing with a quantum system using semi-classical approximation, a classical 
solution should be thought of as being produced by a quantum vertex operator carrying a large charge. Further, since in our framework the vertex operator 
 is replaced by the corresponding wave function, the angle variables 
defined through a classical solution should  be used to describe the wave function
 of the corresponding semiclassical state. 

Now the serious problem is that we do not know the exact saddle point solution for the 
three-point function. The only information we know is that in the vicinity of each  vertex insertion point $z_i$, the exact three-point solution, to be represented  by a $2\times 2$ matrix $\bbY$ given by 
\begin{align}
\bbY &= \matrixii{Z_1}{Z_2}{-\Zbar_2}{\Zbar_1} \comma \qquad 
Z_1 =Y_1+iY_2 \comma \qquad Z_2 = Y_3+iY_4 \comma 
\end{align}
which must be almost identical to the two-point solution produced by the same vertex operator. Let us denote such a solution by $\bbYref$ and call it a reference solution. As we have to normalize the three-point function precisely by such a  two-point function for each 
 leg, what is important is the difference between $\bbY$ and $\bbYref$. 
Note that even if they are produced by the same vertex operator, they are different 
 because $\bbY$ is influenced by the presence of other vertex operators 
 in the three-point function. 

Here and in what follows, the global isometry group $G={\rm SU(2)}_L\times{\rm SU(2)}_R$ and its complexification  $G^c={\rm SL(2,C)}_L\times{\rm SL(2,C)}_R$ play the central roles. 
Being the symmetry groups of the equations of motion (and the Virasoro conditions),  two solutions of the equations of 
motion  are connected by the action of $G$ and/or  $G^c$. The difference between their actions are that (when expressed in terms of the Minkowski worldsheet variables) while $G$ connects a real solution to a real solution, $G^c$
 transforms a real solution to a complex solution. Since the three-point 
 interaction is inherently a tunneling process, the saddle point solution 
for such a process must be complex. 
 Therefore  near $z_i$  the two solutions $\bbY$ and $\bbYref$ 
must be connected  
 by an element of $G^c$ in the manner 
\begin{align}
\bbY = \Vtil \bbYref V \comma \qquad \Vtil \in {\rm SL(2,C)}_L, V \in {\rm SL(2,C)}_R
\end{align}
 This means that the angle variables associated to $\bbY$,  as 
 defined relative to the ones associated to $\bbYref$,  should be computable 
 from the knowledge of the transformation matrices   $\Vtil$ and $V$.
 This connection was made completely  explicit in \cite{KK2} and the  master formulas  giving such shifts  of the angle variables were obtained. 
 Corresponding to 
the solutions $\psi(x)$ and $\psitil(x)$ of the  the right and the left ALP
 respectively, there are right angle variable $\phi_R$ and the left angle 
 variable $\phi_L$. Their shifts are given by\footnote{These equations are obtained from 
 the fundamental formula (3.74) of \cite{KK2} by substituting the definition
 of the function $f(x)$ given in (3.62) of the same reference and 
 noting the expression of the function $h(x)$ shown in \eqref{defh} 
 of this paper. }
\begin{align}
\iDelta \phi_R &= -i \ln \left({ (n \cdot \psi_+(\infty) ) (n \cdot \psiref_-(\infty))
\over (n \cdot \psiref_+(\infty)) (n \cdot \psi_-(\infty)) }\right) 
\comma 
\label{DeltaphiR}  \\
\iDelta \phi_L &= -i \ln \left({ (\ntil \cdot \psitil_+(0) ) (\ntil \cdot \psitilref_-(0))
\over (\ntil \cdot \psitilref_+(0)) (\ntil \cdot \psitil_-(0)) }\right) 
\label{DeltaphiL} 
\end{align}
where $n$ and $\ntil$ are the normalization vectors for the right and the 
left sector  and $\psi_\pm(x)$ and $\psiref_\pm(x)$ are the Baker-Akhiezer eigenvectors  corresponding  to the solutions $\bbY$ and  $\bbYref$ respectively  and  are related by 
\begin{align}
\psi_\pm &= V^{-1} \psiref_\pm \comma \qquad 
\psitil_\pm = \Vtil \psitilref_\pm  \period
\end{align}
How $V$ and $\Vtil$ can be obtained will be described  in detail in subsection \ref{subsec:wave}. 

The remaining problem is to fix the normalization
 vectors $n$ and $\ntil$, relevant for the left and the right sectors. 
In the case of the string which is entirely in $AdS_3$ \cite{KK2}, 
we fixed them by the following argument.  Consider for simplicity 
the wave function corresponding to a conformal primary operator of the gauge theory  sitting at the origin of the boundary of $AdS_5$. Such an operator 
 is characterized by the invariance under the special conformal transformation.  Therefore the corresponding wave function and the angle variables 
comprising it should also be invariant. Explicitly it  requires  that $n\cdot \psi$ and $\ntil \cdot \psitil$ must  be preserved  under the special conformal transformation and this  determined  $n$ and $\ntil$. 

The essence of the argument we shall employ for  the case of a string in \EAdS3 $\times S^3$ studied in the present  work  is  the same. However because the structures of the gauge theory operators and the corresponding string solutions are more complicated, we need to generalize and refine the argument. 
As a result  of this improvement, not only has the determination of the 
normalization vectors become more systematic  but also their physical meaning 
has been identified  more clearly.  Moreover,  the entire procedure of the constructions of the wave functions  for the $S^3$ part and the  \EAdS3 part has become completely parallel and transparent. 
Below we shall begin the analysis first from the gauge theory side. 
%%%%%%%%%%%%%%%%
\subsection{Characterization of the gauge theory operators by symmetry 
 properties\label{subsec:highest}}
%%%%%%%%%%%%%%%%%%
As sketched above, in order to construct the wave functions expressing the effect of the 
 insertion of the vertex operators,  we must study how to characterize the global symmetry properties of the vertex operators and the classical 
 configurations that they produce in their vicinity.

 For this purpose, it is 
 convenient to first look at the symmetry properties of the corresponding gauge theory operators.  The three composite operators 
 $\calO_1(x_1), \calO_2(x_2), \calO_3(x_3)$ making up the three-point functions in the so-called ``SU(2) sector" are composed of the complex scalar fields  $Z\equiv \Phi_1+i\Phi_2$, $X \equiv \Phi_3 + i\Phi_4$ and their 
 complex conjugates $\Zbar$ and $\Xbar$, where $\Phi_I$ $(I=1,2,3,4)$ are four of the six real hermitian fields in the adjoint representation of the gauge group. Under the global symmetry group  ${\rm SO(4)}={\rm SU(2)}_{R}\times  {\rm SU(2)}_{L}$, these fields transform in  the doublet 
 representations of ${\rm SU(2)}_{R}$ and ${\rm SU(2)}_{L}$ 
 with the  right  and the left  charges $\mathcal{R}$ and $\mathcal{L}$ given in Table \hyperlink{table1}{1}:\\\nxt
\begin{minipage}{0.5\hsize}
\begin{center}
\hypertarget{table1}Table 1. The SU(2)${}_R$ and SU(2)${}_L$ charges for the basic scalar fields.
\beq{
\begin{array}{c|c|c}
& \calR & \calL \\ \hline
Z& +1/2 & +1/2 \\ 
\Zbar & -1/2 & -1/2 \\
X& -1/2 & +1/2 \\
\Xbar & +1/2 & -1/2  
\end{array}
\nn}
\end{center}
\end{minipage}
\begin{minipage}{0.5\hsize}
\begin{center}
\hypertarget{table2}Table 2. Roles of the scalar fields for the operators $\calO_i$.
\beq{
\begin{array}{c|cc}
&\text{vacuum} &\text{excitation} \\ \hline
\mathcal{O}_1& Z  &X   \\
\mathcal{O}_2& \bar{Z}&\bar{X}\\
\mathcal{O}_3& Z&\bar{X}
\end{array}
\nn}
\,
\end{center}
\end{minipage}\\
\nxt
These transformation properties are succinctly represented by 
 the $2\times 2$ matrix 
\begin{align}
\Phi  &= \matrixii{Z}{X}{-\Xbar}{\Zbar} \comma 
\end{align}
which gets transformed as $U_L \Phi \, U_R $, where $U_L \in \SUL, U_R \in \SUR$. 
In spite of this SO(4) symmetry,  in the existing literature \cite{Tailoring1}  the operators $\calO_i$ are taken  to be composed  of a special  pair of  fields\footnote{The reason 
 for this choice is that it is the simplest one that can produce 
non-extremal three-point functions. }
 indicated in Table \hyperlink{table2}{2}. 
For example,  $\calO_1$ is of the form $\tr (ZZ\cdots X ZZX\cdots Z)$. 
In the spin-chain interpretation, $Z$ and $X$ represent  the up and the 
 down spin respectively so that $\calO_1$ is a state built upon the all-spin-up 
 vacuum state $\tr Z^l$ on $l$ sites 
 by flipping some of the up-spins  into the 
down-spins which represent  excitations. Therefore at each site  there is an SU(2) group acting 
 on a spin, and according to  Table \hyperlink{table1}{1} it  is  identified with 
${\rm SU(2)}_{\rm R}$ for this case.  For the entire operator $\calO_1$, what is 
 relevant is the total  ${\rm SU(2)}_{\rm R}$, the generator  of which will be denoted by $S^i_R$. 

Let us now characterize the spin-chain states corresponding 
 to the operators of the type $\calO_1$ from the point of view of this total 
 $\SUR$. First, since the constituents $Z$ and $X$ carry definite spin quantum numbers,  
every state of type $\calO_1$ carries a definite right and  left global charges.  
Second, every such state is actually a highest weight state annihilated by the  operator 
 $S^+_R= S^1_R + iS^2_R$. For the vacuum state $\ket{Z^l}= \ket{\!\uparrow^l}$ it is obvious. As for the  excited states, they can be written as 
  the Bethe states $\prod_{i=1} 
B(u_i) \ket{\!\uparrow^l}$, where $B(u_i)$ is   the familiar magnon creation operator  carrying  the spectral parameter  $u_i$. It is well-known\cite{book} that  such a state is  a  highest weight state of the total  $\SUR$  and hence annihilated by the same $S^+_R$, provided that the Bethe state is ``on-shell", namely that  the spectral parameters satisfy the Bethe ansatz equations. Therefore we have 
 found that kinematically all the operators of type $\calO_1$ can be characterized  as the highest weight state of the total ${\rm SU(2)}_{\rm R}$.

Now in order to deal with other operators built upon a  ``vacuum state" different from   $\tr Z^l$, let us introduce  the general linear combinations of 
$\Phi_I$ as $\vec{P} \cdot \vec{\Phi} = \sum_{I=1}^4 P_I \Phi_I$. 
To discuss the transformation property under $\SUR \times \SUL$,   it is more convenient to deal with the matrix 
\begin{align}
\bbP &\equiv  \matrixii{P_1+iP_2}{P_3+iP_4}{-(P_3 -iP_4)}{P_1-iP_2} 
= P_I \Sig_I \comma \\
\Sig_I &\equiv  (1, i\sig_3, i\sig_2, i\sig_1) 
\period
\end{align}
Then, we have the representation
\begin{align}
\vec{P} \cdot \vec{\Phi} &= \half \tr \left( \sig_2 \bbP^t \sig_2 \Phi\right)  \period \label{PPhi}
\end{align}
In this notation, $\bbP$ corresponding to $Z, \Zbar, X, \Xbar$ take the form
$ \bbP_Z = 1-\sig_3, \bbP_\Zbar = 1+\sig_3, \bbP_X =-(\sig_1-i\sig_2), 
\bbP_\Xbar=
\sig_1+i\sig_2$. 

As we argued above, all the on-shell states  built upon a common vacuum are 
 annihilated by the same $S^+_R$. In other words as long as the 
global transformation property is concerned,  the vacuum state can be considered as the representative of all the states built upon it. Further,   since the local spin state is identical at each site for the vacuum state we can characterize the vacuum by the form of the ``annihilation operator"  $s^+_R$ acting on a single spin state. As it will be slightly more convenient, instead of the 
 annihilation operator, we will use  the ``raising operator" 
 $K = \exp (\al s^+_R)$, where $\al$ is any constant. 
The vacuum is then characterized by the form of $K$  that  
 leaves its building block   {\it invariant}. 

 Let us explain this idea concretely for the operator $Z$, which is  the building block for the simplest vacuum state $\tr Z^l$.  In the general notation (\ref{PPhi}), we can express $Z$ as $Z=\half \tr (\sig_2 \bbP_Z^t \sig_2 \Phi )$ with  $\bbP_Z = 1-\sig_3$.  Now let us  look for the raising operators $K_Z$
 and $\tilde{K}_Z$ for  $\SUR$ and $\SUL$  respectively, 
which leave $Z$ invariant. Since $\Phi$ transforms into $\tilde{K}_Z \Phi K_Z$, the invariance condition reads 
\begin{align}
\half \tr \left( \sig_2 \bbP^t_Z \sig_2 \tilde{K}_Z \Phi K_Z \right) 
= \half \tr \left( \sig_2 \bbP^t_Z \sig_2 \Phi\right)  \period
\end{align}
This is equivalent to  the condition
\begin{align}
\bbP_Z = \tilde{K}_Z^{-1}\bbP_Z  K_Z^{-1} \period 
\label{invbbPZ}
\end{align}
It is easy to find the solutions\footnote{The most general solutions 
 are of the form $\matrixii{\alpha}{\be}{0}{\alpha^{-1}} $
 and $\matrixii{\alpha^{-1}}{0}{\tilde{\be}}{\alpha} $. However since we are 
interested in the raising type operators,  it is sufficient to consider 
the operators of the form (\ref{VpZ}). 
}, which read
\begin{align}
K_Z &= \matrixii{1}{\be}{0}{1} = e^{\half \be\sig_+} \comma 
\qquad \tilde{K}_Z =  \matrixii{1}{0}{\tilde{\be}}{1} =e^{\half \tilde{\be} \sig_-} 
\comma \label{VpZ}
\end{align}
where $\be$ and $\tilde{\be}$ are arbitrary constants. 

Next we consider a general case where the vacuum state is given by 
 $\tr (\vec{P} \cdot \vec{\Phi})^l$, with arbitrary $\vec{P}$. 
Since,  in general,  $\vec{P} \cdot \vec{\Phi}$ does not carry a definite set of left and 
 right charges defined as in Table \hyperlink{table1}{1}, this state and the ones built upon it 
by some spin-chain type excitations are not charge eigenstates. Nevertheless, we can characterize this family of states again by the raising operators 
$K$ and $\tilde{K}$ 
 which leave $\vec{P}\cdot \vec{\Phi}$ invariant. Just as in (\ref{invbbPZ}), 
 this condition is expressed as 
\begin{align}
\bbP = \tilde{K}^{-1}\bbP  K^{-1} \period 
\label{invbbP}
\end{align}
where $\bbP$ corresponds to $\vec{P}$. 
Since $\vec{P}\cdot \vec{\Phi}$ can be obtained from $Z$ by an 
$\SUL \times \SUR$ transformation,  $\bbP$ can be  obtained from  $\bbP_Z$ by a corresponding  transformation of the form 
\begin{align}
\bbP &= U_L \bbP_Z U_R   \period \label{relPPZ}
\end{align}
Then combined with  (\ref{invbbP}) we readily obtain 
 the relation  $\bbP_Z = (U_L^{-1} \tilde{K}^{-1}U_L)
 \bbP_Z(U_R  K^{-1}U_R^{-1})$. Comparing this with (\ref{invbbPZ})
 we can express the raising operators $K$ and $\tilde{K}$
in terms of the ones for the operator $Z$ given in (\ref{VpZ})  in the form
\begin{align}
K &= U_R^{-1} K_Z U_R \comma \qquad 
\tilde{K} = U_L \tilde{K}_Z U_L^{-1}  \period  \label{VpVZp}
\end{align}

Now these raising operators can in turn be characterized 
 by the two-component 
vectors  $n$ and $\ntil$, which are left invariant under the following action 
 of $K$ and $\tilde{K}$ respectively\footnote{Intentionally we are using the same letters   $n$ and $\ntil$ for the vectors introduced here 
 as those used previously for the normalization  vectors. This is because 
they will be shown to be identical. }
:
\begin{align}
K^t n &= n \comma \qquad \tilde{K}^t \ntil = \ntil \period 
\label{defnntil}
\end{align}
Since  the overall factor for these vectors are inessential,  we can 
 normalize them to have unit length as $n^\dagger n=\ntil^\dagger \ntil =1$. We shall  refer to them as {\it polarization spinors}, as they characterize, 
so to speak,  the  ``direction of polarization" of the highest weight operator $\vec{P} \cdot \vec{\Phi}$. It should be noted that from the knowledge of $n$ and $\ntil$, one can reconstruct
 $\bbP$ which is invariant under the raising operators, as in (\ref{invbbP}). 
In fact, if we set 
\begin{align}
\bbP &= -2i \sig_2 \ntil n^t \comma  \label{recbbP}
\end{align}
one can easily check that this $\bbP$ satisfies (\ref{invbbP}), 
 with the use of the defining equations (\ref{defnntil}) and a simple 
formula $\sig_2 U^{-1} \sig_2 = U^t$ valid for any  invertible $2\times 2$ matrix 
 $U$ satisfying $\det U=1$. 

Let us illustrate these concepts by computing the polarization spinors 
 for the operators  $Z$ and $\Zbar$ respectively. For the operator 
$Z$ we already computed 
 the right and the left raising operators in (\ref{VpZ}). Then it is easy to see 
 that the corresponding polarization spinors $n_Z$ and $\ntil_Z$ satisfying $K_Z^t n_Z  =n_Z$ and $\tilde{K}_Z^t \ntil_Z =\ntil_Z$ are 
given by 
\begin{align}
n_Z &= \vecii{0}{1} \comma \qquad \ntil_Z =\vecii{1}{0} \period 
\label{nZntilZ}
\end{align}
As a check, from the formula (\ref{recbbP}), we immediately get 
 $\bbP_Z =$  {\small $\matrixii{0}{0}{0}{2} $}, which is the desired form. 
As for the operator $\Zbar$, repeating the similar analysis,  the raising operators leaving $\bbP_\Zbar = 1+\sig_3$ invariant can be readily  obtained to be
\begin{align}
K_\Zbar &= \matrixii{1}{0}{\al}{1} 
\comma \qquad \tilde{K}_\Zbar = \matrixii{1}{\tilde{\al}}{0}{1} \comma 
\end{align}
with  $\al$ and $\tilde{\al}$ being  arbitrary constants. The corresponding polarization spinors can be taken to be 
\begin{align}
n_\Zbar &= \vecii{1}{0} \comma \qquad \ntil_\Zbar =\vecii{0}{1} 
\period \label{nzbar}
\end{align}
Finally consider the normalization spinors for a general operator $\vec{P}\cdot\vec{\Phi}$ which is related to $Z=\vec{P}_Z\cdot \vec{\Phi}$ through  the relation of the form (\ref{relPPZ}). Since the raising operators for such 
an operator are obtained from those for $Z$ in the manner (\ref{VpVZp}), 
 the polarization vectors $n$ and $\ntil$ are expressed in terms of $n_Z$
 and $\ntil_Z$ as
\begin{align}
n &= U_R^t n_Z \comma \qquad \ntil = (U_L^t)^{-1} \ntil_Z \period
\label{nntil}
\end{align}
As an application of this formula, let us re-derive  $n_\Zbar$ and $\ntil_\Zbar$
from this perspective. Since $\bbP_\Zbar = 1+\sig_3$ and $\bbP_Z =1-\sig_3$, 
 it is easy to see that they are related by an $\SUL\times \SUR$ transformation of the form 
\begin{align}
\bbP_\Zbar &= U_L \bbP_Z U_R \comma  \qquad 
U_L = i\sig_2 \comma \quad U_R = -i\sig_2 \period 
\label{relZbarZ}
\end{align}
In fact this transformation realizes  the mapping 
$(Z,X) \rightarrow (\Zbar, -\Xbar)$. 
 Substituting  the forms of 
 $U_L$ and $U_R$  into  the above formula (\ref{nntil}),
 we  obtain 
$U^t_R n_Z = (1,0)^t $ and $(U^t_L)^{-1} \ntil_Z = -(0,1)^t
\propto (0,1)^t$, which 
 agree with (\ref{nzbar}).

Summarizing,  we can  say that, as far as the global 
 symmetry properties are concerned, the operators of 
type $\calO_1$ and $\calO_3$ are characterized by the polarization 
spinors $n_Z$ and $\ntil_Z$, while the operators of type $\calO_2$ are associated with $n_\Zbar$ and $\ntil_\Zbar$. For more general operators 
 built upon the vacuum $\tr (\vec{P}\cdot \vec{\Phi})^l$, the corresponding polarization spinors are obtained from $n_Z$ and $\ntil_Z$ by appropriate 
 transformations which connect $\bbP$ with $\bbP_Z$ as shown  in (\ref{nntil}). 

The importance of the above analysis  is that, as we shall describe below,  precisely the same characterization scheme must be valid for the vertex operators in string theory which correspond  to the gauge theory composite operators like  $\calO_i$. Moreover, it will be shown that the polarization spinors introduced purely from the 
group theoretic point of view above  will be identified with the 
 ``normalization vectors"  that appeared in  (\ref{defh}),  which play  pivotal roles  in the construction of the angle variables and hence the construction of the wave functions describing the contribution  of the vertex operators. 
%%%%%%%%%%%%%%%%%%
\subsection{Wave functions  for the $S^3$ part\label{subsec:wave}}
%%%%%%%%%%%%%%%%%
\subsubsection{Symmetry structure of the vertex operators and the classical 
 solutions}
%%%%%%%%%%%%%%%%%
We now begin the explicit construction of the wave functions contributing to the three-point functions in string theory.  As emphasized in the introduction, an  essential ingredient for the success of the computation of the three-point functions 
is the separation of the kinematical  and the dynamical factors.  Although the dynamics is quite  different between the gauge theory and the corresponding string theory, 
 the kinematical symmetry  properties correspond quite directly between the gauge theory  operators and the vertex operators of string theory. 
Therefore in this subsection we will describe how we can implement the 
scheme of the symmetry characterization of the operators developed 
in the preceding subsection  for the gauge theory operators to the vertex operators and 
the classical solutions produced by them. 
Since the analysis concerning  the each factor of the symmetry group $\SUR \times \SUL$ is completely similar and can be performed independently, 
 after some general discussions we will almost exclusively focus on 
 the $\SUR$ part of the symmetry transformations and various corresponding quantities  for clarity of presentations. 

In the saddle point approximation scheme  we are employing, we cannot directly  deal with the vertex operator: What we can deal with are the classical solutions produced by
 the vertex operators carrying large charges. Therefore we need to  extract the information of the quantum vertex operators indirectly through such classical solutions. 

For definiteness, we first focus on a solution with diagonal $\SUR\times \SUL$ charges describing a two-point function of an operator built on the $\tr (Z^l)$-vacuum ($\calO_1$ and $\calO_3$ in section \ref{subsec:highest}) and its conjugate\fn{What is meant by ``conjugation" is the usual complex conjugation of the fields, $Z\to \bar{Z}$ and $X\to \bar{X}$.}. In what follows, we shall denote 
 such a solution by $\bbYdiag$. Then we can associate a pair of polarization spinors $n_Z$ and $\ntil_Z$ and the raising operators (\ref{VpZ}) to the vertex operator that produces the solution. For convenience, we display them again with appropriate renaming:
\begin{align}
\ndiag&=\vecii{0}{1} \comma \qquad \ntildiag=\vecii{1}{0} \comma 
\label{polspdiag}  \\
K^{\rm diag} (\be)&= \matrixii{1}{\be}{0}{1}  \comma \qquad \tilde{K}^{\rm diag}(\tilde{\be}) =  \matrixii{1}{0}{\tilde{\be}}{1}  
\period 
\end{align}
All the solutions describing a two-point function of mutually conjugate operators, $\langle\mathcal{O} \overline{\mathcal{O}}\rangle$, can be obtained from this basic solution $\bbYdiag$ by an $\SUR\times \SUL$  transformation. 
Since a normalized three-point function in the gauge theory can be obtained by dividing an unnormalized one by $\langle\mathcal{O} \overline{\mathcal{O}}\rangle$-type two-point functions as
\beq{
\frac{\langle \mathcal{O}_i \mathcal{O}_j \mathcal{O}_k\rangle}{\sqrt{\langle\mathcal{O}_i\overline{\mathcal{O}}_i\rangle\langle\mathcal{O}_j\overline{\mathcal{O}}_j\rangle\langle\mathcal{O}_k\overline{\mathcal{O}}_k\rangle}}\comma
}
the aforementioned solutions, to be denoted by $\mathbb{Y}^{\rm ref}$, can be used as {\it reference solutions} to determine the normalization of the wave function.
An important feature of such solutions is that they are real-valued when expressed in terms of the Minkowski worldsheet variables.
 This qualification will be extremely important since the equation of motion is actually invariant under a larger group $\SLCR\times \SLCL$ and its action can produce  ``complex" solutions which signify tunneling. Such a  tunneling process is  necessary for the three-point interactions to take place, as we shall see.

From now on till the end of this subsection, we shall suppress  all the left transformations and display only the  right transformations. The results  for the left transformations will be summarized in subsection \ref{subsubsec:left}. 

Now consider a three-point function produced by vertex operators, corresponding to the gauge theory operators, inserted at $z_i$ on the worldsheet.  
We will take the operators to be those obtained by SO(4) rotations of the operators built on the $\tr(Z^l)$-vacuum. This suffices for the present purpose since such three-point functions include\fn{Note that $\calO_1$ and $\calO_3$ in section \ref{subsec:highest} are built on the $\tr(Z^l)$-vacuum while $\calO_2$ can be obtained from the operator built on $\tr(Z^l)$ by an SO(4) rotation (\ref{relZbarZ}), which effects $(Z, X) \rightarrow (\Zbar, -\Xbar)$.} the ones discussed in section \ref{subsec:highest}.

\begin{figure}[htbp]
  \begin{center}
   \picture{clip,height=4cm}{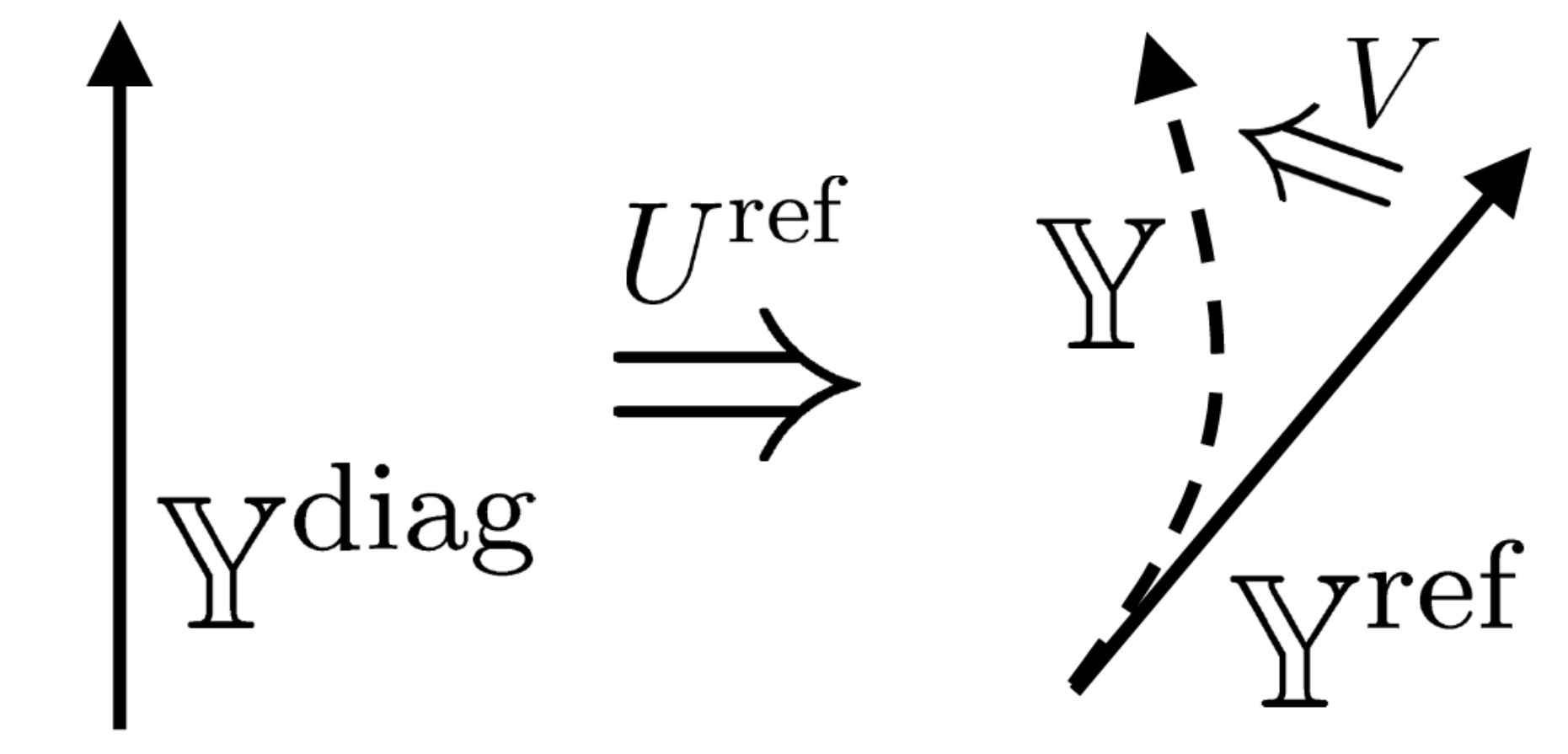}
  \end{center}
\caption{A schematic picture which explains the relation between the local three-point 
 solution $\bbY$ and the two-point solutions $\bbYdiag$ and $\bbYref$. 
$\bbYref$ is obtained  from $\bbYdiag$ by a {\it real} global transformation $U^{\rm ref}$, while $\bbY$ and  $\bbYref$, which are produced by the same vertex operator, are related  through a {\it complexified} global transformation  $V$.} 
\label{fig:rel-Ys}
\end{figure}

Although  the saddle point solution 
 for such a three-point function is so far not available explicitly, 
 let us  denote the solution in the vicinity of $z_i$ by $\bbY$. 
Asymptotically as $z \rightarrow z_i$ such a configuration must be 
well-approximated by a two-point reference solution 
 $\bbYref$, which is produced by the {\it same} vertex operator. 
Even if they are  produced by the same vertex operator, $\bbY$ and $\bbYref$ are 
 different since $\bbY$ is influenced non-trivially and dynamically by the other two vertex operators present. We write the transformation  between them at $z\simeq  z_i$ as\fn{Note that $\bbYref$ is the solution for the two-point function, expressed globally in terms of the cylinder coordinate. Thus we need to express $\bbY$ in terms of the local coordinate $\left( \tau^{(i)}, \sigma^{(i)}\right)$ given in \eqref{localcoordinate} to compare two solutions.} 
\begin{align}
&\bbY (z\simeq z_i) = \bbYref V  \qquad (z\to z_i)\comma\qquad V \in \SLCR \period \label{relYYref}
\end{align}
 This relative difference  is the quantity of interest  
 since  we need to normalize the three-point function by the two-point functions.  In general $V$  belongs to $\SLCR \supset \SUR$, since the three-point interaction is necessarily a tunneling  process. In contrast 
the reference solution  $\bbYref$ can be obtained from  $\bbYdiag$ by 
 a transformation belonging to $\SUR$ in the form 
\begin{align}
\bbYref &= \bbYdiag U^{{\rm ref}} \comma \qquad U^{\rm ref} \in \SUR \period \label{relYrefYdiag}
\end{align}
The relation between $\bbY$, $\bbYref$ and $\bbYdiag$ is sketched in \figref{fig:rel-Ys}. 

Now just as we did already for the solution $\bbYdiag$, 
 we can associate 
 to  the solution $\bbYref$ the polarization spinor $\nref$ 
 and the raising transformation $K^{\rm ref}$ which leaves it invariant. 
Then from the general formula (\ref{nntil}) and  (\ref{VpVZp}) we can express them in terms of the quantities associated to the diagonal solution as
\begin{align}
\nref &= (U^{\rm ref})^t \ndiag \comma  \label{nrefdiag}\\
K^{\rm ref}(\be) &= (U^{\rm ref})^{-1}
K^{\rm diag}(\be) U^{\rm ref} \period \label{Vrefdiag}
\end{align}
By the same token we can associate the polarization spinor $n$ and the 
 raising transformation $K$ to the local solution $\bbY$. However
 since $\bbY$ is produced by the same vertex operator as $\bbYref$, 
we must have $n=\nref$.  As for $K$, just as in \eqref{Vrefdiag},  under the transformation $V$ which produces  $\bbY$ from $\bbYref$, the raising operator $K^{\rm ref}(\be)$ transforms into $K(\be)$ in the manner
$ K(\be) = V^{-1} K^{\rm ref}(\be) V$.  Since this operator must leave $n$ and hence $\nref$ invariant,  we must have 
\begin{align}
V^{-1} K^{\rm ref}(\be) V =K^{\rm ref}(\be')
\end{align}
for some $\be'$. Substituting the relation  (\ref{Vrefdiag}), 
 we get
\begin{align}
\begin{aligned}
&\left( V^{\prime}\right)^{-1} K^{\rm diag}(\be) V^{\prime} 
=K^{\rm diag} (\be') \comma\\
&V^{\prime}\equiv U^{\rm ref} V (U^{\rm ref})^{-1}\period
\end{aligned}
\end{align}
This means that  the operator $V'$ transforms a raising operator  into a raising operator for the diagonal solution. 
It is not difficult to show that the general form of such an operator is 
{\small $\matrixii{a}{b}{0}{a^{-1}} $}. Note that this contains a scale 
transformation which is in $\SLCR$ but not in $\SUR$.  From this result we can solve for 
 $V$ and its inverse and obtain the following useful  representations
\begin{align}
V &= (U^{\rm ref})^{-1} \matrixii{a}{b}{0}{a^{-1}} U^{\rm ref} \comma 
\label{repV} \\
V^{-1}&= (U^{\rm ref})^{-1} \matrixii{a^{-1}}{-b}{0}{a} U^{\rm ref} \period 
\label{repVinv}
\end{align}
At this stage we need not know the actual values of $a$ and $b$ in these 
 formulas. $b$ will turn out to be irrelevant and $a$ will be expressed 
 in terms of certain Wronskians. 
%%%%%%%%%%%%%%%%
\subsubsection{Construction of the wave function for the right sector }
%%%%%%%%%%%%%%%%
We are now ready for the construction of the wave function for the 
 right sector using the formula for the shift of the angle variable $\phi_R$
 given in (\ref{DeltaphiR}). 

First we need to fix the normalization vector $n$ appearing in that  formula. 
As we shall show, the answer is that it coincides  precisely with  the polarization spinor $n$ introduced from the group theoretical point of view in (\ref{defnntil}) in subsection \ref{subsec:highest}.  Recall that  in the formalism developed in \cite{DV1,DV2,Vicedo},  the zeros of  $n \cdot \psi(x)$, where $\psi $ is  the Baker-Akhiezer vector and $n$ is the normalization vector,  determines   the angle variables. When one makes a global  SL(2,C)$_R$ transformation $V_R$ on the string solution $\bbY$ like $\bbY \rightarrow \bbY V_R$, the Baker-Akhiezer vector transforms like $\psi\rightarrow V_R^{-1}\psi$.  In particular, take $V_R$  to be the raising operator $K$ under which the vertex operator  producing the solution $\bbY$ is invariant.  Then  the wave function 
 corresponding to the vertex operator and hence the angle variables comprising it must also be invariant.  This means that the zeros of $n \cdot (K^{-1} \psi)  = (K^t n) \cdot \psi$ must coincide with the zeros of  $n \cdot \psi$ and hence we must have
 $K^t n \propto n$. However since $K$ is 
similar to $K^{\rm diag}$, it is clear that the constant of proportionality can only be unity and $n$ must 
 satisfy $K^t n=n$. This, however, is nothing but the definition of the 
polarization spinor given in (\ref{defnntil}). In other words, the proper 
 choice of the normalization vector for constructing the wave function 
 is precisely the polarization spinor associated to the vertex operator to which 
 the wave function corresponds. 
 
Having found  the proper choice of the normalization vector in the formula 
 (\ref{DeltaphiR}) for the shift of the angle variable $\phi_R$, what remains to be understood is how to evaluate  the inner products  $n \cdot \psi_\pm (\infty)$ and $n \cdot \psiref_\pm(\infty)$.  Corresponding to  the relation \eqref{relYYref}, in the vicinity of $z_i$,  $\psi_\pm$ and $\psiref_\pm$ are related by 
 the constant transformation $V$ as $\psi_\pm(z\simeq z_i) = V^{-1} \psiref_\pm(z\simeq z_i)$.  Now recall the form of the ALP for the 
 right sector given in \eqref{RALP}.  We see that  for  $x=\infty$ the coefficients of  the connections $j_z$ and $j_{\zbar}$ vanish and hence the solutions  $\psi_\pm(x=\infty)$ and $\psiref_\pm(x=\infty)$ themselves become constant. Combining these pieces of information,  we obtain  the relation  
\begin{align}
\psi_\pm(\infty) &= V^{-1} \psiref_\pm(\infty) \period \label{psipmV}
\end{align}
The right hand side can be evaluated using  the representation (\ref{repVinv}) as 
\begin{align}
\psi_\pm(\infty) &=  (U^{\rm ref})^{-1}  \matrixii{a^{-1}}{-b}{0}{a} U^{\rm ref} \psiref_\pm(\infty)
= (U^{\rm ref})^{-1}  \matrixii{a^{-1}}{-b}{0}{a} \psidiag_\pm(\infty) 
\comma \label{psipminf}
\end{align}
where $\psi^{\rm diag}_{\pm}(x)$ is the Baker-Akhiezer vector for $\mathbb{Y}^{\rm diag}$, which is related to $\psi^{\rm ref}_{\pm}(x)$ by
\beq{
\psi^{\rm ref}_{\pm} (x)=\left( U^{\rm ref}\right)^{-1} \psi^{\rm diag}_{\pm} (x)\period
} 

We now need to know $\psidiag_\pm(\infty) $,  which are the eigenstates 
 of the monodromy matrix near $x=\infty$ corresponding to the eigenvalues $e^{\pm ip(x)}$.  For a charge-diagonal solution $\mathbb{Y}^{\rm diag}$, the  monodromy matrix  near $x=\infty$ is diagonal and hence is either of the form 
 (a)\ ${\rm diag}\,(e^{ip(x)}, e^{-ip(x)})$  or (b)\ ${\rm diag}\,(e^{-ip(x)}, e^{ip(x)})$, depending on the solution.   For the case $(a)$
 the eigenvectors are 
 $\psidiag_+(\infty)=(1,0)^t, \psidiag_-(\infty) =(0,1)^t$, while  for the case (b) their forms are swapped. Since $\mathbb{Y}^{\rm diag}$ is produced by 
  the vertex operator with the definite polarization spinor specified in (\ref{polspdiag}), there should  be a definite answer.  
To determine the proper choice of (a) or (b), we need to construct the wave function for each choice and see if it has the same transformation property as the corresponding operator in the gauge theory.  
As it will be checked later in this subsection,  it turned out that the case (b) is the correct choice. Therefore we will take 
\begin{align}
\psidiag_+(\infty) &= \vecii{0}{1} \comma \qquad \psidiag_-(\infty) = \vecii{1}{0} \period \label{choiceb}
\end{align}
Substituting them into (\ref{psipminf}), we  obtain the important relations
\begin{align}
\psi_+(\infty) &=    (U^{\rm ref})^{-1}\left(a \psidiag_+(\infty)-b\psidiag_-(\infty)\right) = a  \psiref_+(\infty) -b\psiref_-(\infty) \comma 
\label{ipirefp} \\
\psi_- (\infty) &=  (U^{\rm ref})^{-1} a^{-1}\psidiag_-(\infty)  = a^{-1} \psiref_-(\infty)  
 \period \label{imiref}
\end{align}
As for the polarization spinor,  observe that by inspection 
the following relation holds:
\begin{align}
\ndiag=  (-i \sig_2) \psidiag_-(\infty) \period
\end{align}
This relation is actually  universal  in the following sense. Let us act 
 $(U^{\rm ref})^t$ from left. Then the relation becomes 
\begin{align}
\begin{aligned}
 \left( \,(U^{\rm ref})^t \ndiag =\,\right)\, \nref & = (U^{\rm ref})^t (-i \sig_2) \psidiag_- (\infty)\\
 &= (-i\sig_2)(U^{\rm ref})^{-1} \psidiag_-(\infty)   \\
&= -i\sig_2 \psiref_-(\infty)
\comma 
\end{aligned}
\label{covreln}
\end{align}
where we used the identity $\sig_2 (U^{\rm ref})^t \sig_2 = (U^{\rm ref})^{-1}$. Thus, 
 exactly the same form of  relation holds for the reference solution and 
in fact for any solution related by an $\SUR$ transformation. Together with 
the formula  (\ref{imiref}) we get  the relation 
\begin{align}
n &= -ia\sig_2 \psi_-(\infty) \comma  \label{nitopsim}
\end{align}
which will be extremely important. 

Let us now recall the formula (\ref{DeltaphiR}) for the shift of the angle variable $\phi_R$.
Displaying it again for convenience, it is of the form  
\begin{align}
\iDelta \phi_R &= -i \ln \left({ (n \cdot \psi_+(\infty) ) (n \cdot \psiref_-(\infty))
\over (n \cdot \psiref_+(\infty)) (n \cdot \psi_-(\infty)) }\right) 
\period\label{again}
\end{align}
From (\ref{ipirefp}) and (\ref{covreln}),  we can write  $n \cdot \psi_+(\infty) = a  n \cdot \psiref_+(\infty)$. 
 As for $n \cdot \psi_-(\infty)$, use of (\ref{imiref}) gives 
$n \cdot \psi_-(\infty) =a^{-1} n \cdot \psiref_-(\infty) $.  
Now due to the relation (\ref{covreln}), 
 the quantity $n \cdot \psi_-(\infty)=\nref \cdot \psi_-(\infty) $, which appears both in the numerator and the denominator of the formula (\ref{again}),  vanishes. Therefore we must first regularize $n$ slightly  to make 
the quantity $n \cdot \psi_-(\infty)$ finite, cancel them in the formula 
and then remove the regularization. As for the same quantity 
appearing in $n \cdot \psi_+(\infty)$, we can safely set it to zero from the beginning since $n \cdot \psiref_+(\infty) $ is non-vanishing. In this way we find 
 that $n \cdot \psiref_\pm(\infty)$'s  all cancel out and we are left with an 
 extremely simple formula for $\iDelta \phi_R$ given by 
\begin{align}
\iDelta \phi_{R} &= -i \ln a^2  \period  \label{phiRasq}
\end{align}
 Note that the shift depends only on the quantity
 $a$, which parametrizes the scale transformation not belonging to  $\SUR$, showing  the tunneling nature of the effect. 

Let us now  write the formula (\ref{nitopsim}) for the operator at $z_i$ 
with a subscript $i$  as $n_i = -ia_i\sig_2 i_-(\infty)$. 
Then, from the definition of the Wronskian we  obtain  $\wrons{n_i}{n_j} = a_i  a_j   \wrons{i_-}{j_-}\big|_\infty$. 
Writing out all the relations of this form and forming appropriate ratios, we can easily extract out each $a_i^2$. The result can be written in a universal form 
 as 
\begin{align}
a_i^2 &= {\wrons{j_-}{k_-} \over \wrons{i_-}{j_-} \wrons{k_-}{i_-}  }\bigg|_\infty  {
\wrons{n_i}{n_j}  \wrons{n_k}{n_i} \over \wrons{n_j}{n_k} }\period 
\end{align}
Then substituting this expression  into the formula (\ref{phiRasq})
 we obtain the shift of the 
angle variable $\phi_R$ at the position $z_i$ as 
\begin{align}
e^{i \Delta \phi_{R,i}} &=  {\wrons{j_-}{k_-} \over \wrons{i_-}{j_-} \wrons{k_-}{i_-}  }\bigg|_\infty  {
\wrons{n_i}{n_j}  \wrons{n_k}{n_i} \over \wrons{n_j}{n_k} }
 \period \label{expshiftR}
\end{align}
 This formula  is remarkable in that it cleanly 
 separates  the kinematical part composed of $\wrons{n_i}{n_j}$ and 
 the dynamical part described by $\wrons{i_-}{j_-}\big|_\infty$. 

As the last step of  the construction of  the wave function, we need to 
pay   attention to  the convention of \cite{Vicedo} that we are adopting. 
In that work, the Poisson bracket is defined to be  $\acom{p}{q}
=1$ for the usual momentum $p$ and the coordinate $q$. In this convention
 the Poisson bracket of the action angle variables was worked out to be
 given by  $\acom{\phi}{S} =1$. In other words  the action variable $S$ 
 corresponds to $q$ and the angle variable $\phi$ corresponds to $p$. Therefore 
upon quantization in the angle variable representation, we must set $S=i \del /\del \phi$. This means that  the wave function that carries  charge  $S$ is  given by $e^{-iS \phi}$, {\it not} by $e^{iS\phi}$. 

Recalling the relation \eqref{SinfS0} between the action variable $S_\infty$ and the right charge $R$, namely $S_\infty =-R$, and employing the formula 
(\ref{expshiftR}),   the contribution to the wave function from the right sector is obtained as 
\begin{align}
\Psi_R^{S^3} &= \exp \left(-i\sum_{i=1}^3 \left(- R_i\right) \iDelta \phi_{R, i}
\right) =\prod_{\{i,j,k\}} \left({\wrons{n_i}{n_j} \over \wrons{i_-}{j_-} \big|_\infty}
\right)^{R_i+R_j -R_k}   \comma \label{PsiR}
\end{align}
where $\{i,j,k\}$ denotes the cyclic permutations  of $\{1,2,3\}$. 

At this stage, let us confirm that the wave function so constructed indeed 
 carries the correct charge. To see this, it suffices to consider the U(1) 
 transformation which corresponds to the diagonal right-charge rotations. 
Let us examine the case of the charge-diagonal operator built upon the $Z$-type vacuum, such as $\calO_1$ or $\calO_3$ in section \ref{subsec:highest}. 
In such a case the reference state is the charge-diagonal state itself, hence  $U^{\rm ref} =1$.  Then if we set $a=e^{i\theta/2}, b=0$ in the formula (\ref{repV}), the $\SUR$ transformation matrix $V$ becomes $\diag (e^{i\theta/2}, e^{-i\theta/2})$,  which is a U(1) transformation under which $Z$ and $\Zbar$, carrying the right charge $1/2$ and $-1/2$ 
respectively,   transform as $Z \rightarrow e^{i\theta/2}Z$
 and $\Zbar \rightarrow e^{-i\theta/2}$.  Now 
according to (\ref{phiRasq}), under such a transformation the wave function 
acquires the phase $e^{ -i (-R) \ln a^2}= e^{iR\theta}$. This shows that the wave function has the same (positive) charge $R$ as the operator of the form $\tr( Z^{2R})$. 
This proves that the choice of $\psidiag_\pm (\infty)$ we made in (\ref{choiceb})
 is the correct one. If we had made the other choice, the wave function would have acquired the phase $e^{-iR\theta}$, which contradicts the fact that the corresponding operator in the gauge theory is built on the $\tr (Z^l)$-vacuum.
Similar argument can be made for the left sector 
and again one can check that the wave function (\ref{PsiR}) carries the correct charges.
%%%%%%%%%%%%%%%%
\subsubsection{Contribution of the left sector and complete wave function for the $S^3$ part\label{subsubsec:left}}
%%%%%%%%%%%%%%%%
We now briefly describe the analysis for the left sector, to complete the construction of the wave function for the $S^3$ part. 

The procedure  is exactly the same as for the right sector but there are a couple of notable differences. First, the transformation matrices act from the left and consequently  in various formulas the matrices are replaced by their inverses.
In particular, the formulas corresponding to (\ref{repV}) and (\ref{psipmV}) 
for the transformation $\Vtil$ that connects three-point solution and the reference solution in the manner  $\bbY = \Vtil \bbYref$ take the form 
\begin{align}
&\Vtil = \tilde{U}^{\rm ref} \matrixii{a}{0}{b}{a^{-1}} (\tilde{U}^{\rm ref})^{-1} \comma  \\
&\psitil_\pm(0) = \Vtil \psitilref_\pm (0) \comma 
\end{align} 
where $\tilde{U}^{\rm ref}\in \SUL$ is the matrix effecting the connection 
 $\bbYref = \tilde{U}^{\rm ref} \bbYdiag$. 
 Second, the raising matrix for the diagonal solution is now lower triangular,  namely
\begin{align}
\tilde{K}^{\rm diag}(\be) &= \matrixii{1}{0}{\be}{1} \period
\end{align}
Thirdly, the polarization spinor for $Z$ is $\ntildiag = (1,0)^t$,  as discussed in (\ref{nzbar}). 
Lastly, because of the form of the ALP for the left sector, the Baker-Akhiezer vector becomes coordinate-independent at  $x=0$ instead of at $x=\infty$. 

Let us now list the basic results for the left sector, omitting the intermediate details. Just as for the right sector, the formulas below are valid for any type of 
 operator. 
\begin{align}
&\psidiag_+= \vecii{0}{1} \comma\qquad \psidiag_- = \vecii{1}{0} \comma  \\
&\psitil_+(0) = a^{-1}\psitilref_+(0) + b\psitilref_-(0)\comma  \qquad 
\psitil_-(0) = a  \psitilref_-(0) \comma  \\
&\ntil= a i\sig_2 \psitil_+(0)  \comma \qquad 
\iDelta \phi_L = -i \ln a^{-2 } \period \label{Zleft}
\end{align}
Using these formulas, we obtain the contribution to the wave function from  the left sector  as 
\begin{align}
&\Psi_L^{S^3}=\exp \left(-i\sum_{i=1}^3  L_i   \iDelta \phi_{L, i}
\right) =\prod_{\{i,j,k\}} \left({\wrons{\ntil_i}{\ntil_j} \over \wrons{i_+}{j_+}\big|_0}
\right)^{L_i+L_j -L_k}  \label{PsiL} \comma
\end{align}
where we used the gauge invariance of the Wronskians and replaced $\sl{\tilde{i}_{+},\tilde{j}_+}$ with $\sl{i_+,j_+}$.
Together with $\Psi_R^{S^3}$ obtained  in (\ref{PsiR}) we now have 
 the complete wave function for the $S^3$ part. It is of the structure
\begin{align}
&e^{\calF_{\rm vertex}} = \Psi^{S^3}_L\Psi^{S^3}_R e^{\calV_{\rm energy}} 
  \comma \nn\\
&\calF_{\rm vertex} = \calV_{\rm kin} +\calV_{\rm dyn} + \calV_{\rm energy} \period \label{strucFvertex}
\end{align}
Let us explain each term \eqref{strucFvertex} in order.
The first term $\calV_{\rm kin}$ stands for 
 the kinematical part composed of the Wronskians $\wrons{n_i}{n_j}$ and 
 $\wrons{\ntil_i}{\ntil_j} $, 
\beq{
&\mathcal{V}_{\rm kin}=\nn\\
&(R_1+R_2-R_3)\ln \sl{n_1,n_2}+(R_2+R_3-R_1)\ln \sl{n_2,n_3}+(R_3+R_1-R_2)\ln \sl{n_3,n_1}\nn\\
&+(L_1+L_2-L_3)\ln \sl{\tilde{n}_1,\tilde{n}_2}+(L_2+L_3-L_1)\ln \sl{\tilde{n}_2,\tilde{n}_3}+(L_3+L_1-L_2)\ln \sl{\tilde{n}_3,\tilde{n}_1}\period\label{Vkin}
}
The second term $\calV_{\rm dyn}$ refers to 
 the dynamical part consisting of the Wronskians 
$\wrons{i_-}{j_-} \big|_\infty$ and  $\wrons{\itil_+}{\jtil_+} \big|_0$,
\beq{
&\mathcal{V}_{\rm dyn}=\nn\\
&-(R_1+R_2-R_3)\ln \sl{1_-,2_-}\big|_{\infty}-(R_2+R_3-R_1)\ln \sl{2_-,3_-}\big|_{\infty}-(R_3+R_1-R_2)\ln \sl{3_-,1_-}\big|_{\infty}\nn\\
&-(L_1+L_2-L_3)\ln \sl{1_+,2_+}\big|_{0} -(L_2+L_3-L_1)\ln \sl{2_+,3_+}\big|_{0}-(L_3+L_1-L_2)\ln \sl{3_+,1_+}\big|_{0}\period\label{Vdyn}
} 
The last term $\mathcal{V}_{\rm energy}$ denotes the contribution involving the worldsheet energy shown in the last term of \eqref{involvingenergy}. Such a term is necessary for the following reason. As explained below \eqref{localcoordinate} and at the beginning of section \ref{subsec:contour},  we evaluate our wave function on the 
 circle defined by $\tau^{(i)} =0$, corresponding to $\ln |z-z_i| = \ln \ep_i$. On the other hand, the wave function introduced through the  state operator mapping in \eqref{stateoperatormap} is defined on the unit circle described by $\ln |z-z_i| =0$. 
The term $\calV_{\rm energy}$ is needed to fill this gap. 
 As the energy of the each external state is given\fn{The energy can be computed from the behavior of the stress-energy tensor around the puncture \eqref{stressbehavior}.} by
$2\sqrt{\lambda}\kappa_i^2$, $\mathcal{V}_{\rm energy}$ can be evaluated explicitly as
\beq{
\mathcal{V}_{\rm energy}=2\sqrt{\lambda}\sum_{i=1}^{3} \kappa_i^2 \ln \epsilon_i\period\label{Venergy}
}

Before ending this subsection, let us make two comments. First, it is not guaranteed at this stage that the wave function thus constructed produces a correctly normalized two-point function. In addition, as discussed in \cite{KK2}, there may be additional contributions which come from the canonical change of variables, $\{ \mathbb{Y}, \del_\tau \mathbb{Y}\} \to \{\phi_i, S_i\}$.  However, in section \ref{subsec:2pt}, it will be checked that our result for the three-point function reproduces the normalized two-point function in an appropriate limit. Therefore we can a posteriori confirm that the wave function is 
properly  normalized and the additional contributions are absent. 
 Second, one recognizes 
 that the power of $\wrons{n_i}{n_j}$, namely  $R_i+R_j -R_k$, is 
 the familiar combination, made out of conformal weights and spins, for the coordinate differences in the three-point  functions of
 a conformal field theory, {\it except for the overall sign}. 
In the next subsection, we will elaborate on this structure of the 
 power from the point of view of the dual gauge theory. 
Also in section \ref{subsec:AdS},  where we construct  the wave function 
for the $E\! AdS_3$ part, the above difference in the overall sign will be explained.  

Summarizing,  the product of (\ref{PsiR}) and (\ref{PsiL}) gives the general form of the wave functions  for the three-point function. 
It is expressed in terms of the two types of Wronskians. 
One type is the Wronskians between the solutions of the ALP around 
 vertex insertion points. They will be evaluated in section \ref{sec:wron}. The others
 are the Wronskians between the polarization spinors associated with the 
 vertex operators, which are of purely kinematical nature and hence should be common  to the string and the gauge theory sides. 
\subsubsection{Correspondence with the gauge theory side \label{subsec:relation}}
We shall now examine our formula for the wave function from 
 the point of view of correspondence with the gauge theory side. 

First consider the question of how to distinguish the different types of gauge theory  operators  $\calO_i$ from their corresponding wave functions  in string theory.  The wave function constructed above is expressed in terms of the polarization spinors, which depend 
 only on the type of the vacuum on which the corresponding gauge
 theory operator is built, the eigenvectors of the ALP in the vicinity of the 
 insertion point $z_i$, and the charges carried by the vertex operators. 
A natural question is how we can distinguish the type of vertex operators involved  from these data.
Operators of $\calO_1$ and $\calO_2$ in section \ref{subsec:highest} can 
be distinguished by the structure of their polarization spinors because the vacuum on which they are built are different. On the other hand, operators of $\calO_1$  and $\calO_3$, which are built on the same type of the vacuum,  are characterized by the same polarization spinors 
and hence it appears that one cannot  distinguish them from the formula 
 for the wave function. Since these operators differ only in the types of 
 excitations, $X$ or $\Xbar$, the question is how this is reflected. The answer is in the relation between the absolute magnitude of the charges $\calR$ and $\calL$, which are given by $R$ and $L$ respectively. Because the charges 
 carried by the operator $X$ are  
$(\calR,\calL) =(-1/2, 1/2)$, the magnitude of the total charges 
 of the type $\calO_1$  operator built upon $Z$-vacuum with $X$ as excitations must satisfy the inequality $R <L$. Similarly, the magnitudes of the total charges for the operator of type  $\calO_2$ also obey  $R <L$. 
On the other hand, for the operator of type $\calO_3$, we have $R >L$. 

 Such distinction is reflected 
 not only on the charges but also on the dynamical property of the 
eigenstates  $i_\pm$ appearing in the wave function formula. As discussed in \eqref{ug1} and \eqref{ul1}, the relative magnitude of $R$ and $L$ for a one-cut solution 
is determined by the  position of the cut in the quasi-momentum $p(x)$: When the real part of  the position of the branch cut is in the interval  $[-1, 1]$ in the spectral parameter space such a solution has $R>L$ and hence corresponds to the operator of type $\calO_3$. Contrarily the operator of type $\calO_1$ 
having $R<L$ corresponds  to a solution with the cut outside the above 
 interval. 
Conceptually this is quite intriguing.  From the spin-chain perspective, 
 $\calO_1$ and $\calO_3$ form distinct types of spin chains, 
 which cannot be transformed into each other by an $\SUR \times \SUL$ transformation. On the other hand, in string theory the solutions corresponding to 
 these distinct  spin chains are described in a more unified way. 
It would be interesting to realize such a unified treatment on the  gauge theory side as well. 

Let us next examine the role and the meaning of the kinematical factor $\mathcal{V}_{\rm kin}$ from the point of view  of the dual gauge theory.  In this regard, 
 note that the quantity $\wrons{n_i}{n_j}$, being a skew product, vanishes when $n_i$ and $n_j $ coincide. This in fact happens for the case of 
 the operators  $\calO_1$ and $\calO_3$ discussed in section \ref{subsec:highest}, which are built upon the same $Z$-vacuum and hence carry the same polarization spinors. There are three possibilities. 
If the power $R_i+R_j -R_k$ is positive, then the wave function and hence the three-point function vanishes. 
This would express a selection rule. On the other hand,  if it is negative, the three-point function diverges. For an internal symmetry such as $\SUR$ this should not occur. The last   possibility is that the power is exactly zero. In this case,   we should  regularize $\wrons{n_i}{n_j}$ slightly away from zero and then
 apply the vanishing power to get the result,  which is unity. 

Let us see which of these cases  is actually realized  for the set of operators $\calO_1, \calO_2$ and $\calO_3$ in section \ref{subsec:highest}. Let $l_i$ be the total length of the operator $\calO_i$  and 
$M_i$ be the number of excitations. The number of ``vacuum fields" is then 
given by   $l_i -M_i$.  There are two obvious conservation laws for these numbers  if all the fields and anti-fields of the set $\{\calO_i\}$ are fully contracted to form propagators. One is the conservation concerning excitations, \ie, the total number of  $X$'s  should  equal the total number of $\Xbar$'s. The other is the 
conservation concerning the vacuum fields, \ie the total number of $Z$'s should equal the total number of $\Zbar$'s. From the structure of $\calO_i$'s it is easy to find  that these two conservation laws 
are expressed as 
\begin{align}
\begin{aligned}
&(i)\quad  M_1 = M_2 + M_3 \comma \\
&(ii)\quad l_1+l_3 -l_2 = 2M_3 \period
\end{aligned}\label{cons-charge}
\end{align}
Now consider the right and the left charges carried by $\calO_i$. From Table \hyperlink{table1}{1}, and the compositions of $\calO_i$, we get, for example, $R_1=\half l_1-M_1, L_1=\half l_1$, etc.. Then, computing the powers  of interest we get
\begin{align}
R_1+R_3 -R_2 &= M_3+M_2 -M_1 =0 \comma \\
L_1 +L_3 -L_2 &= \half (l_1+l_3-l_2) -M_3 =0 \period
\end{align}
Therefore precisely due to the conservation laws, $(i)$ and $(ii)$ above,  of the number of contracting  fields, the  powers that occur for the vanishing Wronskians  $\wrons{n_1}{n_3}$ and  $\wrons{\ntil_1}{\ntil_3}$ are zero. Hence,  in the computation 
of the three-point function of $\calO_i$'s such factors simply produce  unity. 

Up to this point we have obtained the general formulas for the contribution
 of the action part and the wave function part, both of which are expressed in terms of the Wronskians of the form $\wrons{i_\pm}{j_\pm} $. In the next section  we will evaluate these quantities to substantiate the general formulas. 
%%%%%%%%%%%%%%
%%%%%%%%%%%%%%%%%%%%%%%%%%%%%%%%%%
\section{\bfall{Evaluation of the Wronskians}\label{sec:wron}}
%%%%%%%%%%%%%%%%%%%%%%%%%%%%%%%%%%
In the previous two sections, we have shown that both the contribution of the action and that  of the vertex operators are expressible in terms of the Wronskians $\wrons{i_\pm}{j_\pm}$ between the eigenvectors of the monodromy matrices. 
The goal of this section is to evaluate those Wronskians. First, in section \ref{subsec:gen}, we show that certain products of Wronskians are expressed in terms of the quasi-momenta. Next, in sections \ref{subsec:pole} and \ref{subsec:zero}, we determine the analytic properties (\ie poles and zeros) of each Wronskian as a function of the spectral parameter $x$. With such a knowledge, 
we apply,  in section \ref{subsec:WH},  a generalized version of  the Wiener-Hopf decomposition formula to the products of the Wronskians and determine the individual factor. Finally, in section \ref{subsec:sing}, we compute the singular part and the constant part of the Wronskian, which cannot be determined by the Wiener-Hopf method.   
%%%%%%%%%%%%%%%%%%%%%%%%%%%%%%%%%%%%
\subsection{Products of Wronskians in terms of quasi-momenta\label{subsec:gen}}
%%%%%%%%%%%%%%%%%%%%%%%%%%%%%%%%%
To obtain the information of the Wronskian $\wrons{i_\pm}{j_\pm}$ between the eigenvectors  of the ALP at different points, we need some condition 
which governs the global property of such Wronskians.  As we shall see, 
 such a condition is provided by the global consistency condition for the 
 product of  the local monodromy matrices $\Omega_i$ associated with the 
 vertex insertion points $z_i$.  Since the total monodromy must be 
trivial  upon going around the entire worldsheet, we must have 
\beq{
\Omega_1 \Omega_2 \Omega_3 =\id\period\label{eq-5-1}
}
Although this  appears to be a rather weak condition, it is sufficiently  powerful to determine  the forms of  certain products of the Wronskians 
 in terms of the quasi-momenta $p_i(x)$, as discussed  in \cite{JW, KK1}. 
Let us quickly reproduce those expressions.  Take the  basis in which $\Omega_1$ is diagonal, namely 
\beq{
\Omega_1=\pmatrix{cc}{e^{ip_1}&0\\0&e^{-ip_1}}\period\label{eq-5-2}
}
Since the set of eigenvectors $j_\pm$ at $z_j$ form a complete basis, 
 one can expand the eigenvectors $i_{\pm}$ at $z_i$ in terms of 
 them in the following way:
\beq{
i_{\pm} = \sl{i_{\pm} \comma j_{-}}j_{+}-\sl{i_{\pm} \comma j_{+}}j_{-}\period
} 
Making use of  this formula, $\Omega_2$ can be expressed in the $\Omega_1$-diagonal basis as 
\beq{
\Omega_2 = M_{12}\pmatrix{cc}{e^{ip_2}&0\\0&e^{-ip_2}}M_{21}\comma\label{eq-5-3}
}
where the matrix $M_{ij}$, effecting the change of basis, is given by 
\beq{
M_{i j}=\pmatrix{cc}{-\sl{i_-\comma j_+}&-\sl{i_-\comma j_-}\\ \sl{i_+\comma j_+}&\sl{i_+\comma j_-}}\period
}
Now owing to the constraint \eqref{eq-5-1},  $\Omega_1$ and $\Omega_2$ must satisfy the following relation: 
\beq{
\tr \left( \Omega_1 \Omega_2\right)=\tr \Omega_3^{-1}=2\cos p_3\period\label{eq-5-4}
}
Substituting the equations \eqref{eq-5-2} and \eqref{eq-5-3} into \eqref{eq-5-4}, we obtain an  equation for $\sl{1_{\pm}\comma 2_{\pm}}$ of the form
\beq{
&\cos \left(p_1-p_2\right)\sl{1_{+}\comma 2_{+}}\sl{1_{-}\comma 2_{-}}-\cos \left(p_1+p_2\right)\sl{1_{+}\comma 2_{-}}\sl{1_{-}\comma 2_{+}}=\cos p_3\period
}
This equation,  together with the Schouten identity\fn{The general form of the Schouten identity is given by $\sl{i\comma j}\sl{k\comma l}+\sl{i\comma k}\sl{j\comma l}+\sl{i\comma l}\sl{j\comma k}=0$. It can be proven directly from the definition of the Wronskians.} for  $1_{\pm}$ and $2_{\pm}$
given by 
\beq{
\sl{1_{+}\comma 2_{+}}\sl{1_{-}\comma 2_{-}}-\sl{1_{+}\comma 2_{-}}\sl{1_{-}\comma 2_{+}}=\sl{1_{+}\comma 1_{-}}\sl{2_{+}\comma 2_{-}}=1\comma
}
completely determines the products of Wronskians, $\sl{1_{+}\comma 2_{+}}\sl{1_{-}\comma 2_{-}}$ and $\sl{1_{+}\comma 2_{-}}\sl{1_{-}\comma 2_{+}}$. 
In a similar manner,  products of certain other Wronskians can also be obtained, which are summarized as the following set of equations\fn{Note that the equations \eqref{eq-5-5}--\eqref{eq-5-10} appear slightly different in form from those derived in \cite{KK1}. This is because $2_{+}$ in this paper corresponds to $2_{-}$ in \cite{KK1} and $2_{-}$ in this paper corresponds to $-2_{+}$ in \cite{KK1}.}:
\beq{
&\sl{1_+\comma 2_+}\sl{1_-\comma 2_-} = \frac{\sin \frac{p_1+p_2+p_3}{2}\sin \frac{p_1+p_2-p_3}{2}}{\sin p_1 \sin p_2}\comma \label{eq-5-5}\\
&\sl{2_+\comma 3_+}\sl{2_-\comma 3_-} = \frac{\sin \frac{p_1+p_2+p_3}{2}\sin \frac{-p_1+p_2+p_3}{2}}{\sin p_2 \sin p_3}\comma\label{eq-5-6}\\
&\sl{3_+\comma 1_+}\sl{3_-\comma 1_-} = \frac{\sin \frac{p_1+p_2+p_3}{2}\sin \frac{p_1-p_2+p_3}{2}}{\sin p_3 \sin p_1}\comma\label{eq-5-7}\\
&\sl{1_+\comma 2_-}\sl{1_-\comma 2_+} = \frac{\sin \frac{p_1-p_2+p_3}{2}\sin \frac{p_1-p_2-p_3}{2}}{\sin p_1 \sin p_2}\comma \label{eq-5-8}\\
&\sl{2_+\comma 3_-}\sl{2_-\comma 3_+} = \frac{\sin \frac{p_1+p_2-p_3}{2}\sin \frac{-p_1+p_2-p_3}{2}}{\sin p_2 \sin p_3}\comma\label{eq-5-9}\\
&\sl{3_+\comma 1_-}\sl{3_-\comma 1_+} = \frac{\sin \frac{-p_1+p_2+p_3}{2}\sin \frac{-p_1-p_2+p_3}{2}}{\sin p_3 \sin p_1}\period\label{eq-5-10}
}

What we need  for the computation of the three-point functions, however, 
are the individual Wronskians and not just the products given in \eqref{eq-5-5}--\eqref{eq-5-10}.  Such a knowledge will be extracted based on 
  the analytic properties of the Wronskians regarded as 
functions of the complex spectral parameter $x$. We will analyze such 
 properties in the next two subsections. 
%%%%%%%%%%%%%%%%%%%%%%%%%
\subsection{Analytic properties of the Wronskians I: Poles\label{subsec:pole}}%%%%%%%%%%%%%%%%%%%%%%%%%
An individual Wronskian, viewed as a function of $x$, is almost uniquely determined\fn{As we will discuss later, the Wronskian also contains essential singularities at $x=\pm 1$. In addition, an overall proportionality constant cannot be determined by the positions of zeros and poles. These ambiguities will be fixed in section \ref{subsec:sing}.} by its analytic properties, namely the positions of the poles and the zeros.  From the expressions exhibited 
 in \eqref{eq-5-5}--\eqref{eq-5-10}, we know that the products of 
 Wronskians have  poles at $\sin p_i=0$ and zeros at $\sin\left((\pm p_1\pm p_2 \pm p_3)/2 \right)=0$.  Therefore the question is which factor 
of the product is responsible for such  a pole and/or  a zero. 
In this subsection, we will describe how to analyze the structure of the poles. 

To illustrate the basic idea, we will consider the  Wronskians  $\sl{1_{+}\comma 2_{+}}$ and $\sl{1_{-}\comma 2_{-}}$ as examples, for which the 
product  is given by 
\beq{
&\sl{1_+\comma 2_+}\sl{1_-\comma 2_-} = \frac{\sin \frac{p_1+p_2+p_3}{2}\sin \frac{p_1+p_2-p_3}{2}}{\sin p_1 \sin p_2}\period\nn
}
Let us focus on the  pole associated with $\sin p_1 =0$ and denote the position of the pole by $x_{\rm pole}$. There are two types of points at which $\sin p_1$ vanishes,  the branch points and the ``singular points". 
First consider the case where $x_{\rm pole}$ is a singular point, at which the 
two eigenvalues of the monodromy matrix $\Omega_1$ degenerate to 
either $+1$ or $-1$.   This, however, does not mean that $\Omega_1$ is proportional to the unit matrix  for the following reason:  If $\Omega_1 \propto \id$, the monodromy condition $\Omega_1 \Omega_2 \Omega_3=\id$ forces $p_2$ to be equal to $+p_3$ or $-p_3$ modulo $\pi$. However, since 
$p_1$, $p_2$ and $p_3$ can be chosen completely independently, 
 there is no reason for such special relation to hold.  Thus, the only remaining possibility is that the monodromy matrix $\Omega_1$ takes the form of a Jordan-block at $x=x_{\rm pole}$, namely, 
\beq{
\Omega_1 (x_{\rm pole} )\sim \pm \pmatrix{cc}{1&c\\0&1}\period\label{jord}
}
In this case, the eigenvectors  $1_{+}$ and $1_{-}$ degenerate at $x=x_{\rm pole}$ and we have  one eigenvector. 
To see what happens at $x=x_{\rm pole}$ more explicitly, let us study the asymptotic behavior of $1_{\pm}$ near $z_1$. In the vicinity of each puncture, 
 the saddle point solution for the three-point function can be well-approximated  by an appropriate solution for a two-point function.  Consequently, 
the eigenvectors for the three-point function $1_{\pm}$ can also be approximated near $z_1$ by the eigenvectors for the two-point function  $1_{\pm}^{\rm 2pt}$. 
 As shown in \eqref{normcond3}, this structure can be seen most transparently in the Pohlmeyer gauge.  Working out the subleading corrections, we obtain the following expansion for the eigenfunctions $\hat{1}_{\pm}$:
\beq{
&\hat{1}_{+}=\hat{1}_{+}^{\rm 2pt} \left( 1+ c_1(\sigma^{(1)}, x) e^{a_1 \tau^{(1)} }+c_2(\sigma^{(1)}, x) e^{a_2 \tau^{(1)} } +\cdots \right)\comma\label{exp1+}\\
&\hat{1}_{-}=\hat{1}_{-}^{\rm 2pt} \left( 1+ \tilde{c}_1(\sigma^{(1)}, x) e^{\tilde{a}_1 \tau^{(1)} }+\tilde{c}_2(\sigma^{(1)}, x) e^{\tilde{a}_2 \tau^{(1)} } +\cdots\right)\period\label{exp1-}
}
Here  $\tau^{(1)}$ and $\sigma^{(1)} $ are  the local coordinates near $z_1$ given in \eqref{localcoordinate} and $c_k$ and $\tilde{c}_k$ are $2\times 2$ matrices dependent only on $\sigma^{(1)}$ and $x$. The constants $a_k$ in the exponents are such that successive terms are becoming 
smaller by exponential factors as $\tau \rightarrow -\infty$. An important 
observation is that since $\hat{1}_{\pm}^{\rm 2pt}$ are eigenfunctions 
corresponding to a two-point function, they are insensitive to the global 
 monodromy constraint  (\ref{eq-5-1}) on the three-point function and hence
non-degenerate at $x=x_{\rm pole}$. 
An apparent puzzle now is how  exponentially small  corrections can produce  the degeneracy of $\hat{1}_\pm$. 

The answer is the following.  
 Since one of the solutions $\hat{1}_\pm^{\rm 2pt}$ is exponentially increasing (\ie big) and the other is decreasing (\ie small) as $\tau \rightarrow   -\infty$, let us consider the case where $\hat{1}_+^{\rm 2pt}$ is big and 
$\hat{1}_-^{\rm 2pt}$ is small. Now  for $\hat{1}_\pm$ to become 
 degenerate at $x=x_{\rm pole}$, logically there are three  possibilities 
\begin{align}
&(a)\qquad \hat{1}_+ = \al \hat{1}_- \comma \qquad \al=\mbox{finite}\comma \\
& (b) \qquad \hat{1}_+ = \be \hat{1}_- \comma 
\qquad \be \rightarrow \infty \comma \\
& (c)\qquad  \hat{1}_- = \be \hat{1}_+ \comma \qquad \be \rightarrow \infty \period 
\end{align}
First, since $\hat{1}_+^{\rm 2pt}$ is much larger than  $ \hat{1}_-^{\rm 2pt}$ by assumption,  the case $(a)$ cannot  occur.   Now consider the case where $x$ is slightly different  from 
 $x_{\rm pole}$. Then 
$\be$ is large but  finite and the relations $(b)$ or $(c)$ must 
be realized approximately.   But it is obvious  that  $(b)$ is the  only consistent relation since exponentially small solution can appear in the big solution 
but not the other way around.  Therefore we must have the situation 
\begin{align}
\hat{1}_+ &= \hat{1}_+^{\rm 2pt}  + \cdots + \be \hat{1}_- + \cdots \comma \label{smallinbig}
\end{align}
 As $x \rightarrow x_{\rm pole}$, $\be$ diverges 
 and (\ref{smallinbig}) goes over to the relation $(b)$.  The situation is the same  if $\hat{1}_-$ is the big solution: Always the big solution diverges at the degeneration point,  while the small solution remains finite\footnote{
Remark:\ This does not mean of course that there is only one solution at the degeneration point. There must exist another  independent 
 solution of new structure, namely the structure which is different 
 from $\hat{1}_\pm$. However, as long as we stick to this basis, what we see is  that one of the solutions diverges and disappears. }. 

Similar argument can be applied to the other Wronskians, making use 
of the general asymptotic behavior of the eigenvectors in the Pohlmeyer gauge, 
which is of the form
\beq{
\hat{i}_{\pm} \sim e^{\pm q(x)\tau^{(i)}} \quad (z\sim z_i)\period
}
It is clear from this expression that which one of the $\hat{i}_{\pm}$
 diverges as $z\rightarrow z_i$ is governed by the sign of the real 
 part of the quasi-energy $q(x)$. Since the divergence of the eigenfunction 
 produces a pole on the Wronskian containing it, we can determine which 
 Wronskian of the product is responsible for the pole with the following 
 general rule:  At $\sin p_i =0$, the Wronskians  behave as 
\beq{
&{\rm Re}\,q(x) >0 \Rightarrow \sl{i_{+}\comma \ast}= \text{finite}\comma \quad \sl{i_{-}\comma \ast}=\infty\comma\label{result:pole}\\
&{\rm Re}\,q(x) <0 \Rightarrow\sl{i_{+}\comma \ast}=\infty\comma \quad \sl{i_{-}\comma \ast}= \text{finite} \period\label{result:pole2}
}
Hence, for  $\Re q(x) >0$ the pole occurs in $\sl{i_{-}\comma \ast}$, while for  $\Re q(x) <0$ it occurs in $\sl{i_{+}\comma \ast}$. 
%%%%%%%%%%%%%%%%%%%%%%%%
\subsection{Analytic properties of the Wronskians II: Zeros\label{subsec:zero}}
%%%%%%%%%%%%%%%%%%%%%%%%
Having determined the pole structure, let us next discuss the zeros of the Wronskians. 
The determination of the zeros is substantially more difficult since, 
 in contrast to the poles which are local phenomena, the zeros are determined by the global properties on the Riemann surface. 
  As shown in previous works \cite{AMSV,AM,AGM}, the notion 
 of the WKB curve \cite{GMN} is one of the main tools to explore such global properties. However, as its name indicates, the WKB curve is useful only when the 
 leading term in the WKB expansion is sufficiently accurate. 
 For this reason, it is not  powerful enough  to fully determine the zeros of the Wronskians in the whole region of the spectral parameter space. In this subsection we shall introduce an appropriate generalization of the WKB curve, to be called the {\it exact WKB curve}, to overcome this difficulty.  
%%%%%%%%%%%%%%%%%%%%%%%%
\subsubsection{WKB approximation and WKB curves}
%%%%%%%%%%%%%%%%%%%%%%%%%%
In order to motivate the generalized version, we shall first  briefly review the ordinary WKB curves defined in \cite{GMN}.  

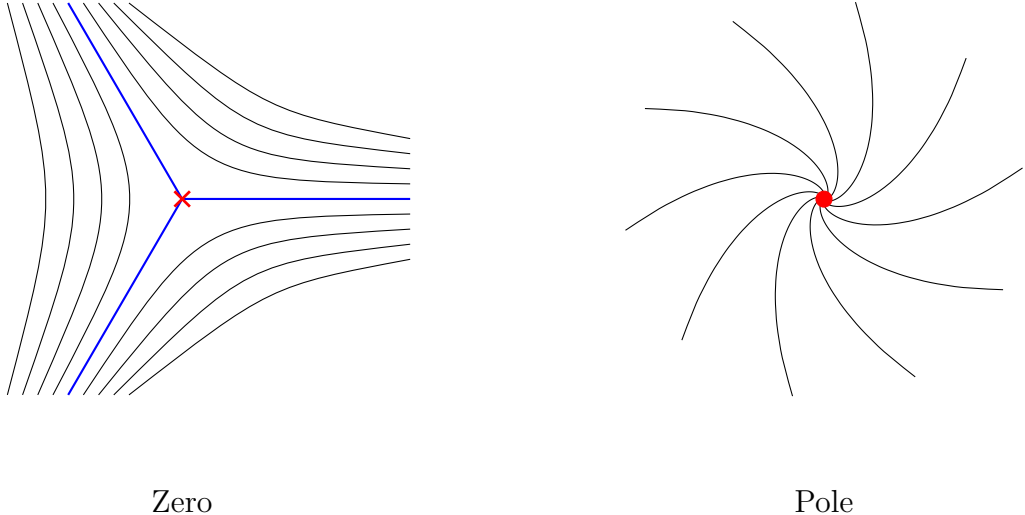
\begin{figure}
\begin{minipage}{0.5\hsize}
\centering
\begin{tikzpicture}
\draw[blue,thick] (0,0) -- (3,0)
(0,0) -- (-3/2,{3/2*sqrt(3)})
(0,0) -- (-3/2,{-3/2*sqrt(3)});
\foreach \x in {2,4,...,8}
{\draw ({-3/2+0.1*\x},{3/2*sqrt(3)}) .. controls ({0.15*\x -0.05},{0.15*\x -0.05}) .. (3,{0.1*\x})
({-3/2+0.1*\x},{-3/2*sqrt(3)}) .. controls ({0.15*\x -0.05},{-0.15*\x +0.05}) .. (3,{-0.1*\x})
({-3/2-0.1*\x},{3/2*sqrt(3)}) .. controls ({-(0.15*\x -0.05)*sqrt(2)},0) .. ({-3/2-0.1*\x},{-3/2*sqrt(3)});
}
\node at (0,0) {\Cross};
\node at (0,-4) {Zero};
\end{tikzpicture}
\end{minipage}
\begin{minipage}{0.5\hsize}
\centering
\begin{tikzpicture}
\foreach \x in {1,...,10}{
\draw[scale=0.55,domain={-pi}:{pi/2},smooth,variable=\t]
plot ({exp(\t)*cos((0.5*\t + \x*pi/5) r)},{exp(\t)*sin((0.5*\t + \x*pi/5) r)});
}
\filldraw[red] (0,0) circle (3pt);
\node at (0,-4) {Pole};
\end{tikzpicture}
\end{minipage}
\caption{Sketch of  WKB curves near a zero ( a red cross in the left figure) and a pole (a red circle in the right figure). There are exactly three WKB curves that emanate from a zero. In contrast,  there are infinitely many WKB curves radiating from a pole in all directions.}
\label{fig-WKB}
\end{figure}

When the expansion parameter $\zeta$ is sufficiently small, the leading term
 of the WKB expansion for the solutions to ALP \eqref{diag-gauge} around $z_i$ is given by 
\beq{
\psihat \sim \exp \left(\pm \frac{1}{\zeta}\int_{z_i^{\ast}}^{z} \sqrt{T}dz \right)\period\label{leadwkb}
}
Of the two independent solutions given above, one is the {\it small solution},  which decreases exponentially as it approaches $z_i$ and the other is the  {\it big solution}, which increases exponentially in the same limit.  
In order to make the variation of the magnitude of the solution more precise, 
one defines the WKB curves as the curves along which the phase of the leading term \eqref{leadwkb} in the WKB expansion is constant. More explicitly, they are characterized  by the equation
\beq{
\Im \left( \frac{\sqrt{T}}{\zeta}dz\right) =0\period\label{def-wkb}
}
By analyzing the structure of \eqref{def-wkb}, one finds the following three  characteristic properties of the WKB curves.  
(i)\, At generic points on the worldsheet, the WKB curves are non-intersecting. 
(ii)\,  At a puncture, the WKB curves radiate in all directions  from the puncture. (iii)\, At a zero of $T(z)$, there are three special WKB curves which radiate from the zero and separate three different regions of the WKB curves. For  details, see \figref{fig-WKB}. 

Along the WKB curve, the magnitude of the leading term in the WKB expansion \eqref{leadwkb} increases or decreases monotonically,  until they reach a zero or a pole of $T(z)$. Thus, if two punctures $z_i$ and $z_j$ are connected by a WKB curve and the spectral parameter $\zeta$ is sufficiently small, the small solution $s_i$ defined around $z_i$ will grow exponentially as it approaches the other puncture $z_j$. In other words, 
 the small solution $s_i$ behaves like the big solution around $z_j$. 
Therefore $s_i$ will be linearly independent of $s_j$  and hence the Wronskian between these two small solutions $\sl{s_i\comma s_j}$ must be non-vanishing.  

With this logic, we conclude that the Wronskians $\sl{i_{\pm}\comma j_{\pm}}$ are non-vanishing  if the following three conditions are satisfied: 
(a)\, Two punctures $z_i$ and $z_j$ are connected by a WKB curve. 
(b)\,  Two eigenvectors $i_{\pm}$ and $j_{\pm}$ are both small solutions.
(c)\,  The leading WKB solutions \eqref{leadwkb} are sufficiently accurate. 
%%%%%%%%%%
%%%%%%%%%%%%%%%%%%%
\subsubsection{Exact WKB curves}
%%%%%%%%%%%%%%%%%%%%
Evidently, the analysis above is valid only in a restricted region of  the spectral parameter plane where the approximation by the leading term of  the WKB expansion is reliable. 
Actually, even if we improve the approximation by going to the next order 
approximation, we still cannot cover the entire spectral parameter plane 
 because such an expansion is only an asymptotic series. It is indeed possible 
 that as we change $x$ the small and the big solutions interchange 
their roles. Such a phenomenon is clearly  non-perturbative  and cannot be captured by the usual expansion. 
So to understand the structure of  the zeros on the whole spectral parameter plane, it is necessary to generalize  the notion of WKB curves in a 
non-perturbative fashion. 

In order to seek such an improvement, we need  to look closely 
 at the general structure of the conventional  WKB expansion. 
Let us denote the components of the solution $\hat{\psi}^d$ to the ALP in the diagonal gauge \eqref{diag-gauge} as
\beq{
&\hat{\psi}^d=\pmatrix{c}{\psi^{(1)}\\\psi^{(2)}}\period\label{sec-6-comp}
}
By  substituting \eqref{sec-6-comp} into the ALP \eqref{diag-gauge}, we obtain the equations for the components $\psi^{(1)}$ and $\psi^{(2)}$. 
Then, upon eliminating $\psi^{(2)}$ in favor of $\psi^{(1)}$, we get a second-order differential equation for $\psi^{(1)}$. To solve this equation, 
we expand  $\psi^{(1)}$ in powers of $\zeta$ in the form 
\beq{
\psi^{(1)}=\sqrt{\frac{\rho}{T}-\frac{\del \gamma}{\sqrt{T}}}\,\exp\left[\int_{z_0}^{z}\left(\frac{W_{-1}}{\zeta}+W_0 + \zeta W_1 +\cdots\right)\right]\period\label{exppsi-6}
}
One can then determine the one-forms $W_n$ order by order recursively. 
This procedure is described in Appendix \ref{apsub:direct}. 
As a result of such a computation, we find that the WKB expansions for two linearly independent solutions to the ALP can be expressed in the following form:
\beq{
 \pmatrix{c}{f_{\pm }^{(1)}\\f_{\pm }^{(2)}}\exp\left(\pm\int_{z_0}^{z}W_{\swkb}(z, \barz ;\zeta )\right)\period\label{WKBP}
}
Here  $W_{\swkb}\equiv W_{\swkb}^z dz +W_{\swkb}^{\barz} d\barz$ is the one-form defined as a power series in  $\zeta$, with the leading term given by $\sqrt{T}dz/(2\zeta)$. On the other hand, the functions $f_{\pm }^{(1)}$ and $f_{\pm }^{(2)}$ are defined in terms of $W_{\swkb}^z$ by
\beq{
&f_{\pm }^{(1)}\equiv k_{\swkb}=\sqrt{\frac{\rho-\sqrt{T}\del \gamma}{T\,W^z_{\swkb}}}\comma \\
&f_{\pm}^{(2)}\equiv \frac{-i}{\sqrt{W_{\swkb}^z}}\left[ \pm W^z_{\swkb}  + \left(\frac{\sqrt{T}}{2\zeta}-\frac{\rho\cos 2\gamma}{\sqrt{T}} +\frac{\del \ln k_{\swkb}}{2} \right)\right]\period
}  

With this structure in mind, we now introduce an improved notion of the WKB curve, to be called  the ``exact WKB curve", 
 by writing the exact  solutions to the ALP in the form 
\beq{
\hat{\psi}^d = \pmatrix{c}{f_{\rm ex}^{(1)}\\f_{\rm ex}^{(2)}}\exp\left(\int_{z_0}^{z}W_{\rm ex}(z, \barz ;\zeta )\right) \comma \label{exactP}
}
where $f_{\rm ex}^{(1)}$ and $f_{\rm ex}^{(2)}$ are given by
\beq{
&f_{\rm ex}^{(1)} = \sqrt{\frac{\rho-\sqrt{T}\del \gamma}{T\,W^z_{\rm ex}}} \comma\quad f_{\rm ex}^{(2)} = \frac{-i}{\sqrt{W_{\rm ex}^z}}\left[  W^z_{\rm ex}  + \left(\frac{\sqrt{T}}{2\zeta}-\frac{\rho\cos 2\gamma}{\sqrt{T}} +\frac{\del \ln f_{\rm ex}^{(1)}}{2} \right)\right]\period
}
Note that the expression \eqref{exactP} is identical in form to \eqref{WKBP} with the plus sign chosen. However,  there is an essential  difference. 
 While $W_{\swkb}$ is given by the  asymptotic series in powers of  $\zeta$ and is hence ambiguous non-perturbatively,   $W_{\rm ex}$ on the other hand  is unambiguous as it is defined directly by the exact solution  $\hat{\psi}$. 
Of course, if we expand $W_{\rm ex}$ perturbatively in powers of  $\zeta$, 
the series  will coincide with $W_{\rm WKB}$.  In this sense, $W_{\rm ex}$ can be regarded as the non-perturbative completion of $W_{\swkb}$. Now 
one of the virtues of the expression \eqref{exactP} is that we can easily construct another solution satisfying $\sl{\hat{\psi}^d \comma \hat{\psi}'{}^d}=1$ by choosing  the 
 opposite the signs  as 
\beq{
 \hat{\psi}'{}^d = \pmatrix{c}{f'_{\rm ex}{}^{(1)}\\f'_{\rm ex}{}^{(2)}}\exp\left(-\int_{z_0}^{z}W_{\rm ex}(z, \barz ;\zeta )\right) \comma \label{exactP2}
}
where $f'_{\rm ex}{}^{(1)}$ and $f'_{\rm ex}{}^{(2)}$ are given by
\beq{
&f'_{\rm ex}{}^{(1)} = \sqrt{\frac{\rho-\sqrt{T}\del \gamma}{T\,W^z_{\rm ex}}} \comma\,\, f'_{\rm ex}{}^{(2)} = \frac{-i}{\sqrt{W_{\rm ex}^z}}\left[ - W^z_{\rm ex}  + \left(\frac{\sqrt{T}}{2\zeta}-\frac{\rho\cos 2\gamma}{\sqrt{T}} +\frac{\del \ln f'_{\rm ex}{}^{(1)}}{2} \right)\right]\period
}

Using the definition \eqref{exactP}, let us now discuss the generalization of the WKB curves.  The quantity $\sqrt{T}dz/\zeta$ used to define the original WKB curves  is proportional to the leading term in the expansion of $W_{\swkb}$. Therefore the most natural generalization of the WKB curves would be to use $W_{\rm ex}$, which is a non-perturbative completion of $W_{\swkb}$, to  define them  as 
\beq{
\Im \left( W_{\rm ex}(z;\zeta) \right)=0\period
}
Unfortunately, there is a problem with this definition. Since there are many 
 exact solutions to the ALP, a different choice of the solution $\hat{\psi}^d$ leads to a different $W_{\rm ex}$ and thus to different curves.  We can avoid this  problem by defining  the curves in terms of the small solution $s_i$ (for a general  value of $\zeta$) near each puncture $z_i$.  We shall call them  the {\it exact WKB curves} and denote them by  EWKB$_{(i)}$. 

The precise definition is given as follows:\,  
The exact WKB curves associated to the puncture $z_i$ are defined as  the 
 curves satisfying the equation
 \beq{
 \Im \left( W^{(i)}_{\rm ex}(z;\zeta) \right) =0\comma\label{exWKB}
 }
where $W^{(i)}_{\rm ex}$ is the exponential factor for the solution 
 $s_i$, which is the smaller of the 
 two eigenvectors  $i_{+}$ and $i_-$. Explicitly, it is defined through the 
expression
\beq{
s_i \propto \pmatrix{c}{f_{\rm ex}^{(1)}\\f_{\rm ex}^{(2)}}\exp\left(\int_{z_0}^{z}W_{\rm ex}^{(i)}(z, \barz ;\zeta )\right)\period\label{si}
}

Let us now make several comments. First, it is easy to see that this definition of the exact WKB curves reduces  to that of the ordinary WKB curves when $\zeta$ is sufficiently small.  Second,  as in \eqref{exactP2},   with a flip of sign in the exponent, we can obtain another solution
\beq{
b_i \equiv  \pmatrix{c}{f'_{\rm ex}{}^{(1)}\\f'_{\rm ex}{}^{(2)}}\exp\left(-\int_{z_0}^{z}W_{\rm ex}^{(i)}(z, \barz ;\zeta )\right)\comma\label{bi}
}
which is big near the puncture $z_i$ and satisfies $\sl{s_i \comma b_i}=1$. 
Such a solution $b_i$,  however,  is not guaranteed to be an eigenvector 
since the eigenvector distinct  from $s_i$ is  in general  given by a linear combination of the form $b_i + c s_i$. 

Now the definition of EWKB$_{(i)}$ given above refers to a specific 
puncture  from which the curves emanate. In order for the notion of 
the exact WKB curve to be valid for the entire worldsheet, we must guarantee
 that the  definitions of EWKB$_{(i)}$'s  for $i=1,2,3$ are consistent in the region  where they overlap. To check this, let us consider the behavior of the 
 small solution $s_i$ as we follow an EWKB$_{(i)}$. Along such a curve 
 the phase of the exponential factor of $s_i$ stays constant, while its magnitude  increases monotonically\fn{Strictly speaking the small eigenvector \eqref{si} also contains a prefactor in front of the exponential. This prefactor, however, does not play a significant role in our discussion since it drops out if we consider the ratio of two solutions $s_i/b_i$. It is in fact sufficient to know the ratio in order to identify the small solution and the big solution.},  until it reaches some endpoint. Consider the case in which this endpoint is the puncture at $z_j$. 
In such a case, we know that  $s_i$  grows  exponentially as it approaches $z_j$ and in fact behaves like a big solution $b_j$,  up to an admixture of 
the exponentially  small solution $s_j$. Thus,  with sufficient accuracy,  $s_i$ 
 can be expressed  in the small neighborhood of $z_j$ as 
\beq{
s_i \propto b_j = \pmatrix{c}{f'_{\rm ex}{}^{(1)}\\f'_{\rm ex}{}^{(2)}}\exp\left(-\int_{z_0}^{z}W_{\rm ex}^{(j)}(z, \barz ;\zeta )\right)\period \label{nearzj}
} 
But since the exponent of the small solution $s_j$, which is used to 
 define EWKB$_{(j)}$,  is the same as that of $b_j$ except for the sign, we see that  by definition  the curve we have been following  becomes an  EWKB$_{(j)}$  curve in the vicinity of $z_j$, when $z_i$ and $z_j$ are connected 
by such a curve.    Therefore  the definitions of EWKB$_{(i)}$ and EWKB$_{(j)}$ are indeed globally  consistent. 

Let us now make use of the exact WKB curves to determine the analytic properties  of the Wronskians.  
First,  by following exactly the same logic as in the case of the ordinary WKB curves, we can  immediately conclude that the Wronskian involving two small solutions  $s_i$ and $s_j$  must be nonzero if two punctures $z_i$ and $z_j$ are connected by some exact WKB curves. Although this is an extremely useful 
information, the problem seems to be  that, unlike the ordinary WKB curves, we do not  know the configurations of the EWKB curves since the 
 exact solutions  to the ALP are not available. 

Nevertheless, we shall show below that by making use of a characteristic quantity defined locally around each puncture for the EWKB curves,  it is possible 
 to fully classify the topology (connectivity) of the curves on the entire worldsheet. 
The quantity in question is the ``number density" of the EWKB curves emanating  from a puncture at $z_i$. To motivate its definition, consider two such curves which emanate from $z_i$ and end at $z_j$ and let the constant 
 phase of $W^{(i)}_{\rm ex}$ along the two curves be $\phi_1$ and $\phi_2$. Evidently the magnitude of the difference $|\phi_1-\phi_2|$ is the same around $z_i$ 
 and around $z_j$,  that is,  it is conserved.  If there is no singularity in 
the region between these lines, we can draw in more EWKB curves  connecting $z_i$ and $z_j$. Because of the property of the constancy of the phase difference noted above,  it is quite natural  to draw the curves in such a way that  the difference of the phases of the adjacent  curves is some fixed unit  angle.  Going around $z_i$ and counting the number of such lines, we can define  the number (density) of the EWKB$_{(i)}$ curves as\fn{In \eqref{numex}, we have chosen a convenient normalization of $N_i$.} 
\beq{
N_i\equiv \frac{1}{2\pi}\oint_{\mathcal{C}_i}|\Im \, W_{\rm ex}^{(i)}|\comma \label{numex}
}
where $\calC_i$ is an infinitesimal circle around $z_i$.  Although $N_i$ is not 
 an integer in general, we will call it ``a number of lines". 
Actually we can express $N_i$ in a more explicit way. From the 
 asymptotic behavior of $i_{\pm}$ \eqref{normcond3}, 
 we can obtain the  form of  $W_{\rm ex}^{(i)}$ near $z_i$ as 
\beq{
&W_{\rm ex}^{(i)} \sim \pm \left( q_i(x) d\tau^{(i)} + ip_i(x)d\sigma^{(i)}\right)  \quad \text{as }z\to z_i \period
\label{appex}
}
Here  $( \tau^{(i)}\comma \sigma^{(i)})$ is the local coordinate defined in \eqref{localcoordinate}, and $+$ or $-$ sign is chosen depending on which of the solutions $i_{\pm}$ is small.  Substituting \eqref{appex} into the definition \eqref{numex}, we obtain a simple expression
\beq{
N_i \equiv \left|\Re p_i(x)\right|\period\label{defni}
}
Since the phase around the puncture is governed  by the local monodromy, 
 it is natural that $N_i$ can be expressed in terms of $p_i(x)$. 

Before we make use of the concept of $N_i$ in a more global context, 
let us derive  two important properties of the EWKB$_{(i)}$'s which will be necessary for determining their configurations. 

The first property 
 will be termed the {\it non-contractibility}.  It can be stated as follows:
\begin{itemize}
	\item[] ``All the exact WKB curves which start and end at the same puncture are non-contractible."
\end{itemize}
  In other words, such curves must go around a different puncture at least once.  
The proof is simple. Recall that the Wronskians between small solutions should be nonzero if two punctures are connected by an exact WKB curve. 
If we apply this statement  to the same puncture $z_i$ connected by an EWKB curve, we would conclude that $\wrons{s_i}{s_i}$ is non-zero, which is clearly false. The only way to be consistent with the general assertion above 
 is that the curve is non-contractible and the solution gets transformed 
 by the non-trivial monodromy $\Omega$ as it goes around other punctures. In this case  the Wronskian is of the form $\wrons{s_i}{\Omega s_i}$,  which need not vanish.  

The next property is concerned with   the endpoints of the exact WKB curves. 
It can be stated as follows: 
\begin{itemize}
	\item[] `` All but finitely many exact WKB curves terminate at punctures. " 
\end{itemize}
The proof can be given as follows.  As in the 
 case of the ordinary WKB curves, the possible endpoints  are the zeros or the poles
 of  $W_{\rm ex}^{(i)}$. Concerning  the former,  the number of  exact WKB curves flowing into a zero is always finite,  as shown in \figref{fig-WKB}.  
On the other hand, a pole can be the endpoint of  infinitely many curves 
and thus plays a crucial role in the study of the analyticity of the Wronskians. 
Now there are three different types of poles for  $W_{\rm ex}^{(i)}$. 
The first is a puncture, at which the vertex operator is inserted. 
The second type of a pole corresponds to the situation  where  the small eigenvector 
 $s_i$ develops a singularity at a position different from the puncture. 
Since we only consider the worldsheet without additional singularities as mentioned in section \ref{subsec:gen-three},
 such a singularity in $s_i$ should not occur. 
The last type of divergence for $W_{\rm ex}^{(i)}$ occurs  when 
$s_i$ develops a zero. Indeed,  $s_i$ in general has several zeros on the Riemann surface.  However,  such points cannot be the endpoints of the exact WKB curves for the following reason: 
At the zeros of $s_i$, the ratio $s_i/b_i$ of the small and the big solutions must also
 vanish\fn{The big solution  $b_i$ cannot  vanish at such points so as  to ensure the normalization condition $\sl{s_i\comma b_i}=1$.}.  But  this contradicts the basic property of the exact WKB curve that such a ratio, determined by the exponential factor in \eqref{si}, monotonically increases along the exact WKB curve as we move away from $z_i$.   From these considerations, we find  that apart from a finite number of curves 
 which can flow into zeros of $W_{\rm ex}^{(i)}$, the rest of the infinitely many exact   WKB curves must end at the  punctures. 

The two properties we have proved above are extremely important 
for  the following reason. They provide certain global restrictions  for the 
EWKB curves  for all values of the spectral parameter,  about which we 
  only  know the local behaviors explicitly  in the vicinity of the punctures. 
Below, they allow us to show that there are essentially two distinct classes of 
configurations for the exact WKB curves. 

\begin{figure}
\begin{minipage}{0.5\hsize}
\centering
\begin{tikzpicture}
\draw (0,0) -- (2,{2*sqrt(3)});
\draw (0,0) to [out=80, in=220] (2,{2*sqrt(3)});
\draw (0,0) to [out=40, in=260]  (2,{2*sqrt(3)});
\draw (4,0) -- (2,{2*sqrt(3)});
\draw (4,0) to [out=100, in=320] (2,{2*sqrt(3)});
\draw (4,0) to [out=140, in=280] (2,{2*sqrt(3)});
\draw (0,0) -- (4,0);
\draw (0,0) to [out=20, in=160] (4,0);
\draw (0,0) to [out=340, in=200] (4,0);
\filldraw[red] (0,0) circle (3pt)
(2,{2*sqrt(3)}) circle (3pt)
(4,0) circle (3pt);
\node[below,left] at (-0.2,-0.2) {$z_2$};
\node[above] at (2, {2*sqrt(3)+0.2}) {$z_1$};
\node[below,right] at (4.2,-0.2) {$z_3$};
\node[below] at (2,-1) {(a) Symmetric case ($N_i + N_j-N_k>0$)};
\end{tikzpicture}
\end{minipage}
\begin{minipage}{0.5\hsize}
\centering
\begin{tikzpicture}
\draw (0,0) -- (2,{2*sqrt(3)});
\draw (0,0) to [out=80, in=220] (2,{2*sqrt(3)});
\draw (0,0) to [out=40, in=260]  (2,{2*sqrt(3)});
\draw (4,0) -- (2,{2*sqrt(3)});
\draw (4,0) to [out=100, in=320] (2,{2*sqrt(3)});
\draw (4,0) to [out=140, in=280] (2,{2*sqrt(3)});
\draw (2,{2*sqrt(3)}) to [out=270, in=180] (4.4,-0.8) to [out=0, in=270] (5.2,0.5) to [out=90, in=0] (2.7,3.7) to [out=180, in=40] (2,{2*sqrt(3)});
\filldraw[red] (0,0) circle (3pt)
(2,{2*sqrt(3)}) circle (3pt)
(4,0) circle (3pt);
\node[below,left] at (-0.2,-0.2) {$z_2$};
\node[above] at (2, {2*sqrt(3)+0.2}) {$z_1$};
\node[below,right] at (4.2,-0.2) {$z_3$};
\node[below] at (2,-1) {(b) Asymmetric case ($N_2 + N_3-N_1<0$)};
\end{tikzpicture}
\end{minipage}
\caption{Two distinct classes for the connectivity of exact WKB curves. 
When all the triangle inequalities are satisfied (``symmetric" case), each 
exact WKB curve connects  two different punctures. On the other hand, 
 when some  of the triangle inequalities are  violated (``asymmetric" case), 
there must exist non-contractible curve(s) which starts and ends at the same 
 puncture.}
\label{fig-connection}
\end{figure}
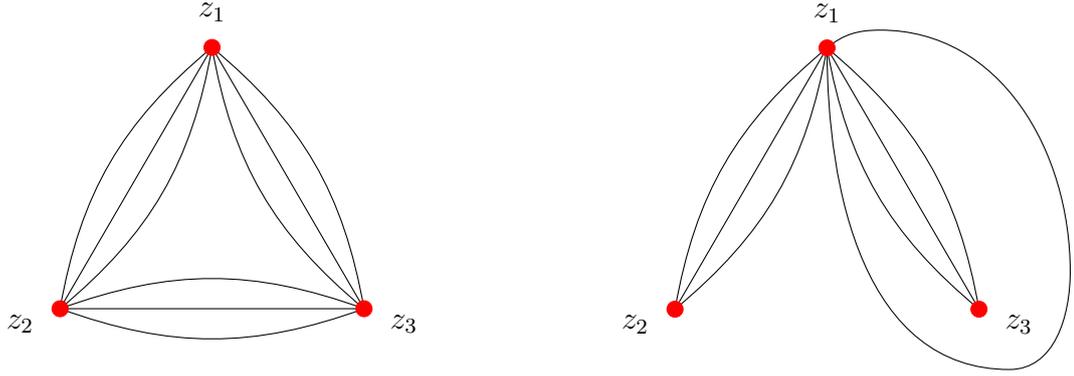

These two classes are distinguished by whether the number of lines $N_i$ 
fully  satisfy the  triangle inequalities  or not\footnote{In the case of the 
 usual WKB curves, $W_{\rm ex} \sim \sqrt{T(z)}dz$ and hence $N_i$ is proportional to  $\kappa_i$. Classification by the triangle inequalities for $\kappa_i$ already appeared in \cite{KK1}. }.  When  $N_i$'s  satisfy the 
relations 
\beq{
N_i + N_j-N_k>0\comma \label{triineq}
}
for all possible combinations of distinct $i,j,k$, we refer to such a configuration 
 as  {\it symmetric}.  It is easy to show that if \eqref{triineq} is 
 satisfied  the number of lines connecting $z_i$ and $z_j$ 
 cannot be zero.  As this holds for all the interconnecting lines,  the three punctures 
 must be  piece-wise connected to each other as in the left figure of \figref{fig-connection}. 

On the other hand, in the second case, which we shall call {\it asymmetric}, not all the triangle inequalities are satisfied. For example, one is violated like 
\beq{
N_2+N_3-N_1<0\period \label{TIv}
}
In this case, one can readily convince oneself  that,  while all the curves emanating 
 from $z_2$ and $z_3$ end at $z_1$, there must  exist  a
non-contractible curve connecting  $z_1$ to itself.  This is depicted 
 in the right figure of \figref{fig-connection}. 

In this way, we can completely classify the configurations  of the exact WKB curves from the local information  $N_i =|\Re p_i(x)|$.  Note that $N_i$ depends on $x$. In fact it happens that as $x$ changes a symmetric configuration 
 can turn into an asymmetric configuration and vice versa. In an application 
 of the present idea to the classical  three-point function in Liouville theory\cite{DH}, it was checked that such a transition must be taken into account 
in order to obtain the correct result. 
 Below, we will see explicitly how the patterns of the configurations of the exact WKB curves analyzed above can be used to determine the zeros of the Wronskians. 
%%%%%%%%%%%%%%%%%%%%%%%
\subsubsection{Determination of the zeros of the Wronskians }
%%%%%%%%%%%%%%%%%%%%%%%
As an example, let us focus on the factor 
\beq{
\sin \left( \frac{p_1 +p_2 +p_3}{2}\right) \comma\label{plplpl}
}
and determine which  Wronskians  develop  a zero when this factor 
 vanishes. (The logic below applies to all the other cases straightforwardly.) 
From the relations \eqref{eq-5-5}--\eqref{eq-5-10}, we find  
 that the products of  Wronskians that  become zero  are 
\beq{
\sl{1_{+}\comma 2_{+}}\sl{1_{-}\comma 2_{-}}\comma \quad \sl{2_{+}\comma 3_{+}}\sl{2_{-}\comma 3_{-}}\comma \quad\sl{3_{+}\comma 1_{+}}\sl{3_{-}\comma 1_{-}}\period\label{group}
}
For convenience, let us define  the following two sets  of eigenvectors, namely the set  $\calS_+ \equiv \{1_{+},2_{+},3_{+}\}$ and 
 the set $\calS_- \equiv \{1_{-},2_{-},3_{-}\}$. 
An important feature of the quantities shown in \eqref{group} is that 
 only the Wronskians of the eigenstates in the same group,  $\calS_+$ or $\calS_-$,  appear.  This is in fact a general feature and holds  also for other situations. 

Now, let us present two theorems, which will be useful in the determination of the zeros. The first theorem is the following assertion,  which we have already proved: 
\begin{itemize}
\item[]\underline{\it Theorem 1.}\ \ 
When two punctures $z_i$ and $z_j$ are connected by an  exact WKB curve, the Wronskian between the two small eigenvectors $\sl{s_i \comma s_j}$ is non-vanishing.
\end{itemize}
The second theorem classifies the possibilities of the patterns of the zeros 
 and is stated as follows:
\begin{itemize}
\item[]\underline{\it Theorem 2.}\ \ 
There are only two distinct possibilities concerning the zeros of the Wronskians in \eqref{group}: Either (a) all the Wronskians among the members of $\calS_+$ are zero  and those among $\calS_-$ are nonzero, or (b) all the Wronskians among $\calS_-$ are zero and those among  $\calS_+$ are nonzero. 
\end{itemize}
The proof is as follows. 
 Let us first note that in each product of two Wronskians
 appearing in \eqref{group}, only one of them vanishes. In fact if 
both factors become zero simultaneously, the product  develops a double 
 zero, which contradicts the fact that the zeros of \eqref{plplpl} are all simple 
 zeros.  This property implies that in the list  given in \eqref{group}, at least two of the individual Wronskians which actually vanish must be between the members belonging to the same set, which can be  $\calS_+$ or $\calS_-$.  Suppose  they belong to $\calS_+$. Since 
 $\wrons{i_+}{j_+}=0$ means that $i_+$ and $j_+$   are parallel to each other, vanishing of  two such different Wronskians between the states of  $\calS_+$ implies that in fact all the three states in $\calS_+$ are proportional to 
 each other. Therefore the third Wronskian from the set $\calS_+$ must also 
 vanish. Obviously the same logic applies to the $\calS_-$ case. 
This proves the theorem. 

We can  now analyze the zeros of the Wronskians using these theorems. 
First consider the symmetric case. Since one of the states $i_\pm$ must 
 be a small solution,  either $\calS_+$ or $\calS_-$ must contain two
  small solutions.  For a symmetric configuration, they must be 
 connected by an exact WKB curve. Then by theorem 1 the Wronskian between them must be non-vanishing. Theorem 2 further asserts  that all the Wronskians for the members of that set are non-vanishing, while the ones for 
 elements of the other set all vanish. 

Next, consider the asymmetric case. For simplicity, let us 
assume that $N_1>N_2 + N_3$ is satisfied\fn{Generalization to other cases is straightforward.}.  In such a case, there exist exact WKB curves which start from $z_1$, go around $z_2$ (or $z_3$),  and return to $z_1$. 
To make use of the existence of such a curve, consider the following Wronskians:
\beq{
\sl{1_+ \comma \Omega_2 1_+} \comma \quad \sl{1_{-}\comma \Omega_2 1_{-}}\period\label{asymwron}
}
To compute them,  we first note that $1_\pm$ can be expressed  in terms of $2_{\pm}$ in the following manner
\beq{
1_{\pm} = \sl{1_{\pm}\comma 2_{-}} 2_{+}-\sl{1_{\pm}\comma 2_{+}} 2_{-}\period\label{1intermsof2}
}
Then, applying $\Omega_2$ to \eqref{1intermsof2} and substituting them to \eqref{asymwron}, we can express \eqref{asymwron} in terms of the ordinary Wronskians as
\beq{
&\sl{1_+ \comma \Omega_2 1_+} = 2i \sin p_2 \sl{1_{+}\comma 2_{-}}\sl{1_{+}\comma 2_{+}}\comma\label{1om1}\\
&\sl{1_- \comma \Omega_2 1_-}= 2i \sin p_2 \sl{1_{-}\comma 2_{-}}\sl{1_{-}\comma 2_{+}}\period
}
Consider the case where $1_+$ is the small solution. 
Since $\Omega_2 1_{+}$ can be obtained by parallel-transporting $1_{+}$ along the exact WKB curve which starts and ends at $z_1$, 
$\Omega_2 1_{+}$ must behave  as the big solution around $z_1$. 
Therefore, the Wronskian $\sl{1_+ \comma \Omega_2 1_+} $ is non-vanishing  in this case. 
Then from \eqref{1om1}  it follows that $\sl{1_{+}\comma 2_{+}}$ must also be non-vanishing. Applying the theorem 2, we conclude that 
the Wronskians between the members of $\calS_+$ are  non-vanishing 
 and those of $\calS_-$ all vanish. In an entirely similar manner, 
 when $1_{-}$ is the small eigenvector, we obtain the result where 
 the roles of $\calS_+$ and $\calS_-$ are interchanged.

Performing  similar analyses  for the other cases, we obtain the  general rules
summarized below. 
%%%%%%%%
\begin{itemize}
\item[]\underline{Rule 1:}\  {\it Decomposition of the eigenvectors into two groups.}\\
When a factor of the form $\sin \left( \sum_i \epsilon_i p_i/2\right)$ vanishes, the Wronskians which vanish are the ones among $\{1_{\epsilon_1}, 2_{\epsilon_2},3_{\epsilon_3} \}$ or  the ones among $\{1_{-\epsilon_1}, 2_{-\epsilon_2},3_{-\epsilon_3} \}$.
\item[]\underline{Rule 2:}\  {\it Symmetric case.}\\When the configuration of the exact WKB curves is symmetric, the Wronskians from the group which contains two or more small solutions are nonzero whereas the Wronskians from the other group are zero.
\item[]\underline{Rule 3:}\  {\it Asymmetric case.}\\When the configuration of the exact WKB curves is asymmetric and $N_i$'s satisfy $N_i>N_j+N_k$, the Wronskians from the group which contains the smaller of the two solutions $i_{\pm}$ are nonzero whereas the Wronskians from the other group are zero.
\end{itemize}
In the next subsection, we will utilize these rules to evaluate the individual Wronskians.
%%%%%%%%%%%%%%%%%%%%%%%%%%%%%%%%
\subsection{Individual Wronskian from the Wiener-Hopf decomposition\label{subsec:WH}}
%%%%%%%%%%%%%%%%%%%%%%%%%%%%%%%%%%%%
Making use of the data for the analyticity of the Wronskians obtained in the previous subsection, we  now set up and solve a Riemann-Hilbert problem 
to decompose the product of Wronskians and extract the individual Wronskians.The standard method for such a procedure is known as the Wiener-Hopf decomposition, which extracts from a complicated function a part regular on the upper half plane and  the part regular on the  lower half plane.  The typical set up is as follows. 
Suppose $F(x)$ is  a function which decreases sufficiently fast at infinity
and can be written as a sum of two components $F(x) =F_{\uparrow}(x)+F_{\downarrow}(x) $,   where $F_{\uparrow}(x)$ is  regular on the upper half plane  while  $F_{\downarrow}(x)$ is regular on the lower half plane. 
Then, each component, in the region where it is regular,  can be extracted from $F(x)$ as 
\beq{
&F_{\uparrow}(x)= \int_{-\infty}^{\infty} \frac{dx^{\prime}}{2\pi i} \frac{1}{x^{\prime}-x}F(x^{\prime})& (\Im x>0)\comma\label{WH1}\\
&F_{\downarrow}(x)=-\int_{-\infty}^{\infty} \frac{dx^{\prime}}{2\pi i} \frac{1}{x^{\prime}-x}F(x^{\prime}) & (\Im x<0)\period\label{WH2}
}
These equations  can be easily proven by first substituting 
$F(x')=F_{\uparrow}(x^{\prime})+F_{\downarrow}(x^{\prime})$ on the 
 right hand side and then closing the integration contour for  $F_{\uparrow}(x^{\prime})$ ($F_{\downarrow}(x^{\prime})$) on the upper (lower) half plane. 
Now when the argument $x$ is not in the region specified  in \eqref{WH1} and \eqref{WH2}, we need to analytically continue the above formulas. 
For instance, $F_{\uparrow}(x)$ in the region where $\Im x<0$ should  be expressed as 
\beq{
F_{\uparrow}(x)= F(x) - F_\downarrow(x) = F(x) +\int_{-\infty}^{\infty} \frac{dx^{\prime}}{2\pi i} \frac{1}{x^{\prime}-x}F(x^{\prime})\period
\label{addF}
}
Note that the first term $F(x)$ on the right hand side can be thought of 
 as due to the integral along a small circle around $x^{\prime}=x$. 

To apply this method to the case of our interest, namely to the equations 
\eqref{eq-5-5}--\eqref{eq-5-10}, we take the logarithm and represent  them in a general form  as 
\beq{
\ln  \sl{i_{\epsilon_i}\comma j_{\epsilon_j}} + \ln \sl{i_{-\epsilon_i}\comma j_{-\epsilon_i}}=&\ln \sin\left(\frac{\epsilon_i p_i+ \epsilon_j p_j + p_k}{2}\right)+\ln \sin\left(\frac{\epsilon_i p_i +\epsilon_j p_j - p_k}{2}\right)\nn\\ & -\ln \sin p_i -\ln \sin p_j\period \label{eq-46}
}
Here $\epsilon_i$ denotes a $+$ or $-$ sign. In this process, we have neglected the contributions of the form $\ln (-1)$, since they only contribute to 
 the overall phase of the three-point functions.  Our aim will be to 
express each of the  terms  on the left hand side of \eqref{eq-46}  in terms of some  convolution integrals of the functions on the right hand side. 
To put it in another way, we wish to decompose each term on the right hand side 
into contributions coming from each term on the left hand side. 
Since the quasi-momentum $p_i(x)$ is defined on a Riemann surface with 
 branch cuts, we need to generalize the Wiener-Hopf decomposition formula in an appropriate way, as discussed below. 
%%%%%%%%%%%%%%%%%%%%%%
\subsubsection{Separation of  the poles}
%%%%%%%%%%%%%%%%%%%%%%%
Let us first decompose the terms of the form  $-\ln \sin p_i$,  which give rise to poles of the Wronskians. 
As shown in the previous section, which Wronskian develops a pole 
 is determined purely by  the sign of the real part of the quasi-momentum $q_i(x)$.  Therefore, we should be able to  decompose the quantity $-\ln \sin p_i$  by using a convolution  integral along the curve  defined by $\Re q_{i}=0$.  For the ordinary Wiener-Hopf decomposition, the convolution kernel is given simply by $1/ (x-x^{\prime})$. In the present case, however, 
we have a two-sheeted Riemann surface and hence we must make sure that
 the kernel has the simple pole only when $x$ and $x'$ coincide on the same
 sheet. When they are  on top of each other on different sheets, no singularity
 should occur. 
The appropriate kernel with this property is given by 
\beq{
\widehat{\mathcal{K}}_{i}(x^{\prime};x)\equiv \frac{1}{2(x^{\prime}-x)}\left(\sqrt{\frac{(x-u_i)(x-\bar{u}_i)}{(x^{\prime}-u_i)(x^{\prime}-\bar{u}_i)}} +1\right) \period\label{conv0} 
}
When $x$ and $x'$ get close to each other but on different sheets, 
 the square root factor tends to $-1$ canceling the $+1$ term and hence
 the kernel is indeed regular. 
 Furthermore, in the limit that $x'$ tends to  $\infty$,  the kernel $\widehat{\mathcal{K}}_{i}(x^{\prime};x)$ decreases 
 like $\left( x^{\prime}\right)^{-2}$, which is sufficiently fast for our purpose. 

With such a convolution kernel, we can carry out the Wiener-Hopf decomposition in the usual way.  Namely the term $-\ln \sin p_i(x)$ can be 
 decomposed into the contributions of  $\sl{i_+\comma j_{\ep_j}}(x)$
 and $\sl{i_-\comma j_{-\ep_j}}(x)$ as 

\beq{
&\sl{i_+\comma j_{\ep_j}}(x)\ni  \oint_{\Gamma_{i_+}} \widehat{\mathcal{K}}_i\ast(-\ln \sin p_i )\comma\label{eq-47}\\
&\sl{i_-\comma j_{-\ep_j}}(x)\ni  \oint_{\Gamma_{i_-}} \widehat{\mathcal{K}}_i\ast(-\ln \sin p_i)\comma\label{eq-48}
}
where the convolution integral is defined as
\beq{\int A*B \equiv 
\int \frac{dx^{\prime}}{2\pi i} A(x^{\prime}; x)B(x^{\prime})\period
}
As for the contours of integration, 
$\Gamma_{i_+}$ is  defined by $\Re \, q_i=0$ and  $\Gamma_{i_-}$ stands for  $-\Gamma_{i_+}$.  
The direction of the contour $\Gamma_{i_+}$ is defined such that $\sl{i_+\comma  j_{\ep_j}}(x)$ does not contain poles in the region to the left of  the contour\fn{A typical form of the contour is depicted in \figref{fig:typical} in section \ref{sec:compare}, where we study explicit examples.}. 

Now note that under the holomorphic involution $x\to \hat{\sigma}x$, 
 the quasi-momentum $p_i(x)$ and the square-root contained in \eqref{conv0} simply flip sign. Making use of  this property,  we can re-express the convolution integrals  \eqref{eq-47} and \eqref{eq-48} as integrals only on the first 
(or the upper) sheet.:
\beq{
&\sl{i_+\comma  j_{\ep_j}}(x)\ni -\oint_{\Gamma^{u}_{i_+}} \mathcal{K}_i\ast\ln \sin p_i \comma\label{eq-49}\\
&\sl{i_-\comma j_{-\ep_j}}(x)\ni -\oint_{\Gamma^{u}_{i_-}} \mathcal{K}_i\ast\ln \sin p_i\period\label{eq-50}
}
Here,   $\Gamma_{i_{\pm}}^u$ denotes the portion of $\Gamma_{i_{\pm}}$ on the upper-sheet  of the spectral curve and the kernel $\mathcal{K}_i(x^{\prime};x)$ (without a hat) is defined by
\beq{
\mathcal{K}_i(x^{\prime};x)\equiv \frac{1}{x^{\prime}-x}\sqrt{\frac{(x-u_i)(x-\bar{u}_i)}{(x^{\prime}-u_i)(x^{\prime}-\bar{u}_i)}}\period
}
Again we have neglected the factors of the form $\ln (-1)$  arising from the sign flip of $p_i(x)$, as they only modify the overall phase of the Wronskians and the three-point functions. 

It is important to note that  \eqref{eq-47} and \eqref{eq-48} are valid only when $x$ is on the left hand side of the contours, just as in the case of the ordinary Wiener-Hopf decomposition. 
When the argument $x$ is on the right hand side of the contour $\Gamma_{i_{\pm}}$, we must add $-\ln\sin p_i$ to \eqref{eq-47} and \eqref{eq-48},   as explained in \eqref{addF}.  Such effects can be taken into account also in \eqref{eq-49} and \eqref{eq-50}, if $x$ is on the upper  sheet, by adding a small circle encircling $x^{\prime}=x$ counterclockwise to the integration contours.   In what follows, such contributions will be referred to as  {\it contact terms}. 
%%%%%%%%%%%%%%%%%%%%%
\subsubsection{Separation of  the zeros}
%%%%%%%%%%%%%%%%%%%%%
Next we shall discuss the decomposition of the first two terms on the right hand side of \eqref{eq-46}, which are responsible for the zeros of the Wronskians. 
To perform the decomposition,  again we need to determine the appropriate convolution kernel and the integration contour. 

Let us first discuss the convolution kernel. As the terms of our focus 
depend on all the quasi-momenta $p_i(x)$'s,  the appropriate convolution kernel must be a function on the Riemann surface which contains all the branch cuts of the $p_i(x)$'s.  Such a kernel can be easily written down as a generalization of the expression \eqref{conv0} and is given by 
\beq{
\widehat{\mathcal{K}}_{\rm  all}\equiv\frac{1}{8(x^{\prime}-x)}\prod_{i=1}^{3}\left(\sqrt{\frac{(x-u_i)(x-\bar{u}_i)}{(x^{\prime}-u_i)(x^{\prime}-\bar{u}_i)}}+1\right) \period\label{conv1}
}
Since there are two choices of sign for each square root factor on the right hand side of \eqref{conv1}, $\widehat{\mathcal{K}}_{\rm all}$ is properly 
 defined  on the eightfold  cover of the complex plane. 
In what follows, we distinguish these eight sheets  as  $\{\ast,\ast,\ast\}$-sheet, where the successive entry $\ast$ is either ``$u$" 
denoting upper sheet or ``$l$" denoting lower sheet,  referring to  the two sheets for $p_1(x)$, $p_2(x)$ and $p_3(x)$ respectively. It is clear that 
the kernel \eqref{conv1} has a pole with a residue $+1$ at $x^{\prime}=x$ only when two-points are on the same sheet. Therefore it has a desired property for the Wiener-Hopf decomposition.

Let us next turn to  the contour of integration. As discussed in the previous section, the zeros of the Wronskians are determined by the following two properties: (i) The connectivity of the exact WKB-curves and (ii) the relative magnitude  of the eigenvectors $i_{\pm}$. Therefore,  curves  across  which these two properties change can be the possible integration contours. Corresponding to the   properties (i) and (ii) above, there are two types of integration contours; the curves defined by $\Re q_i(x)=0$ and the curves defined by $N_i=N_j+N_k$.  An important point to bare in mind is that in general 
only some portions of these curves will be the proper integration contours,  since in some cases the analyticity of the Wronskians does not change even when  we cross these curves. In order to determine the correct  integration contours explicitly, we need to apply the general rules derived in the previous section. 
However, as the form of the contours determined through such a procedure depends  on the specific details of the choice of the external states, 
we will postpone such an analysis  until  section \ref{sec:compare}, where we work out 
 some  specific examples.  Thus, in what follows we will denote the integration contours without specifying their explicit forms as $\mathcal{M}_{\pm\pm\pm}$, where $\mathcal{M}_{\epsilon_1\epsilon_2\epsilon_3}$ denotes the contour we use to determine the contribution of the factor $\sin \left(\sum_i \epsilon_i p_i \right)$ to $\sl{i_{\epsilon_i}\comma j_{\epsilon_j}}$. They are defined such that they flip the orientation if we flip the signs of  three indices, for example  $\mathcal{M}_{+++}=-\mathcal{M}_{---}$

Employing the kernel and the contours given above, let us perform 
the decomposition of the product of Wronskians,  taking that of $\sl{1_{+}\comma 2_{+}}$ and $\sl{1_{-}\comma 2_{-}}$ as a representative example. 
Applying the Wiener-Hopf decomposition to the relation \eqref{eq-46} 
with $i=1, j=2$ and $\ep_1=+, \ep_2=+$, we obtain 
\beq{
&\sl{1_{+}\comma 2_{+}} \ni\nn\\
& \oint_{\mathcal{M}_{+++}}\widehat{\mathcal{K}}_{\rm all}\ast \ln \sin \left(\frac{p_1 + p_2 +p_3}{2}\right)+\oint_{\mathcal{M}_{++-}}\widehat{\mathcal{K}}_{\rm all}\ast \ln \sin \left(\frac{ p_1 + p_2- p_3}{2}\right)\comma\label{eq-51}\\
&\sl{1_-\comma 2_-}\ni  \nn\\
& \oint_{\mathcal{M}_{---}}\widehat{\mathcal{K}}_{\rm all}\ast \ln \sin \left(\frac{p_1+p_2+p_3}{2}\right)+\oint_{\mathcal{M}_{--+}}\widehat{\mathcal{K}}_{\rm all}\ast \ln \sin \left(\frac{p_1+p_2-p_3}{2}\right)\period\label{eq-52}
}
As in the case of the ordinary Wiener-Hopf decomposition, the expressions \eqref{eq-51} and \eqref{eq-52} are valid only when $x$  is located to the left  of the integration contour.   Additional terms, to be discussed shortly,  are needed  when $x$ 
is on the other side of the contour. 

Let us now show that the kernel $\widehat{\mathcal{K}}_{\rm all}$ 
used  in \eqref{eq-51} and \eqref{eq-52} can be effectively 
 replaced by simpler combinations of the form  $\left( \mathcal{K}_i +\mathcal{K}_j\right)/8$.  To explain the idea,  consider the following integral as an example:
\beq{
\oint_{\mathcal{M}_{+++}}\frac{dx^{\prime}}{2\pi i}\widehat{\mathcal{K}}_{\rm all}(x^{\prime};x) \ln \sin \left( \frac{p_1 +p_2 +p_3}{2}\right)(x^{\prime})\period\label{origmppp}
}
As the first step,  we  make a change of integration variable from 
$x^{\prime}$ to  $\hat{\sigma}_3 x^{\prime}$, where $\hat{\sigma}_i$ 
 denotes the holomorphic involution with respect to $p_i$, namely the 
 operation that  exchanges the two sheets associated with $p_i$. 
Although this clearly leaves  the value of the integral intact,  the form 
 of the integral changes. One can easily verify that the following 
transformation formulas for the integrand and the contours hold:
\beq{
&\ln \sin \left( \frac{p_1+p_2 +p_3}{2}\right)(\hat{\sigma}_3 x^{\prime})=\ln \sin \left( \frac{p_1+p_2 -p_3}{2}\right)(x^{\prime})\comma\label{3flip1}\\
&\widehat{\mathcal{K}}_{\rm all}(\hat{\sigma}_3 x^{\prime};x)= \widehat{\mathcal{K}}_{\rm all}^{(3)}( x^{\prime};x)\comma\label{3flip2}\\
&\oint_{\mathcal{M}_{+++}}d(\hat{\sigma}_3 x^{\prime})=\oint_{\mathcal{M}_{++-}}dx^{\prime}\period\label{3flip3}
}
In the second line \eqref{3flip2}, the ``sign-flipped kernel"  $\widehat{\mathcal{K}}^{(3)}_{\rm  all}$ 
is defined by 
\beq{
\widehat{\mathcal{K}}^{(3)}_{\rm  all}\equiv\frac{1}{8(x^{\prime}-x)}\left(-\sqrt{\frac{(x-u_3)(x-\bar{u}_3)}{(x^{\prime}-u_3)(x^{\prime}-\bar{u}_3)}}+1\right)\prod_{\ell = 1,2}\left(\sqrt{\frac{(x-u_\ell )(x-\bar{u}_\ell )}{(x^{\prime}-u_\ell )(x^{\prime}-\bar{u}_\ell )}}+1\right)\period\label{conv2}
}
Making such transformations,   we can re-express the integral \eqref{origmppp} 
as 
\beq{
\oint_{\mathcal{M}_{++-}}\frac{dx^{\prime}}{2\pi i}\widehat{\mathcal{K}}^{(3)}_{\rm all}(x^{\prime};x) \ln \sin \left( \frac{p_1 +p_2 -p_3}{2}\right)(x^{\prime})\period
}
Performing similar analysis for all the possible sign-flips, we obtain $2^3$ different expressions for \eqref{origmppp}. Then averaging over all the $2^3$ expressions, we find that the final expressions are given in terms of the kernels $\mathcal{K}_i$ as follows:
\beq{
&\sl{1_{+}\comma 2_{+}}\ni \nn\\
&\frac{1}{16}\left(\oint_{\mathcal{M}_{+++}}\hspace{-20pt}\left( \mathcal{K}_1+\mathcal{K}_2\right)\ast \ln \sin \left( \frac{p_1 +p_2+p_3}{2}\right) +\oint_{\mathcal{M}_{++-}}\hspace{-20pt}\left( \mathcal{K}_1+\mathcal{K}_2\right)\ast \ln \sin \left( \frac{p_1 +p_2-p_3}{2}\right)\right.\nn\\
&\left.+\oint_{\mathcal{M}_{+-+}}\hspace{-20pt}\left( \mathcal{K}_1-\mathcal{K}_2\right)\ast \ln \sin \left( \frac{p_1 -p_2+p_3}{2}\right) +\oint_{\mathcal{M}_{-++}}\hspace{-20pt}\left( -\mathcal{K}_1+\mathcal{K}_2\right)\ast \ln \sin \left( \frac{-p_1 +p_2+p_3}{2}\right)\right)\comma\label{12ppzero}\\
&\sl{1_{-}\comma 2_{-}}\ni \nn\\
&\frac{1}{16}\left(\oint_{\mathcal{M}_{---}}\hspace{-20pt}\left( \mathcal{K}_1+\mathcal{K}_2\right)\ast \ln \sin \left( \frac{p_1 +p_2+p_3}{2}\right) +\oint_{\mathcal{M}_{--+}}\hspace{-20pt}\left( \mathcal{K}_1+\mathcal{K}_2\right)\ast \ln \sin \left( \frac{p_1 +p_2-p_3}{2}\right)\right.\nn\\
&\left.+\oint_{\mathcal{M}_{-+-}}\hspace{-20pt}\left( \mathcal{K}_1-\mathcal{K}_2\right)\ast \ln \sin \left( \frac{p_1 -p_2+p_3}{2}\right) +\oint_{\mathcal{M}_{+--}}\hspace{-20pt}\left( -\mathcal{K}_1+\mathcal{K}_2\right)\ast \ln \sin \left( \frac{-p_1 +p_2+p_3}{2}\right)\right)\period\label{12mmzero}
}
Just as before, we neglected the contributions of the form $\ln(-1)$ as leading 
 to pure phases. Also, the same remarks  made below  
 equations \eqref{eq-51} and \eqref{eq-52} on the position of $x$ relative 
 to the contour lines apply to the expressions \eqref{12ppzero} and \eqref{12mmzero} above. 

Finally, for later  convenience,  let us further re-write  the above 
 expressions  as integrals  performed  purely on the $\{u,u,u\}$-sheet. Each  contour $\calM_{\ep_1\ep_2\ep_3}$ has parts on the eight different  sheets denoted  by  $\calM_{\ep_1\ep_2\ep_3}^{u,u,u}, \calM_{\ep_1\ep_2\ep_3}^{u,u,l}$, etc., where the superscripts indicate 
 the relevant sheet in an obvious way.  Consider for example 
 the first integral in   \eqref{12ppzero} along the contour $\calM_{+++}$. 
The  form as given is  for the portion $\calM_{+++}^{uuu}$. For the portion   denoted by $\calM_{+++}^{ulu}$ for example, if we wish to express 
 its contribution in terms of an integral on the  $\{u,u,u\}$-sheet,  we need  to change the sign of $\calK_2$ and $p_2$. Then  the integral  becomes  identical to that of the first term in the second line of \eqref{12ppzero}, except along $\calM_{+-+}^{uuu}$. 
In similar fashions we can re-express the contributions from the eight 
parts of $\calM_{+++}$ in terms of the integrals on the  $\{u,u,u\}$-sheet. After repeating the same procedure for the 
 rest of the three terms in \eqref{12ppzero},  one finds that the net effect is that each term of \eqref{12ppzero} is multiplied 
 by a factor of eight, with each contour restricted to the  $\{u,u,u\}$-sheet.  In this way we obtain the representations
\beq{
&\sl{1_{+}\comma 2_{+}}\ni \nn\\
&\frac{1}{2}\left(\oint_{\mathcal{M}^{{uuu}}_{+++}}\hspace{-20pt}\left( \mathcal{K}_1+\mathcal{K}_2\right)\ast \ln \sin \left( \frac{p_1 +p_2+p_3}{2}\right) +\oint_{\mathcal{M}^{{uuu}}_{++-}}\hspace{-20pt}\left( \mathcal{K}_1+\mathcal{K}_2\right)\ast \ln \sin \left( \frac{p_1 +p_2-p_3}{2}\right)\right.\nn\\
&\left.+\oint_{\mathcal{M}^{{uuu}}_{+-+}}\hspace{-20pt}\left( \mathcal{K}_1-\mathcal{K}_2\right)\ast \ln \sin \left( \frac{p_1 -p_2+p_3}{2}\right) +\oint_{\mathcal{M}^{{uuu}}_{-++}}\hspace{-20pt}\left( -\mathcal{K}_1+\mathcal{K}_2\right)\ast \ln \sin \left( \frac{-p_1 +p_2+p_3}{2}\right)\right)\comma\label{12ppzero2}\\
&\sl{1_{-}\comma 2_{-}}\ni \nn\\
&\frac{1}{2}\left(\oint_{\mathcal{M}^{{uuu}}_{---}}\hspace{-20pt}\left( \mathcal{K}_1+\mathcal{K}_2\right)\ast \ln \sin \left( \frac{p_1 +p_2+p_3}{2}\right) +\oint_{\mathcal{M}^{{uuu}}_{--+}}\hspace{-20pt}\left( \mathcal{K}_1+\mathcal{K}_2\right)\ast \ln \sin \left( \frac{p_1 +p_2-p_3}{2}\right)\right.\nn\\
&\left.+\oint_{\mathcal{M}^{{uuu}}_{-+-}}\hspace{-20pt}\left( \mathcal{K}_1-\mathcal{K}_2\right)\ast \ln \sin \left( \frac{p_1 -p_2+p_3}{2}\right) +\oint_{\mathcal{M}^{{uuu}}_{+--}}\hspace{-20pt}\left( -\mathcal{K}_1+\mathcal{K}_2\right)\ast \ln \sin \left( \frac{-p_1 +p_2+p_3}{2}\right)\right)\period\label{12mmzero2}
}
The results obtained in this subsection and the previous subsection are both expressed in terms of certain convolution integrals on the spectral curve. Thus, in what follows, we will denote their sum by $\text{\sf Conv}\sl{i_{\pm},j_{\pm}}$.

Before ending this subsection, let us make one important remark. Although each convolution integral obtained so far is divergent at $x=\pm 1$, the divergence cancels\fn{One can confirm this by expanding the convolution integrals around $x=\pm1$.} in the sum $\text{\sf Conv}\sl{i_{\pm},j_{\pm}}$. Thus the contribution singular at $x=\pm 1$ must be separately taken into account as we will do in the next subsection.
%%%%%%%%%%%%%%%%%%%%%%%%%%%%%%%
\subsection{Singular part and constant part of the Wronskians\label{subsec:sing}}
%%%%%%%%%%%%%%%%%%%%%%%%%%%%
In addition to the main non-trivial parts determined by the Wiener-Hopf decomposition described above, there are two further contributions to the 
 Wronskians.  One is the contribution  singular at $x=\pm 1$, 
coming from such structure in  the connections  used  in  ALP.  
The other is the possibility of adding a constant function on the spectral 
 curve.  In this subsection, we will determine these two contributions.

Let us first focus on terms singular at $x=1$. To determine such terms, we will need the WKB expansions around $x=1$ for all the Wronskians, not just the ones that were discussed in section \ref{subsec:WKB}, namely $\sl{i_+\comma j_+}$ and $\sl{i_-\comma j_-}$.  This is because of the following reason:
 Although the formulas we obtained for the contribution of the action and 
 that of the wave function appear to contain Wronskians of the type 
$\sl{i_+\comma j_+}$ and  $\sl{i_-\comma j_-}$ only,  we must understand  their behavior when they are followed into the second sheet as well in order to know the analyticity property on the entire Riemann surface. 
As shown in \eqref{change-sheet},  when  we cross the branch cut 
associated with $p_i(x)$ into the lower sheet,  the eigenfunctions $i_+$ and $i_-$ behave like  $i_-$ and $-i_+$ on the upper sheet, respectively,  . Therefore  the behavior of $\sl{i_+\comma j_+}$ on the $\{ u,l,*\}$-sheet can be obtained  from  the behavior of $\sl{i_+\comma j_-}$ on the $\{u,u,*\}$-sheet, etc. 

Now the WKB expansions of the Wronskians of the type $\sl{i_+\comma j_-}$ can be obtained from those of  $\sl{i_+\comma j_+}$  by the use of the following Schouten identities:
\beq{
\sl{i_+\comma j_-}\sl{j_+\comma k_+}+\sl{i_+\comma j_+}\sl{j_-\comma k_+}+\sl{i_+\comma k_+}\sl{j_-\comma j_+}=0\period\label{eq-60}
}
Indeed these identities can be regarded  as the equations for  the six unknown Wronskians of the form $\sl{i_+\comma j_-}$. 
If we consider all the combinations of $i,j$ and $k$ in \eqref{eq-60}, 
 we obtain three independent equations.  Combining them with the equations 
\eqref{eq-5-8}--\eqref{eq-5-10} for the products of the Wronskians,  we can completely determine $\sl{i_+\comma j_-}$'s in terms of $\sl{i_+\comma j_+}$ in the following form:
\beq{
&\sl{1_+\comma 2_-}=e^{-i(p_1+p_2-p_3)/2} \frac{\sin \left(\frac{p_1-p_2-p_3}{2} \right)}{\sin p_2}\frac{\sl{3_+\comma 1_+}}{\sl{2_+\comma 3_+}}\comma\label{eq-61}\\
&\sl{1_-\comma 2_+}=e^{i(p_1+p_2-p_3)/2} \frac{\sin \left(\frac{p_1-p_2-p_3}{2} 
\right)}{\sin p_1}\frac{\sl{2_+\comma 3_+}}{\sl{3_+\comma 1_+}}\comma\label{eq-62}\\
&\sl{2_+\comma 3_-}=e^{-i(-p_1+p_2+p_3)/2} \frac{\sin \left(\frac{-p_1+p_2-p_3}{2} \right)}{\sin p_3}\frac{\sl{1_+\comma 2_+}}{\sl{3_+\comma 1_+}}\comma\label{eq-63}\\
&\sl{2_-\comma 3_+}=e^{i(-p_1+p_2+p_3)/2} \frac{\sin \left(\frac{p_1+p_2-p_3}{2} 
\right)}{\sin p_2}\frac{\sl{3_+\comma 1_+}}{\sl{1_+\comma 2_+}}\comma\label{eq-64}\\
&\sl{3_+\comma 1_-}=e^{-i(p_1-p_2+p_3)/2} \frac{\sin \left(\frac{-p_1-p_2+p_3}{2} \right)}{\sin p_1}\frac{\sl{2_+\comma 3_+}}{\sl{1_+\comma 2_+}}\comma\label{eq-65}\\
&\sl{3_-\comma 1_+}=e^{i(p_1-p_2+p_3)/2} \frac{\sin \left(\frac{-p_1+p_2+p_3}{2} 
\right)}{\sin p_3}\frac{\sl{1_+\comma 2_+}}{\sl{2_+\comma 3_+}}\period\label{eq-66}
}
From these expressions, we can obtain the WKB-expansion for every Wronskian using the results for $\sl{i_+\comma j_+}$. 

The singular term of the Wronskians is given simply by the leading term in the WKB expansion. For instance, the singular terms for $\sl{i_+\comma j_+}$ and $\sl{i_-\comma j_-}$ at $x=1$ on the $\{u,u,u\}$-sheet is determined from the expansion \eqref{++wkb} and \eqref{--wkb} as
\beq{
\ln \sl{1_+\comma 2_{+}} \overset{x\sim 1}{\sim} \frac{2}{1-x} \int_{\ell_{21}}\sqrt{T}dz\comma \qquad \ln \sl{1_-\comma 2_{-}} \overset{x\sim 1}{\sim} \frac{2}{1-x} \int_{\ell_{12}}\sqrt{T}dz\comma\label{eq-67}\\
\ln \sl{2_+\comma 3_{+}} \overset{x\sim 1}{\sim} \frac{2}{1-x} \int_{\ell_{23}}\sqrt{T}dz\comma \qquad \ln \sl{2_-\comma 3_{-}} \overset{x\sim 1}{\sim} \frac{2}{1-x} \int_{\ell_{32}}\sqrt{T}dz\comma\label{eq-68}\\
\ln \sl{3_+\comma 1_{+}} \overset{x\sim 1}{\sim} \frac{2}{1-x} \int_{\ell_{\hat{3}1}}\sqrt{T}dz\comma\qquad \ln \sl{3_-\comma 1_{-}} \overset{x\sim 1}{\sim} \frac{2}{1-x} \int_{\ell_{1\hat{3}}}\sqrt{T}dz\period\label{eq-69}
}
Then by using \eqref{eq-61}--\eqref{eq-66} we can determine 
 the singular terms for $\sl{i_+\comma j_-}$  on the $\{u,u,u\}$-sheet as 
\beq{
&\ln \sl{1_{+}\comma 2_{-}} \overset{x\sim 1}{\sim}\frac{2\pi i (\kappa_1+\kappa_2 -\kappa_3)}{1-x}+\frac{2}{1-x} \int_{\ell_{\hat{2}\hat{3}}+\ell_{\hat{3}1}}\sqrt{T}dz\comma \label{eq-73}\\
&\ln \sl{1_{-}\comma 2_{+}} \overset{x\sim 1}{\sim}\frac{2\pi i (-\kappa_1 -\kappa_2 +\kappa_3)}{1-x}+ \frac{2}{1-x} \int_{\ell_{1\hat{3}}+\ell_{\hat{3}\hat{2}}}\sqrt{T}dz\comma\label{eq-74}\\
&\ln \sl{2_{+}\comma 3_{-}} \overset{x\sim 1}{\sim}\frac{2\pi i (-\kappa_1+\kappa_2 +\kappa_3)}{1-x}+ \frac{2}{1-x} \int_{\ell_{21}+\ell_{1\hat{3}}}\sqrt{T}dz\comma \label{eq-75}\\
&\ln \sl{2_{-}\comma 3_{+}} \overset{x\sim 1}{\sim}\frac{2\pi i (\kappa_1 -\kappa_2 -\kappa_3)}{1-x}+ \frac{2}{1-x} \int_{\ell_{\hat{3}1}+\ell_{12}}\sqrt{T}dz\comma\label{eq-76}\\
&\ln \sl{3_{+}\comma 1_{-}} \overset{x\sim 1}{\sim}\frac{2\pi i (\kappa_1 -\kappa_2 +\kappa_3)}{1-x}+ \frac{2}{1-x} \int_{\ell_{12}+\ell_{23}}\sqrt{T}dz\comma \label{eq-77}\\
&\ln \sl{3_{-}\comma 1_{+}} \overset{x\sim 1}{\sim} \frac{2\pi i (-\kappa_1 +\kappa_2 -\kappa_3)}{1-x}+ \frac{2}{1-x} \int_{\ell_{32}+\ell_{21}}\sqrt{T}dz\period\label{eq-78}
}

In order to determine the singular terms completely, we also need to understand the singular behavior on other sheets. As already described, this can be
 done by utilizing the fact that  $i_{+}$ and $i_-$ transform  into $i_-$ and $-i_+$ respectively as  one crosses a branch cut associated to $p_i(x)$. 
For instance, applying this rule  we can easily find that the  singular term for $\sl{1_+\comma 2_{+}}$ must behave in the following way on each 
 sheet:
\beq{
&\sl{1_{+}\comma 2_{+}}\overset{x\sim 1}{\sim} \frac{2}{1-x}\int_{\ell_{21}}\sqrt{T}dz &(\text{on the $\{u\comma u\comma\ast\}$-sheet})\comma\\
&\sl{1_{+}\comma 2_{+}}\overset{x\sim 1}{\sim} \frac{2\pi i (\kappa_1+\kappa_2 -\kappa_3)}{1-x}+\frac{2}{1-x} \int_{\ell_{\hat{2}\hat{3}}+\ell_{\hat{3}1}}\sqrt{T}dz &(\text{on the $\{u\comma {l}\comma\ast\}$-sheet})\comma\\
&\sl{1_{+}\comma 2_{+}}\overset{x\sim 1}{\sim}\frac{2\pi i (-\kappa_1 -\kappa_2 +\kappa_3)}{1-x}+ \frac{2}{1-x} \int_{\ell_{1\hat{3}}+\ell_{\hat{3}\hat{2}}}\sqrt{T}dz &(\text{on the $\{{l}\comma u\comma\ast\}$-sheet})\comma\\
&\sl{1_{+}\comma 2_{+}}\overset{x\sim 1}{\sim} \frac{2}{1-x}\int_{\ell_{12}}\sqrt{T}dz &(\text{on the $\{{l}\comma {l}\comma\ast\}$-sheet})\period
}

Combining all these results, it is possible to write down the expression 
on the entire Riemann surface which gives the correct singular behavior
 on the respective sheet. It is given by 
\beq{
&\textsf{Sing}_{+}\left[\sl{1_{+}\comma 2_{+}}\right]=\frac{1}{1-x}\sqrt{\frac{(x-u_1)(x-\bar{u}_1)}{(1-u_1)(1-\bar{u}_1)}}\left(\pi i(\kappa_1 +\kappa_2 -\kappa_3) + 2\int_{\ell_{\hat{1}\hat{2}}+\ell_{\hat{2}\hat{3}}+\ell_{\hat{3}1}}\sqrt{T}dz \right)\nn\\
&\hspace{44pt}+\frac{1}{1-x}\sqrt{\frac{(x-u_2)(x-\bar{u}_2)}{(1-u_2)(1-\bar{u}_2)}}\left(\pi i(-\kappa_1 -\kappa_2 +\kappa_3) + 2\int_{\ell_{23}+\ell_{3\hat{1}}+\ell_{\hat{1}\hat{2}}}\sqrt{T}dz \right)\period\label{sing12++1}
}
Here and hereafter, we will use the notation  $\textsf{Sing}_{\pm}\left[f(x)\right]$ to denote  the  singular term of  $f(x)$ around $x=\pm 1$. 
In an entirely  similar manner, we can determine the terms  singular at $x=-1$ as 
\beq{
&\textsf{Sing}_{-}\left[\sl{1_{+}\comma 2_{+}}\right]=\frac{1}{1+x}\sqrt{\frac{(x-u_1)(x-\bar{u}_1)}{(1-u_1)(1-\bar{u}_1)}}\left(\pi i(-\kappa_1 -\kappa_2 +\kappa_3) + 2\int_{\ell_{\hat{1}\hat{2}}+\ell_{\hat{2}\hat{3}}+\ell_{\hat{3}1}}\sqrt{\bar{T}}d\barz \right)\nn\\
&\hspace{44pt}+\frac{1}{1+x}\sqrt{\frac{(x-u_2)(x-\bar{u}_2)}{(1-u_2)(1-\bar{u}_2)}}\left(\pi i(\kappa_1 +\kappa_2 -\kappa_3) + 2\int_{\ell_{23}+\ell_{3\hat{1}}+\ell_{\hat{1}\hat{2}}}\sqrt{\bar{T}}d\barz \right)
\period\label{sing12++2}
}
Singular terms for other Wronskians at $x=\pm 1$ can be determined in a similar manner.

The remaining issue is the ambiguity of adding a  constant function to the logarithm of the Wronskian. 
Such an ambiguity can be fixed by once more utilizing the property that  $i_{\pm}$ that $i_{+}$ ($i_{-}$) transforms  into $i_{-}$ ($-i_{+}$) as  it crosses the branch cut of $p_i$. 
This leads to the following constraint for the Wronskians
\beq{
 \sl{i_{+}\comma j_{+}}(\hat{\sigma}_i\hat{\sigma}_j x)= \sl{i_{-}\comma j_{-}}(x)\period\label{flipwron}
}
It turns out that all the results obtained so far satisfy \eqref{flipwron}. 
Since this property gets lost upon adding a constant to the logarithm 
of the Wronskian,  it shows that our results  are already complete and we should not add any constant functions. 
%%%%%%%%%%%%%%%%%%%%%%%
%%%%%%%%%%%%%%%%%%%
\section{\bfall{Complete three-point functions at strong coupling}\label{sec:3pt}}
%%%%%%%%%%%%%%%%%%%
Up to the last section, we have developed  necessary methods and 
acquired the knowledge of the various parts that make up the three-point functions of our interest.  Now we are  ready to put  them together and see that they combine in a non-trivial fashion to produce a rather remarkable
 answer. 

First  in subsection \ref{subsec:S}, we obtain the complete result for the $S^3$ part  by putting together the contribution of the action and 
that of the vertex operators. These two contributions combine nicely to produce  a simple expression in terms of integrals on the spectral curve. 
Then, adapting the methods developed for the $S^3$ part, we evaluate in subsection \ref{subsec:AdS} the $E\!AdS_3$ part of the three-point function. 
Our focus will be on the differences between the $S^3$ and $E\!AdS_3$ contributions. Finally in subsection \ref{subsec:full}, we present the full answer by 
combining the contributions of the $S^3$ part and the $E\!AdS_3$ part. We will see  that  the structure of the final answer closely resembles that of the weak coupling result.  Detailed comparison for certain specific  cases will be performed in section \ref{sec:compare}. 
%%%%%%%%%%%%%%%%
\subsection{The $S^3$ part\label{subsec:S}}
%%%%%%%%%%%%%%%%
Before we begin the actual computations, let us  summarize the structure of the contributions 
 from the $S^3$ part to the logarithm of the three-point function, which 
 we denote by $F_{S^3}$.  As was already indicated in section \ref{subsec:gen-three},   $F_{S^3}$ consists of the contribution 
 of the action and that of the vertex operators, namely 
\begin{align}
F_{S^3} &= \calF_{\text{action}} + \calF_{\text{vertex}} \period
\end{align}
Each contribution can be further split into several different pieces  as 
\beq{
&\calF_{\text{action}}=\frac{\sqrt{\lambda}}{6}+\easy + \hard\comma
&\calF_{\text{vertex}}=\kin+\dyn + \calV_{\rm energy} \period
}
Among these terms,  $\easy$, $\kin$ and $\mathcal{V}_{\rm energy}$ have  already been evaluated respectively in \eqref{Api}, \eqref{Vkin} and \eqref{Venergy}.
Thus, our main task will be to compute $\hard$ and $\dyn$. As shown in \eqref{Aeta} and \eqref{Vdyn}, $\hard$ is given by the normal ordered derivatives of the Wronskians, $\no{\del_x \ln \sl{i_{+}\comma j_{+}}}_{\pm}$, whereas $\dyn$ is given by the Wronskians evaluated at $x=0$ and $x=\infty$, $\left.\ln\sl{i_{+}\comma j_{+}}\right|_{\infty}$ and $\left.\ln\sl{i_{-}\comma j_{-}}\right|_{0}$.
From the discussion in section \ref{sec:wron}, we know the Wronskians are comprised of two different parts, the convolution-integral part $\textsf{Conv}\left[\ln\sl{i_\ast\comma j_\ast}\right]$ and the singular part $\textsf{Sing}_{\pm}\left[\ln\sl{i_\ast\comma j_\ast}\right]$. They both contribute to $\hard$ and $\dyn$. In what follows, we examine these two parts separately and evaluate their contributions to $\hard$ and $\dyn$.
%%%%%%%%%%%%%%%%%%%
\subsubsection{Contributions from the convolution integrals}
%%%%%%%%%%%%%%%%%%%%%
We begin with the computation of  the convolution integrals. To illustrate the basic idea,  let us study $\left.\textsf{Conv}\left[\ln\sl{2_+\comma 1_+}\right]\right|_{\infty}$, $\left.\textsf{Conv}\left[\ln\sl{2_-\comma 1_-}\right]\right|_{0}$ and  $\no{\del_x\textsf{Conv}\left[\ln\sl{2_{+}\comma 1_{+}}\right]}_{\pm}$ as representative  examples. 

To compute  the first two quantities,  we need to know on which side of the integration contours the points $x=0$ and $x=\infty$ are located.  
This is because the convolution integrals derived in subsection \ref{subsec:WH} are valid only when $x$ is on the left hand side of the contours.  When $x$ is on the right hand side of the contours, we must include the contact terms, which originate from the integration around $x^{\prime}=x$.  Unfortunately,  the form of the contours depend on the specific details of the solutions we use and hence we cannot give a general discussion. 
We will therefore postpone the discussion of the contact terms until 
 we study several explicit examples in the next section. 
 
 Apart from such contact terms, $\left.\textsf{Conv}\left[\sl{2_+\comma 1_+}\right]\right|_{\infty}$ and  $\left.\textsf{Conv}\left[\sl{2_-\comma 1_-}\right]\right|_{0}$ can be obtained directly from \eqref{eq-49}, \eqref{eq-50}, \eqref{12ppzero2} and \eqref{12mmzero2} by setting the value of $x$ in the convolution kernels $\mathcal{K}_i(x^{\prime};x)$ to be $0$ and $\infty$ respectively. 

Next, consider the evaluation of the normal-ordered derivative  $\no{\del_x\textsf{Conv}\left[\ln\sl{2_{+}\comma 1_{+}}\right]}_{\pm}$. This quantity 
does not  receive contributions from the contact terms since the integration contours pass right through $x=\pm1$ and we can compute $\no{\del_x\textsf{Conv}\left[\ln\sl{2_{+}\comma 1_{+}}\right]}_{\pm}$ always on the left hand side of the contour. In addition, since the convolution integrals are nonsingular at $x=\pm 1$,  as discussed at the end of section \ref{subsec:WH}, the normal ordering is in fact unnecessary. Thus, $\no{\del_x\textsf{Conv}\left[\ln\sl{2_{+}\comma 1_{+}}\right]}_{\pm}$ 
can be obtained from \eqref{eq-49} and \eqref{12ppzero2}  by simply replacing $\mathcal{K}_i(x^{\prime};x)$ with their derivatives $\left.\del_x\mathcal{K}_i(x^{\prime};x)\right|_{x=\pm 1}$.

Applying similar analyses to other Wronskians and using the formulas \eqref{Aeta} and \eqref{Vdyn}, we can obtain  the contributions of the convolution integrals to $\hard$ and $\dyn$, which will be  denoted by  $\textsf{Conv}\left[\hard\right]$ and $\textsf{Conv}\left[\dyn\right]$. They are given by 
\beq{
\textsf{Conv}\left[\hard\right] = &\sqrt{\lambda}\left[\int_{\mathcal{M}^{uuu}_{---}}\sym{\kappa_i\ev{\del_x\mathcal{K}_i}_{+}-\kappa_i\ev{\del_x \mathcal{K}_i}_{-}}_{123} \ast \ln \sin \left(\frac{p_1+p_2+p_3}{2} \right)\right.
\nn\\&+\int_{\mathcal{M}^{uuu}_{--+}}\sym{\kappa_i\ev{\del_x\mathcal{K}_i}_{+}-\kappa_i\ev{\del_x\mathcal{K}_i}_{-}}_{12}^{3} \ast \ln \sin \left(\frac{p_1+p_2-p_3}{2} \right)\nn\\
&+\int_{\mathcal{M}^{uuu}_{-+-}}\sym{\kappa_i\ev{\del_x\mathcal{K}_i}_{+}-\kappa_i\ev{\del_x\mathcal{K}_i}_{-}}_{13}^{2} \ast \ln \sin \left(\frac{p_1-p_2+p_3}{2} \right)\nn\\
&+\int_{\mathcal{M}^{uuu}_{+--}}\sym{\kappa_i\ev{\del_x\mathcal{K}_i}_{+}-\kappa_i\ev{\del_x\mathcal{K}_i}_{-}}_{23}^{1} \ast \ln \sin \left(\frac{-p_1+p_2+p_3}{2} \right)\nn\\
&\left. -2\sum_{j=1}^{3}\int_{\Gamma_{j_-}^u} \left(\kappa_j\ev{\del_x\mathcal{K}_j}_{+}-\kappa_j\ev{\del_x\mathcal{K}_j}_{-}\right) \ast \ln \sin p_j\right]\comma\label{Aomega}
}
\beq{
\textsf{Conv}\left[\dyn\right]= &\int_{\mathcal{M}^{uuu}_{---}}\sym{S_\infty^{i}\left.\mathcal{K}_i\right|_{\infty}+S_0^{i}\left.\mathcal{K}_i\right|_{0}}_{123} \ast \ln \sin \left(\frac{p_1+p_2+p_3}{2} \right)
\nn\\&+\int_{\mathcal{M}^{uuu}_{--+}}\sym{S_\infty^{i}\left.\mathcal{K}_i\right|_{\infty}+S_0^{i}\left.\mathcal{K}_i\right|_{0}}_{12}^{3} \ast \ln \sin \left(\frac{p_1+p_2-p_3}{2} \right)\nn\\
&+\int_{\mathcal{M}^{uuu}_{-+-}}\sym{S_\infty^{i}\left.\mathcal{K}_i\right|_{\infty}+S_0^{i}\left.\mathcal{K}_i\right|_{0}}_{13}^{2} \ast \ln \sin \left(\frac{p_1-p_2+p_3}{2} \right)\nn\\
&+\int_{\mathcal{M}^{uuu}_{+--}}\sym{S_\infty^{i}\left.\mathcal{K}_i\right|_{\infty}+S_0^{i}\left.\mathcal{K}_i\right|_{0}}_{23}^{1} \ast \ln \sin \left(\frac{-p_1+p_2+p_3}{2} \right)\nn\\
& -2\sum_{j=1}^{3}\int_{\Gamma_{j_-}^u} \left(S_\infty^{j}\left.\mathcal{K}_j\right|_{\infty}+S_0^{j}\left.\mathcal{K}_j\right|_{0}\right) \ast \ln \sin p_j\period\label{convVdyn}
}
To simplify the expressions, we have introduced 
 the double bracket notation $\sym{\star }$, to denote   sum of 
 three terms with designated  combinations of signs,  defined as 
\beq{
\sym{a_i}_{123}=a_1 +a_2 +a_3\comma \quad \sym{a_i}_{12}^3=a_1 +a_2 -a_3\comma \quad \text{etc.}\comma\label{def-db}
}
Also, we have employed  the abbreviated symbols $\ev{\del_{x}\mathcal{K}_i}_{\pm}$, $\left.\mathcal{K}_i\right|_{\infty}$ and $\left.\mathcal{K}_i\right|_{0}$, 
which are defined by 
\beq{
\ev{\del_{x}\mathcal{K}_i}_{\pm}\equiv \left.\del_x \mathcal{K}_i(x^{\prime};x)\right|_{x=\pm 1}\comma \quad \left.\mathcal{K}_i\right|_{\infty}\equiv \mathcal{K}_i(x^{\prime};\infty)\comma \quad \left.\mathcal{K}_i\right|_{0}\equiv \mathcal{K}_i(x^{\prime};0)\period\label{pminf0}
}

It turns out that the two contributions  \eqref{Aomega} and \eqref{convVdyn}
combine  to give a remarkably simple  expression displayed below. 
This is due to the crucial relation of the form 
\beq{
\sqrt{\lambda}\kappa_i\ev{\del_x \mathcal{K}_i}_{+}-\sqrt{\lambda}\kappa_i\ev{\del_x\mathcal{K}_i}_{-}+S_{\infty}^i\left.\mathcal{K}_i\right|_{\infty}+S_0^{i}\left.\mathcal{K}_i\right|_{0} = z(x^{\prime}) \frac{dp_i(x^{\prime})}{dx^{\prime}}\comma \label{nice}
}
where   $z(x)$ on the right hand side is  the  Zhukovsky variable,  defined  
 in \eqref{Zhukvar}.  
Although this equality  can be verified by a direct computation using the explicit form of $p_i(x)$ for the one-cut solutions given in  \eqref{onecutpx}, it is 
 important to give a more intuitive and essential understanding.  Note that the right hand side of \eqref{nice}  is proportional to the integrand of the filling fraction given in \eqref{fillfrac}. 
Therefore when integrated over  appropriate  $a$-type cycles, it produces 
 the corresponding conserved charges. In other words, it is characterized by the singularities associated with such charges.  Now observe that the left hand side 
precisely consists of terms which provide such singularities. The first two 
terms are responsible for  the singularities at $x=\pm 1$, while the last two terms  contain the poles at $x=\infty$ and $x=0$ associated with the 
 charges $S_\infty^i$ and $S_0^i$ respectively. Furthermore, it should be emphasized 
 that the formula above {\it unifies} the contributions in two sense of the word. First,   it  unites the contributions from the action, 
represented by the first two terms, and those from the vertex operators, 
represented by the last two terms. Only when they are put together one can 
 reproduce all the singularities of the right hand side.  Second, the expression obtained 
 on the right hand side is universal in that all the specific data shown on the left hand side, namely $\kappa_i, S^i_\infty$ and $S^i_0$, are contained in one quantity $p_i(x)$.  As we shall discuss  in section \ref{subsec:AdS}, this feature allows us to write down the same form of the result (except for an overall sign) given by 
 the right hand side of \eqref{nice} for the contributions from the $E\!AdS_3$ part, using 
the quasi-momentum for that part of the string.

Now,  applying \eqref{nice}  we can rewrite the sum  $\mathcal{T}_{\rm conv}\equiv \textsf{Conv}\left[\hard\right] + \textsf{Conv}\left[\dyn\right] $ into the following 
 compact expression:
\beq{
\mathcal{T}_{\rm conv}= &\int_{\mathcal{M}^{uuu}_{---}}\frac{z(x)\left( dp_1 +dp_2 +dp_3\right)}{2\pi i} \ln \sin \left(\frac{p_1+p_2+p_3}{2} \right)
\nn\\&+\int_{\mathcal{M}^{uuu}_{--+}}\frac{z(x)\left( dp_1 +dp_2 -dp_3\right)}{2\pi i} \ln \sin \left(\frac{p_1+p_2-p_3}{2} \right)\nn\\
&+\int_{\mathcal{M}^{uuu}_{-+-}}\frac{z(x)\left( dp_1 -dp_2 +dp_3\right)}{2\pi i} \ln \sin \left(\frac{p_1-p_2+p_3}{2} \right)\nn\\
&+\int_{\mathcal{M}^{uuu}_{+--}}\frac{z(x)\left( -dp_1 +dp_2 +dp_3\right)}{2\pi i} \ln \sin \left(\frac{-p_1+p_2+p_3}{2} \right)\nn\\
& -2\sum_{j=1}^{3}\int_{\Gamma_{j_-}^u} \frac{z(x) \,dp_j}{2\pi i} \ln \sin p_j + \textsf{Contact}\period\label{totalconv}
}
In the last line, we included  the possible contributions from the contact terms, 
denoted by \textsf{Contact}.
%%%%%%%%%%%%%%%%%
\subsubsection{Contributions from the singular part of the Wronskians}
%%%%%%%%%%%%%%%%%%
We now turn to the computation of the  singular part $\textsf{Sing}_{\pm}\left[\ln\sl{i_\ast\comma j_\ast}\right]$. By substituting the expressions
 for  the singular part of the Wronskians, such as \eqref{sing12++1} and \eqref{sing12++2}, into the formulas \eqref{Aeta} and \eqref{Vdyn}, we can  evaluate the contributions of  the singular part in a straightforward manner. 
From this calculation, we find that a part of the terms contribute only to 
 the  overall phase of  the three-point functions.  For instance, the first and the third term in \eqref{sing12++1}, which are proportional to $\pm \pi i ( \kappa_1 + \kappa_2 - \kappa_3)$, will only yield an  overall phase owing to the  factor of $\pi i$. Just as before,  we will ignore such   contributions 
 in this work. 
 Then the contributions of  $\textsf{Sing}_{+}\left[\ln \sl{i_{\ast}\comma j_{\ast}}\right]$ to $\hard$ and $\dyn$,   denoted by $\textsf{Sing}_{+}\left[\hard\right]$ and $\textsf{Sing}_{+}\left[\dyn \right]$, are obtained as 
\beq{
\textsf{Sing}_{+}\left[\hard\right]= \sqrt{\lambda}&\left[ \sym{\kappa_i\no{\del_x\mathcal{K}_i(1;x)}_{+}-\kappa_i\left.\del_x \mathcal{K}_i(1;x)\right|_{-}}_{12}^3\int_{\ell_{21}}\varpi\right.\nn\\
& + \sym{\kappa_i\no{\del_x\mathcal{K}_i(1;x)}_{+}-\kappa_i\left.\del_x \mathcal{K}_i(1;x)\right|_{-}}_{23}^{1}\int_{\ell_{23}}\varpi\nn\\
&+\left.\sym{\kappa_i\no{\del_x\mathcal{K}_i(1;x)}_{+}-\kappa_i\left.\del_x \mathcal{K}_i(1;x)\right|_{-}}_{13}^{2}\int_{\ell_{\hat{3}1}}\varpi\right]\comma\label{A-sing+}
}
and
\beq{
\textsf{Sing}_{+}\left[\dyn\right]= &\left[\sym{S_\infty^{i}\left.\mathcal{K}_i\right|_{\infty}+S_0^{i}\left.\mathcal{K}_i\right|_{0}}_{12}^3 \int_{\ell_{21}}\varpi +\sym{S_\infty^{i}\left.\mathcal{K}_i\right|_{\infty}+S_0^{i}\left.\mathcal{K}_i\right|_{0}}_{23}^1 \int_{\ell_{23}}\varpi\right.\nn\\
&\left.\left. +\sym{S_\infty^{i}\left.\mathcal{K}_i\right|_{\infty}+S_0^{i}\left.\mathcal{K}_i\right|_{0}}_{13}^2 \int_{\ell_{\hat{3}1}}\varpi\right]\right|_{x^{\prime}=+1}\period\label{V-sing+}
}
Note that  in the present case, in contrast to the case of $\no{\del_x\textsf{Conv}\left[\ln\sl{i_{\ast}\comma j_{\ast}}\right]}_{\pm}$ discussed previously,  the normal ordering in  $\no{\del_x\mathcal{K}_i(1;x)}_{+}$ is necessary  since $\del_x \mathcal{K}_i(1;x)$ is singular at $x= 1$.
In an entirely similar manner,  the contributions of $\textsf{Sing}_{-}\left[\ln \sl{i_{\ast}\comma j_{\ast}}\right]$ to $\hard$ and $\dyn$,  denoted by $\textsf{Sing}_{-}\left[\hard\right]$ and $\textsf{Sing}_{-}\left[\dyn \right]$, are 
computed as 
\beq{
\textsf{Sing}_{-}\left[\hard\right]= -\sqrt{\lambda}&\left[ \sym{\kappa_i\left.\del_x\mathcal{K}_i(-1;x)\right|_{+}-\kappa_i\no{\del_x \mathcal{K}_i(-1;x)}_{-}}_{12}^3\int_{\ell_{21}}\bar{\varpi} \right.\nn\\
&+ \sym{\kappa_i\left.\del_x\mathcal{K}_i(-1;x)\right|_{+}-\kappa_i\no{\del_x \mathcal{K}_i(-1;x)}_{-}}_{23}^{1}\int_{\ell_{23}}\bar{\varpi}\nn\\
&+\left.\sym{\kappa_i\left.\del_x\mathcal{K}_i(-1;x) \right|_{+}-\kappa_i\no{\del_x \mathcal{K}_i(-1;x)}_{-}}_{13}^{2}\int_{\ell_{\hat{3}1}}\bar{\varpi}\right]\comma\label{A-sing-}
}
and
\beq{
\textsf{Sing}_{-}\left[\dyn\right]= -&\left[\sym{S_\infty^{i}\left.\mathcal{K}_i\right|_{\infty}+S_0^{i}\left.\mathcal{K}_i\right|_{0}}_{12}^3 \int_{\ell_{21}}\bar{\varpi}+\sym{S_\infty^{i}\left.\mathcal{K}_i\right|_{\infty}+S_0^{i}\left.\mathcal{K}_i\right|_{0}}_{23}^1 \int_{\ell_{23}}\bar{\varpi}\right.\nn\\
&\left.\left.\sym{S_\infty^{i}\left.\mathcal{K}_i\right|_{\infty}+S_0^{i}\left.\mathcal{K}_i\right|_{0}}_{13}^2 \int_{\ell_{\hat{3}1}}\bar{\varpi}\right]\right|_{x^{\prime}=-1}\period\label{V-sing-}
}

Now just as we did for  $\textsf{Conv}\left[\hard\right] + \textsf{Conv}\left[\dyn\right] $, we can make use of the relation \eqref{nice} to 
 rewrite the sum $\textsf{Sing}_{\pm}\left[\hard\right]+\textsf{Sing}_{\pm}\left[\dyn\right]$ into  much simpler forms. The results are 
\beq{
\textsf{Sing}_{+}\left[\hard\right]+\textsf{Sing}_{+}\left[\dyn\right]=&   \no{z(x)\left(\frac{dp_1}{dx}+\frac{dp_2}{dx}-\frac{dp_3}{dx}\right)}_{+}\int_{\ell_{21}}\varpi   \nn\\
& + \no{z(x)\left(\frac{dp_1}{dx}-\frac{dp_2}{dx}+\frac{dp_3}{dx}\right)}_{+}\int_{\ell_{\hat{3}1}}\varpi\nn\\
&+\no{z(x)\left(-\frac{dp_1}{dx}+\frac{dp_2}{dx}+\frac{dp_3}{dx}\right)}_{+}\int_{\ell_{23}}\varpi\comma \label{totalsing+}
}
and 
\beq{
\textsf{Sing}_{-}\left[\hard\right]+\textsf{Sing}_{-}\left[\dyn\right]=& - \no{z(x)\left(\frac{dp_1}{dx}+\frac{dp_2}{dx}-\frac{dp_3}{dx}\right)}_{-}\int_{\ell_{21}}\bar{\varpi}\nn\\ 
&- \no{z(x)\left(\frac{dp_1}{dx}-\frac{dp_2}{dx}+\frac{dp_3}{dx}\right)}_{-}\int_{\ell_{\hat{3}1}}\bar{\varpi}\nn\\
& -\no{z(x)\left(-\frac{dp_1}{dx}+\frac{dp_2}{dx}+\frac{dp_3}{dx}\right)}_{-}\int_{\ell_{23}}\bar{\varpi}\period\label{totalsing-}
}
The expressions  $\no{z(x)\,dp_i/dx}_{\pm}$ in \eqref{totalsing+} and \eqref{totalsing-} above can be evaluated using the explicit form of the quasi-momentum,   given in \eqref{onecutpx}, as\fn{Definitions of $\Lambda_i$ and $\bar{\Lambda}_i$ are given in \eqref{gi} and \eqref{bargi}.}
\beq{
\no{z(x)\frac{dp_i}{dx}}_{+}=-2\pi\kappa_i -\pi \kappa_i \Lam_i\comma \quad \no{z(x)\frac{dp_i}{dx}}_{-}=2\pi \kappa_i + \pi \kappa_i \bar{\Lam}_i\period\label{explicit}
}
This provides  fairly explicit forms for the expressions $\textsf{Sing}_{\pm}\left[\hard\right]+\textsf{Sing}_{\pm}\left[\dyn\right]$. 
%%%%%%%%%%%%%%
\subsubsection{Result for the $S^3$ part}
%%%%%%%%%%%%%%%
We can now combine the results obtained so far and obtain the net contribution of  the $S^3$ part. Recall that the general structure of the $S^3$ part of the three-point functions we have computed is of the form 
\beq{
F_{S^3}=&\frac{\sqrt{\lambda}}{6}+2\sqrt{\lambda}\sum_{i=1}^3\kappa_i^2 \ln \epsilon_i +\easy +\kin + \textsf{Conv}\left[\hard\right]+ \textsf{Conv}\left[\dyn\right]\nn\\
&+\textsf{Sing}_{+}\left[\hard\right]+\textsf{Sing}_{+}\left[\dyn\right]+\textsf{Sing}_{-}\left[\hard\right]+\textsf{Sing}_{-}\left[\dyn\right]\period\label{s3full1}
}
Among the various terms  shown above, those which can be expressed in terms of  the contour integrals of $\varpi$ or $\bar{\varpi}$ can be combined and evaluated 
 using the explicit form of $\no{z\,dp_i/dx}_{\pm}$ given in \eqref{explicit}. 
The result is 
\beq{
\mathcal{T}_{\rm sing}\equiv &\easy +\textsf{Sing}_{+}\left[\hard\right]+\textsf{Sing}_{+}\left[\dyn\right]+\textsf{Sing}_{-}\left[\hard\right]+\textsf{Sing}_{-}\left[\dyn\right]\nn\\
=&-\frac{\sqrt{\lambda}}{2}\left[(\kappa_1 +\kappa_2 -\kappa_3)\int_{\ell_{21}}(\varpi +\bar{\varpi})+(\kappa_1 -\kappa_2 +\kappa_3)\int_{\ell_{\hat{3}1}} (\varpi +\bar{\varpi})\right.\nn\\
&\left.+(-\kappa_1 +\kappa_2 +\kappa_3)\int_{\ell_{23}} (\varpi +\bar{\varpi})\right]\period\label{Tlam}
}
Since $\varpi$ and $\bar{\varpi}$ behave near the punctures as
\beq{
\varpi \to \frac{\kappa_i}{z-z_i}\comma \quad \bar{\varpi} \to \frac{\kappa_i}{\barz-\barz_i}\comma\quad (z\to z_i)\quad \text{for }i=1,\bar{2},3\comma
}
the expression \eqref{Tlam} diverges in the following fashion as the regularization parameters $\epsilon_i$'s tend  to zero:
\beq{
\mathcal{T}_{\rm sing}\to -2\sqrt{\lambda}\sum_{i=1}^{3} \kappa_i^2 \ln \epsilon_i=-\mathcal{V}_{\rm energy}\period 
}
Notice, however, that this divergence  is precisely canceled by the second term 
of  \eqref{s3full1}. Therefore, the quantity \eqref{s3full1} as a whole is finite 
in the limit $\epsilon_i \rightarrow 0$.  This is as expected for correctly normalized three-point functions.

Let us summarize the  final result for the logarithm of the three-point functions
 coming from  the $S^3$ part.  It can be written in the form 
\beq{
F_{S^3}=\frac{\sqrt{\lambda}}{6}+\mathcal{V}_{\rm energy} + \mathcal{T}_{\rm sing}+\kin +\mathcal{T}_{\rm conv}\comma\label{s3result}
}
where $\kin$ is the kinematical factor depending only on the normalization 
 vectors given in \eqref{Vkin},  $\mathcal{T}_{\rm conv}$ is  the sum of the contributions from the convolution integrals  \eqref{totalconv}, and  $\mathcal{T}_{\rm sing}$, which is  given  in \eqref{Tlam},  represents  the sum of $\mathcal{A}_{\varpi}$  defined in \eqref{Api}  and the contributions from the singular parts of the Wronskians.
%%%%%%%%%%%%%%%%%%%%
\subsection{The $E\!AdS_3$ part\label{subsec:AdS}}
%%%%%%%%%%%%%%%%%%%
We now discuss the contributions from the $E\!AdS_3$ part. 
Since the logic of the evaluation  is almost entirely similar,  we will not repeat the long analysis we performed for the $S^3$ part. In fact  it suffices to 
explain  which part of the analysis for the $S^3$ part can be ``copied" 
and which part has to  be  modified. 
%%%%%%%%%%%%%%%%%%%%%%%%
\subsubsection{Contribution  from the action}
%%%%%%%%%%%%%%%%%%%%%%%
Let us begin with the contribution from the action integral. 
Since $E\!AdS_3$ and $S^3$ are formally quite similar, the computation 
 of the action integral can be performed in exactly the same manner. 
There is, however, a simple but crucial difference. It is the overall 
 sign of the integral.  For $E\!AdS_3$, the counterpart of the matrix $\bbY$ 
 shown in \eqref{bbY} is  given by 
\begin{align}
\bbX &\equiv \matrixii{X_+}{X}{X_-}{\Xbar} \comma 
\end{align}
where 
\begin{align}
X_\pm &\equiv X^{-1} \pm  X^4\comma \comma \qquad X \equiv X^1+iX^2\comma \qquad \Xbar \equiv X^1-iX^2 \period
\end{align}
The right current is then defined as 
\begin{align}
\jhat &\equiv \bbX^{-1} d\bbX  = \jhat_z dz + \jhat_\zbar d\zbar \period
\end{align}
Now compare the expressions of the stress tensors  and 
 the action integrals  for $S^3$ and $E\!AdS_3$,  expressed in terms of 
 the respective right current. They are given by 
\begin{align}
T(z) &\equiv T_{AdS}(z) =  \half \tr (\jhat_z \jhat_z) = 
\kappa^2  \comma 
\quad T_{S}(z) = -\half\tr(j_z j_z) = -\kappa^2 \comma \\
S_{E\! AdS_3} &= {\sqrt{\lam} \over 2\pi} 
\int d^2z \,\tr(\jhat_z \jhat_\zbar) \comma \qquad 
S_{S^3} = -{\sqrt{\lam} \over 2\pi} 
\int d^2z \,\tr(j_z j_\zbar) \period
\end{align}
This shows that while we have the equality $\tr (\jhat_z \jhat_z) = \tr(j_z j_\zbar) =\kappa^2$, the signs in front of the action integrals 
are opposite. Therefore all the results for the action integral are formally 
 the same as those for the $S^3$ case, but with opposite signs. 
This will lead to various cancellations with the contributions from the $S^3$ 
 part, as we shall see shortly. 
%%%%%%%%%%%%%%%%%%%%%%
\subsubsection{Contribution from the wave function }
%%%%%%%%%%%%%%%%%%%%%%%%
As for the evaluation of the contribution from the wave function,  the basic  logic of the formalism developed in section \ref{sec:vertex} for the  $S^3$  still applies. 
However, there are a few important modifications, as we shall explain below. 

%%%%%%%%%%%%%%%
As  discussed in our previous work \cite{KK2},   in the case of a string in $E\! AdS_3$ 
  the global symmetry group is $\SLCR \times \SLCL$ and hence the 
 the raising operators with respect to which we define the highest weight state 
are the left and the right special conformal transformations given by 
\begin{align}
V_R^{\rm sc} &= \matrixii{1}{0}{\be_R}{1} \comma \qquad 
V_L^{\rm sc} = \matrixii{1}{\be_L}{0}{1}  \comma 
\end{align}
where $\be_R$ and $\be_L$ are constants.  
Applying our general argument for the determination of the polarization 
 spinors, we readily find 
\begin{align}
(V_R^{\rm sc})^t \ndiag &= \ndiag \comma \qquad \ndiag = \vecii{1}{0} \comma  \label{nAdS}\\
(V_L^{\rm sc})^t \ntildiag &= \ntildiag \comma \qquad \ntildiag 
 = \vecii{0}{1} \period \label{ntilAdS}
\end{align}
It should be noted that, compared to the $S^3$ case  given in (\ref{nZntilZ}), 
 $\ndiag$ here for the right sector is the same as  $\ntil^Z$ for the left sector there 
and similarly  $\ntil$ for the left sector in the present case is 
 identical to $n^Z$ for the right sector for the $S^3$ case.
Now the algebraic manipulations for the construction of the wave 
 functions are the same as for the $S^3$ case up to the computation of the factor 
 $e^{i \Delta \phi}$. Therefore, for the right sector, we get the same result 
 for the $Z$-type operator in the left sector, given in (\ref{Zleft}). For example at $z_1$ we have 
\begin{align}
e^{i\Delta \phi_{R,1}} &= a_1^{-2} = {\wrons{1_+}{2_+} 
\wrons{3_+}{1_+} 
\over \wrons{2_+}{3_+}}\bigg|_\infty  { \wrons{n_2}{n_3}
\over \wrons{n_1}{n_2} \wrons{n_3}{n_1}} 
\end{align}
This is the {\it inverse} of the result  for $S^3$ obtained in (\ref{expshiftR})
with $i_-$ replaced by $i_+$. 
The  result for the left sector is similar. What this means is that 
 the wave function for the $E\! AdS_3$ is obtained from the one for the $S^3$ case 
by (i) reversing the sign of the powers and (ii) exchanging $i_+$ and  $i_-$.  
Abusing the same notations for the 
polarization spinors and the eigenvectors as in the $S^3$ case, 
 we get
\begin{align}
\Psi_R^{E\! AdS_3} &= \prod_{i \ne j \ne k} \left({\wrons{n_i}{n_j} \over \wrons{i_+}{j_+}\big|_\infty}
\right)^{-(R_i+R_j -R_k)}  \comma \label{AdSR}\\
\Psi_L^{E\! AdS_3}&=\prod_{i \ne j \ne k} \left({\wrons{\ntil_i}{\ntil_j} \over \wrons{\itil_-}{\jtil_-}\big|_0}
\right)^{-(L_i+L_j -L_k)} \comma  \label{AdSL}
\end{align}
where $R_i$ and $L_i$ here are the combinations of the conformal dimension
 $\Delta_i$ and the spin $S_i$ given by 
\begin{align}
R_i &= \frac{1}{2}\left( \Delta_i-S_i\right) \comma \qquad L_i = \frac{1}{2}\left( \Delta_i +S_i\right)
\period
\end{align}
This reversal of power relative to the $S^3$ case is what is desired. 
Effectively it is equivalent to employing  $e^{+iS \phi}$ as the form of the 
wave function, which is what we adopted in the previous work\cite{KK2} for 
 the three-point function of the GKP string in \EAdS3 and lead to 
the power structure given in (\ref{AdSR}) and (\ref{AdSL}). As we shall show below, correctness of this power structure becomes obvious when we relate the Wronskian $\wrons{n_i}{n_j} $ to  the difference of  the coordinates
 $x_i$ and $x_j$,  where $x_i$ is the position of the $i$-th vertex  operator on the boundary of $E\! AdS_3$.

Recall that  the  embedding coordinates of $E\! AdS_3$ are taken to be   $X^\mu$ $(\mu = -1,1,2,4)$, which is a vector of $SO(1,3)$ with 
signature  $(-,+,+,+)$, while  the  Poincar\'e coordinates are given by $z=1/(X^{-1} + X^4)$, $x^r = z X^r$, $(r=1,2)$, with which  $X^{-1} -X^4$ 
is expressed as $z+(\vecx^2/z)$.  Consider approaching a point on the boundary $z=0$ with finite values of $x^r$.  Then the term $z$ in $X^{-1} -X^4$ becomes  negligible compared to  $\vecx^2/z$ and  $X^\mu$ approaches a null vector,  with large components. Such a vector can be parametrized, up to an overall scale,  by the boundary coordinates  $\vecx = (x^1, x^2)$  as 
\begin{align}
X^{-1} &= \half (1+\vecx^2)\comma \qquad X^4 = \half (1-\vecx^2)
\comma   \qquad (X^1, X^2) = \vecx \comma \\
\vecx^2 &= x^r \eta_{rs} x^s = x^r x_r\comma \qquad \eta_{rs} =(+,+)
\comma \quad r,s=1,2  \period
\end{align}
As usual, one can map $X^\mu$  to the matrix $X^\mu \Sighat_\mu$, with $\Sighat_\mu 
 = (1,  \sig_1, \sig_2, \sig_3)$, which transforms  from left under  $SL(2,C)$ and from right under $SL(2,C)^\ast$. 
Then, it is well-known that  for a null vector $X^\mu$  the matrix elements of $X^\mu \Sighat_\mu$   can be written as a product of spinors 
(or twistors) as 
\begin{align}
(X^\mu \Sighat_\mu)_{\al\aldot}
 &  =\matrixii{1}{x}{\bar{x}}{\vecx^2} = (\sig_1 \ntil)_\al n_\aldot 
 \comma 
\end{align}
where 
\begin{align} 
&x \equiv x^1+ix^2\comma \quad \bar{x} \equiv  x^1-ix^2   \comma \\
&n = \vecii{1}{x} \comma \qquad 
\ntil = \vecii{\xbar}{1} \period
\end{align}
These spinors can be identified precisely as the polarization spinors characterizing a vertex operator  which is placed  at $\vecx$ on the boundary for the following reasons. 
 First they 
 transform in the correct way: Under the global transformation 
$X^\mu \Sighat_\mu  \rightarrow V_L( X^\mu \Sighat_\mu) V_R$, 
we have 
$(\sig_1 \ntil )_\al \rightarrow (V_L \sig_1\ntil)_\al$ and $n_\aldot 
 \rightarrow (n V_R)_\aldot$. This is equivalent to $\ntil \rightarrow V_L^t \ntil$ and $n \rightarrow V_R^t n$, which are the right transformation laws. 
Second, these spinors coincide with the polarization spinors given in 
(\ref{nAdS}) and (\ref{ntilAdS}) when we bring the point $\vecx$ to the origin of the boundary by the translation by the vector $-\vecx$. This is 
 effected by the right and the left  translation matrices given by 
\begin{align}
V_R^{\rm tr}(-x) &= \matrixii{1}{-x}{0}{1}  \comma \qquad 
V_L^{\rm tr}(-\bar{x}) = \matrixii{1}{0}{-\bar{x}}{1} 
\period
\end{align}
Then we get
\begin{align}
(V_R^{\rm tr})^t(-x) n &= \vecii{1}{0} \comma \qquad 
(V_L^{\rm tr})^t(-\xbar) \ntil  = \vecii{0}{1}  
\period
\end{align}
Therefore  $n$ and $\ntil$ can be identified with 
  the polarization spinors for the vertex operator at $\vecx$ on the boundary. Now let $n'$ and $\ntil'$ be similar polarization spinors corresponding to 
 a vertex operator at $\vec{x'}$ on the boundary. Then we immediately get 
\begin{align}
\wrons{n}{n'} &=  x'-x 
\comma  \qquad 
\wrons{\ntil}{\ntil'} =\xbar' -\xbar \comma \\
\wrons{n}{n'}\wrons{\ntil}{\ntil'} &= (x'-x)( \xbar' -\xbar) =(x'-x)^2 \period
\end{align}
In this way, for the $E\! AdS_3$  the Wronskians formed by the polarization spinors produce the difference of the boundary  position vectors. Therefore 
the relevant  part of the wave function becomes 
\begin{align}
& \prod_{\{ i,j,k\}} \wrons{n_i}{n_j}^{- (R_i+R_j -R_k)} 
\wrons{\ntil_i}{\ntil_j}^{- (L_i+L_j -L_k)}  \nn\\
&=\prod_{\{ i,j,k \}} (x_i-x_j)^{- (R_i+R_j -R_k)}
(\xbar_i-\xbar_j)^{- (L_i+L_j -L_k)} \period
\end{align}
In particular, for the case of spinless configurations that we are considering, 
this becomes
\begin{align}
\prod_{\{ i,j,k\}} {1\over |x_i -x_j|^{\Delta_i + \Delta_j -\Delta_k}}
\comma \label{kinads}
\end{align}
which exhibits the familiar coordinate dependence for the three-point function in such a case.  
%%%%%%%%%%%%%%%%%%%%%%%%%
\subsubsection{Total contribution from the  $E\!AdS_3$ part}
%%%%%%%%%%%%%%%%%%%%%%%%%%
As we have seen, the structure of the contribution from the $E\!AdS_3$ part 
 is essentially the same as that from the $S^3$ case, except for 
 the important reversal of signs in the powers in the contributing factor (or the terms contributing  to the the logarithm of the three-point coupling.) 
This change of sign occurred both for the action and for the wave function. 
As we compute the basic  Wronskians in exactly the same way as before 
and use them to compute  the contributions to the logarithm of the three-point  function from  the action part and the wave function part, we again 
 obtain  the expression of the form of the left hand side of \eqref{nice}, with 
 the overall sign reversed. Therefore, we can use the identity \eqref{nice}
 again to obtain the result  $-z(x') d\hat{p}_i(x') /dx'$, 
 where $\hat{p}_i$ denotes the quasi-momentum for the $E\!AdS_3$ part 
 of the string.  One can check that in fact this rule of correspondence, 
 namely $p_i(x) \rightarrow \hat{p}_i(x)$ and the reversal of sign for the convolution integrals, applies  to all the contributions. Thus, combining 
 all the results for the AdS part,  the contribution to the logarithm of 
 the three-point function is given by the following expression:
\beq{
F_{E\!AdS_3}=-\frac{\sqrt{\lambda}}{6}+\hat{\mathcal{V}}_{\rm energy} + \hat{\mathcal{T}}_{\rm sing}+\hat{\calV}_{\rm kin}  +\hat{\mathcal{T}}_{\rm conv}\period\label{ads3result}
}
Here, $\hat{\mathcal{V}}_{\rm energy}$ and $\hat{\mathcal{T}}_{\rm sing}$ are equal to $-\mathcal{V}_{\rm energy}$ and $-\mathcal{T}_{\rm sing}$ respectively,  $\hat{\calV}_{\rm kin}$
is the kinematical factor given in \eqref{kinads}, and $\hat{\mathcal{T}}_{\rm conv}$  is the convolution integrals obtained from the unhatted counterpart 
 for the $S^3$ case with the substitution rule described above. 
%%%%%%%%%%%%%%%%%%%%%%%%%
\subsection{Complete  expression  for the three-point function \label{subsec:full}}
%%%%%%%%%%%%%%%%%%%%%%%%%
We are finally ready to put together the contributions from the 
 $S^3$ part summarized in \eqref{s3result} and those from the $E\!AdS_3$ part given in \eqref{ads3result} and present  the full answer for the three-point function.  As we have already discussed, the divergent terms cancel with 
 each other for the $S^3$ part and the $E\!AdS_3$ part  separately. 
On the other hand, the constant terms proportional to $\sqrt{\lam}/6$ 
cancel between $S^3$ and $E\!AdS_3$ contributions. Thus we are left with 
the kinematical factors and  the contributions from the convolution integrals 
which are of the same structure except for the overall sign. 
Therefore, factoring the kinematical structure  as 
\beq{
\begin{aligned}
\langle \mathcal{V}_1 \mathcal{V}_2 \mathcal{V}_3  \rangle = &\frac{1}{N}\frac{C_{123}}{|x_1-x_2|^{\Delta_1 +\Delta_2-\Delta_3}|x_2-x_3|^{\Delta_2 +\Delta_3-\Delta_1}|x_3-x_1|^{\Delta_3 +\Delta_1-\Delta_2}}\\
&\times \sl{n_1\comma n_2}^{R_1 +R_2-R_3}\sl{n_2\comma n_3}^{R_2 +R_3-R_1}\sl{n_3\comma n_1}^{R_3 +R_1-R_2}\\
&\times\sl{\tilde{n}_1\comma \tilde{n}_2}^{L_1 +L_2-L_3}\sl{\tilde{n}_2\comma \tilde{n}_3}^{L_2 +L_3-L_1}\sl{\tilde{n}_3\comma \tilde{n}_1}^{L_3 +L_1-L_2}\comma
\end{aligned}
\label{finalkin}
}
the logarithm of the structure constant $C_{123}$ is finally given by 

{\footnotesize 
\beq{
&\ln C_{123}=\nn\\
&\int_{\mathcal{M}^{uuu}_{---}}\hspace{-20pt}\frac{z(x)\left( dp_1 +dp_2 +dp_3\right)}{2\pi i} \ln \sin \left(\frac{p_1+p_2+p_3}{2} \right)
+\int_{\mathcal{M}^{uuu}_{--+}}\hspace{-20pt}\frac{z(x)\left( dp_1 +dp_2 -dp_3\right)}{2\pi i} \ln \sin \left(\frac{p_1+p_2-p_3}{2} \right)\nn\\
&+\int_{\mathcal{M}^{uuu}_{-+-}}\hspace{-20pt}\frac{z(x)\left( dp_1 -dp_2 +dp_3\right)}{2\pi i} \ln \sin \left(\frac{p_1-p_2+p_3}{2} \right)+\int_{\mathcal{M}^{uuu}_{+--}}\hspace{-20pt}\frac{z(x)\left( -dp_1 +dp_2 +dp_3\right)}{2\pi i} \ln \sin \left(\frac{-p_1+p_2+p_3}{2} \right)\nn\\
&-\int_{\hat{\mathcal{M}}^{uuu}_{---}}\hspace{-20pt}\frac{z(x)\left( d\hat{p}_1 +d\hat{p}_2 +d\hat{p}_3\right)}{2\pi i} \ln \sin \left(\frac{\hat{p}_1+\hat{p}_2+\hat{p}_3}{2} \right)
-\int_{\hat{\mathcal{M}}^{uuu}_{--+}}\hspace{-20pt}\frac{z(x)\left( d\hat{p}_1 +d\hat{p}_2 -d\hat{p}_3\right)}{2\pi i} \ln \sin \left(\frac{\hat{p}_1+\hat{p}_2-\hat{p}_3}{2} \right)\nn\\
&-\int_{\hat{\mathcal{M}}^{uuu}_{-+-}}\hspace{-20pt}\frac{z(x)\left( d\hat{p}_1 -d\hat{p}_2 +d\hat{p}_3\right)}{2\pi i} \ln \sin \left(\frac{\hat{p}_1-\hat{p}_2+\hat{p}_3}{2} \right)-\int_{\hat{\mathcal{M}}^{uuu}_{+--}}\hspace{-20pt}\frac{z(x)\left( -d\hat{p}_1 +d\hat{p}_2 +d\hat{p}_3\right)}{2\pi i} \ln \sin \left(\frac{-\hat{p}_1+\hat{p}_2+\hat{p}_3}{2} \right)\nn\\
& -2\sum_{j=1}^{3}\int_{\Gamma_{j_-}^u} \frac{z(x) \,dp_j}{2\pi i} \ln \sin p_j+2\sum_{j=1}^{3}\int_{\hat{\Gamma}_{j_-}^u} \frac{z(x) \,d\hat{p}_j}{2\pi i} \ln \sin \hat{p}_j + \textsf{Contact}\comma  \label{final3pt}}
}
where \textsf{Contact} stands for  the contribution from the contact terms. 
We find it truly remarkable that, in spite of the complexity of both the analysis and 
 the intermediate expressions,  the final  answer takes such a simple form. 
Moreover, it  exhibits essential similarity to the  form of the weak coupling result \cite{Tailoring3, Kostov1,Kostov2, Serban}
even before taking any further limits. 
 In the next section, we shall  evaluate 
the structure constant \eqref{final3pt} more explicitly,  including 
the quantity \textsf{Contact},  for several important examples and 
 compare with the weak coupling results more closely. 

%%%%%%%%%%%%%%%%
\section{\bfall{Examples and comparison with the weak coupling result}\label{sec:compare}}
The results obtained in the previous section are quite general and applicable to three-point functions of arbitrary one-cut solutions on $E\!AdS_3\times S^3$.
 In this section we focus on several explicit examples, make some basic checks  and discuss the relation with the results at weak coupling. 

In subsection \ref{subsec:setup}, we first explain the basic set-up, which will be used throughout this section. Then, in subsection \ref{subsec:BPS}, we study the correlation functions of three BPS operators and see that the contributions from the $S^3$ part and the $E\!AdS_3$ part completely cancel out in this case. The results thus obtained fully agree with the results obtained  in the gauge theory. 
In subsection \ref{subsec:2pt}, we study the behavior of the three-point function under the limit where the charge of one of the operators becomes negligibly small while  the other two operators become identical. We confirm that the result reduces to that of the two-point function, as expected. 
Next, in subsection \ref{subsec:nonBPS}, we study three-point functions of one non-BPS and two BPS operators, which were studied on the gauge theory side in \cite{Tailoring3}. We will focus on certain explicit examples and show that  the full three-point functions can be expressed in terms of  simple integrals which resemble the semi-classical limit of the results at weak coupling \cite{Tailoring3, Kostov1, Kostov2,Serban}.  Then, in subsection \ref{subsec:FT}, we discuss the Frolov-Tseytlin limit of such three-point functions.  In this limit, the integrands in the final expression approximately agree with the ones in the weak coupling,  whereas the integration contours are rather different.  Lastly, we discuss the possible origin and the implication of this mismatch.
%%%%%%%%%%%%%%%%%%%%%%%%%
\subsection{Basic set-up\label{subsec:setup}}
%%%%%%%%%%%%%%%%%%%%%%%
Before starting the detailed analysis, let us clarify the basic set-ups to be used in this section.

The three-point functions studied extensively on the gauge theory side 
are those of the following three types of operators (see also Table \hyperlink{table2}{2}.):
\beq{
\mathcal{O}_1 = \tr \left( Z^{l_1-M_1}X^{M_1} \right) +\cdots\comma\quad \mathcal{O}_2 = \tr \left( \bar{Z}^{l_2-M_2}\bar{X}^{M_2} \right) +\cdots\comma \quad 
\mathcal{O}_3 = \tr \left( Z^{l_3-M_3}\bar{X}^{M_3} \right) +\cdots\period\nn
}
As explained in section \ref{subsec:relation}, such three-point functions vanish unless the conservation laws\fn{As we have shown in section \ref{sec:vertex}, such conservation laws can be derived also on the string theory side.} for the charges, \eqref{cons-charge}, are satisfied. Due to these  conservation laws, one cannot in general take the operators to be simple BPS states, such as $\tr ( Z^{l})$ or $\tr ( \bar{Z}^{l})$, which are the highest-weight vectors of the global SU(2)$_R\times$SU(2)$_L$ symmetry. Instead, we need to use descendants of the global symmetry to satisfy the conservation laws when we study three-point functions involving BPS operators\cite{Tailoring1,Tailoring3}. While this can be done without problems  on the gauge theory side, it leads to certain difficulty on the string theory side. This is because all the classical solutions of string are known to (or believed to) correspond to some highest-weight states.  To circumvent this difficulty, below we will utilize the global transformations to make all three operators to be built on different ``vacua".  On the string theory side, this corresponds to taking the polarization vectors of the  three operators, $n_i$'s and $\tilde{n}_i$'s, to be all distinct.  Then  no conservation laws will be imposed and we can safely take the limit where some of the operators become BPS while keeping them to be of highest-weight. 
Since the correlation functions involving descendants can be obtained from the correlation functions involving the highest-weight states by simple group theoretical manipulations,   knowledge of the  three-point functions for the highest weight states is  sufficient.  In addition, replacing the highest-weight operator with its descendant only modifies the kinematical factor, $\mathcal{V}_{\rm kin}$, of our result and the dynamical parts of three-point functions, which are main subjects of study in this section, will not be affected. 

After making the global transformations, the operators  $\mathcal{O}_1$, $\mathcal{O}_2$ and $\mathcal{O}_3$ can be treated  almost on the same footing.  However, there is an important difference between $\mathcal{O}_3$ and the other two in string theory:  As explained in section \ref{subsec:relation}, the quasi-momenta for the operators $\mathcal{O}_1$ and $\mathcal{O}_2$  contain branch cuts in the  $|\Re x|>1$ region,  whereas the quasi-momentum for the operator $\mathcal{O}_3$ contains a branch cut in the $|\Re x|<1$ region. This difference is  important in the analysis  to follow,  since the position of the branch cuts affects the contours for  the convolution integrals.
%%%%%%%%%%%%%%%%%%%%%%%%%%%%%%
\subsection{Case of three BPS operators\label{subsec:BPS}}
%%%%%%%%%%%%%%%%%%%%%%%%%%%%%%
Let us first study the correlation functions of three BPS operators. 
In order to apply the general formula for the three-point functions 
 of one-cut solutions obtained in the previous section, we need the explicit forms of $p(x)$ and $q(x)$ for the BPS operators,  which in particular determine the integration contours.  
Within the bosonic sector, the characteristic feature of a BPS state 
  is that, as it should correspond to a supergravity mode,  it is ``point-like",  meaning  that its  two-point function is $\sig$-independent.  In the language of the 
 spectral curve, it  means the absence of a branch cut, since a branch cut 
corresponds to a non-trivial string mode with $\sig$-dependence. 

Now in fixing  the forms of $p(x)$ and $q(x)$, there is a subtle  problem with the configuration without a branch cut.  In the case of one-cut solutions corresponding to non-BPS operators, the constant parts  of $p(x)$ and $q(x)$ 
 are  fixed in such a way that they   vanish at the branch points.
Obviously, for configurations without a branch cut, this prescription cannot be applied.
  One natural  remedy would be to start with a non-BPS solution, apply the usual method  above to fix the constants and then shrink the cut to obtain a BPS solution. 
This idea, however, still does not cure the  problem since  the resultant  $p(x)$ and $q(x)$  depend on the points on the spectral curve at which we shrink the branch cut. 
The existence of such an ambiguity possibly implies that the semi-classical three-point functions are affected by the presence of infinitesimal branch cuts. 
Although such an assertion sounds counter-intuitive,   it is not totally inconceivable since similar effects were already observed in the study of ``heavy-heavy-light" three-point functions\fn{In \cite{Tailoring2}, such effects were called {\it back reactions}.} in \cite{Tailoring2}. 

Below we shall fix  the ambiguity  by  employing  a prescription which is quite natural from the viewpoint of the correspondence with the spin chain on the 
 gauge theory side. The prescription is to shrink the branch 
 cuts  either at $x=0$ or at $x=\infty$ in producing  BPS operators. 
This choice is based on the following fact: In gauge theory, 
 adding a small number of Bethe roots at $x=0$ or $x=\infty$ correspond to performing a small global transformation and keeps the operator to be BPS,  whereas adding a small number of Bethe roots at generic points on the spectral curve creates nontrivial magnon excitations and makes the operator non-BPS.

Having identified the classical solutions corresponding to BPS operators, let us now determine the integration contours. First we focus on the $S^3$-part of three-point functions.  As discussed in section \ref{subsec:relation}, for $\mathcal{O}_1$ and $\mathcal{O}_2$, 
$p_i(x)$ and $q_i(x)$ can have branch cuts only in the  the $|\Re x|>1$ region and 
hence we take the infinitesimal branch cut to be placed at $x=\infty$. 
Then from the general form of the one-cut solution given in \eqref{onecutpx} 
and \eqref{onecutqx}, we get 
\beq{
&p_i(x)=-2\pi \kappa_i\left( \frac{1}{x-1}+\frac{1}{x+1}\right)\comma \quad q_i(x)=-2\pi \kappa_i\left( \frac{1}{x-1}-\frac{1}{x+1}\right)\comma\label{pq12BPS}
}
which vanish at $x=\infty$, as desired. 
On the other hand, for $\mathcal{O}_3$, since the  branch cuts can only be in the $|\Re x|<1$ region,  we place an infinitesimal branch cut  at $x=0$. 
Then from \eqref{onecutpx} and \eqref{onecutqx} we get 
\beq{
p_3(x)&=-2\pi \kappa_3\left( \frac{x}{x-1}-\frac{x}{x+1}\right) 
= -2\pi \kappa_3\left( \frac{1}{x-1}+\frac{1}{x+1}\right)  \\
 q_3(x)&=-2\pi \kappa_3\left( \frac{x}{x-1}+\frac{x}{x+1}\right)\period\label{pq3BPS}
}
These expressions vanish at $x=0$.

\begin{figure}[htbp]
\begin{minipage}{0.5\hsize}
  \begin{center}
   \picture{clip,height=5cm}{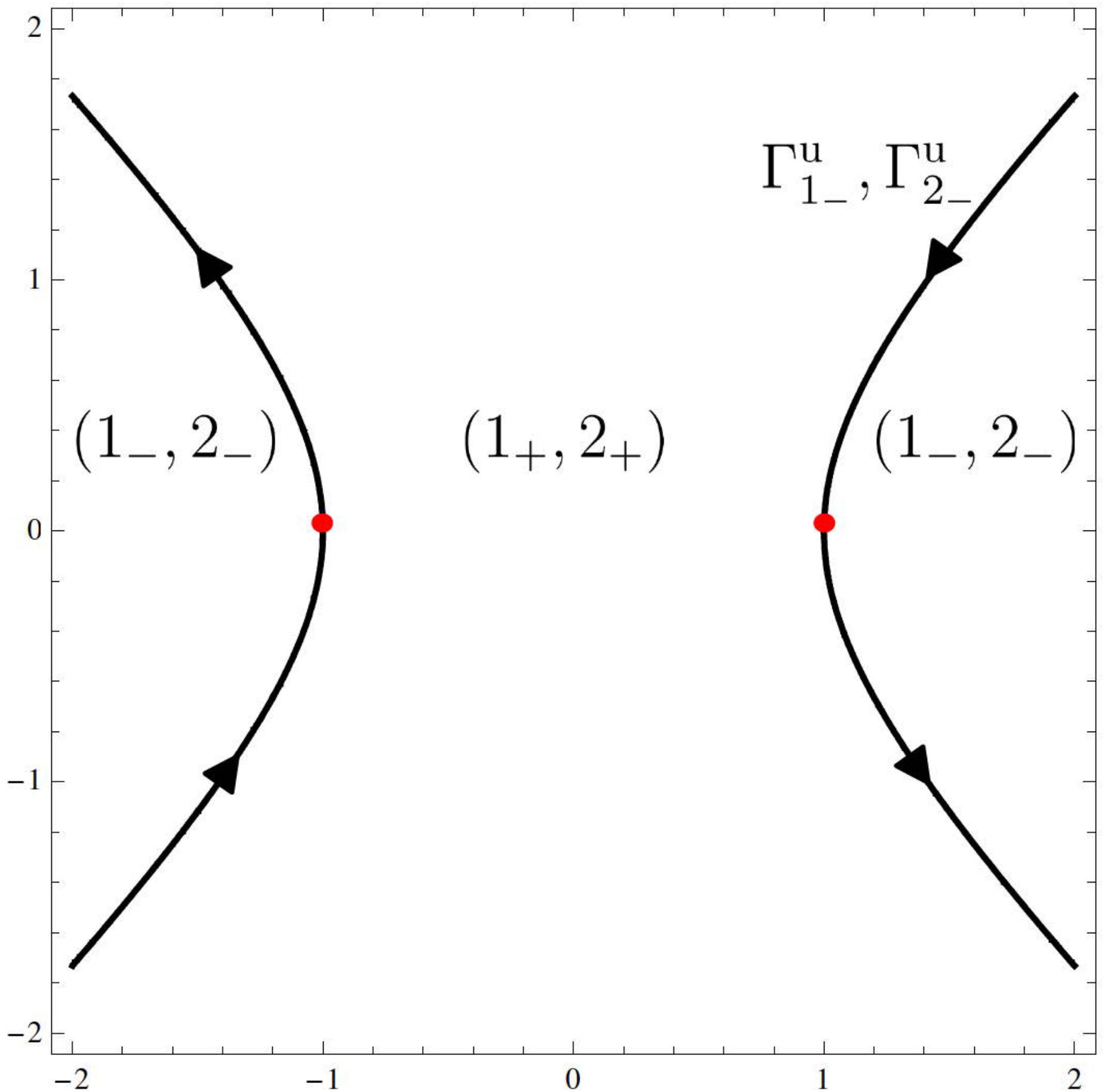}
  \end{center}
 \end{minipage}
 \begin{minipage}{0.5\hsize}
  \begin{center}
   \picture{height=5cm,clip}{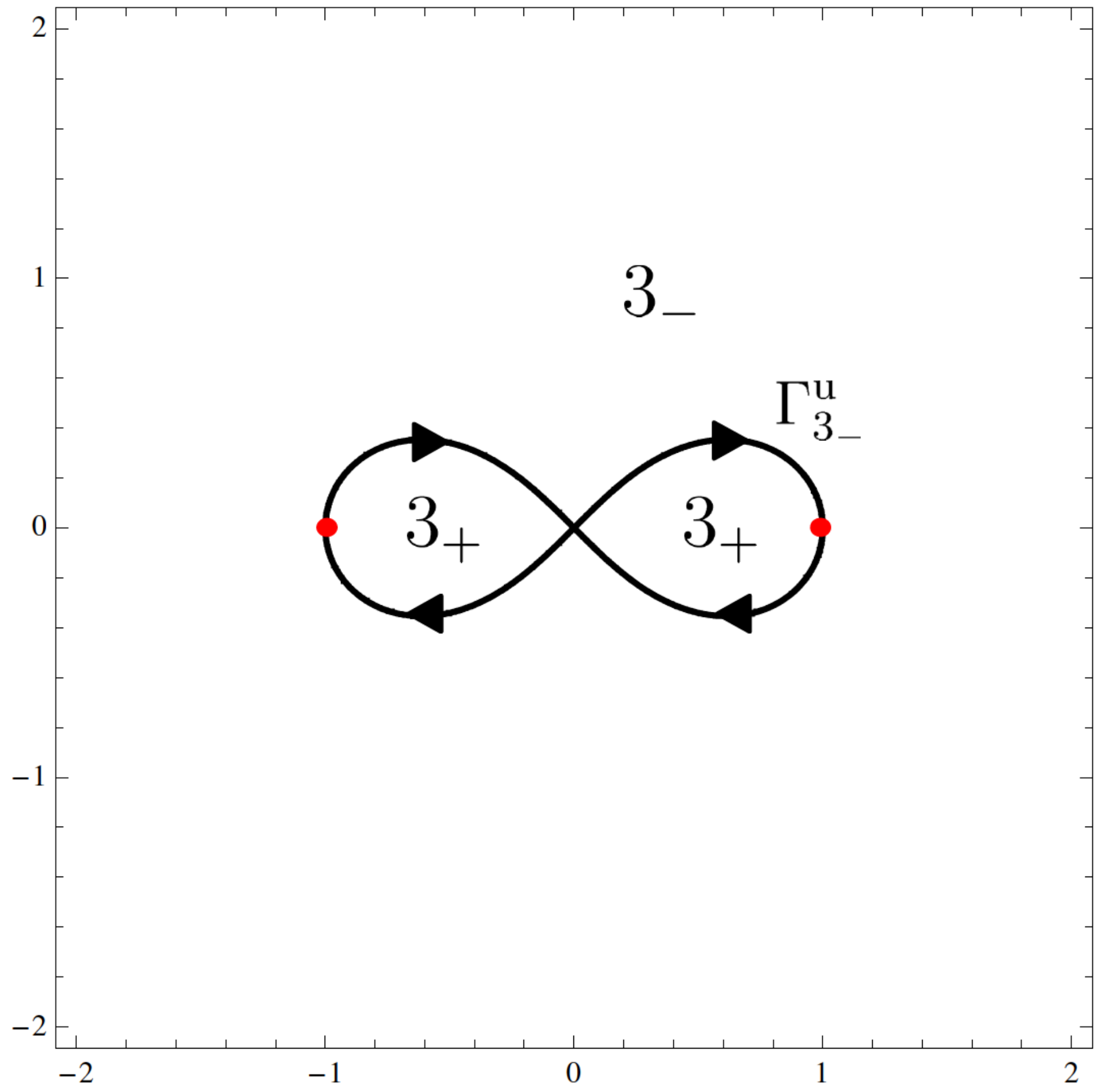}
  \end{center}
 \end{minipage}
\caption{The contours $\Gamma_{i_-}^u$, defined by $\Re q_i=0$. In each region, we showed which of the eigenvectors is the small solution.} 
\label{fig:BPS-curves}
\end{figure}

As discussed in detail in section \ref{sec:wron}, the contours for the convolution integrals consist of two types of  curves. The first type are those defined by $\Re  q_i(x)=0$, across which the relative magnitude of  $i_+$ and $i_-$  changes. They 
determine the integration contours $\Gamma_{i_-}^u$ defined in section \ref{subsec:WH} and are  depicted in \figref{fig:BPS-curves}. Note
 that in the present case, the contours $\Gamma_{1_{-}}^u$ and $\Gamma_{2_{-}}^u$ coincide since $q_1(x) =q_2(x)$. 
The second type are the curves defined by $N_i=N_j+N_k$, across which the connectivity of the exact WKB curves changes.  Now for a BPS  operator, $N_i = |\Re p_i(x)|$ is given by a common function  $|\Re ((x+1)^{-1} + (x-1)^{-1})|$  times the factor $-2\pi\kappa_i$,  as shown above.  Since  $\kappa_i$'s satisfy the triangular inequalities, this means that $N_i=N_j+N_k$ cannot be satisfied. Hence the 
 second  type of curves are absent and the integration contours are determined solely  by the first type of  curves.  

With this knowledge, we can now apply the general rules given  at the end 
 of  section \ref{subsec:zero}  to  determine the integration contours  $\mathcal{M}_{\pm\pm\pm}^{uuu}$.  As an example, consider the contour $\calM_{---}^{uuu}$,  
which is used for the convolution integral involving $\sin\half (-p_1(x)-p_2(x)-p_3(x))$. 
From the Rule 1, either Wronskians among $\calS_-=\{1_-, 2_-, 3_-\}$  vanish or those   among $\calS_+=\{1_+, 2_+, 3_+\}$ vanish.  Then we must apply Rule 2, since the triangle inequalities are satisfied in the present case. It states  that 
if two of the members of $\calS_-$ (resp. $\calS_+$)  are small solutions, then the Wronskians  for the members of $\calS_+$ (resp. $\calS_-$)  vanish. 
Now consider the curve $\Ga^u_{1-}$. From its definition, it is along $\Re q_1(x)=0$ with the direction such that  to the left of this curve $1_-$ is the small solution. The curve $\Ga^u_{2-}$ is identical, as we already 
remarked. These curves are depicted in the left figure of \figref{fig:BPS-curves}, together with the states which are small in the three regions 
 separated by these curves.  
Together with the rules mentioned above, we see explicitly that the analyticity of Wronskians change across such curves and hence  we can identify $\Gamma_{1_-}^u(=\Gamma_{2_-}^u)$
 as the contour $\mathcal{M}_{---}^{uuu}$. 
Similarly, the curve $\Ga^u_{3-}$, identified as  $\mathcal{M}_{+--}^{uuu}$, is shown in the right figure of \figref{fig:BPS-curves}. In this way,  we find the contours $\mathcal{M}_{\pm\pm\pm}^{uuu}$ to be 
given by 
\beq{
\begin{aligned}
&\mathcal{M}_{---}^{uuu}=\Gamma_{1_-}^u(=\Gamma_{2_-}^u)\comma\quad\mathcal{M}_{+--}^{uuu}=\Gamma_{3_-}^u\comma\\
&\mathcal{M}_{-+-}^{uuu}=\Gamma_{3_-}^u\comma \quad \mathcal{M}_{--+}^{uuu}=\Gamma_{1_-}^u(=\Gamma_{2_-}^u)\period
\end{aligned}
}

\begin{figure}[htbp]
\begin{minipage}{0.5\hsize}
  \begin{center}
   \picture{clip,height=2cm}{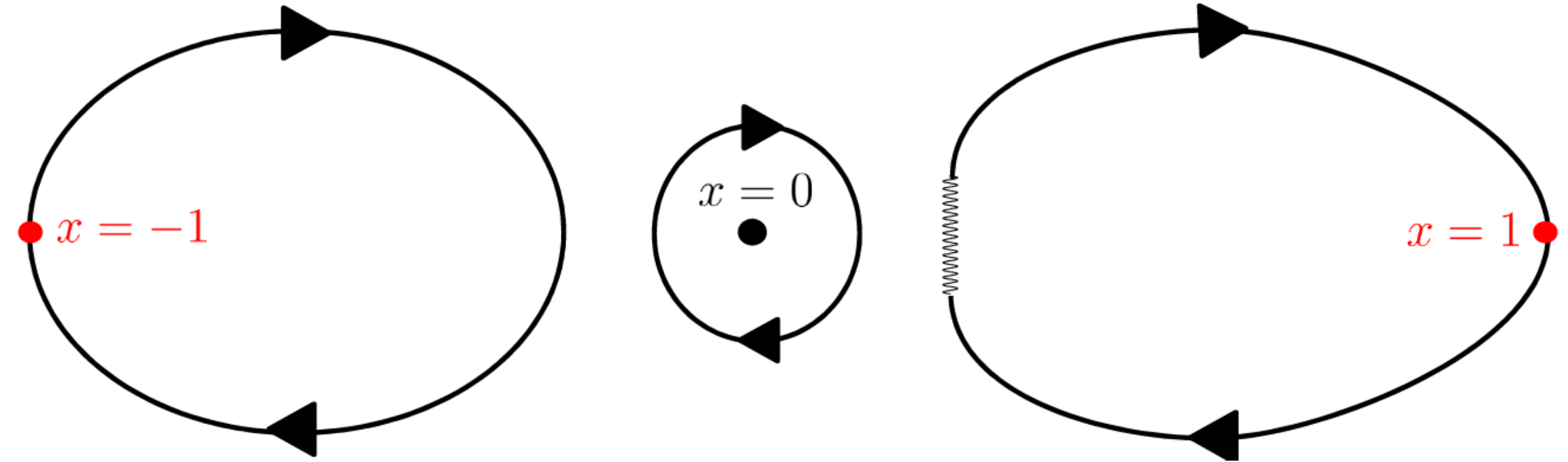}\\
(a) Putting a small branch cut \\ away from $x=0$.
\end{center}
 \end{minipage}
 \begin{minipage}{0.5\hsize}
  \begin{center}
   \picture{height=2cm,clip}{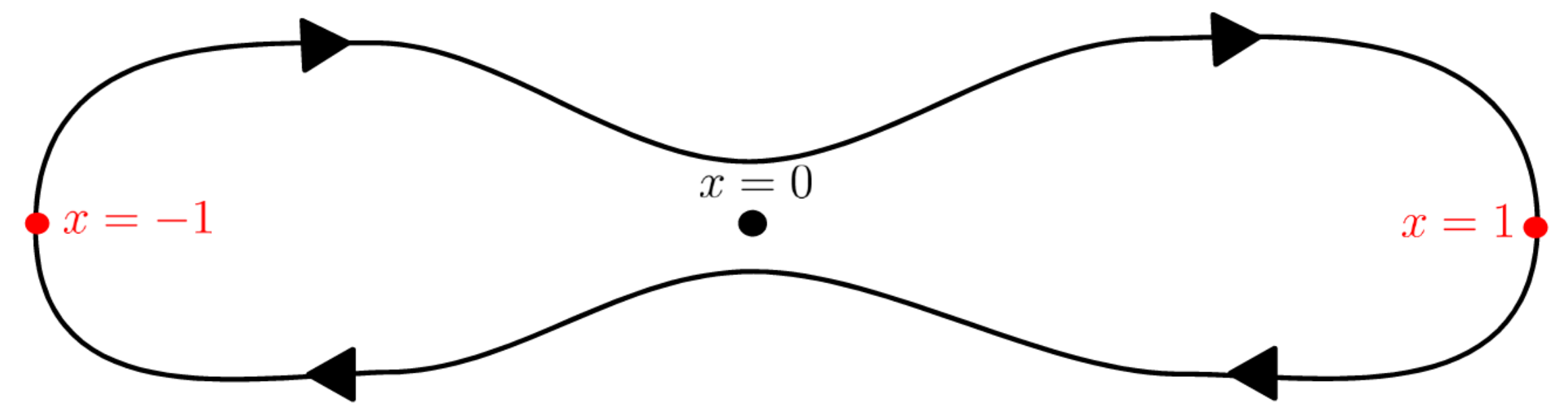}\\
(b) Shrinking the cut and \\ deforming the contour.
\end{center}
 \end{minipage}
\caption{An example of the contour deformation. The contour depicted in (b) can be further deformed into the contour on the unit circle.} 
\label{fig:deformation}
\end{figure}

Let us next consider the effects of the contact terms. As argued  in section \ref{sec:3pt}, such contribution must be taken into account  when $x=0$ ($x=\infty$) is on the left (right) hand side of the integration contours.  The effect is 
most conveniently done by adding a small circle around $x=0$ ($x=\infty$) to the contour for each integration in \eqref{totalconv}. However, in the case of BPS operators, the integration contours  terminate right at $x=0$ or $x=\infty$. Therefore we need to first regularize them by putting a small branch cut slightly away from $x=0$ or $x=\infty$ and then take the limit where the branch cut shrinks to $x=0$ or $x=\infty$.  
An example of such a procedure is depicted in \figref{fig:deformation}. Since the sine-functions in the convolution integrals \eqref{totalconv} turn out to vanish only on the real axis in the case of BPS operators, we can further deform the contours into those on the unit circle.  
As a result, we find that the $S^3$-part of the three-point function  is given by%
\beq{
&\oint_{U}\frac{z\left( dp_1 +dp_2 +dp_3\right)}{2\pi i} \ln \sin \left(\frac{p_1+p_2+p_3}{2} \right)
+\oint_{U}\frac{z\left( dp_1 +dp_2 -dp_3\right)}{2\pi i} \ln \sin \left(\frac{p_1+p_2-p_3}{2} \right)\nn\\
&+\oint_{U}\frac{z\left( dp_1 -dp_2 +dp_3\right)}{2\pi i} \ln \sin \left(\frac{p_1-p_2+p_3}{2} \right)+\oint_{U}\frac{z\left( -dp_1 +dp_2 +dp_3\right)}{2\pi i} \ln \sin \left(\frac{-p_1+p_2+p_3}{2} \right)\nn\\
&-2\sum_{j=1}^{3}\int_{U} \frac{z \,dp_j}{2\pi i} \ln \sin p_j\comma\nn
}
where $U$ denotes the contour  which goes around  the unit circle clockwise. 

Next consider the $E\!AdS_3$-part of the three-point function.  The quasi-momenta and the quasi-energies for the operators without spin in $AdS$ are given 
in \cite{JW} by\fn{The spectral parameter $x$ used in \eqref{sec6-adspq} is related to the spectral parameter $\xi$ used in \cite{JW} by $\xi = (x-1)/(x+1)$.} 
\beq{
&\hat{p}_i(x) =-2\pi \kappa_i\left( \frac{1}{x-1}+\frac{1}{x+1}\right)\comma \quad
\hat{q}_i(x)=-2\pi \kappa_i\left( \frac{1}{x-1}-\frac{1}{x+1}-1\right)\period\label{sec6-adspq}
} 
Then, performing a similar analysis as in the case of $S^3$-part, we find that the result is again given by the integrals  along the unit circle. As the quasi-momenta $p_i(x)$  for the $S^3$-part and the ones $\hat{p}_i(x)$ for the  $E\!AdS_3$-part coincide in the case of BPS operators,  we see from 
 the general formula \eqref{final3pt} that 
the contributions form these  two parts cancel each other completely.  Therefore, the three-point function  for three BPS operators is  given purely by the kinematical factors as 
\beq{
\begin{aligned}
\langle \mathcal{V}_1 \mathcal{V}_2 \mathcal{V}_3  \rangle = &\frac{1}{|x_1-x_2|^{\Delta_1 +\Delta_2-\Delta_3}|x_2-x_3|^{\Delta_2 +\Delta_3-\Delta_1}|x_3-x_1|^{\Delta_3 +\Delta_1-\Delta_2}}\\
&\times \sl{n_1\comma n_2}^{R_1 +R_2-R_3}\sl{n_2\comma n_3}^{R_2 +R_3-R_1}\sl{n_3\comma n_1}^{R_3 +R_1-R_2}\\
&\times\sl{\tilde{n}_1\comma \tilde{n}_2}^{L_1 +L_2-L_3}\sl{\tilde{n}_2\comma \tilde{n}_3}^{L_2 +L_3-L_1}\sl{\tilde{n}_3\comma \tilde{n}_1}^{L_3 +L_1-L_2}\comma
\end{aligned}
}
This is consistent with the result  in the gauge theory that the three-point functions of BPS operators are tree-level exact and have no dependence on the 't Hooft coupling constant $\lambda$.
%%%%%%%%%%%%%%%%%%%%%%%%%
\subsection{Limit producing two-point function\label{subsec:2pt}}
%%%%%%%%%%%%%%%%%%%%%%%%%%
Having seen that the BPS three-point functions are correctly reproduced from our general formula,  let us next discuss the limit where the three-point functions are expected to reduce to two-point functions. As an example,  we take two of the operators  $\mathcal{O}_1$ and $\mathcal{O}_2$  to have  identical quasi-momenta and quasi-energy,  while  $\mathcal{O}_3$ is a  BPS operator 
 with vanishingly small charge\fn{Although the case  considered  here appears  similar to the one studied in the gauge theory \cite{Tailoring2} with  $\mathcal{O}_3$ taken to be small but nonvanishing, there is a difference: In \cite{Tailoring2}, $\mathcal{O}_1$ and $\mathcal{O}_2$ must have slightly different quasi-momenta in the presence of  $\mathcal{O}_3$,  due to the conservation law for the magnons.  In the present case, however,   as we performed the global transformation,  no conservation law is imposed and we can take $\mathcal{O}_1$ and $\mathcal{O}_2$ to have  identical quasi-momenta.}. 
\begin{figure}[htbp]
\begin{minipage}{0.32\hsize}
  \begin{center}
   \picture{clip,height=5cm}{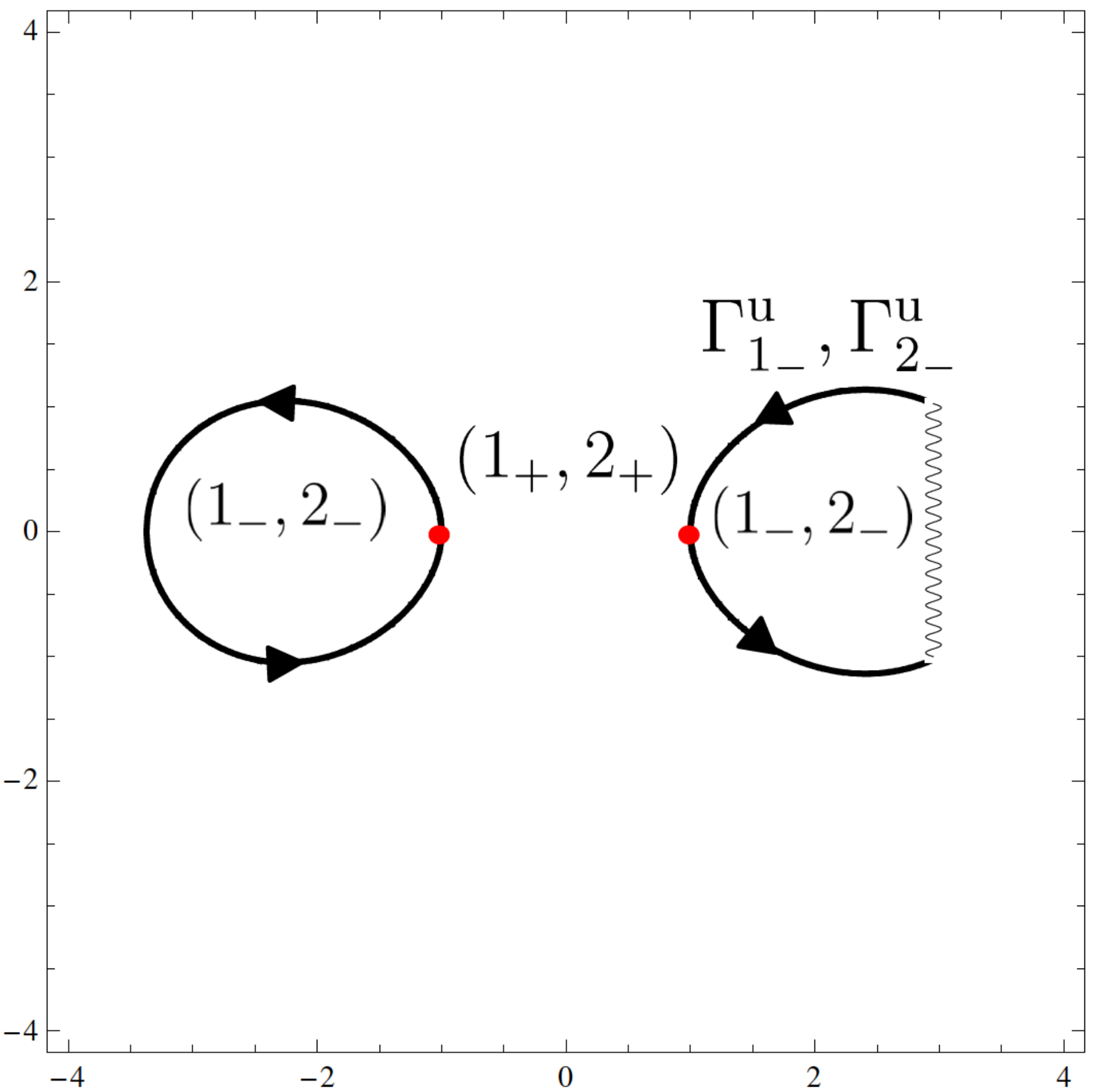}\\
   $(a)$
  \end{center}
 \end{minipage}
\begin{minipage}{0.32\hsize}
  \begin{center}
   \picture{clip,height=5cm}{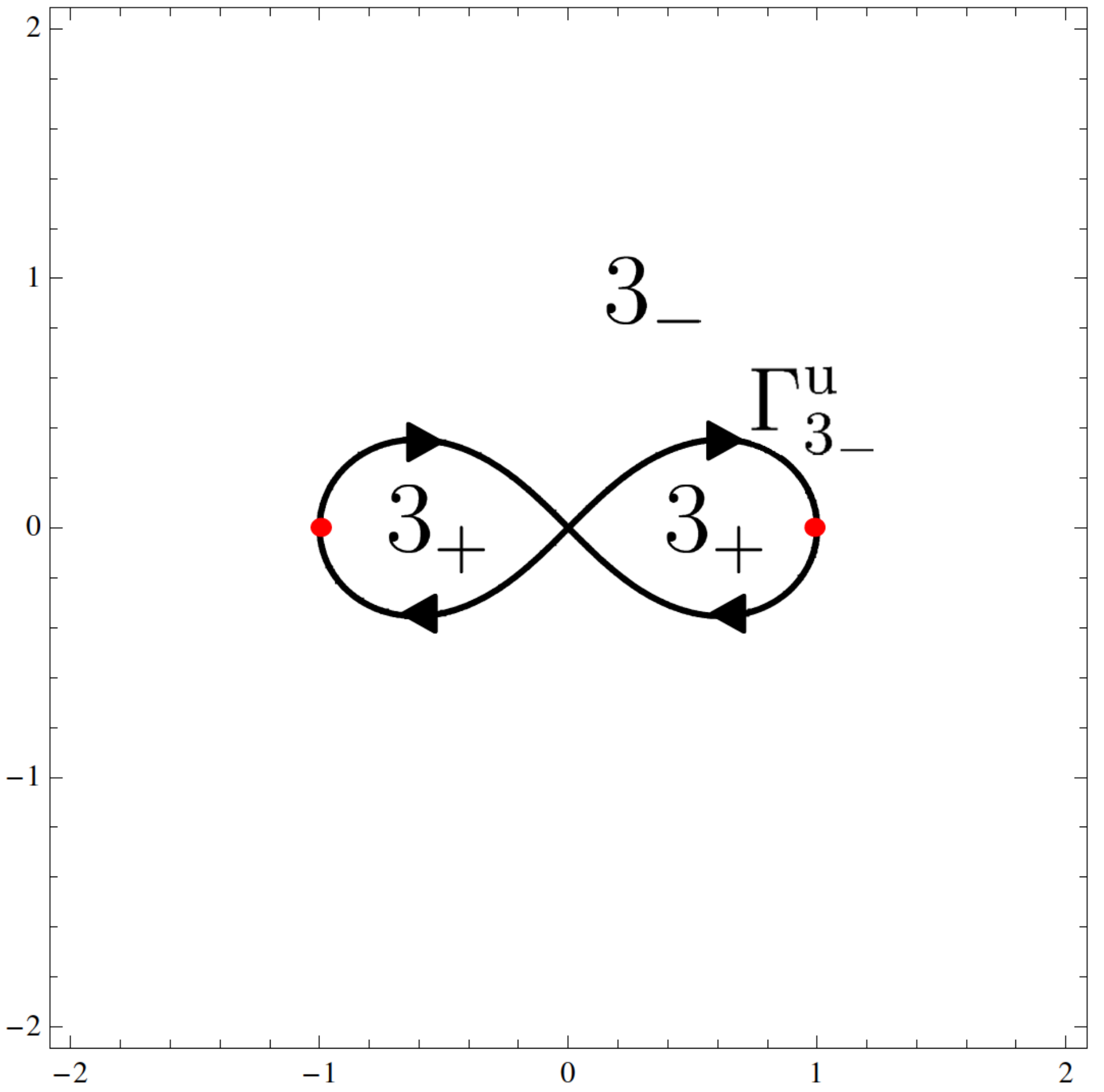}\\
   $(b)$
  \end{center}
 \end{minipage}
 \begin{minipage}{0.32\hsize}
  \begin{center}
   \picture{height=5cm,clip}{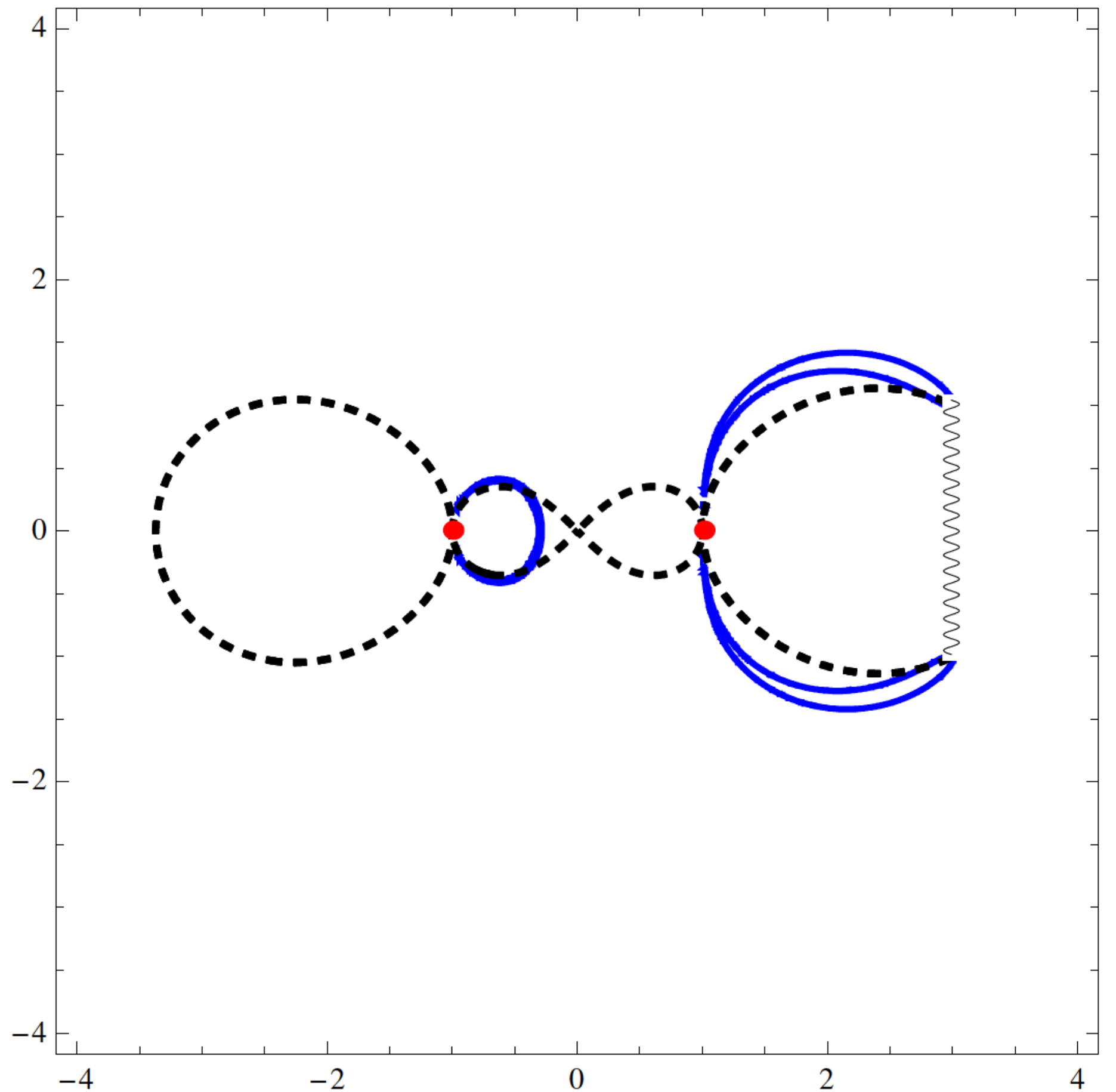}\\
   $(c)$
  \end{center}
 \end{minipage}
\caption{The curves which determine the integration contours in the limit where three-point functions reduce to two-point functions. In the left and the middle figures, the contours $\Gamma_{i_-}^u$, determined by $\Re\, q_i(x)=0$ are depicted. The segment represented by a wavy line is the branch cut. In the rightmost figure, the curve defined by $N_3=N_1+N_3$ is drawn  in blue. For convenience, we redisplayed the curves in figures $(a)$ and $(b)$ as 
dotted lines.} 
\label{fig:2-pnt}
\end{figure}

To understand what happens in such a limit, let us draw the two types of curves, namely $\Re\, q_i=0$ and $N_i=N_j+N_k$. The first type of curves are depicted in the first and the second figures of \figref{fig:2-pnt}. 
As for the second type,   the only curve we need to consider is the curve given by $N_3=N_1+N_2$. This is because the inequalities  $N_1+N_3\geq N_2$ and $N_2+N_3\geq N_1$ are always satisfied since $N_1=N_2$ in the present case. When the operator $\mathcal{O}_3$ is sufficiently small, 
the curve defined by $N_3=N_1+N_2$ almost vanishes  and  we can practically ignore the effects of such a curve.   Thus the integration contours are given purely by $\Re\, q_1=\Re\, q_2 =0$.  Applying the rules given in the previous section and taking into account the contact terms, we find that the convolution integrals for the $S^3$-part are given by 
\beq{
&\int_{\Gamma_{1_-}^u+\mathcal{C}_{\infty}}\hspace{-20pt}\frac{z\left( dp_1 +dp_2 +dp_3\right)}{2\pi i} \ln \sin \left(\frac{p_1+p_2+p_3}{2} \right)
+\int_{\Gamma_{1_-}^u+\mathcal{C}_{\infty}}\hspace{-20pt}\frac{z\left( dp_1 +dp_2 -dp_3\right)}{2\pi i} \ln \sin \left(\frac{p_1+p_2-p_3}{2} \right)\nn\\
&+\int_{\Gamma_{3_-}^u+\mathcal{C}_{0}}\hspace{-20pt}\frac{z\left( dp_1 -dp_2 +dp_3\right)}{2\pi i} \ln \sin \left(\frac{p_1-p_2+p_3}{2} \right)+\int_{\Gamma_{3_-}^u}\hspace{-10pt}\frac{z\left( -dp_1 +dp_2 +dp_3\right)}{2\pi i} \ln \sin \left(\frac{-p_1+p_2+p_3}{2} \right)\nn\\
&-2\sum_{j=1}^{2}\int_{\Gamma_{j_-}^u+\mathcal{C}_{\infty}} \frac{z \,dp_j}{2\pi i} \ln \sin p_j-2\int_{\Gamma_{3_-}^u+\mathcal{C}_{0}} \frac{z \,dp_3}{2\pi i} \ln \sin p_3\comma\label{2ptlimits3}
}
where $\mathcal{C}_{\infty}$ is the contour encircling $x=\infty$ counterclockwise and $\mathcal{C}_{0}$ is the contour encircling $x=0$ clockwise. 
Setting $p_1=p_2$ and $p_3=0$ in this formula, we  see that in this limit 
all the terms  in \eqref{2ptlimits3} completely cancel out with each other.  Similar cancellation occurs also for the $E\!AdS_3$-part. Therefore  the structure constant $C_{123}$ of the three-point function in this limit becomes unity and the result correctly reproduces the correctly normalized two-point function given by 
\beq{
\frac{\sl{n_1\comma n_2}^{2R}\sl{\tilde{n}_1\comma \tilde{n}_2}^{2L}}{|x_1-x_2|^{2\Delta}}\period
}
Here,  $\Delta$,  $R$ and $L$ are, respectively,  the conformal dimension, the (absolute values of the) right  and the left global charges, which are common to $\mathcal{O}_1$ and $\mathcal{O}_2$.
%%%%%%%%%%%%%%%%%%%%%%%%%%%
\subsection{Case of one non-BPS and two BPS operators\label{subsec:nonBPS}}
\begin{figure}[htbp]
\begin{minipage}{0.5\hsize}
  \begin{center}
   \picture{clip,height=5cm}{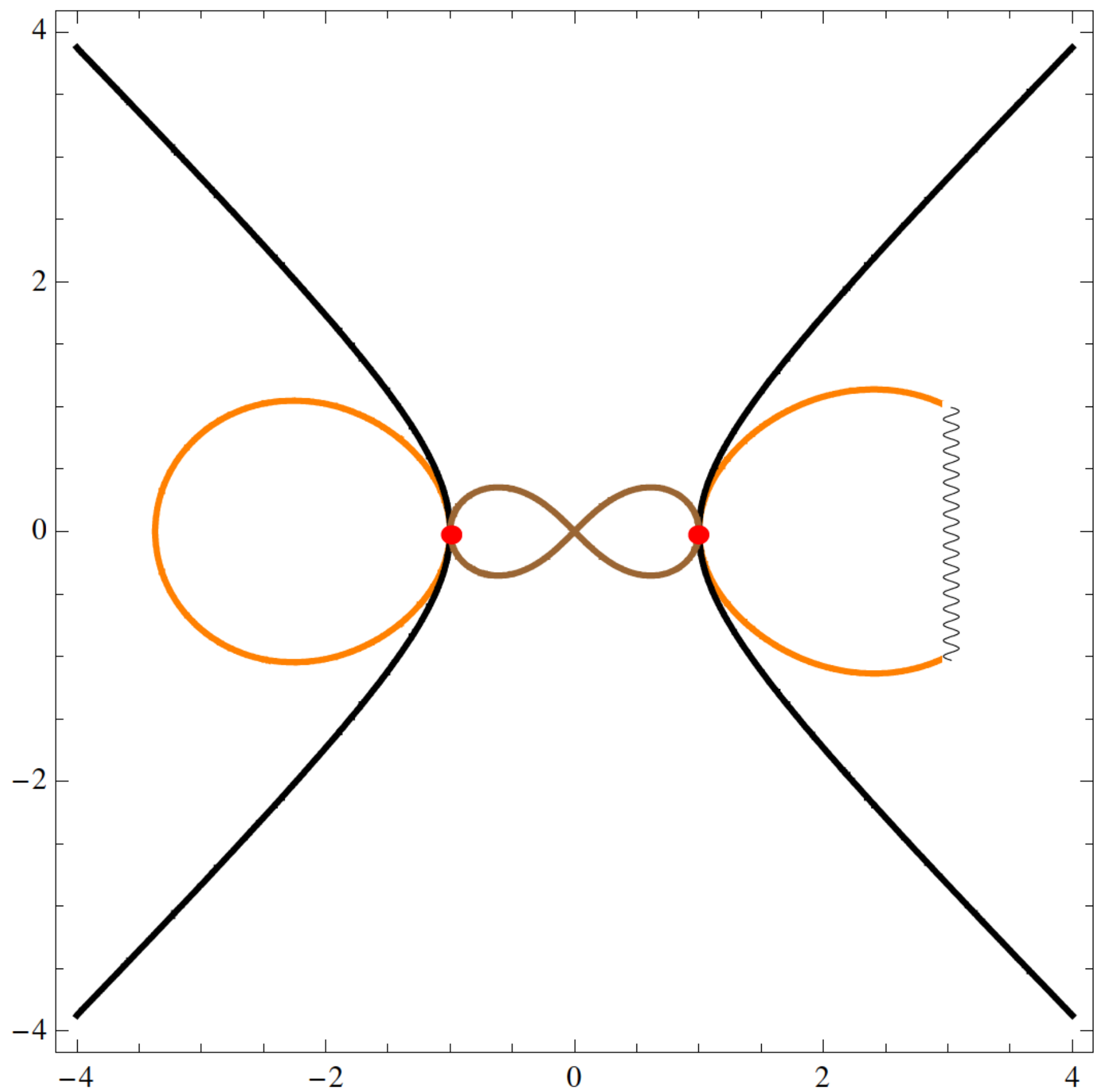}
  \end{center}
 \end{minipage}
\begin{minipage}{0.5\hsize}
  \begin{center}
   \picture{clip,height=5cm}{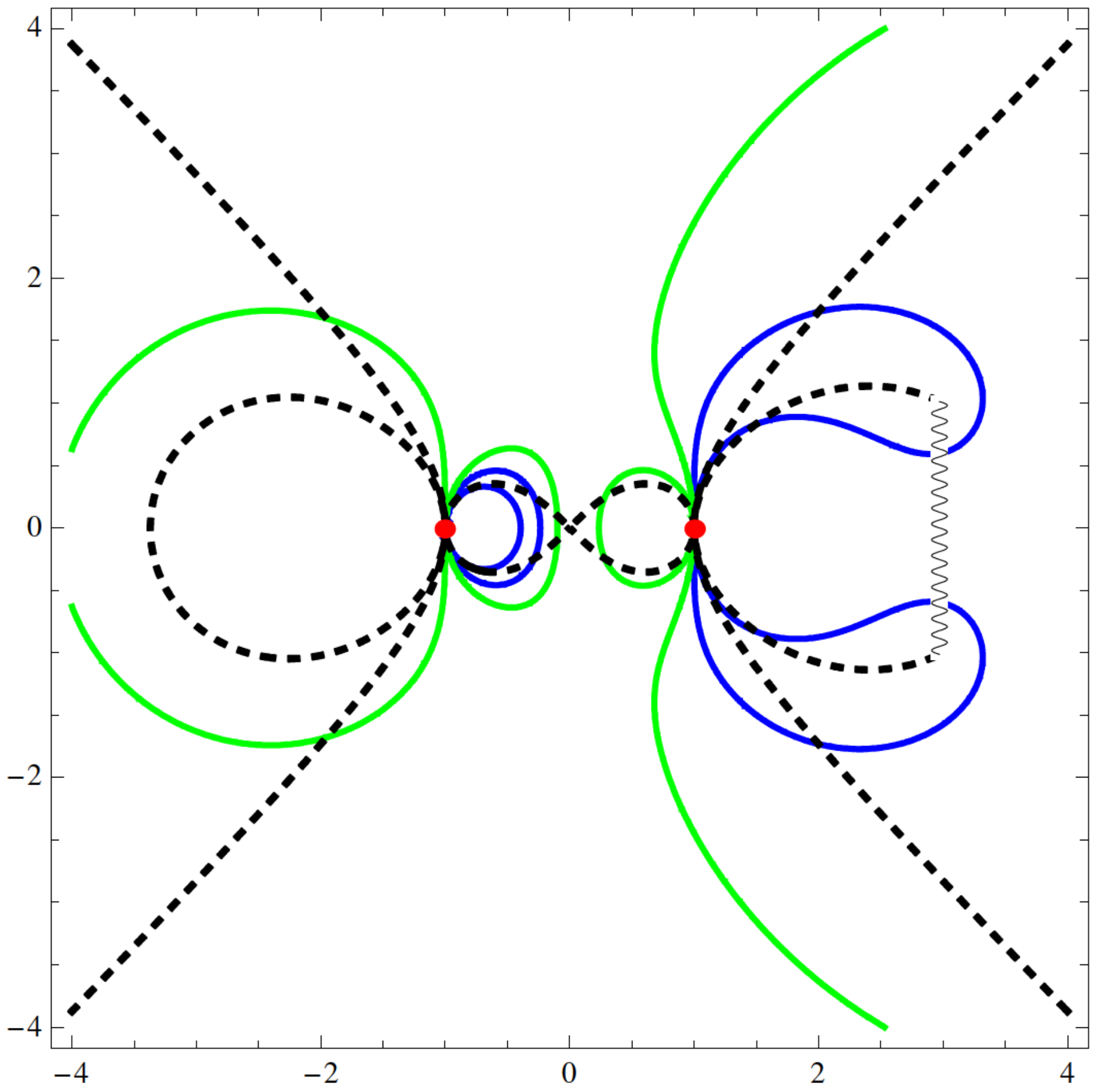}
  \end{center}
 \end{minipage}
 \caption{Typical configuration of the curves produced by the conditions $\Re q_i=0$ and $N_i=N_j+N_k$, for the three-point functions of one non-BPS operator and two BPS operators. In the left figure, $\Re q_1=0$, $\Re q_2=0$ and $\Re  q_3=0$ are drawn respectively in black, orange and brown. In the right figure, $N_1=N_2+N_3$ is drawn in blue and $N_2=N_1+N_3$ is drawn in green.} 
\label{fig:typical}
\end{figure}

%%%%%%%%%%%%%%%%%%%%%%%%%%%%
Having checked that our formula correctly reproduces the known results 
 in simple limits,  let us now study more nontrivial examples.  In this 
subsection, we take up the three-point functions of  one non-BPS  and two BPS operators, which were studied on the gauge-theory side in \cite{Tailoring3}. 
As in \cite{Tailoring3}, we take $\mathcal{O}_2$ to be non-BPS and $\mathcal{O}_1$ and $\mathcal{O}_3$ to be BPS. 
In this case,  the typical forms of the curves corresponding to $\Re q_i=0$ and $N_i=N_j+N_k$,  are given in \figref{fig:typical}. 

\begin{figure}[htbp]
\begin{minipage}{0.5\hsize}
  \begin{center}
   \picture{clip,height=5cm}{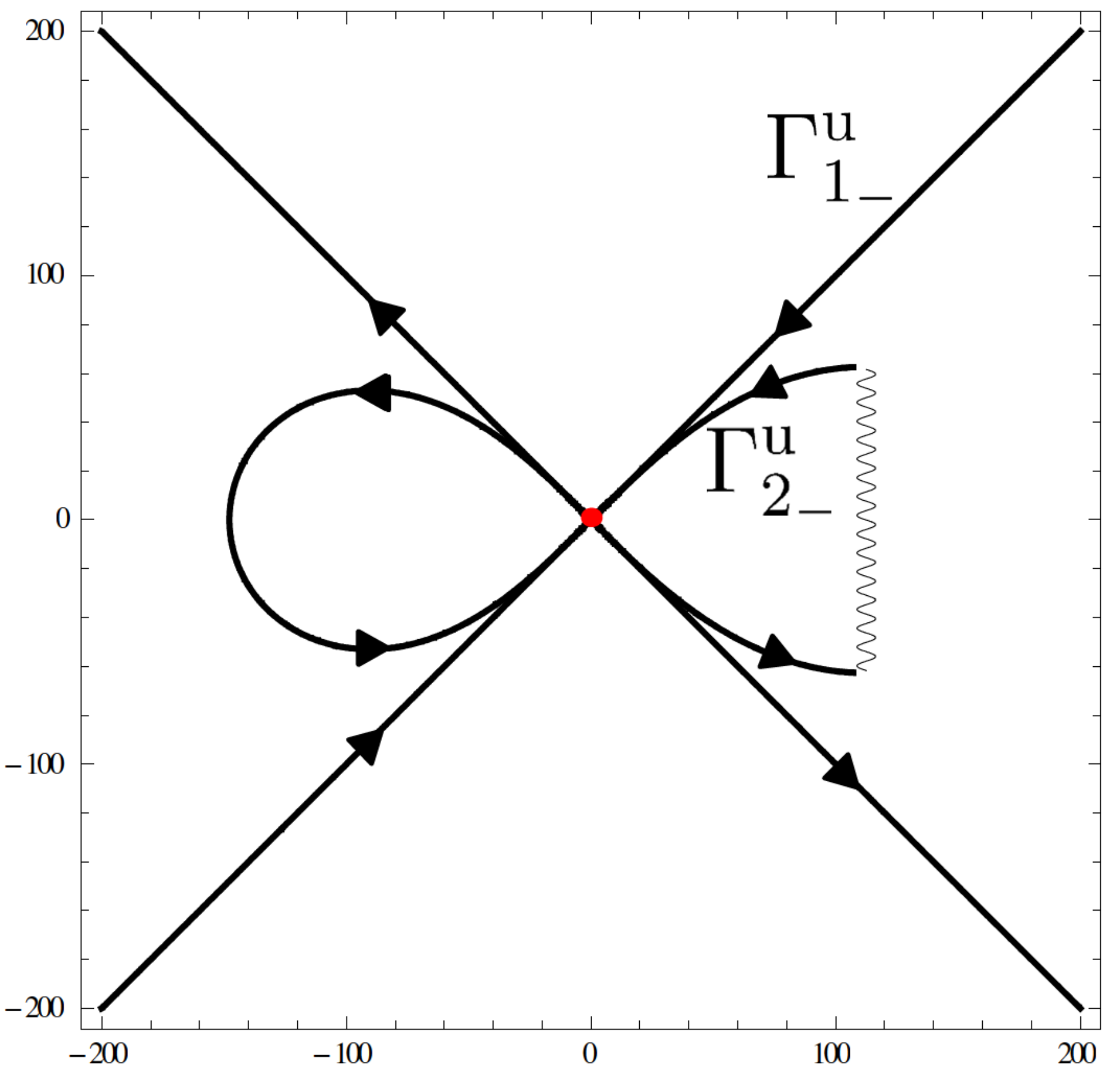}
  \end{center}
 \end{minipage}
\begin{minipage}{0.5\hsize}
  \begin{center}
   \picture{clip,height=5cm}{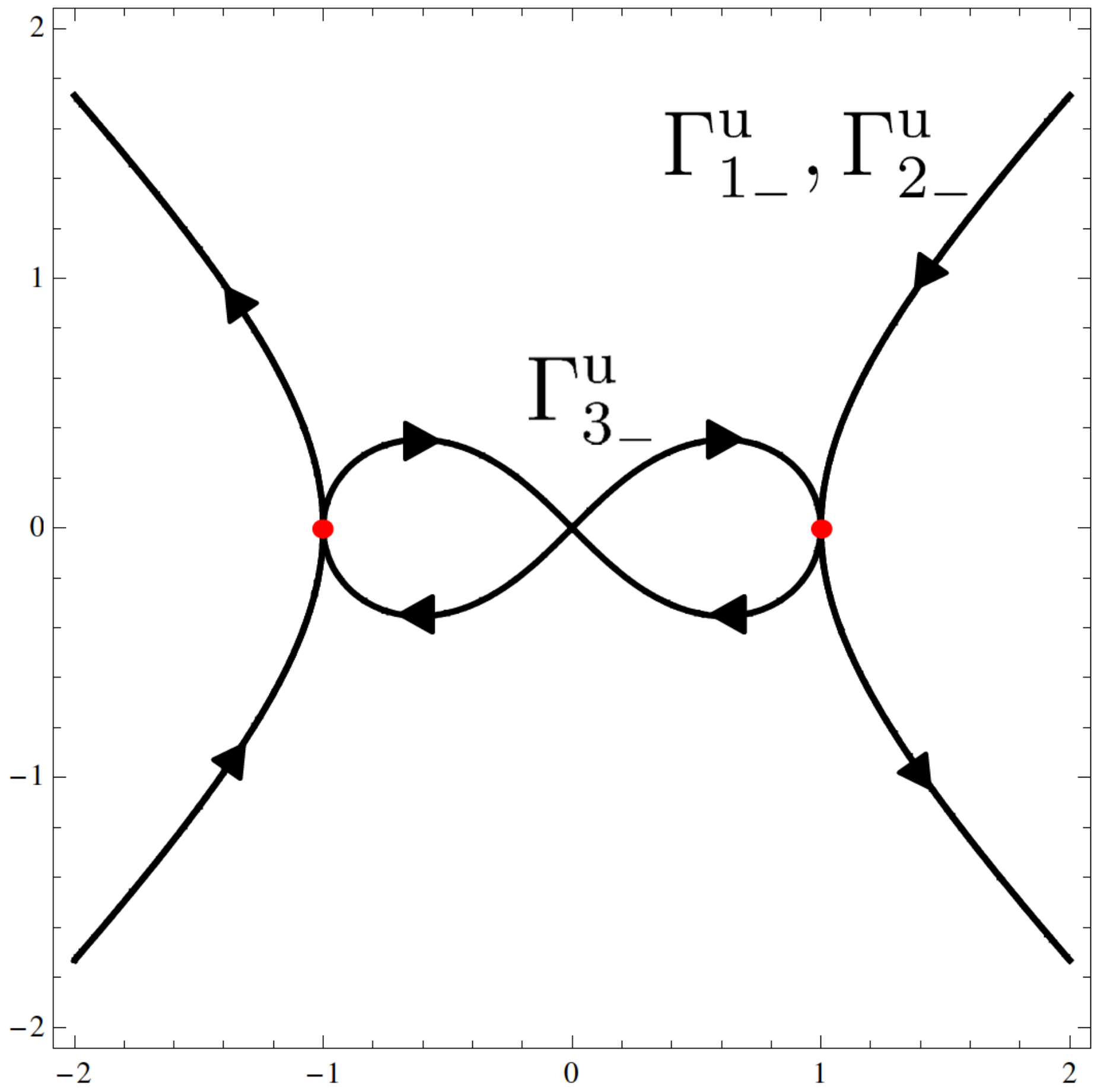}
  \end{center}
 \end{minipage}
 \caption{The contours $\Gamma_{i_-}^u$, defined by $\Re q_i=0$. The left figure shows the configuration of contours in the $|x|\gg 1$ region,  where as the right figure depicts  the configuration of contours in the $|x|<1$ region.} 
\label{fig:non-BPS1}
\end{figure}

\begin{figure}[bthp]
\begin{minipage}{0.32\hsize}
  \begin{center}
   \picture{clip,height=5cm}{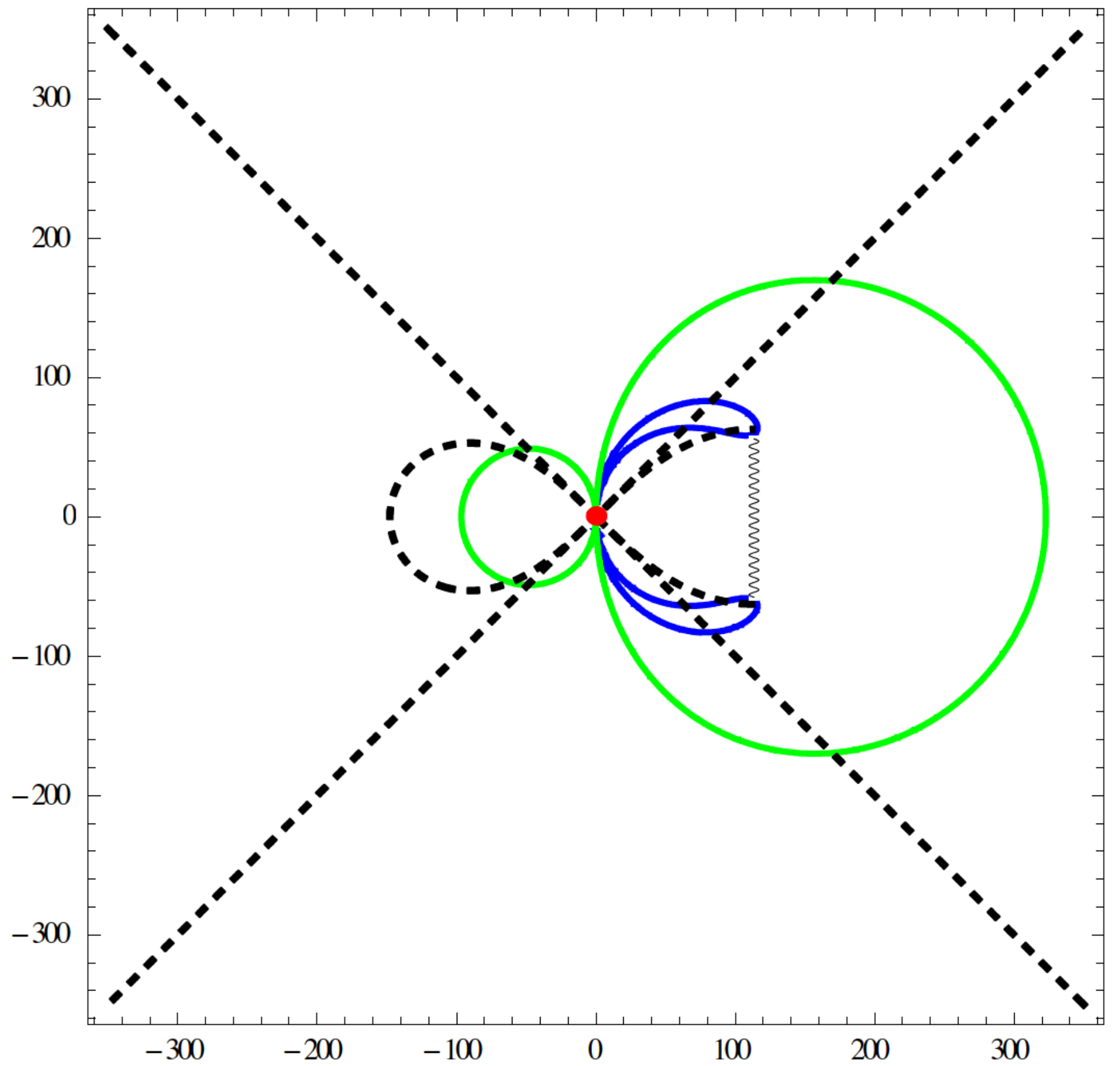}\\
   $(a)$
\end{center}
\end{minipage}
\begin{minipage}{0.32\hsize}
  \begin{center}
   \picture{clip,height=5cm}{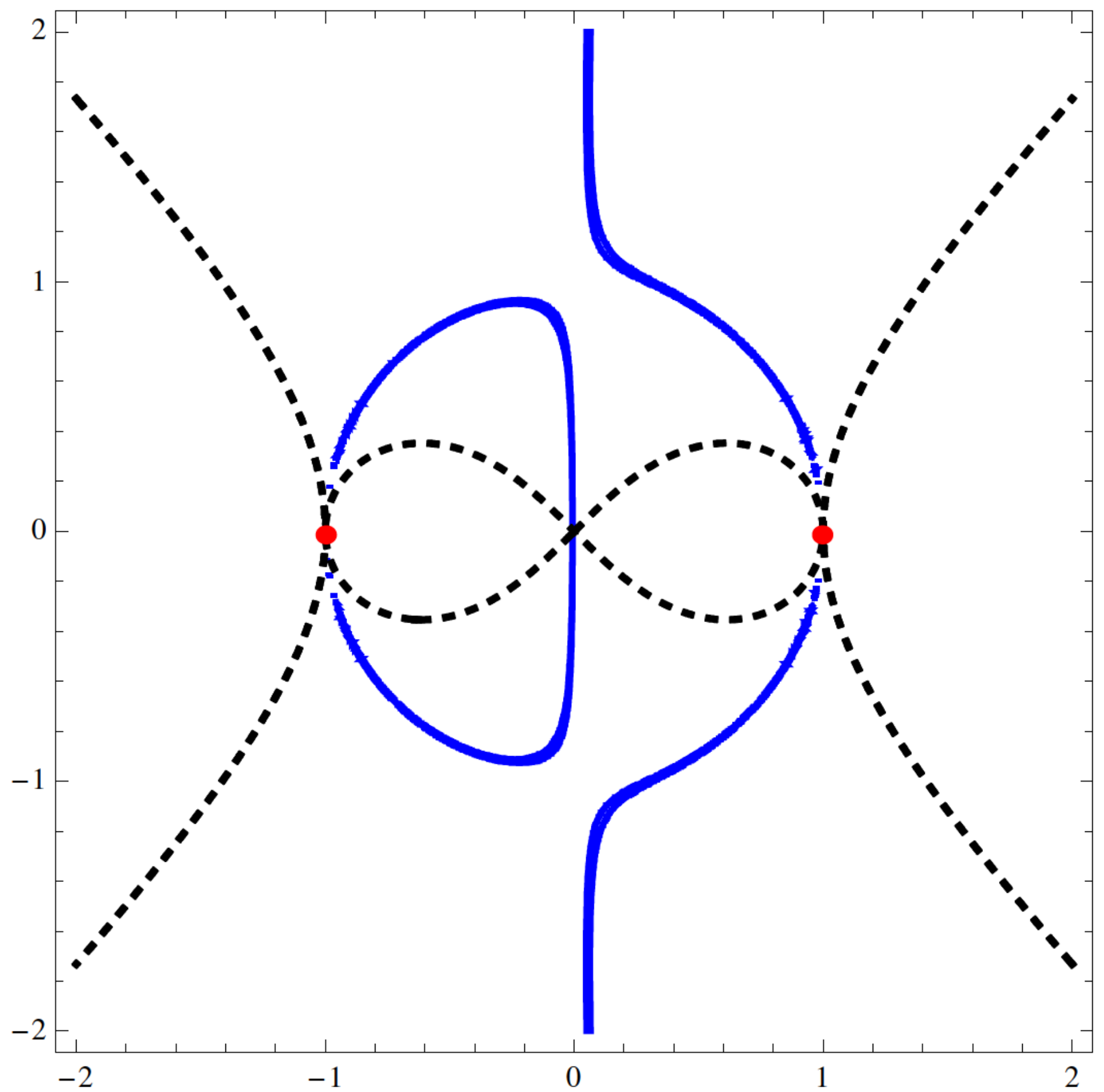}\\
   $(b)$
\end{center}
\end{minipage}
\begin{minipage}{0.32\hsize}
  \begin{center}
   \picture{clip,height=5cm}{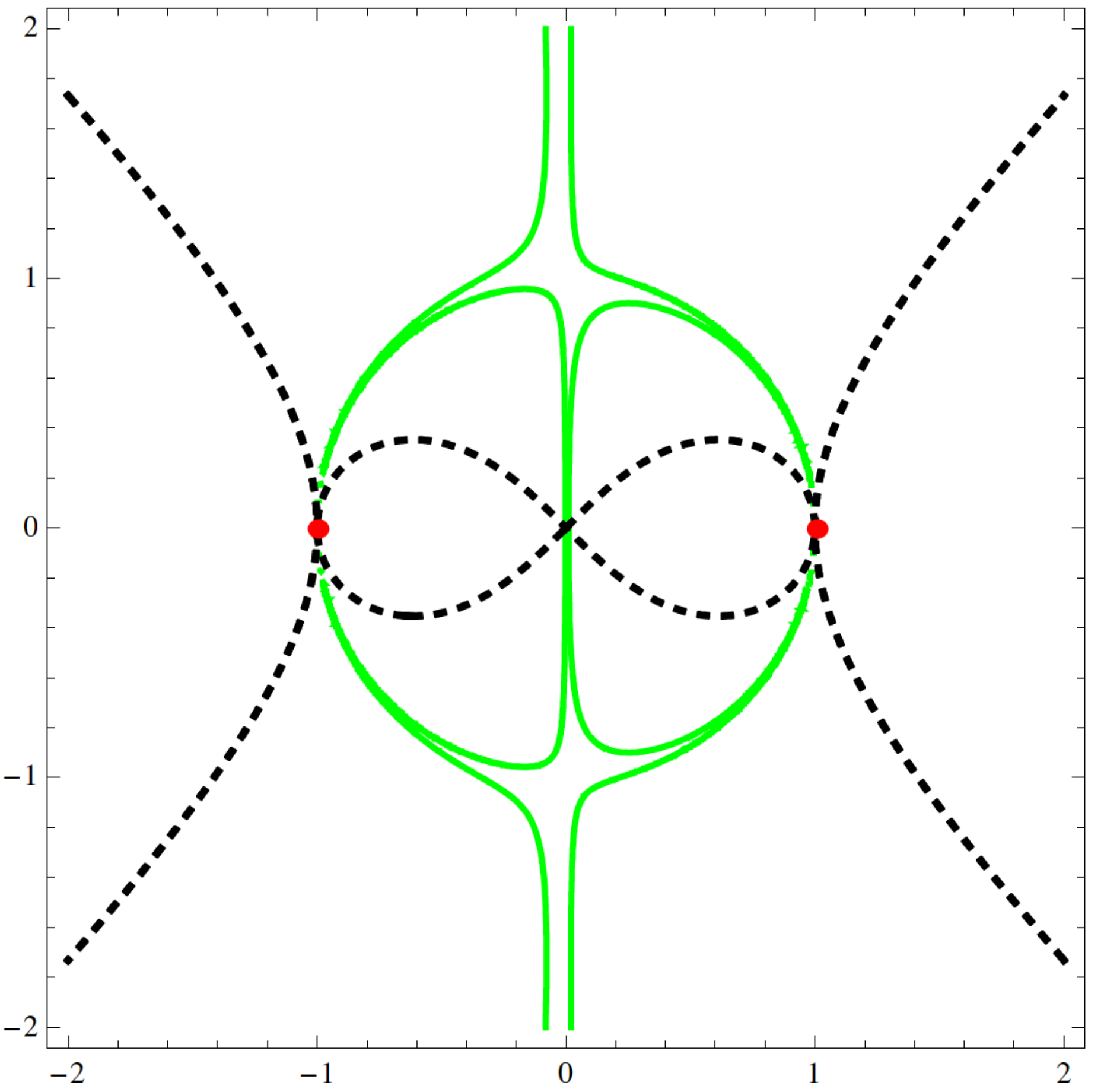}\\
$(c)$
\end{center}
\end{minipage}
 \caption{The curves defined by $N_i=N_j+N_k$. The  figure $(a)$  shows the configuration of curves in $|x|\gg 1$ region where as the figure  $(b)$ shows the configuration of $N_1=N_2+N_3$ in $|x|<1$ region and the figure $(c)$  shows the configuration of $N_2=N_1+N_3$ in $|x|<1$ region. In the present case, the curve $N_3=N_1+N_2$ does not exist.} 
\label{fig:non-BPS2}
\end{figure}

To perform a more detailed analysis, we need to specify the properties of the operators more explicitly, since the precise form of the integration contours depend on such details. As we wish to analyze the so-called Frolov-Tseytlin limit  and make a comparison with the results in the gauge theory in the next subsection, we will take as a representative example the following 
 set of operators carrying large conformal dimensions:
\beq{
\begin{aligned}
\mathcal{O}_1:&\text{ BPS}\comma \quad 2\pi \kappa_1 =2500\comma\\
\mathcal{O}_2:&\text{ non-BPS}\comma\quad 2\pi \kappa_2 = 3250\comma\\
&\text{ }  p(u)-p(\infty^{+})=-16\pi\comma \quad p(0^{+})-p(\infty^{+})=-2\pi\comma\\
\mathcal{O}_3:&\text{ BPS}\comma \quad 2\pi \kappa_3 = 3000\period
\end{aligned}
}
Here $u$ denotes  the position of an  end of the branch cut for the non-BPS operator $\mathcal{O}_2$. 
For these operators, the curves defined by $\Re \, q_i=0$  and those  defined by $N_i=N_j+N_k$ are depicted respectively in \figref{fig:non-BPS1} and \figref{fig:non-BPS2}.  

As in the case of the three BPS operators, we must now apply the general rules of section \ref{sec:wron} to determine the integration contours. As an example, 
 consider the contour $\mathcal{M}_{\rm ---}^{uuu}$ in the region where $|\Re x| \gg 1$. Focus first on the left figure of \figref{fig:non-BPS1}. 
Compared to the typical configuration shown in the left figure of \figref{fig:typical}, the curve determined by $\Re q_3=0$ (shown in brown in \figref{fig:typical}) is depicted here as a point in the middle since we are 
 considering the region where $|\Re x|\gg 1$.   Since the inside of the  
  shrunken region is where $3_+$ is small,  we  have $3_-$ as  the small solution everywhere in this figure.  From the direction of the curves $\Ga^u_{1-}$  and $\Ga^u_{2-}$, we can easily tell which of the states $1_\pm$  and $2_\pm$ are the small solutions in  each of the region separated by 
 these curves.
%\red{We want figures 7.5 and 7.6 here}
\begin{figure}[htbp]
  \begin{center}
   \picture{clip,height=5cm}{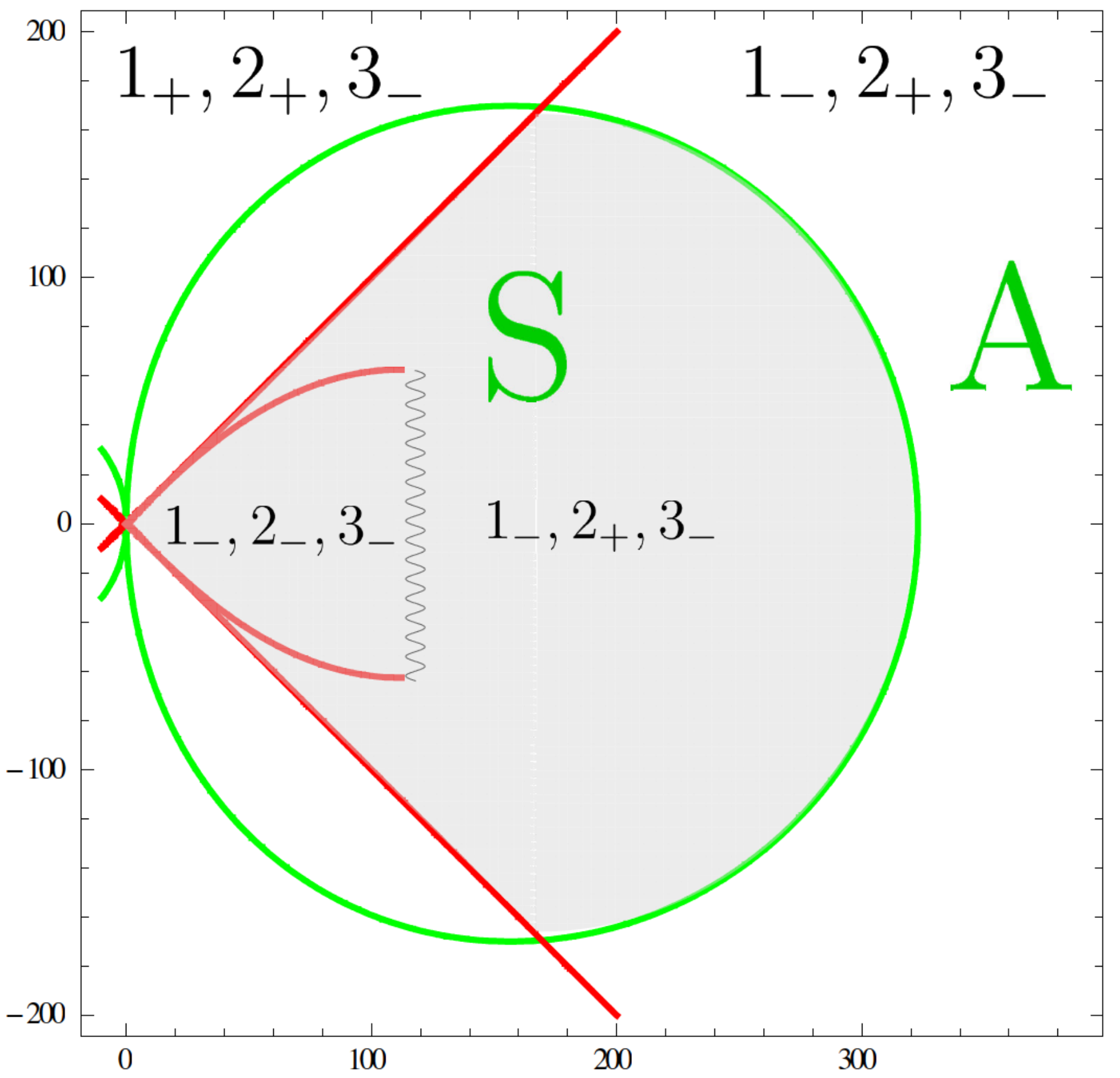}
  \end{center}
\caption{Magnified view of a  part of  the figure $(a)$ of \figref{fig:non-BPS2},  with data necessary  for determining the  contour of integration.  In each region separated by lines and/or the cut (wavy line), the set of ``small" eigenvectors are indicated. The green circle separates the symmetric (S) and the assymetric (A) regions, to which  different rules of analysis apply.  The result is that across the boundary of the shaded area, the analyticity of the Wronskian changes. 
For details of the analysis using this figure,  see the explanation in the main text.
}
\label{fig:new}
\end{figure}

\begin{figure}[tbp]
\begin{minipage}{0.24\hsize}
  \begin{center}
   \picture{clip,height=4cm}{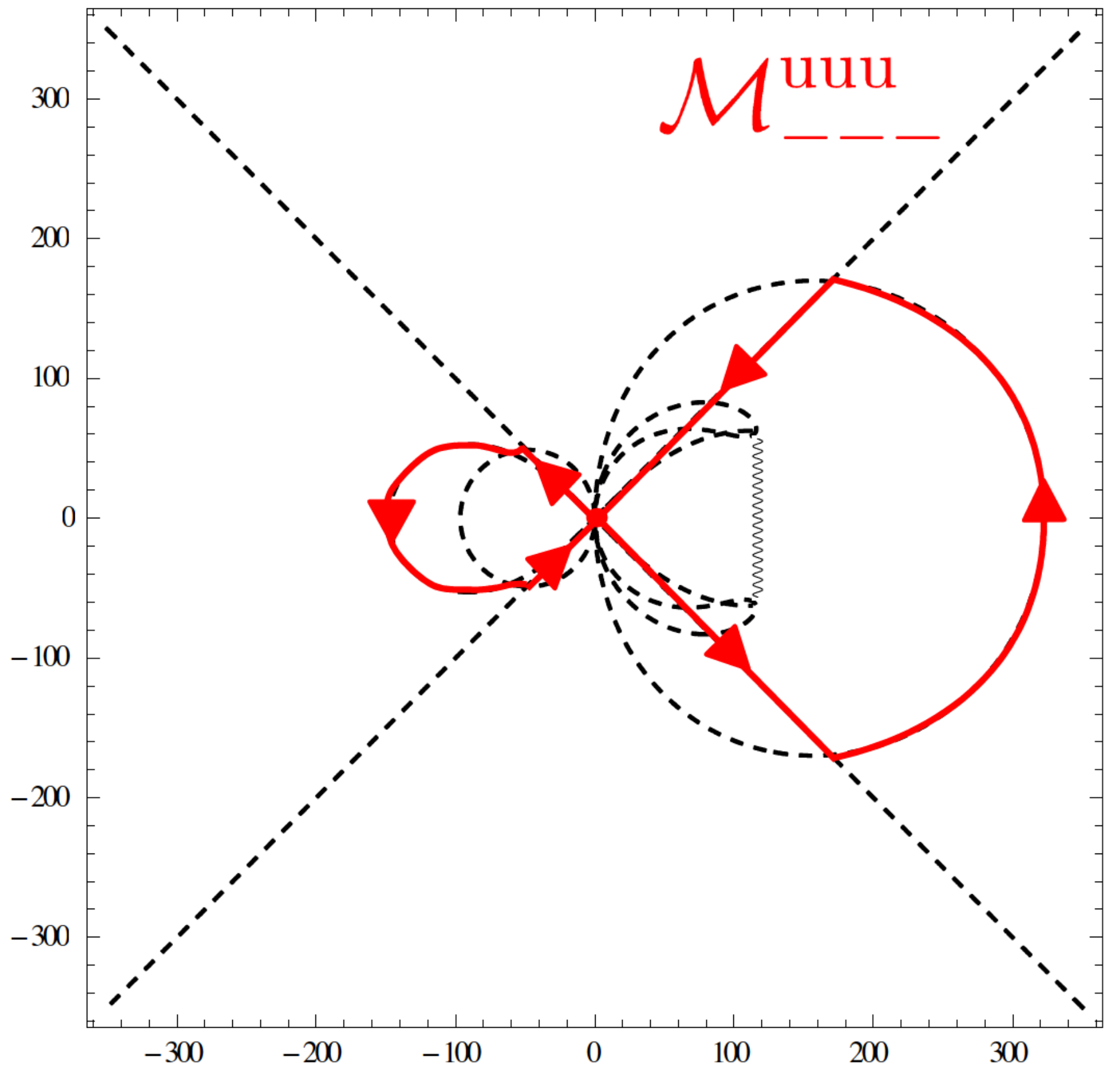}
  \end{center}
 \end{minipage}
\begin{minipage}{0.24\hsize}
  \begin{center}
   \picture{clip,height=4cm}{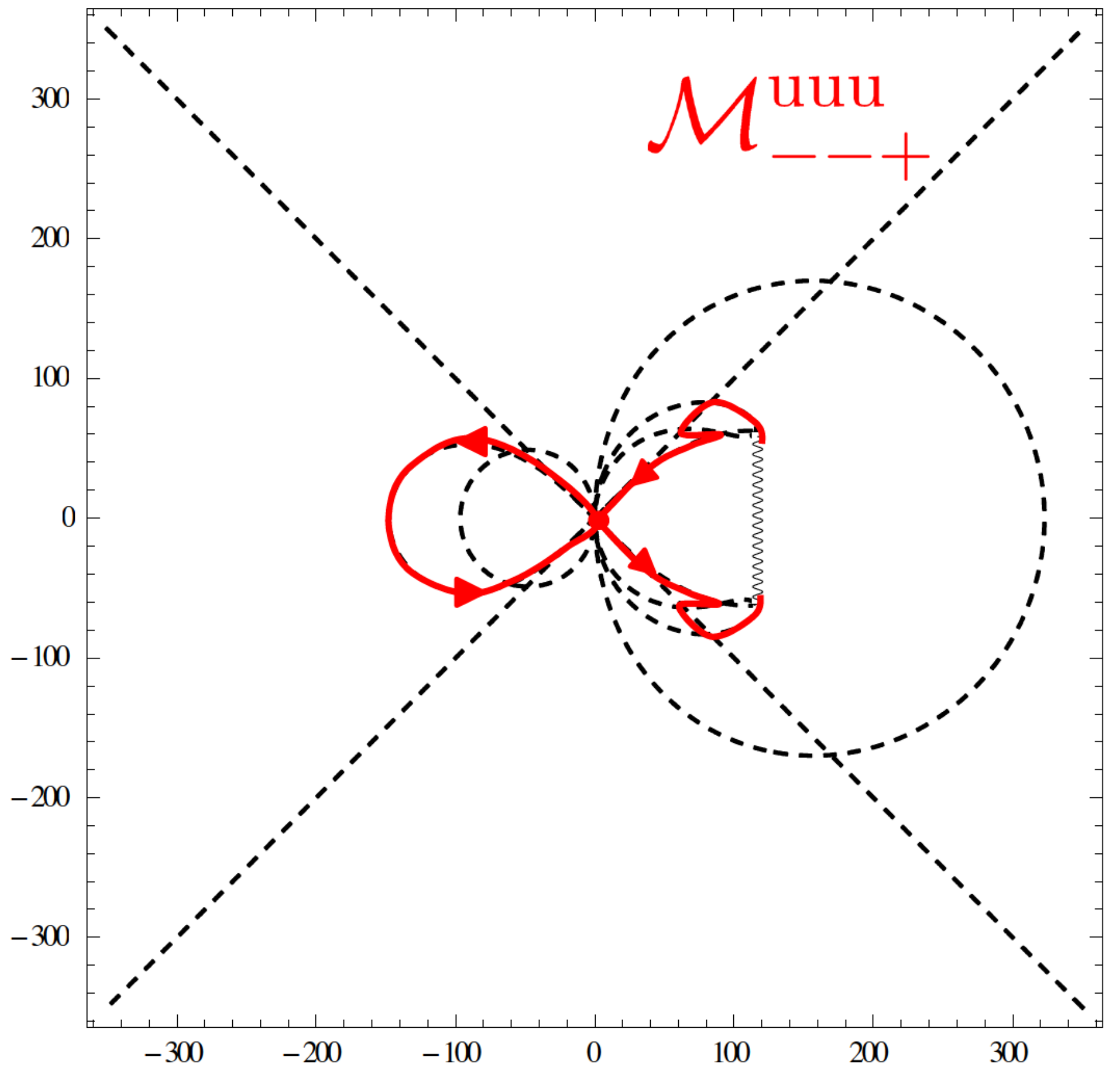}
   \end{center}
 \end{minipage}
\begin{minipage}{0.24\hsize}
  \begin{center}
   \picture{clip,height=4cm}{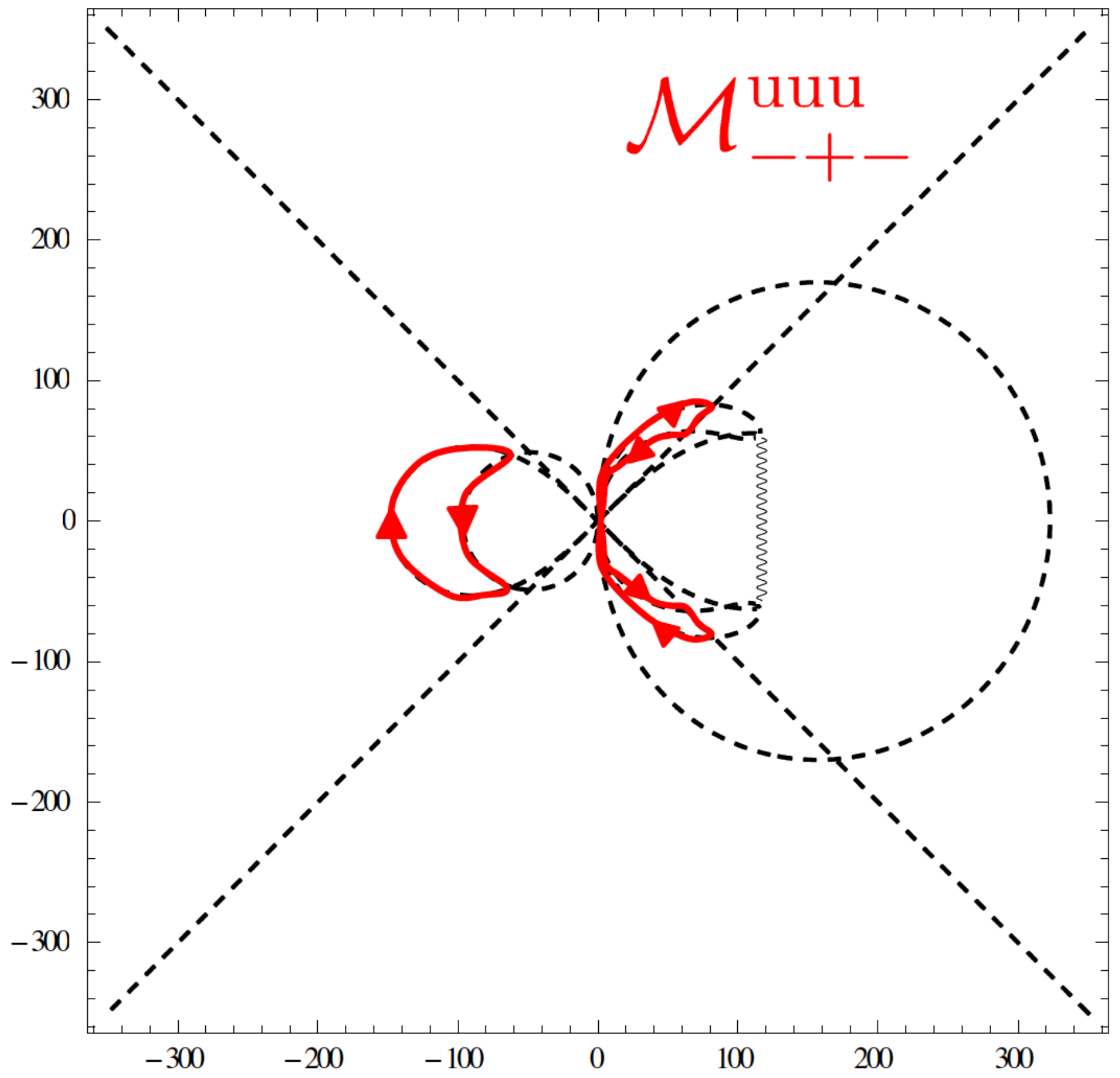}
  \end{center}
 \end{minipage}
\begin{minipage}{0.24\hsize}
  \begin{center}
   \picture{clip,height=4cm}{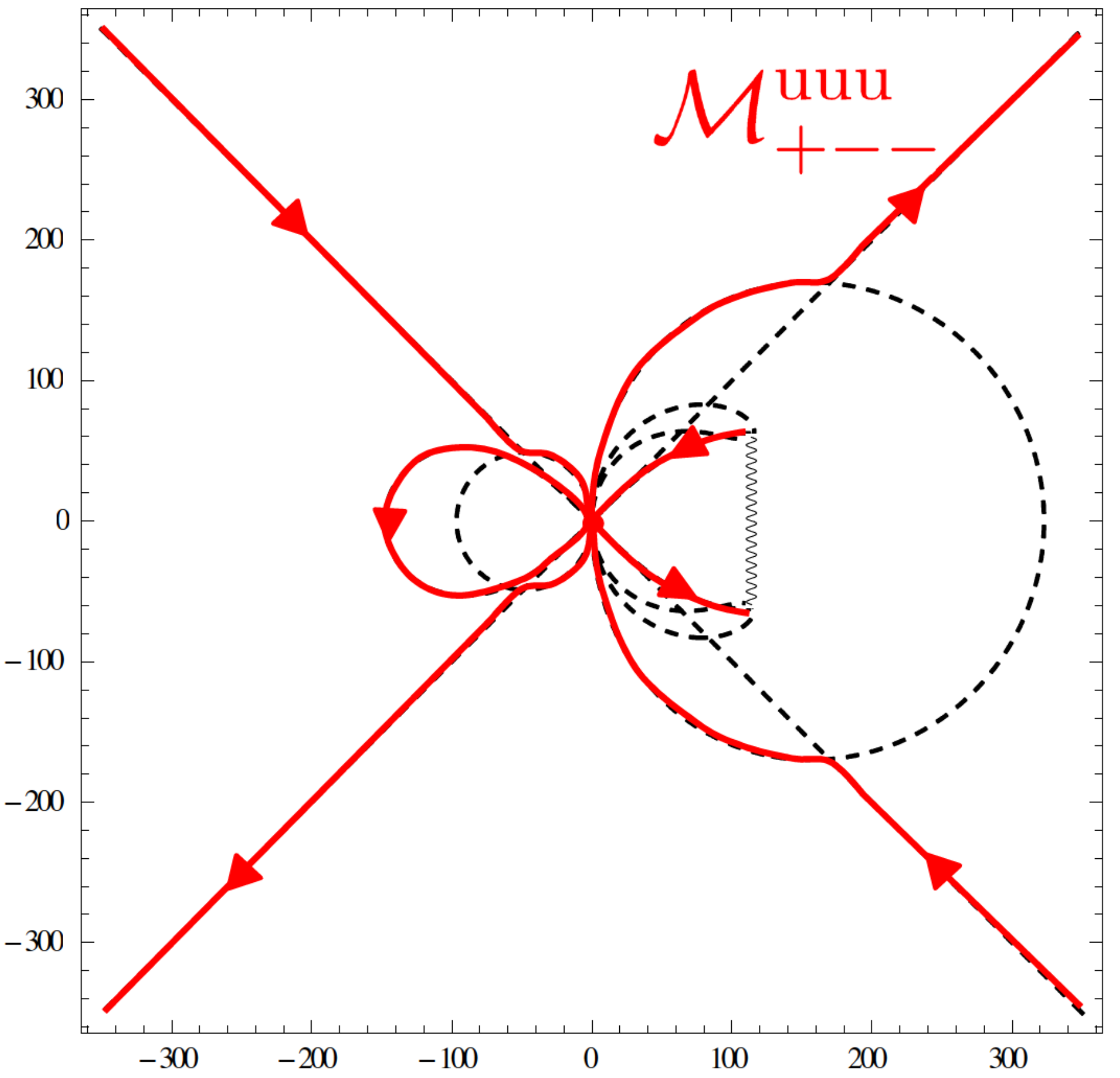}
  \end{center}
 \end{minipage}
 \caption{The integration contours $\mathcal{M}_{\pm\pm\pm}^{uuu}$ in the region $|x|\gg 1$. From left to right, $\mathcal{M}_{\rm ---}^{uuu}$, $\mathcal{M}_{\rm --+}^{uuu}$, $\mathcal{M}_{\rm -+-}^{uuu}$ and $\mathcal{M}_{\rm +--}^{uuu}$.} 
\label{fig:non-BPS3}
\end{figure}

\begin{figure}[tbp]
\begin{minipage}{0.24\hsize}
  \begin{center}
   \picture{clip,height=4cm}{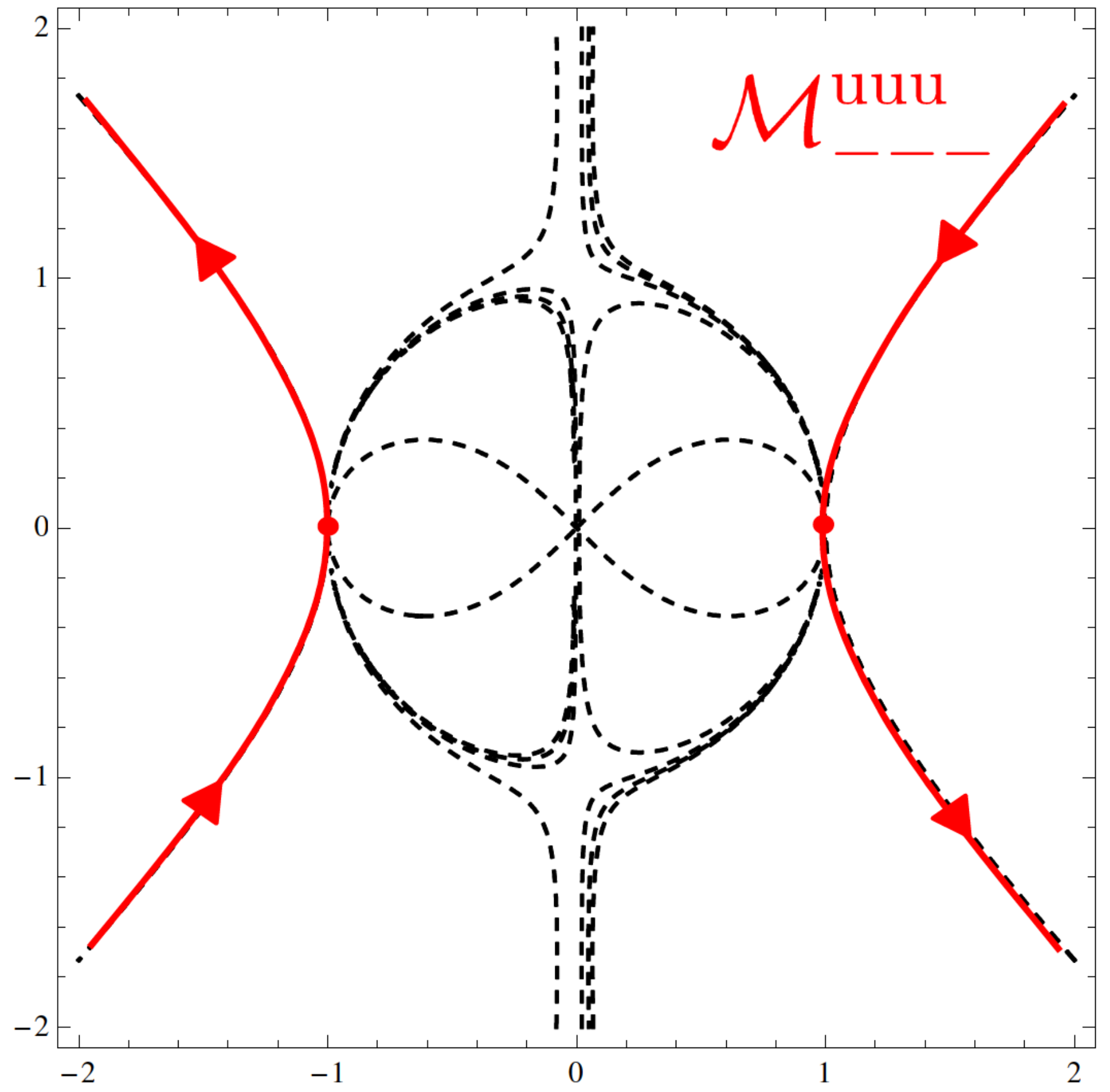}
  \end{center}
 \end{minipage}
\begin{minipage}{0.24\hsize}
  \begin{center}
   \picture{clip,height=4cm}{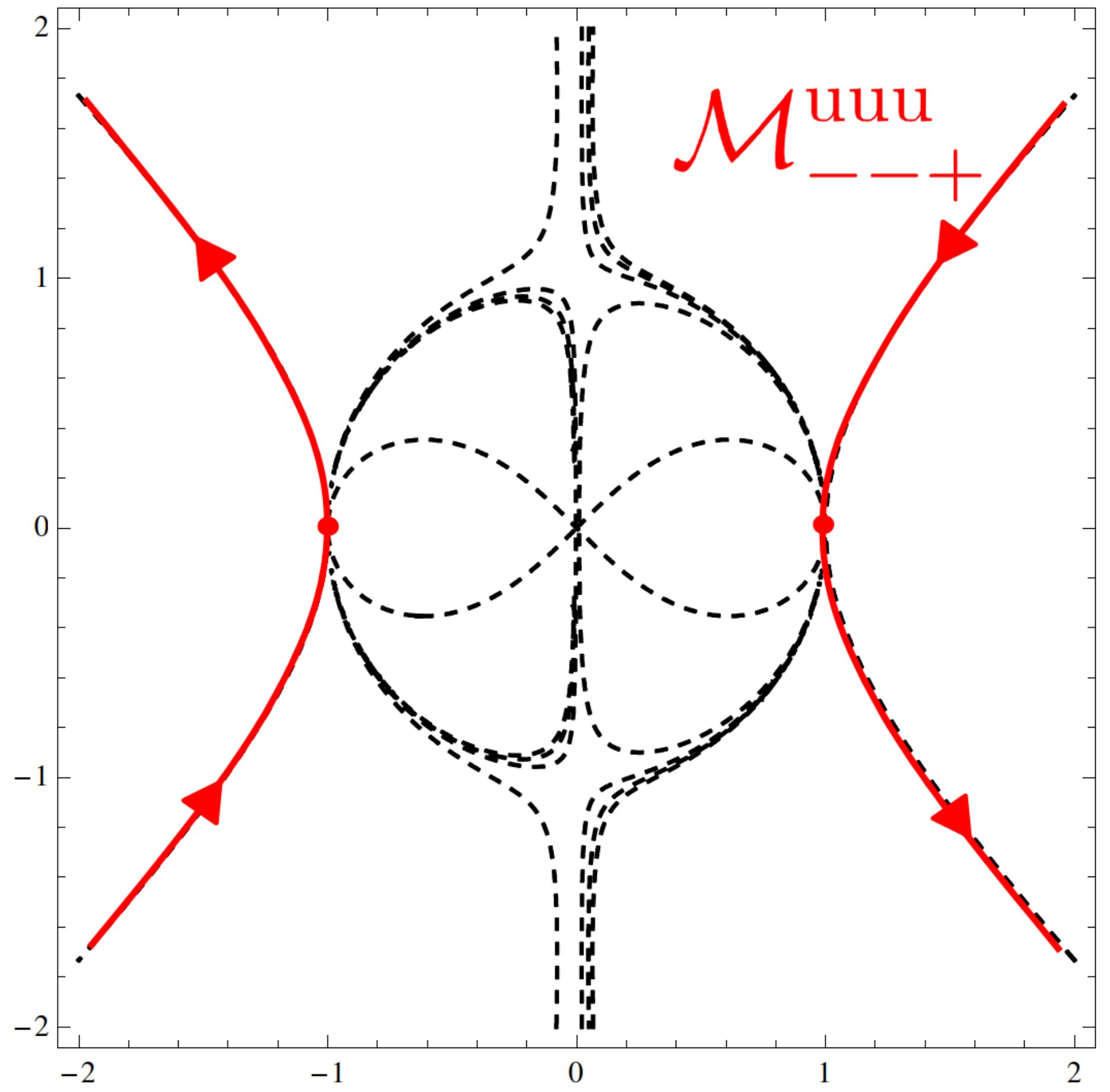}
  \end{center}
 \end{minipage}
\begin{minipage}{0.24\hsize}
  \begin{center}
   \picture{clip,height=4cm}{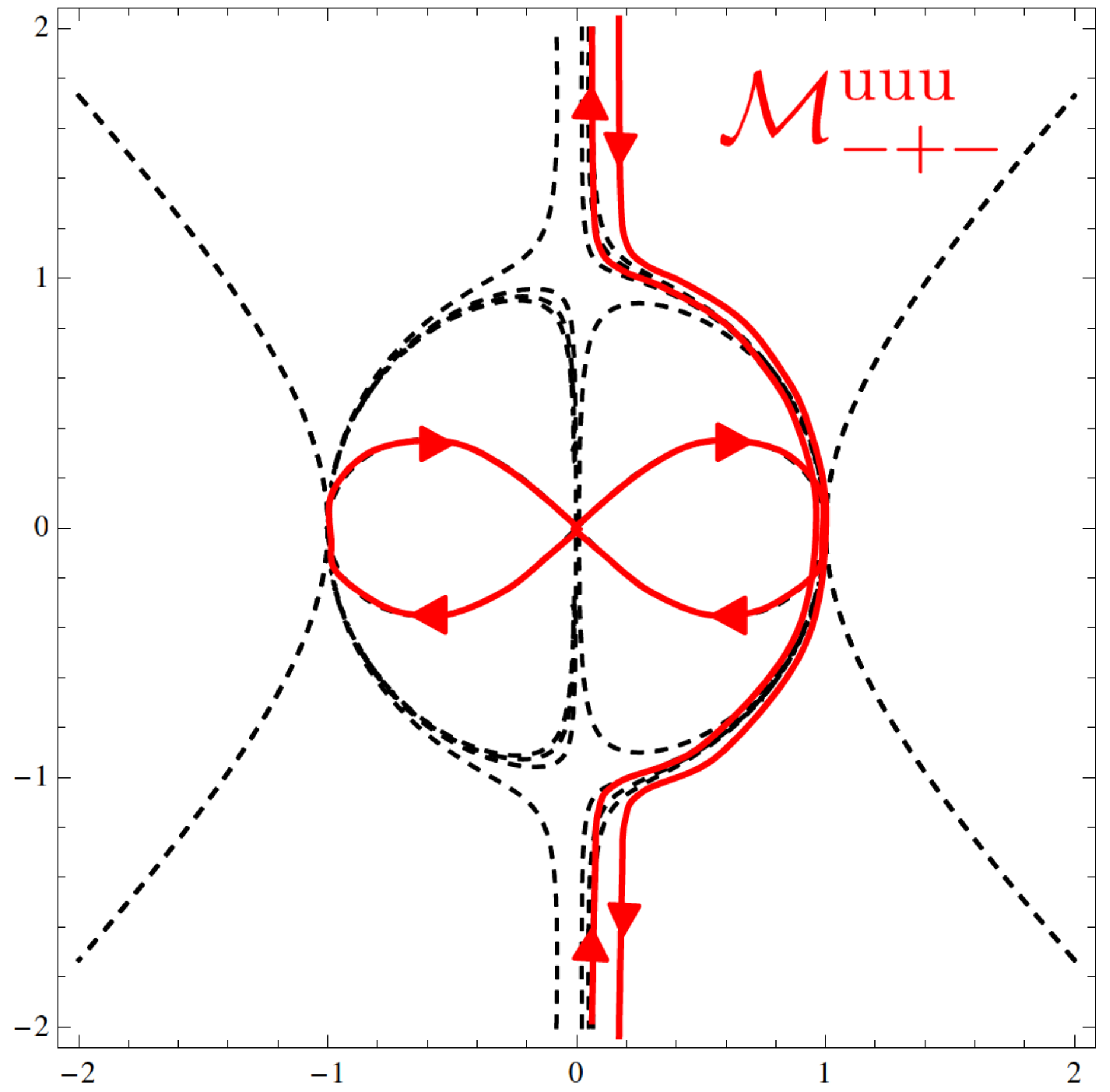}
  \end{center}
 \end{minipage}
\begin{minipage}{0.24\hsize}
  \begin{center}
   \picture{clip,height=4cm}{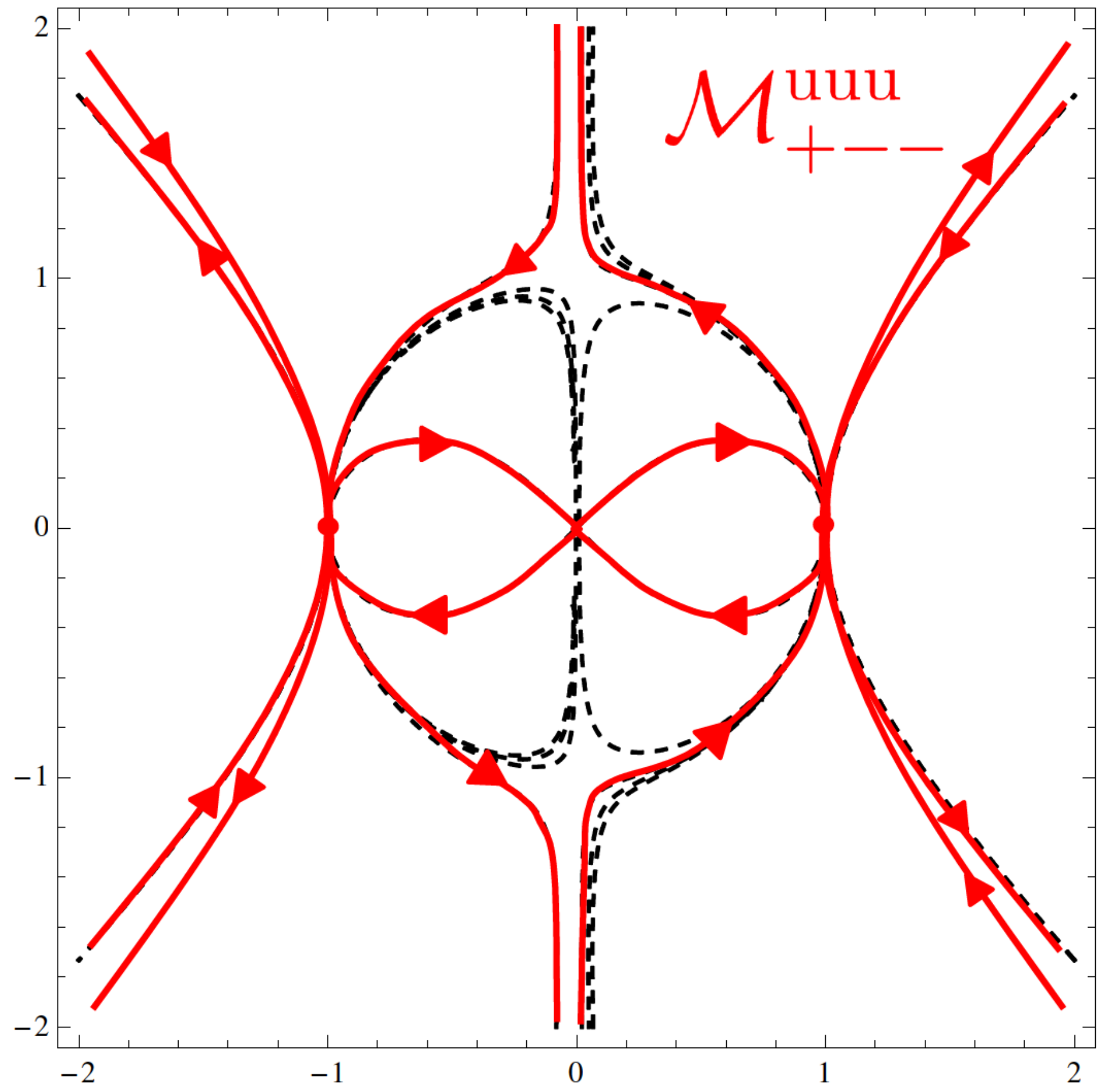}
  \end{center}
 \end{minipage}
 \caption{The integration contours $\mathcal{M}_{\pm\pm\pm}^{uuu}$ in the region $|x|< 1$. From left to right, $\mathcal{M}_{\rm ---}^{uuu}$, $\mathcal{M}_{\rm --+}^{uuu}$, $\mathcal{M}_{\rm -+-}^{uuu}$ and $\mathcal{M}_{\rm +--}^{uuu}$.} 
\label{fig:non-BPS4}
\end{figure}

Now, in distinction to the case of three BPS operators, 
we must also take into account the possible change of the analyticity of the 
 Wronskians as we cross  the lines defined by $N_i=N_j+N_k$. Thus, we 
 must analyze relevant curves drawn  in \figref{fig:non-BPS2} $(a)$, where the one in green corresponds to $N_2=N_1+N_3$ and the one in 
 blue represents $N_1=N_2+N_3$. Across these lines  the configuration 
 changes from symmetric to  asymmetric.  Accordingly, 
 the rule to find the non-vanishing set of Wronskians changes from  Rule 2 to  Rule 3.  Let us focus on the green curve, which is re-drawn in \figref{fig:new}, with additional information. 
It turns out that the configuration is symmetric inside the green circles and asymmetric outside, indicated by the letters S and A respectively.  Now in the region outside of the arc of the large green circle bordered by the lines representing $\Ga^u_{1-}$,  shown in \figref{fig:new} by  the red straight lines,   $1_-, 2_+, 3_-$ are the small solutions, as indicated in the figure.  As this is the asymmetric region  we apply the Rules 1 and 3 and conclude that the Wronskians among the states $\{1_+, 2_+, 3_+\}$ are non-vanishing.
 As we cross the arc into the shaded region inside of the green circle where the configuration is symmetric, still $1_-, 2_+, 3_-$ are the small solutions  but now we must apply the Rules 1 and 2. 
Then we learn that the Wronskians among the 
states $\{1_-, 2_-, 3_-\}$ are non-vanishing instead. In other words, the 
analyticity property of the Wronskians change across this portion of the green curve and hence it serves as a part of the contour for the convolution integral. This explains the portion of the contour along the arc of the large circle 
 shown in the left-most figure in  \figref{fig:non-BPS3}.  Now 
 consider what happens when this contour meets the $\Ga^u_{1-}$ 
 line. Across this line, the small solution changes from $1_-$ to $1_+$. 
Thus when we cross this line from inside the large circle, the set of small 
 solutions change from $1_-$, $2_+$ and $3_-$ to $1_+$, $2_+$ and $3_-$ as shown in \figref{fig:new}. 
As we are still in the symmetric region, the Rules 1 and 2 apply and hence
 we learn that set of non-vanishing Wronskians change across this line. 
Therefore this portion must constitute a part of the contour. This explains 
 the straight red line starting from the the point of intersection with the large circle.  In this fashion, we can uniquely obtain the integration contour 
$\mathcal{M}_{- - -}^{uuu}$,  shown in the leftmost figure of \figref{fig:non-BPS3},  across which the analyticity property of the Wronskians change. 
All the other contours  $\mathcal{M}_{\pm\pm\pm}^{uuu}$ can also be
 determined in an entirely similar manner,  the result  of which are depicted  in \figref{fig:non-BPS3} and  \figref{fig:non-BPS4}.
%%%%%%%%
%

\begin{figure}[htbp]
\begin{minipage}{0.5\hsize}
  \begin{center}
   \picture{clip,height=5cm}{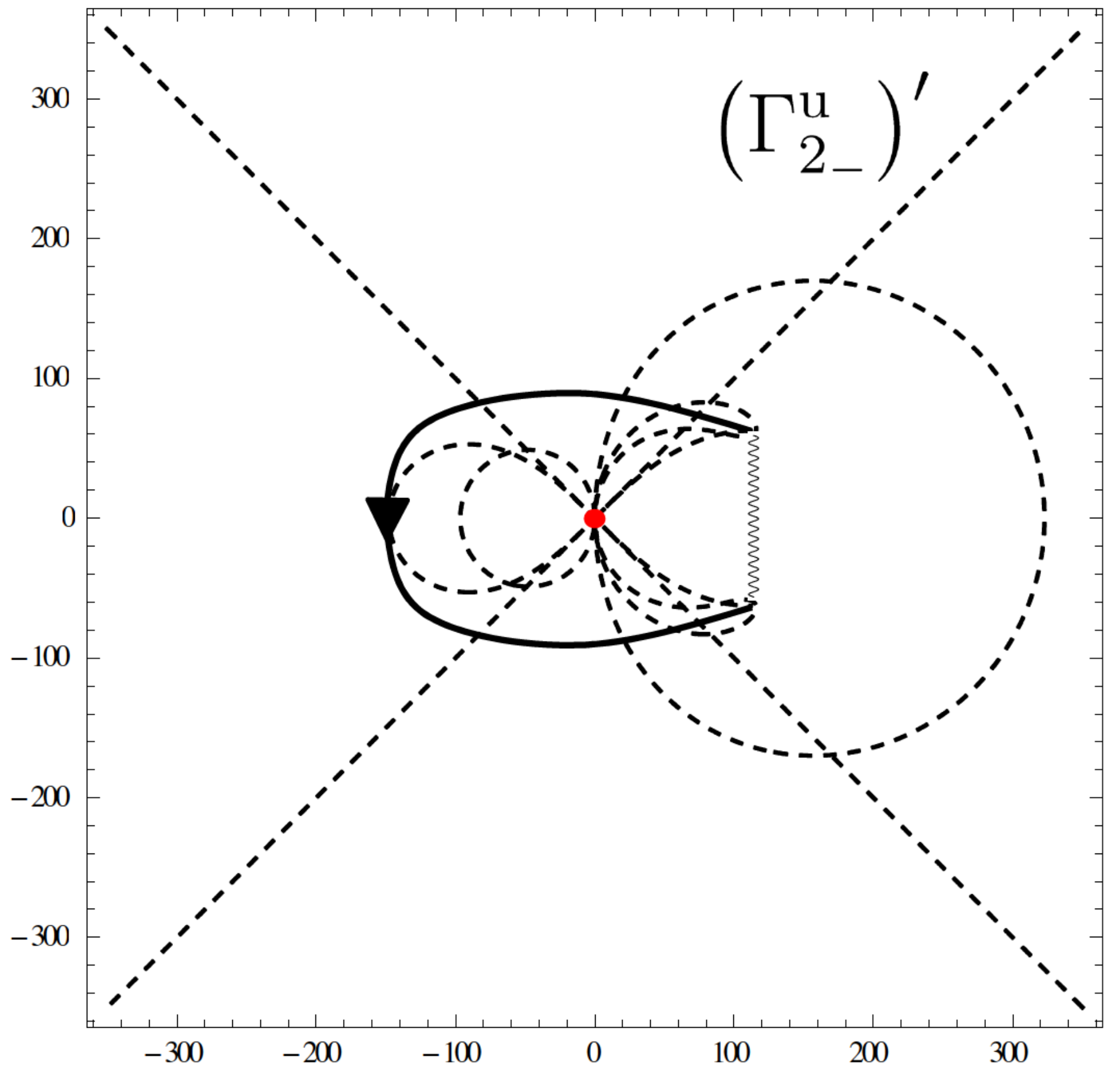}
  \end{center}
 \end{minipage}
\begin{minipage}{0.5\hsize}
  \begin{center}
   \picture{clip,height=5cm}{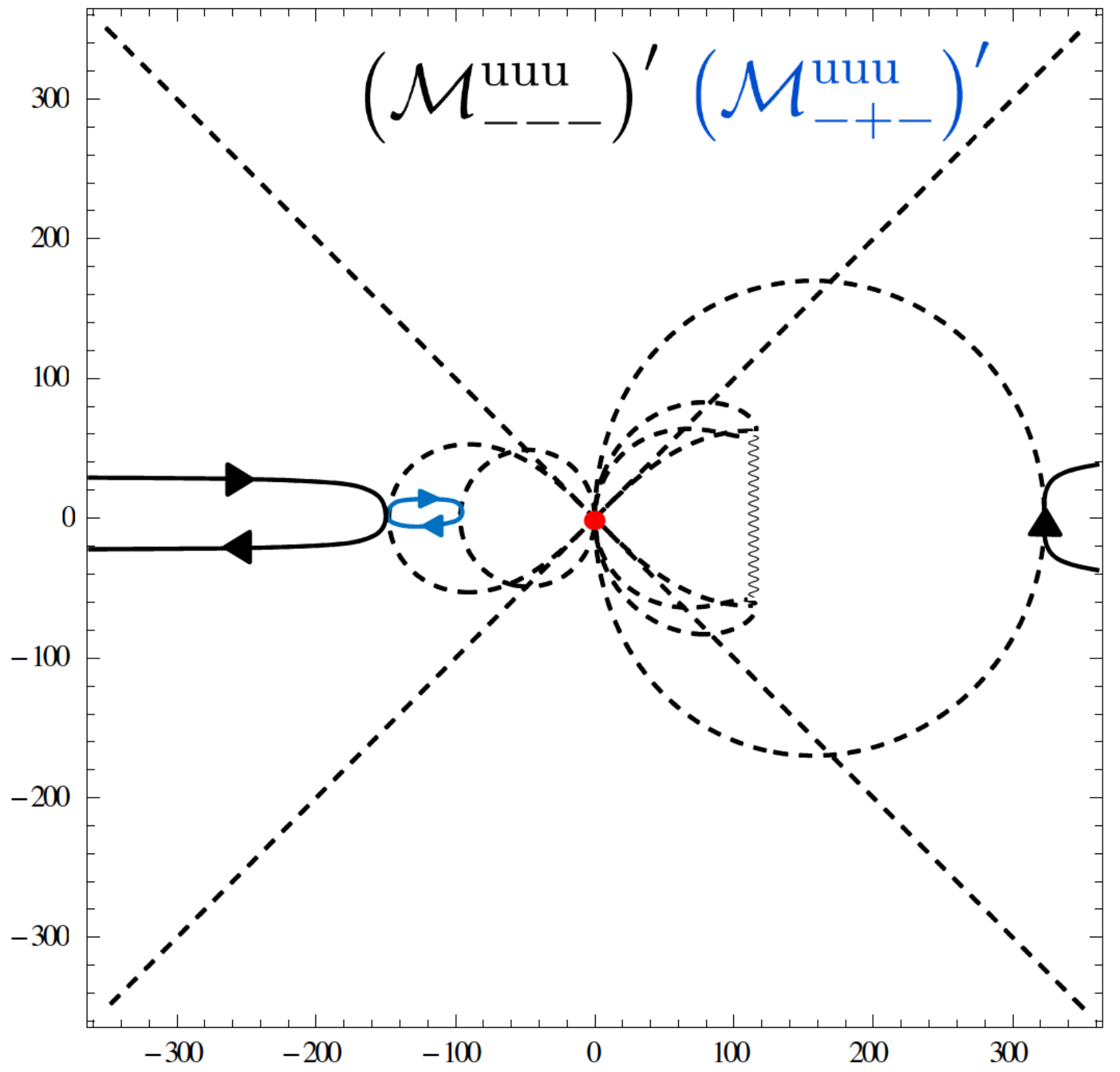}
  \end{center}
 \end{minipage}
 \caption{The contours obtained after the deformation. On the left figure, we depicted $\left( \Gamma_{2_-}^u\right)^{\prime}$. On the right figure, we depicted $\left( \mathcal{M}_{---}^{uuu}\right)^{\prime}$ in black and $\left( \mathcal{M}_{-+-}^{uuu}\right)^{\prime}$ in blue.} 
\label{fig:deform}
\end{figure}

The contours shown in \figref{fig:non-BPS3} and \figref{fig:non-BPS4}  can be simplified by continuous deformation as long as  we do not make them  pass through the singularities of the integrands. 
We can determine the positions of the singularities  numerically and  find that most of the singularities lie on the real axis.  Avoiding them, 
 we can deform each contour into a sum of  the  contour  along the unit circle and  the one which is  far from the unit circle. The results of this deformation 
are summarized as 
\beq{
\begin{aligned}
&\mathcal{M}_{---}^{uuu}\longmapsto \left( \mathcal{M}_{---}^{uuu}\right)^{\prime}+U\comma\quad
\mathcal{M}_{--+}^{uuu}\longmapsto \left( \Gamma_{2_-}^u\right)^{\prime}+U\comma\quad \\
&\mathcal{M}_{-+-}^{uuu}\longmapsto \left( \mathcal{M}_{-+-}^{uuu}\right)^{\prime}+U\comma\quad\mathcal{M}_{+--}^{uuu}\longmapsto \left( \Gamma_{2_-}^u\right)^{\prime}+U\comma\\
&\Gamma_{1_-}^u\longmapsto U\comma\quad \Gamma_{2_-}^u\longmapsto \left(\Gamma_{2_-}^u\right)^{\prime}+U\comma \quad \Gamma_{3_-}^u\longmapsto U\comma
\end{aligned}
}
where, as before, $U$ denotes the unit circle and the primed contours 
are as depicted in \figref{fig:deform}.

Let us make a remark on the separation of the integration contours into the unit circle and the large contours. It is intriguingly reminiscent of the expressions for the one-loop correction to the spectrum of a classical string \cite{GV}.
 In that context, the integration along the unit circle is interpreted as giving the dressing phase and the finite size corrections. Since our results do not include one-loop corrections, it is not at all clear whether our results can be interpreted in a similar way. However, the apparent structural similarity calls for further study. 
%%%%%%%%%%%%%%%%%%%%%%%
\subsection{Frolov-Tseytlin limit and comparison with the weak coupling result\label{subsec:FT}}
%%%%%%%%%%%%%%%%%%%%%%%%%%
\subsubsection{Frolov-Tseytlin limit of the three-point function}
We are now ready to discuss the Frolov-Tseytlin limit of the three-point function and 
 compare it with the weak coupling result.  Let us briefly  recall  how such a limit 
 arises. As shown  in \cite{Kruczenski}, the dynamics of the fluctuations around a fast-rotating  string on $S^3$ can be mapped to the dynamics of the Landau-Lifshitz model, which arises as a coherent state description of the XXX spin chain. In such a 
situation,  the angular momentum $J$ of the $S^3$ rotation can be taken to be  so large that the ratio $\sqrt{\lambda}/J$ becomes vanishingly small, even when 
 $\lam$ is large. For the spectral problem,  it has been demonstrated that such a limit is quite useful in comparing the strong coupling result with the 
weak coupling counterpart.  We would like to see if it applies also to the three 
 point functions.  For this purpose, we need to know how such a limit is taken 
 at the level of the quasi-momenta. Since the SO(4) charges of the external 
 states are proportional  to $\kappa_i$,  the appropriate limit is 
 to scale  all the $\kappa_i$ to infinity  while keeping the mode numbers 
 $\oint_{b_i}dp$ finite.  As already indicated,  we have chosen the example  in the previous subsection to be such that we can readily take such a limit. 

Upon taking the  Frolov-Tseytlin limit,  two simplifications occur in our formula. First, since the branch points are far away from the unit circle, we can approximate $p_2(x)$ on the unit circle by a quasi-momentum for a BPS operator, namely 
\beq{
p_2 (x)\simeq p_2^{\rm BPS}(x)=-2\pi \kappa_2\left( \frac{1}{x-1}+\frac{1}{x+1}\right)\period
}
Now recall that the contribution from the $E\!AdS_3$ part is such that 
 it precisely canceled the $S^3$ part in the case of the three BPS operators. 
Since the $E\!AdS_3$ part is unchanged  for the present case, again the 
 same exact cancellation takes place as far as the integrals over the unit circles 
 are concerned. Therefore we can drop such integrals and obtain 

{\small 
\beq{
&\int_{\left( \mathcal{M}_{---}^{uuu}\right)^{\prime}+\mathcal{C}_{\infty}}\hspace{-30pt}\frac{z\left( dp_1 +dp_2 +dp_3\right)}{2\pi i} \ln \sin \left(\frac{p_1+p_2+p_3}{2} \right)
+\int_{\left( \Gamma_{2_-}^u\right)^{\prime}+\mathcal{C}_{\infty}}\hspace{-30pt}\frac{z\left( dp_1 +dp_2 -dp_3\right)}{2\pi i} \ln \sin \left(\frac{p_1+p_2-p_3}{2} \right)\nn\\
&+\int_{\left( \mathcal{M}_{-+-}^{uuu}\right)^{\prime}+\mathcal{C}_{\infty}}\hspace{-30pt}\frac{z\left( dp_1 -dp_2 +dp_3\right)}{2\pi i} \ln \sin \left(\frac{p_1-p_2+p_3}{2} \right)+\int_{\left( \Gamma_{2_-}^u\right)^{\prime}+\mathcal{C}_{\infty}}\hspace{-30pt}\frac{z\left( -dp_1 +dp_2 +dp_3\right)}{2\pi i} \ln \sin \left(\frac{-p_1+p_2+p_3}{2} \right)\nn\\
&-2\int_{\left( \Gamma_{2_-}^u\right)^{\prime}+\mathcal{C}_{\infty}} \frac{z \,dp_2}{2\pi i} \ln \sin p_2\comma\label{ftfinal}
}
}
Second simplification occurs because on the large contours  the integration variable $x$ is of order $\kappa_i$.  This is precisely the situation where we can approximate the quasi-momenta of the classical strings by the corresponding quantities for  the spin-chains.  Indeed, as explained in \cite{KMMZ},  the quasi-momentum for the string can be identified with that of the Landau-Lifshitz model,  which describes the spin-chain on the gauge theory side in the above limit. 
More precisely, we can use the following identification of the quasi-momenta on the large contour: 
\beq{
&p^{\rm string} (x)\simeq p^{\rm spin}(z(x)) \period 
}
The use of the Zhukovsky variable $z(x)$ on the right hand side is motivated by 
 the fact that  in the all-loop asymptotic Bethe ansatz
equation \cite{BDS,Beisert2}, the rapidity of the spin-chain on the
gauge theory side is identified with the Zhukovsky variable on the
string theory side. 
In the present situation, however, since $z(x) \simeq x$ for large $x$,  the quasi-momenta in \eqref{ftfinal} can be replaced simply with the quasi-momenta for the corresponding spin-chain states at the same value of $x$. 

With such a replacement, the expression \eqref{ftfinal} already appears  rather 
 similar to the weak-coupling result. To make the resemblance more 
conspicuous,  we can  regard  the integral  of $\sin \left((-p_1+p_2+p_3)/2 \right)$
along $(\Ga^u_{2-})'$ on the upper sheet as the integral  of $\sin \left((p_1+p_2-p_3)/2 \right)$ along the reversed contour on the lower sheet  for $p_2$, which we denote by $(\Ga^{l}_{2-})'$. Combining this with the integral of 
 $\sin \left((p_1+p_2-p_3)/2 \right)$ along  $(\Ga^u_{2-})'$
already present and defining $\left( \Gamma_{2_-}\right)^{\prime}$ 
to be the sum of $(\Ga^u_{2-})'$ and $(\Ga^{l}_{2-})'$, 
 we can write 
\eqref{ftfinal} as 
\beq{
&\int_{\left( \Gamma_{2_-}\right)^{\prime}+\mathcal{C}_{\infty}}\hspace{-30pt}\frac{z\left( dp_1 +dp_2 -dp_3\right)}{2\pi i} \ln \sin \left(\frac{p_1+p_2-p_3}{2} \right) -\int_{\left( \Gamma_{2_-}\right)^{\prime}+\mathcal{C}_{\infty}} \frac{z \,dp_2}{2\pi i} \ln \sin p_2 +\text{\sf Mismatch}\comma\label{ftfinal2}
}
where   {\sf Mismatch} is given by
\beq{{\sf Mismatch}=&
\int_{\left( \mathcal{M}_{---}^{uuu}\right)^{\prime}+\mathcal{C}_{\infty}}\hspace{-30pt}\frac{z\left( dp_1 +dp_2 +dp_3\right)}{2\pi i} \ln \sin \left(\frac{p_1+p_2+p_3}{2} \right)\nn\\
&+\int_{\left( \mathcal{M}_{-+-}^{uuu}\right)^{\prime}+\mathcal{C}_{\infty}}\hspace{-30pt}\frac{z\left( dp_1 -dp_2 +dp_3\right)}{2\pi i} \ln \sin \left(\frac{p_1-p_2+p_3}{2} \right)\period\label{mismatch}
}
Now the corresponding weak-coupling result  obtained in \cite{Tailoring3} can be re-cast into the following form by the use of integration by parts, 
\beq{
&\int_{-\mathcal{A}_2}\frac{z\left( dp^{\rm spin}_1 +dp^{\rm spin}_2 -dp^{\rm spin}_3\right)}{2\pi i} \ln \sin \left(\frac{p^{\rm spin}_1+p^{\rm spin}_2-p^{\rm spin}_3}{2} \right) -\int_{-\mathcal{A}_2} \frac{z \,dp^{\rm spin}_2}{2\pi i} \ln \sin p^{\rm spin}_2 \comma \label{weakresult}
}
where $\mathcal{A}_2$ is the contour which encircles the branch cut of $p_2$ counterclockwise. Comparing \eqref{ftfinal2} and \eqref{weakresult}, 
one notes the following: (i)\ The terms denoted by {\sf Mismatch} in the 
strong coupling result are not present in the weak coupling expression. 
(ii)\ The integrands of the rest of the terms are precisely of the same form
as for the weak coupling result,  but the contours of integrations are different. 
This makes a difference in the answer since in deforming the contours  from those for the strong coupling  to those for the weak coupling  picks up non-vanishing  contributions from the singularities of the integrands. Concerning the three-point functions, there is no firm argument that the Frolov-Tseytlin limit must be universal for all the observables. Therefore the discrepancies 
that we found above do not immediately imply the breakdown of the duality. 
However, it is certainly of importance to clarify  the origin of these 
 differences. As a part of the possible understanding, below we shall offer a natural mechanism which can change the  contours of integration. 

\subsubsection{A mechanism for modifying the contours \label{subsubsec:mech}}
%%%%%%%%%%%%%%%%%
The mechanism that we wish to point out  is based on the possibility of having extra singularities on the worldsheet. To see this, let us first recall that in the derivation of the important rules which determine  the analyticity of the Wronskians, we have made an important assumption that the only singularities on the worldsheet of the solutions of the ALP occur at the positions of the vertex insertion points. This in turn means that if there exist extra singularities this assumption breaks 
 down and affects the rules for  determining  the contours of  the convolution integrals\fn{A similar mechanism of changing the integration contour by the extra singularities is discussed in the context of the so-called ODE/IM correspondence \cite{Bazhanov}.}.  Depending on the number and the positions of 
 the extra singularities, the contours can be modified in various ways and 
 it might  be possible to obtain the contour  which appear in the weak coupling result.  

Now we can provide some arguments which indicate that indeed the 
 existence of additional singularities is not uncommon. First, recall that the usual finite gap method is capable of constructing solutions which correspond 
 to the saddle point configurations for two-point functions.  As such they 
 contain only two singularities, normally placed at $\tau =\pm \infty$ in the 
 cylinder coordinates.  In such a formalism designed to deal with two-point 
 functions,  description of three-point solutions would  require 
additional singularities.  In our treatment, due to the inability to construct 
  genuine three-point saddle solutions, we describe the effect of the three vertex operators separately except for imposing the global monodromy condition
that reflects the essence  of their interaction.  However, as already emphasized in our previous work\cite{KK2}, if we wish to deal properly with the three- and higher- point functions using  algebraic curve setup  for a theory with infinite degrees of freedom,  one should actually start from the infinite gap solutions and then consider the limits where  the infinite number of cuts on the spectral curve degenerate to zero size. This process is rather non-trivial and it should be possible to produce some extra singularities on the worldsheet. Although we cannot 
 demonstrate this phenomenon explicitly for  the three-point solution, 
 we  know that already at the level of two-point solution such a mechanism 
exists, as  discussed in some detail in section \ref{subsubsec:multi}. 
There we saw explicitly that a
``one-cut" solution obtained from a multi-cut solution  in a certain 
 degeneration limit  can produce extra singularities without affecting the 
 infinite number of conserved  charges carried by the solution.  It is certainly 
 expected that such a mechanism would exist also in the case of higher-point 
 solutions.  An interesting question is which of the saddle points, those with extra singularities or those without, describe the correlator of the gauge-theory operators. In any case,  further studies  are  definitely needed to clarify this 
issue. 

%%%%%%%%%%%%%%%%%

%%%%%%%%%%%%%%%%%%
\section{Discussions\label{sec:discussion}}
%%%%%%%%%%%%%%
In this paper, we have succeeded in computing the three-point functions 
at strong coupling  of certain non-BPS states with large charges  corresponding
 to the composite operators in the SU(2) sector of the $\mathcal{N}=4$ super Yang-Mills 
 theory. As we have already given a summary of the main result 
 in section \ref{subsec:intro-mot}, we shall not repeat it here. Instead,   below we would like to 
give some comments  and indicate some important issues to be clarified 
 in the future. 

One conspicuous feature of our result is that even for  rather general 
 external states the integrands  of the integrals expressing  the 
structure constant exhibit structures quite similar to the corresponding result at weak coupling. This is quite non-trivial since the weak coupling result in the 
relevant semi-classical regime is obtained from the determinant formula
 for the inner product of the Bethe states, which is so different from the 
method employed for strong coupling. This suggests that we should seek 
  better understanding by reformulating the weak coupling computation in 
 a more ``physical" way.  As a step toward such a goal, an attempt was made in \cite{KKN}, where the inner product of the Bethe states is re-expressed 
 in terms of an integral over the separated dynamical variables. As the notion 
 of the wave function is clearly visible in this formulation, it may give a hint 
for the common feature of the strong and the weak coupling regimes, if 
 an efficient method to identify the semi-classical saddle point 
 can be developed. 

In contrast to the similarity of the integrands, there is a rather clear difference in  the contours of the integrals expressing the three-point coupling in the  weak and the strong coupling computations. This is not just a quantitative difference  but rather a qualitative one.  Reflecting the fact that the determinant formula
 deals with the Bethe roots, the contour of integration in the weak coupling 
 case is around a cut formed by the condensation of such Bethe roots. Information of such a cut is contained in the quasi-momentum $p(x)$. On the other hand, the principal quantity which determines the integration contour is 
 the real part of the quasi-energy $q(x)$, which is conjugate to the worldsheet time $\tau$. Apparently, this notion is not present in the weak coupling formulation. Together with the possible extra singularities on the worldsheet discussed in section \ref{subsubsec:mech}, the question of the contour requires better understanding. 

There are a couple of further interesting questions that one should study concerning  our result.  One is about the limit of our formula where one of the operators is much smaller than the other two.  Such three-point functions were first studied on the string-theory side in \cite{Zarembo, Costa, RT} assuming that the light operator does not change the saddle-point configuration of the other two operators. However, a systematic study on the gauge-theory side \cite{Tailoring2} reveals that the light operator in some cases modifies the saddle-point substantially. By examining the limit of our formula, it would be possible to understand in detail when and how such a ``back-reaction" occurs. 
Another important problem is to understand  the physical meaning of the integration along the unit circle in our formula and clarify if it can be interpreted as the contribution from the dressing phase and the finite size correction as in the case of the one-loop spectrum of a classical string \cite{GV}.

Finally, let us go back once again to the rather simple structure of the 
integrand we found,  similar to the weak coupling result. The simplicity 
 of such a result suggests that there should perhaps be a better more intrinsic 
 formulation  for computing the three-point functions. In the existing literature, including this work, the calculation of the three-point function in the strong coupling regime is divided into the computation 
 of the contribution of the action part and that of the vertex 
 operator (wave function) part.   As we have seen in section \ref{sec:3pt}, in the process 
 of putting these separate contributions together  there occurs a substantial 
simplification,  besides the usual cancellation of divergences.  
This strongly indicates that such a separation is not 
essential and one should rather seek relations which reflect the 
structure of the entire three-point function based on some dynamical 
 symmetry of the theory  including the integrable structure. 
This is of utmost importance since the true understanding of  the AdS/CFT duality lies not  just in the comparison of the calculations of various physical quantities in the strong and the weak coupling regimes itself but rather in identifying the common principle behind such computations and agreements. 

To make the above  remark somewhat more concrete, let us recall that the most important ingredient in the computation performed in this paper at strong coupling 
is the global consistency relations for the monodromy
of the solutions of the auxiliary linear problem around three 
 vertex insertion points. Together with the analyticity property in the 
spectral parameter, the important quantity $\langle i_\pm, j_\pm \rangle$, 
which relates the behavior of the solutions around different insertion points, 
 is extracted and serves as the building block for  the 
three-point coupling. On the other hand, in the weak coupling computations 
 so far performed,  the computation of the three-point coupling 
 is reduced to  those of the inner products  of the Bethe states 
 and their combinations.  Although this is an efficient method,
 it is based basically on the picture of the two-point function and 
not on some principle which governs the entire three-point function. 
Therefore we believe that an extremely important problem is to find 
 some functional equations (or differential  equations)  satisfied by the three-point function,  from which one can determine the coupling constant  more or less directly.  We hope  to discuss  this type of formulation elsewhere.

%%%%%%%%%%%%%
%%%%%%%%%%%%%%%%%%%%%%%
%\newpage
\par\bigskip\noindent
{\large\bf Acknowledgment}\par\smallskip\noindent
%%%%%%%%%%%%%%%%%%%%%%%
We would like to thank Y.~Jiang, I.~Kostov, D.~Serban and P.~Vieira for discussions. S.K. would like to acknowledge the hospitality of the Perimeter Institute for Theoretical Physics, where part of this work was done.
The research of Y.K. is supported in part by the 
 Grant-in-Aid for Scientific Research (B) 
No.~25287049, while that of S.K. is supported in part 
 by JSPS Research Fellowship for Young Scientists, from the Japan 
 Ministry of Education, Culture, Sports, Science and Technology. 

%%%%%%%%%%%%%%%%%%%%%%%%%%%%%%%
\appendix
%%%%%%%%%%%%%%%%%
\section{Details on the one-cut solutions\label{apsec:1-cut}}
In this appendix, we will provide some further details on the one-cut solutions.
\subsection{Parameters of one-cut solutions in terms of the position of the cut\label{apsubsec:cut}}
In section \ref{subsec:one-cut} we have given generic expressions for the parameters 
 which characterize the one-cut solutions in terms of the integrals 
 involving  $p(x)$ and $q(x)$.  If we now use  the explicit forms of $p(x)$ and $q(x)$ given in (\ref{onecutpx}) and (\ref{onecutqx}),  one can evaluate the parameters $\nu_i$, $m_i$ and $\theta_0$  in terms of the position 
 of the cut specified by $u$.  The results take several different forms depending on the region where the cut is located. 
 It is convenient to express  them in universal forms by introducing  two additional sign factors  $\eta_1$ and $\eta_{0,1}$. 
 Together with the factor $\ep$ already introduced in \eqref{defep}, we give their definitions in the following table:
\begin{center}
Table 3. \ Sign factors to distinguish between the positions of the cut.
\begin{align}
\begin{array}{c|c |c|c|c}
& \Re u <-1 & -1<\Re u<0 & 0 <\Re u<1 & 1< \Re u \\ \hline 
\ep & + & - &- & +\\ 
\eta_1  & + & +& +& - \\
\eta_{0,1} & +& +&- &+ \nn
\end{array}
\end{align} 
\end{center}
Then, $\nu_1$ and $\nu_2$ are obtained as 
\begin{align}
\nu_1 &= \kappa \left[-{\eta_1 + \eta_{0,1} |u| 
\over |u-1|} + \ep {\eta_1-\eta_{0,1}|u| \over |u+1|}\right] \comma \\
\nu_2 &= \kappa\left[{\eta_1 - \eta_{0,1} |u| 
\over |u-1|} - \ep {\eta_1+\eta_{0,1}|u| \over |u+1|} \right]
= \ep \nu_1(u\rightarrow -u)  \period 
\end{align}
As for $m_i$, we can immediately obtain them form $\nu_i$  
by the substitution $\ep \rightarrow -\ep$, because, as seen in \eqref{onecutpx} and \eqref{onecutqx}, this interchanges  $q(x)$ and $p(x)$:
\begin{align}
m_1 &= \nu_1 (\ep \rightarrow -\ep) \comma \\
m_2 &= \nu_2 (\ep \rightarrow -\ep) \period
\end{align}
Now  $\cos^2 (\theta_0/2)$ and $\sin^2 (\theta_0/2)$ can be deduced from the Virasoro condition (\ref{Virtwo}) as 
\begin{align}
\cos^2{\theta_0 \over 2} &= { |u| - \eta_1\eta_{0,1} \Re u 
\over 2|u|} \comma \qquad 
\sin^2{\theta_0 \over 2} = { |u| + \eta_1\eta_{0,1} \Re u 
\over 2|u|} \period
\end{align}

The right and the left charges are obtained from (\ref{pxinf}) and (\ref{pxzero}) 
 to be 
\begin{align}
R &= -\frac{\kappa \sqrt{\lam} \eta_1}{2} \left({\Re u-1 \over |u-1|}  + \ep {\Re u +1 \over |u+1|}    \right) \comma \\
L &= \frac{\kappa \sqrt{\lam} \eta_{0,1}}{2 |u|}  \left({|u|^2 -\Re u \over |u-1|}  + \ep {|u|^2 +\Re u  \over |u+1|}     \right) \period
\end{align}
From the definition of $R$ and $L$ as the Noether charges, they must be 
 expressed in terms of the parameters $\nu_i$ and $\theta_0$ in a universal 
 manner independent of the position of the cut. Indeed by using the 
 formulas already obtained for the parameters and the charges in terms of $u$, we can check the universal expressions
\begin{align}
{R\over \sqrt{\lam}} &= \frac{1}{2}\left( -\nu_1 \cos^2{\theta_0 \over 2} +\nu_2\sin^2{\theta_0 \over 2}\right) \comma \\
{L\over \sqrt{\lam}} &= \frac{1}{2}\left(-\nu_1 \cos^2{\theta_0 \over 2} -\nu_2\sin^2{\theta_0 \over 2} \right)\period
\end{align}
Finally, let us discuss the signs and the relative magnitudes of the parameters and 
 the charges. 
 The signs  and the relative magnitude  of $\nu_i$ depend on $u$. From the formulas for $\nu_i$ we can check that %
\begin{align}
|\Re u|>1: \qquad &  \nu_2 < \nu_1 < 0 \comma  \label{magnuone}\\ 
|\Re u|<1: \qquad & \nu_1 < 0 < \nu_2 \comma ( |\nu_1| < \nu_2) 
\period \label{magnutwo}
\end{align}
As for the angles, we always have
\begin{align}
\cos^2{\theta_0\over 2} > \sin^2{\theta_0\over 2} \period
\end{align}
The signs of $R$ and $L$ can be checked to be always positive. (
$R$  for the case $|\Re u| >1$ and $L$ for the case $|\Re u|<1$ are 
 somewhat non-trivial  to check.)

The relative magnitude of $R$ and $L$ can be deduced easily from 
the difference
\begin{align}
{1\over \sqrt{\lam}} (R -L) &= 2\nu_2 \sin^2{\theta_0 \over 2} \period
\end{align}
As the sign of $\nu_2$ has  already been 
 obtained in (\ref{magnuone}) and (\ref{magnutwo}), we immediately get
\begin{align}
&R<L \qquad \mbox{for}\quad |\Re u|>1 \comma \\
& R>L \qquad \mbox{for}\quad |\Re u|<1 \period 
\end{align}
\subsection{Pohlmeyer reduction for one-cut solutions\label{apsubsec:pohl}}
Let us next consider the variables appearing in the Pohlmeyer reduction, $\rho$, $\tilde{\rho}$ and $\gamma$ for one-cut solutions. From their definitions, we can express them in terms of the parameters of the one-cut solution as
\beq{
&\cos 2\gamma = \frac{\nu_1^2-m_1^2}{4\kappa^2}=\frac{\nu_2^2-m_2^2}{4\kappa^2}\comma\label{gam1}\\
&\rho = \frac{1}{8}\cos \frac{\theta_0}{2} \sin \frac{\theta_0}{2}\left((\nu_1+m_1)^2-(\nu_2+m_2)^2\right)\comma\label{rho1}\\
&\tilde{\rho}=\frac{1}{8}\cos \frac{\theta_0}{2} \sin \frac{\theta_0}{2}\left((\nu_1-m_1)^2-(\nu_2-m_2)^2\right)\comma\label{rho2}
}
where we used $z=\tau+i\sigma$ coordinate when we compute these quantities\fn{Note that $\gamma$ is invariant under the coordinate change $z\to z^{\prime}=f(z)$, whereas $\rho$ and $\tilde{\rho}$ transform respectively as $\rho \to \rho^{\prime}=\rho/(\del f)^2$ and $\tilde{\rho} \to \tilde{\rho}^{\prime}=\tilde{\rho}/(\delbar f)^2$.}. 

Using the results in the previous subsection, we can re-express \eqref{gam1}, 
\eqref{rho1} and \eqref{rho2} in terms of the branch points $u$ and $\bar{u}$. 
They are given by
\beq{
&\cos 2\gamma = \epsilon \frac{|u|^2-1}{|u^2-1|}\comma\quad \sin 2\gamma = \frac{2 \Im\, u}{|u^2-1|}\comma\\
&\rho = -\kappa^2 \frac{\Im \, u}{|u-1|^2}\comma\quad \tilde{\rho} =\kappa^2 \frac{\Im \, u}{|u+1|^2}\period
}
The ALP in the Pohlmeyer gauge can be solved in a similar manner and the result is given in \eqref{psi1} and \eqref{psi2}.

In the case of three-point functions, we can compute these quantities separately for each puncture as
\beq{
&\gamma_i = \frac{1}{2}\arcsin \left(\frac{2 \Im\, u_i}{|u_i^2-1|}\right)\comma\label{defgami}\\
&\rho_i = -\kappa^2 \frac{\Im \, u_i}{|u_i-1|^2}\comma\label{defrhoi}\\
&\tilde{\rho}_i =\kappa^2 \frac{\Im \, u_i}{|u_i+1|^2}\period\label{defrhotili}
}
They will be used in the computation of three-point functions.

\subsection{Computation of various integrals\label{apsubsec:landd}}
Using the above results, let us compute various integrals which appear in \Local
 and \Double in section \ref{sec:action}. Around a puncture, one can approximate the behavior of the world-sheet by that of the two-point functions. Thus, when three string states are semi-classically described 1-cut solutions, we expect the following asymptotic behavior of the one-forms:
\beq{
&\lambda \overset{z\to z_i}{\sim} \kappa _i dw_i\comma\quad \omega\overset{z\to z_i}{\sim} -\frac{\kappa_i \cos 2\gamma_i}{2}d\bar{w_i} + \frac{2\rho_i^2}{\kappa_i^3}dw_i\comma\label{asymp}
}
where $w_i$ is the local coordinate $w_i \equiv \tau^{(i)}+i\sigma^{(i)}$ around the puncture $z_i$.   

Using \eqref{asymp}, one can evaluate various integrals. First, the contour integrals of $\lambda$ and $\omega$ along $\mathcal{C}_i$'s are given by
\beq{
&\oint_{\mathcal{C}_i}\lambda=2\pi i\kappa_i\comma \quad \oint_{\mathcal{C}_i}\omega=2\pi i\left(\frac{\kappa_i\cos 2\gamma_i}{2}+\frac{2\rho_i^2}{\kappa_i^3}\right)\qquad i=1 ,\bar{2},3\period\label{ci}
}
On the other hand, the double contour integral, which appears in {\sf Double} can be computed as follows:
\beq{
\oint_{\mathcal{C}_i}\omega \int^{z}_{z^{\ast}_i}\lambda &=\int_{\sigma=0}^{\sigma =2\pi} \left(-\frac{\kappa_i \cos 2\gamma_i}{2}d\bar{w_i} + \frac{2\rho_i^2}{\kappa_i^3}dw_i\right)\int_{\sigma^{\prime}=0}^{\sigma^{\prime}=\sigma} \kappa_i dw^{\prime}_{i}\nn\\
&=-\int_{0}^{2\pi}d\sigma \left( \frac{\kappa_i \cos 2\gamma_i}{2} + \frac{2\rho_i^2}{\kappa_i^3}\right)\kappa_i\sigma\nn\\
&=-2\pi^2\left( \frac{\kappa_i \cos 2\gamma_i}{2} + \frac{2\rho_i^2}{\kappa_i^3}\right)\kappa_i\period\label{dc}
}
These results are used in section \ref{subsec:contour} to explicitly evaluate \Local and \Double.
%%%%%%%%%%%%%%%%%%%%%%%%%%%%%%%%
%%%%%%%%%%%%%%%%%%%%%%%%%%%%%%%%%%%5
\section{Pohlmeyer reduction\label{sec:pohl}}
%%%%%%%%%%%%%%%%%%%%%%%%
In this appendix, we will give some details of the Pohlmeyer reduction for the string on $S^3$. 

In terms of the embedding coordinate $Y_{I}$ $(I=1,\ldots,4)$, $S^3$ is realized as a hypersurface in $R^4$ satisfying $\sum_I Y_{I}^2=1$. The basic idea of the Pohlmeyer reduction is to describe the dynamics of the string in terms of a {\it moving frame} in $R^4$ consisting of four basis vectors $\{Y_I, \del Y_I ,\delbar Y_I, N_I\}$, which satisfy the following properties:
\beq{
N^{I}N_{I}=1\comma \quad N^{I}Y_{I} = N^{I}\del Y_{I}=N^{I} \delbar Y_{I}=0\period
}
Then, using the equation of motion, $\del \delbar Y^{I}+\left(\del Y^{J}\delbar Y_{J}\right) Y^{I}=0$ and the Virasoro constraints, $\del Y^I \del Y_I =-T(z)$ and $\delbar Y^I \delbar Y_I=-\bar{T}(\barz)$,
we can express the derivatives of these basis vectors, $\del N^{I}\comma \del^2 Y^{I}\comma$ {\it etc}. again in terms of the basis vectors:
\beq{
&\del N^{I}=\frac{2\rho}{T\sin ^2 2\gamma}\del Y^{I}+\frac{2\cos 2\gamma \rho }{\sqrt{T\bar{T}}\sin ^2 2\gamma}\delbar Y^{I}\comma\label{apeqe-3} \\
&\delbar N^{I}=\frac{2\rho}{\bar{T}\sin ^2 2\gamma}\delbar Y^{I} + \frac{2 \cos 2\gamma \tilde{\rho}}{\sqrt{T\bar{T}}\sin ^2 2\gamma}\del Y^{I}\comma\label{apeqe-4}\\
&\del^2 Y = T Y^{I} + \frac{\del \ln\left( T\bar{T}\sin ^2 2\gamma\right)}{2}\del Y^{I}+\sqrt{\frac{\bar{T}}{T}}\frac{2\del \gamma}{\sin 2\gamma}\delbar Y^{I} + 2\rho N^{I}\comma \label{apeqe-5}\\
&\delbar ^2 Y = \bar{T}Y+\frac{\delbar \ln\left( T\bar{T}\sin ^2 2\gamma\right)}{2}\delbar Y^{I} +\sqrt{\frac{T}{\bar{T}}}\frac{2\delbar \gamma}{\sin 2\gamma}\del Y^{I}+ 2\tilde{\rho} N^{I}\comma\label{apeqe-6}\\
&\del \delbar Y = -\sqrt{T\bar{T}}\cos 2\gamma Y\label{apeqe-7}\comma
}
where $\rho$, $\tilde{\rho}$ and $\gamma$ are defined by
\beq{
&\del Y^{I}\delbar Y_{I}= \sqrt{T\bar{T}}\cos 2\gamma \comma \quad
\rho\equiv \frac{1}{2}N^{I}\del^2 Y_{I}\comma \quad \tilde{\rho}\equiv \frac{1}{2}N^{I}\delbar^2 Y_{I}\period\label{aprhogam}
}
Using the equation of motion, one can also show that $\gamma$, $\rho$ and $\tilde{\rho}$ satisfy the generalized sin-Gordon equation, which is given in \eqref{singordon}.

Let us next derive a flat connection associated with the system of equations \eqref{apeqe-3}--\eqref{apeqe-7}. For this purpose, it is convenient to introduce the following orthonormal basis:
\beq{
&q_1\equiv Y \comma \quad q_2 \equiv -\frac{i}{\sin 2\gamma}\left[\frac{e^{i\gamma }}{\sqrt{T}}\del Y+ \frac{e^{-i\gamma }}{\sqrt{\bar{T}}}\delbar Y\right]\comma\\
&q_3\equiv\frac{i}{\sin 2\gamma}\left[\frac{e^{i\gamma }}{\sqrt{\bar{T}}}\delbar Y+ \frac{e^{-i\gamma }}{\sqrt{T}}\del Y\right]\comma \quad q_4 \equiv N\comma
}
which satisfy the following normalization conditions:
\beq{
q_1^2 = q_4^2=1\comma \quad q_2 q_3 =-2\period
}
With these orthonormal vectors, \eqref{apeqe-3}--\eqref{apeqe-7} can be re-expressed as the following set of equations,
\beq{
&\del q_1 = \frac{\sqrt{T}}{2}\left[e^{i\gamma}q_2+e^{-i\gamma}q_3\right]\comma \\
&\del q_2 =  e^{-i\gamma}\sqrt{T}q_1 +i\del \gamma q_2  - \frac{2i\rho}{\sqrt{T}\sin 2\gamma}e^{i\gamma}q_4\comma \\
&\del q_3 = e^{i\gamma }\sqrt{T}q_1 - i\del \gamma q_3 + \frac{2i\rho}{\sqrt{T}\sin 2\gamma}e^{-i\gamma}q_4\comma \\
&\del q_4 = \frac{i\rho e^{-i\gamma }}{\sqrt{T}\sin 2\gamma}q_2 - \frac{i\rho e^{i\gamma}}{\sqrt{T}\sin 2\gamma} q_3 \comma\\
&\delbar q_1 = -\frac{\sqrt{\bar{T}}}{2}\left[e^{-i\gamma}q_2+e^{i\gamma}q_3\right]\comma \\
&\delbar q_2 = - e^{i\gamma }\sqrt{\bar{T}}q_1 - i\delbar \gamma q_2 - \frac{2i\tilde{\rho}}{\sqrt{\bar{T}}\sin 2\gamma}e^{-i\gamma}q_4\comma \\
&\delbar q_3 = - e^{-i\gamma }\sqrt{\bar{T}}q_1 +i\delbar \gamma q_3 + \frac{2i\tilde{\rho}}{\sqrt{\bar{T}}\sin 2\gamma}e^{i\gamma}q_4\comma \\
&\delbar q_4 = \frac{i\tilde{\rho} e^{i\gamma }}{\sqrt{\bar{T}}\sin 2\gamma}q_2 + \frac{i\tilde{\rho} e^{-i\gamma}}{\sqrt{\bar{T}}\sin 2\gamma} q_3 \period
}
By expressing the basis in a matrix form,
\beq{
W\equiv \frac{1}{2}\pmatrix{cc}{q_1 + i q_4 & q_2\\q_3&q_1-iq_4}\comma
}
we can convert the above equations into the following form:
\beq{
&\del W +B_z^{L} W +W B_z^{R}=0\comma\quad \delbar W + B_{\barz}^{L}W+W B_{\barz}^{R}=0\comma\label{apeqe-8}
}
where $B_{z,\barz}^{L,R}$ are matrices defined by
\beq{
&B_{z}^{L}\equiv \pmatrix{cc}{-\frac{i\del \gamma}{2}&\frac{\rho e^{i\gamma}}{\sqrt{T}\sin 2\gamma}-\frac{\sqrt{T}}{2}e^{-i\gamma}\\\frac{\rho e^{-i\gamma}}{\sqrt{T}\sin 2\gamma} -\frac{\sqrt{T}}{2}e^{i\gamma}&\frac{i\del \gamma}{2}}\comma \label{BLz}\\
&B_{z}^{R}\equiv \pmatrix{cc}{\frac{i\del \gamma}{2}&-\frac{\rho e^{i\gamma}}{\sqrt{T}\sin 2\gamma}-\frac{\sqrt{T}}{2}e^{-i\gamma}\\-\frac{\rho e^{-i\gamma}}{\sqrt{T}\sin 2\gamma}-\frac{\sqrt{T}}{2}e^{i\gamma}&-\frac{i\del \gamma}{2}}\comma\\
&B_{\barz}^{L}\equiv \pmatrix{cc}{\frac{i\delbar \gamma}{2}&\frac{\tilde{\rho} e^{-i\gamma}}{\sqrt{\bar{T}}\sin 2\gamma}+\frac{\sqrt{\bar{T}}}{2}e^{i\gamma}\\\frac{\tilde{\rho} e^{i\gamma}}{\sqrt{\bar{T}}\sin 2\gamma}+\frac{\sqrt{\bar{T}}}{2}e^{-i\gamma}&-\frac{i\delbar \gamma}{2}}\comma\\
&B_{\barz}^{R}\equiv \pmatrix{cc}{-\frac{i\delbar \gamma}{2}&-\frac{\tilde{\rho} e^{-i\gamma}}{\sqrt{\bar{T}}\sin 2\gamma}+\frac{\sqrt{\bar{T}}}{2}e^{i\gamma}\\-\frac{\tilde{\rho} e^{i\gamma}}{\sqrt{\bar{T}}\sin 2\gamma}+\frac{\sqrt{\bar{T}}}{2}e^{-i\gamma}&\frac{i\delbar \gamma}{2}}\period \label{BRzbar}
}
\eqref{apeqe-8} is equivalent to the flatness conditions of the connections $B^{L}$ and $B^{R}$,
\beq{
\del B_{\barz}^{L}-\delbar B_{z}^{L}+[B_z^{L}\comma B_{\barz}^{L}]=0\comma \qquad \del B_{\barz}^{R}-\delbar B_{z}^{R}+[B_z^{R}\comma B_{\barz}^{R}]=0\period
}
Owing to the classical integrability of the string sigma model, we can ``deform" the above connection without spoiling the flatness by introducing a spectral parameter $\zeta =(1-x)/(1+x)$ as 
\beq{
&B_z (\zeta)\equiv \frac{\Phi_z}{\zeta} +A_z \comma \quad B_{\barz}(\zeta) \equiv \zeta \Phi_{\barz} + A_{\barz}\period\label{apeqe-9}
}
 $\Phi$'s and $A$'s are defined by\fn{\eqref{apeqe-9} is equivalent in form to the SL(2)-Hitchin system. However, the boundary conditions we impose around the punctures are different from the ones used in the usual analysis of the Hitchin system.}
\beq{
&\Phi_z \equiv \pmatrix{cc}{0&-\frac{\sqrt{T}}{2}e^{-i\gamma}\\-\frac{\sqrt{T}}{2}e^{i\gamma}&0}\comma\quad \Phi_{\barz} \equiv \pmatrix{cc}{0&\frac{\sqrt{\bar{T}}}{2}e^{i\gamma}\\\frac{\sqrt{\bar{T}}}{2}e^{-i\gamma}&0}\comma\\
&A_z \equiv \pmatrix{cc}{-\frac{i\del \gamma}{2}&\frac{\rho e^{i\gamma}}{\sqrt{T}\sin 2\gamma}\\\frac{\rho e^{-i\gamma}}{\sqrt{T}\sin 2\gamma}&\frac{i\del \gamma}{2}}\comma \quad A_{\barz}\equiv \pmatrix{cc}{\frac{i\delbar \gamma}{2}&\frac{\tilde{\rho} e^{-i\gamma}}{\sqrt{\bar{T}}\sin 2\gamma}\\\frac{\tilde{\rho} e^{i\gamma}}{\sqrt{\bar{T}}\sin 2\gamma}&-\frac{i\delbar \gamma}{2}}\period \label{Azzbar}
}
The deformed connection \eqref{apeqe-9} evaluated at $\zeta=1$ or $\zeta=-1$ is related to the original connection $B^{L,R}$ in the following way:
\beq{
B^{L}=B(\zeta =1)\comma \quad \left( B^{R}\right) ^{t}=\sigma_2 B(\zeta = -1)\sigma_2\period\label{apeqe-12}
}
Furthermore \eqref{apeqe-9} is related to the usual left/right connection by an appropriate gauge transformation as will be shown in Appendix \ref{apsec:relation}.
%%%%%%%%%%%

%%%%%%%%%%%%%%%%%%%%%%%%%%%%%%%%%%%%
\section{Relation between the Pohlmeyer reduction 
 and the sigma model formulation\label{apsec:relation}}
%%%%%%%%%%%%%%%%%%%%%%%%%%
In this appendix, we explain how the Pohlmeyer reduction 
 and the sigma model formulation are related. 
%%%%%%%%%%%%%%%%%
\subsection{Reconstruction formula for the Pohlmeyer reduction\label{apsubsec:pohl-recon}}
%%%%%%%%%%%%%%%%%%%%
 In section \ref{subsec:one-cut}  we presented the simple formulas  \eqref{reconsteqR} and \eqref{reconsteqL}  which reconstruct the solution  $\bbY$ of the equations of motion  from the eigenfunctions of the ALP in  the sigma model formulation.  We now describe a similar 
 formula for the Pohlmeyer reduction and by comparing such reconstruction 
formulas we can relate the two formulations. 
Consider the left and the right ALP associated with the Pohlmeyer reduction,
\beq{
\left( d + B^{L}\right) \psi^{L} =0\comma \quad \left( d+ B^{R} \right)\psi^R=0\comma \label{eq:b1}
}
and let  $\psi^{L,R}_{1}$ and $\psi^{L,R}_{2}$ be two linearly independent solutions satisfying the normalization conditions 
\beq{
\det \left(\psi^{L}_1 \comma\, \psi^{L}_2 \right)=1 \comma \quad \det \left( \psi^{R}_1 \comma\, \psi^{R}_2\right)=1\period 
} 
Then, similarly to the sigma model case,  the embedding coordinates  $\bbY$  can be reconstructed by the formula 
\beq{
\mathbb{Y}=q_1=\pmatrix{cc}{Z_1 &Z_2\\-\bar{Z}_2&\bar{Z}_1}=\left( \Psi^{L}\right)^{t}\Psi^{R}\comma\label{recon2}
}
where $\Psi^{L,R}$ are $2\times 2$ matrices with a unit determinant, defined by
\beq{
\Psi^{L}\equiv \left(\psi^{L}_1 \comma\,\psi^{L}_2 \right)\comma \qquad \Psi^{R}\equiv \left(\psi^{R}_1 \comma\,\psi^{R}_2 \right)\period 
}
Concerning the property under the global symmetry transformations,
 we should note the following. Since the Pohlmeyer connections $B^L$ and $B^R$ in the equation \eqref{eq:b1} are invariant,  $\Psi^L$ and $\Psi^R$ must also be invariant under such transformations acting from left. However, as for transformations from right, they may transform non-trivially. In fact, as we shall see shortly, 
 they must transform covariantly from right so that the solutions of the ALP 
 for the Pohlmeyer and the sigma model formulations are 
 connected consistently by a gauge transformation. 

Furthermore, one can check that the quantities $q_2$ and $q_3$, which 
 consist of the derivatives of $\bbY$, can be reconstructed  as 
\beq{
q_2 = \left(\Psi^{L}\right)^{t}\pmatrix{cc}{0&2\\0&0}\Psi^R\comma\qquad  q_3 = \left( \Psi^{L}\right)^{t}\pmatrix{cc}{0&0\\2&0}\Psi^{R}\period
}
From these formulas the derivatives of $\bbY$ can be obtained as 
\beq{
\del \mathbb{Y}=\frac{\sqrt{T}}{2}\left[e^{i\gamma}q_2+e^{-i\gamma}q_3 \right]\comma\qquad\delbar \mathbb{Y}=-\frac{\sqrt{\bar{T}}}{2}\left[e^{-i\gamma}q_2+e^{i\gamma}q_3 \right]\period\label{deldelbY}
}
Note that,  in distinction to the case of the sigma model,   the reconstruction formulas for the Pohlmeyer reduction does not use the eigenvectors 
 of the monodromy matrices,  namely $\hat{\psi}_{\pm}$. 
The solutions $\psi^{L,R}_i$ used are simply two linearly independent solutions to the ALP, which are not necessarily the eigenvectors of $\Omega$. 
%%%%%%%%%%%%%%%%%%%%%%%%
\subsection{Relation between the connections and the eigenvectors\label{apsubsec:con-eigen}}
%%%%%%%%%%%%%%%%%%%%%%%%%
We now discuss the relation between the connections and the eigenvectors of the  the Pohlmeyer reduction and those of the sigma model. 

First consider the relation to the right connection of the sigma model. 
From the formulas for $\del \bbY$ and $\delbar \bbY$ given in \eqref{deldelbY}, we can  form the right connection $j$  as 
\beq{
j_z= \sqrt{T}\left( \Psi^{R}\right)^{-1}\pmatrix{cc}{0&e^{i\gamma}\\e^{-i\gamma}&0}\Psi^{R}\comma \quad j_{\barz}=-\sqrt{\bar{T}}\left( \Psi^{R}\right)^{-1}\pmatrix{cc}{0&e^{-i\gamma}\\e^{i\gamma}&0}\Psi^{R}\period\label{apeqe-10}
}
Then, comparing \eqref{apeqe-10} with \eqref{apeqe-9}--\eqref{Azzbar}, we find that the following gauge transformation connects  the  flat connections
 of the two formulations:
\begin{align}
\frac{1}{1-x}j_z &= \calG^{-1} B_z(\zeta) \calG + \calG^{-1} \del \calG 
\comma \\
\frac{1}{1+x}j_{\barz} &= \calG^{-1} B_\zbar(\zeta) \calG + \calG^{-1} \delbar \calG \comma 
\end{align}
where 
\begin{align}
\calG &= i\sig_2 \Psi^R \period \label{defcalG}
\end{align}
The eigevectors $\psi_\pm$ of the sigma model formulation and those 
of the Pohlmeyer reduction, denoted by  $\psihat_\pm$, are 
related as 
\beq{
\psi_{\pm}=\calG^{-1}  \hat{\psi}_{\pm}\period\label{eip1}
}
Note that the factor of  $i$  in \eqref{defcalG} is  needed  to reproduce the correct normalization condition  $\sl{\psi_{+}\comma \psi_{-}}=1$. 
Under the global $\SUR$ transformation $U_R$,  $\psi_\pm$ transform 
 as $\psi_\pm \rightarrow U_R^{-1} \psi_\pm$. From the above formulas 
\eqref{defcalG} and \eqref{eip1} we see that this corresponds to 
the transformation  $\Psi^R \rightarrow \Psi^R U_R$, as remarked 
previously.  

In an exactly similar manner, we can construct the left current $l$'s by
\beq{
l_z =\sqrt{T}\left( \Psi^{L}\right)^{t}\pmatrix{cc}{0&e^{i\gamma}\\e^{-i\gamma}&0}\left[\left(\Psi^{L}\right)^{t}\right]^{-1}\comma \quad l_{\barz}=-\sqrt{\bar{T}}\left( \Psi^{L}\right)^{t}\pmatrix{cc}{0&e^{-i\gamma}\\e^{i\gamma}&0}\left[\left(\Psi^{L}\right)^{t}\right]^{-1}\comma\label{apeqe-11}
}
Comparing \eqref{apeqe-11} with  \eqref{apeqe-9}--\eqref{Azzbar}, we find that the following gauge transformation connects the two connections:
\begin{align}
\frac{x}{1-x}l_z &= \tilde{\calG}^{-1} B_z(\zeta) \tilde{\calG} + 
\tilde{\calG}^{-1} \del \tilde{\calG} \comma \\
-\frac{x}{1+x}l_{\barz} &= \tilde{\calG}^{-1} B_z(\zeta) \tilde{\calG} + 
\tilde{\calG}^{-1} \delbar \tilde{\calG} \comma 
\end{align}
where 
\begin{align}
\tilde{\calG} &= \left[  (\Psi_L)^t (-i\sig_2)\right]^{-1} = i\Psi_L \sig_2 \period\label{defcalGtil}
\end{align}
The eigenvectors are related as 
\beq{
\psitil_{\pm}=\tilde{\calG}^{-1} \hat{\psi}_{\pm}\period\label{eip2}
}
Using \eqref{eip1} and \eqref{eip2}, one can show the equivalence between the reconstruction formulas \eqref{reconsteqR}, \eqref{reconsteqL} and \eqref{recon2}.

%%%%%%%%%%%%%%%%%%%%%%%%%%%
\section{Details of the WKB expansion\label{ap:WKB}}
%%%%%%%%%%%%%%%%%%%%%%%%%%%
In this appendix, we explain the details of the WKB expansion for the solutions to the ALP. We will describe two approaches, each of which 
 has its own merit. 
First in subsection \ref{apsub:direct}, we will perform a direct expansion in 
 the small parameter $\zeta$,  which is useful for clarifying the general structure of the expansion. This method, however, turned out to be not quite suitable   for deriving the explicit formulas for the expansion of the Wronskians. 
Therefore, in subsection \ref{apsub:born}, we take a slightly different approach  based on the Born series expansion. This allows us to derive the expressions for the Wronskians up to the $O(\zeta^1)$ terms with relative ease,  with the results  given in  \eqref{++wkb}, \eqref{--wkb}, \eqref{++wkb2} and \eqref{--wkb2}. 
%%%%%%%%%%%%%%%%%%%%%%%%%%%%%%%%
\subsection{Direct expansion of the solutions to the ALP\label{apsub:direct}}
%%%%%%%%%%%%%%%%%%%%%%%%%%%%%%%%
In this subsection, we will perform a direct expansion of the ALP in the ``diagonal gauge"  introduced in section \ref{subsec:WKB}.  In this gauge the ALP equations 
 become 
\beq{
\left(\del + \frac{1}{\zeta}\Phi^d_z + A^d_z\right) \hat{\psi}^d=0\comma \qquad \left(\delbar + \zeta\Phi^d_{\barz} + A^d_{\barz}\right)\hat{\psi}^d=0\period\label{apdiagg}
}
Denoting the components of  $\hat{\psi}^d$ as 
\beq{
\hat{\psi}^d \equiv \pmatrix{c}{\psi^{(1)}\\\psi^{(2)}}\comma
}
and substituting the expressions for $\Phi^d_z, A^d_z$, etc.  given 
 in \eqref{diag-conn}, the ALP equations above take the form 
\beq{
&\del \psi^{(1)} + \frac{\sqrt{T}}{2\zeta}\psi^{(1)} - \frac{\rho}{\sqrt{T}}\cot 2\gamma \psi^{(1)}+i\left( \frac{\rho}{\sqrt{T}}-\del \gamma\right)\psi^{(2)}=0\comma\label{del-1}\\
&\del \psi^{(2)} - \frac{\sqrt{T}}{2\zeta}\psi^{(2)} + \frac{\rho}{\sqrt{T}}\cot 2\gamma \psi^{(2)}-i\left( \frac{\rho}{\sqrt{T}}+\del \gamma\right)\psi^{(1)}=0\comma\label{del-2}
}
and
\beq{
&\delbar \psi^{(1)} - \zeta\frac{\sqrt{\bar{T}}\cos 2\gamma}{2}\psi^{(1)} - \frac{\tilde{\rho}}{\sqrt{\bar{T}}\sin 2\gamma} \psi^{(1)}+i\frac{\sqrt{\bar{T}}\sin 2\gamma}{2}\psi^{(2)}=0\comma\label{delbar-1}\\
&\delbar \psi^{(2)} + \zeta\frac{\sqrt{\bar{T}}\cos 2\gamma}{2}\psi^{(2)} + \frac{\tilde{\rho}}{\sqrt{\bar{T}}\sin 2\gamma} \psi^{(2)}-i\frac{\sqrt{\bar{T}}\sin 2\gamma}{2}\psi^{(1)}=0\period\label{delbar-2}
}

Let us examine the first two equations  \eqref{del-1} and \eqref{del-2}. To perform the  WKB expansion, it is useful to introduce a coordinate $w$ defined by
\beq{
dw=\sqrt{T}dz\period
}
By this coordinate transformation we can absorb the factor $\sqrt{T}$ and bring the equations to  the simplified  form 
\beq{
&\del_w \psi^{(1)} + \frac{1}{2\zeta}\psi^{(1)} - \frac{\rho}{T}\cot 2\gamma \psi^{(1)}+i\left( \frac{\rho}{T}-\del_w \gamma\right)\psi^{(2)}=0\comma\label{delw-1}\\
&\del_w \psi^{(2)} - \frac{1}{2\zeta}\psi^{(2)} + \frac{\rho}{T}\cot 2\gamma \psi^{(2)}-i\left( \frac{\rho}{T}+\del_w \gamma\right)\psi^{(1)}=0\period\label{delw-2}
}
Let us  express $\psi^{(2)}$ in terms of $\psi^{(1)}$ using \eqref{delw-1}. We get
\beq{
\psi^{(2)}=-i\left(\frac{\rho}{T} -\del_w \gamma\right)^{-1}\left[ \del_w \psi^{(1)} + \left(\frac{1}{2\zeta}-\frac{\rho}{T}\cos 2\gamma \right)\psi^{(1)}\right]\period\label{2intermsof1}
}
Substituting \eqref{2intermsof1} into \eqref{delw-2}, we obtain a second order differential equation for $\psi^{(1)}$ of the form
\beq{
\del_w^2 \psi^{(1)} -\del_w \ln \left(\frac{\rho}{T}-\del_w \gamma \right)\del_w\psi^{(1)} -A\psi^{(1)}=0\comma \label{holo2nd}
}
where  $A$ is given by
\beq{
A=&\left(\frac{1}{2\zeta}-\frac{\rho}{T}\cot 2\gamma \right)^2 +\del_w \left(\frac{\rho}{T}\cot 2\gamma \right)+\del_w \ln \left(\frac{\rho}{T}-\del_w \gamma \right)\left(\frac{1}{2\zeta}-\frac{\rho}{T}\cot 2\gamma \right)\nn\\
&+\left( \del_w \gamma\right)^2 -\left( \frac{\rho}{T}\right)^2\period
}
We now make the WKB expansion of $\psi^{(1)}$ in powers of $\zeta$ in the form,
\beq{
\psi^{(1)}=\sqrt{\frac{\rho}{T}-\del_w \gamma}\,\exp\left[\frac{W_{-1}}{\zeta}+W_0 + \zeta W_1 +\cdots\right]\comma\label{exppsi1}
}
and substitute it  into \eqref{holo2nd}.  Then, at order $\zeta^{-2}$, 
we get the equation 
\beq{
\left( \del_{w}W_{-1}\right)^2 = \frac{1}{4}\comma 
} 
with the solutions given by   $\del_{w}W_{-1}=\pm1/2$.  At the next order, we get the equation 
\beq{
\del_w^2 W_{-1} + 2\del_w W_{-1}\del_w W_0 = \frac{1}{2}\del_w \ln \left(\frac{\rho}{T}-\del_w \gamma \right)-\frac{\rho}{T}\cot 2\gamma\period\label{order-1}
}
From this $\del_w W_0$ is determined as
\beq{
\del_w W_0 =\pm\left[ \frac{1}{2}\del_w \ln \left(\frac{\rho}{T}-\del_w \gamma \right)-\frac{\rho}{T}\cot 2\gamma\right]\comma \label{W0}
}
where the plus sign is for $\del_w W_{-1}=+1/2$ and the minus sign is for $\del_w W_{-1}=-1/2$.
Similarly, we can determine $\del_w W_1$ as 
\beq{
\del_w W_1 =& \pm \left[(\del_w\gamma)^2 -\left(\frac{\rho}{T}\right)^2 +\del_w \left( \frac{\rho}{T}\cot 2\gamma\right)-\frac{1}{2}\del_w^2 \ln \left(\frac{\rho}{T}-\del_w \gamma \right)\right]\nn\\
&-\frac{1}{2}\del_w^2 \ln \left(\frac{\rho}{T}-\del_w \gamma \right)
\comma \label{W1}
}
where the choice of the sign should  be the same as in 
 \eqref{W0}. 
Continuing in this fashion  using \eqref{delbar-1} and \eqref{delbar-2},
  we can determine $\delbar W_{-1}$, $\delbar W_{0}$ and $\delbar W_{1}$  to be 
\beq{
\begin{aligned}
&\delbar W_{-1}=0\comma \quad \delbar W_{0}=\pm \left[\frac{1}{2}\delbar \ln \left( \frac{\rho}{T}-\del_w \gamma\right)-\frac{\tilde{\rho}}{\sqrt{\bar{T}}\sin 2\gamma}\right]\comma\\
&\delbar W_{1}=\pm\left[\frac{\eta}{2}-\frac{1}{2}\delbar \del_w \ln \left(\frac{\rho}{T}-\del_w \gamma \right)\right]-\frac{1}{2}\delbar \del_w \ln \left(\frac{\rho}{T}-\del_w \gamma \right)\period
\end{aligned}
}

The results obtained above can be reorganized into a compact form. 
In fact we can write the expansion  \eqref{exppsi1} as
\beq{
\psi^{(1)} = \exp \left[W_{\rm odd} + W_{\rm even} \right]\comma\label{oddeven}
}
where $W_{\rm odd}$ (resp. $W_{\rm even}$)  denotes  terms which (do not) change sign under the sign-flip of $\del_w W_{-1} $. 
 Then, by substituting \eqref{oddeven} into \eqref{holo2nd} and extracting the terms odd under the above flip of sign, we can obtain the following simple equation expressing $W_{\rm even}$ in terms of $W_{\rm odd}$:
\beq{
W_{\rm even}=-\frac{1}{2}\ln \del_w W_{\rm odd}\period\label{weven}
}
As is clear from the analysis above, the WKB expansion of $W_{\rm odd}$ is given in terms of the integrals of certain functions of the worldsheet variables, such as $\gamma$, $\rho$ and $\tilde{\rho}$. On the other hand, the even part  $W_{\rm even}$, which depends only on the derivatives of $W_{\rm odd}$, is expressed purely in terms of  the local values of the worldsheet variables. 
With  such classifications, we can recast the WKB expansion of the two linearly independent solutions of the ALP  into the following form:
\beq{
\hat{\psi}^d= \pmatrix{c}{f_{\pm }^{(1)}\\f_{\pm }^{(2)}}\exp\left(\pm\int_{z_0}^{z}W_{\swkb}(z, \barz ;\zeta )\right)\period
}
Here we renamed $W_{\rm odd}$ to $W_{\swkb}$ and the functions $f_{\pm }^{(1)}$ and $f_{\pm }^{(2)}$ are defined in terms of $W_{\swkb}^z$ by
\beq{
&f_{\pm }^{(1)}\equiv k_{\swkb}=\sqrt{\frac{\rho-\sqrt{T}\del \gamma}{T\,W^z_{\swkb}}}\comma \\
&f_{\pm}^{(2)}\equiv \frac{-i}{\sqrt{W_{\swkb}^z}}\left[ \pm W^z_{\swkb}  + \left(\frac{\sqrt{T}}{2\zeta}-\frac{\rho\cos 2\gamma}{\sqrt{T}} +\frac{\del \ln k_{\swkb}}{2} \right)\right]\period
}  

%%%%%%%%%%%%%%%%%%%%%%%%%%%%%%%%%%
\subsection{Born series expansion of the Wronskians\label{apsub:born}}
%%%%%%%%%%%%%%%%%%%%%%%%%%%%%%%%%%%
In this subsection, we will derive the explicit form of the expansion for the Wronskians up to $O(\zeta^1)$ using the Born series method, which 
 turned out to be more convenient compared to the direct expansion 
 described above. 
 In particular,  with this method it is much easier to take into account the normalization conditions of the eigenvectors $i_{\pm}$ given in  \eqref{normcond3}. 
Although the method has been described in Appendix B of \cite{CT}, 
we will spell out the details of the derivation since several additional considerations are necessary in our case.

To illustrate the basic idea, let us take the Wronskian $\sl{2_{+}\comma 1_{+}}$ as an example and discuss its expansion. To compute $\sl{2_{+}\comma 1_{+}}$, we need to parallel-transport the eigenvector $1_{+}$, which is defined originally in the neighborhood of $z_1$, to the neighborhood of $z_2$  using the flat connection and compute the Wronskian with $2_{+}$. 
In the diagonal gauge, this procedure can be implemented   in the following
 way: 
\beq{
\sl{\hat{2}^d_{+}\comma \hat{1}^d_{+}}=\sl{ \hat{2}^d_{+}(z_2^{\ast})\comma \textrm{P}\exp\left[-\int_{0}^{1}dt \left( \frac{1}{\zeta}H_0(t)+ V(t) \right)\right] \hat{1}^d_{+}(z_1^{\ast})}\period\label{dyson}
}
In this expression   $t$ parametrizes the curve joining  $z_1^{\ast}$ (at $t=0$) and  $z_2^{\ast}$ (at $t=1$) and $H_0$ and $V$ are defined in terms of the connection in the diagonal gauge,  given in \eqref{diag-conn},  as
\beq{
H_0(t) \equiv \tilde{\Phi}_z \dot{z}\comma \quad V(t) \equiv \tilde{A}_{z} \dot{z} + \tilde{A}_{\barz}\dot{\barz}+ \zeta \tilde{\Phi}_{\barz} \dot{\barz}\comma 
}
with  $\dot{z}$ and $\dot{\barz}$ standing for  $dz/dt$ and $d\barz/dt$ respectively. The equation 
\eqref{dyson} is similar in form to the transition amplitude in quantum mechanics,  where $H_0(t)/\zeta$ is the unperturbed Hamiltonian and $V(t)$ is the time-dependent perturbation. 
Therefore we can  derive the expansion of \eqref{dyson} by applying the 
familiar Born series expansion. 

As the  first step toward this goal, let us determine the expansion of the ``initial states", $\hat{1}^d_{+}(z_1^{\ast})$ and $\hat{2}^d_{+}(z_2^{\ast})$. As explained in section \ref{subsec:gen-three}, the eigenvectors can be well-approximated near the puncture by those of the corresponding two-point functions. Thus, the expansion of the initial states can be obtained from the explicit form of $\hat{i}_{\pm}^{\rm 2pt}$ given in \eqref{psi1} and \eqref{psi2} as
\beq{
\hat{1}^d_{+}(z_1^{\ast}) \sim \hat{1}_{+}^{{\rm 2pt},d} = \pmatrix{c}{O(\zeta^1)\\ 1+O(\zeta^2 )}\comma \qquad \hat{2}^d_{+}(z_2^{\ast})\sim \hat{2}_{+}^{{\rm 2pt},d}= \pmatrix{c}{1+O(\zeta^2 )\\ O(\zeta^1)}\period \label{Born-initial}
}

Let us now study  the leading terms (\ie the $O(V^0)$ terms) in the Born series expansion of \eqref{dyson}. They can be expressed as 
\beq{
1_{+}^{(2)}(z_1^{\ast})2_{+}^{(1)}(z_2^{\ast})\bra{\text{\bf e}_2}e^{-\int_{0}^{1}H_0dt/\zeta}\ket{\text{\bf e}_2}-1_{+}^{(1)}(z_1^{\ast})2_{+}^{(2)}(z_2^{\ast})\bra{\text{\bf e}_1}e^{-\int_{0}^{1}H_0dt/\zeta}\ket{\text{\bf e}_1}\comma\label{leadborn0}
}
where $\ket{\text{\bf e}_1}$ and $\ket{\text{\bf e}_2}$ stand for 
 the unit vectors 
\beq{
\ket{\text{\bf e}_1}=\pmatrix{c}{1\\0}\comma \quad \ket{\text{\bf e}_2}=\pmatrix{c}{0\\1}\comma
}
and $i_{\pm}^{(1)}$ and $i_{\pm}^{(2)}$ are  the upper and the lower component of $\hat{i}_{\pm}^d$ respectively,  which can be expressed as 
\beq{
\hat{i}_{\pm}^d = i_{\pm}^{(1)}\ket{\text{\bf e}_1} + i_{\pm}^{(2)}\ket{\text{\bf e}_2}\period
}
Using \eqref{Born-initial}, we can evaluate the expression \eqref{leadborn0} explicitly as
\beq{
\left(1 +O(\zeta^2 )\right)\exp\left( \int_{\ell_{12}}{1 \over \zeta}
\varpi \right) - O(\zeta^2) \exp\left( -\int_{\ell_{12}}{1 \over \zeta}
\varpi \right)\comma\label{leadingborn}
} 
where $\ell_{12}$ is the contour that connects  $z_1^{\ast}$ and $z_2^{\ast}$,  defined in section \ref{subsec:contour}. Note that the second term in \eqref{leadingborn}, which has an overall $O(\zeta^2)$ factor can be safely neglected only when $\Re\left( \int_{\ell_{12}}\varpi/\zeta\right)$ is positive so that  the exponential  $\exp\left( -\int_{\ell_{12}}\varpi/\zeta\right)$  becomes vanishingly small. The positivity of $\Re\left(\int_{\ell_{12}}\varpi/\zeta\right)$ is guaranteed when the following two conditions are satisfied:
\begin{enumerate}
\item The eigenvectors, $1_{+}$ and $2_{+}$, are small solutions.
\item $z_1$ and $z_2$ are connected by a {\it WKB curve} $z(s)$ 
defined to be satisfying the condition 
\beq{
\Im  \left( \sqrt{T} \frac{dz}{ds}\right) =0\comma\label{defWKB}
}
where $s$ parameterizes the curves.
\end{enumerate}
This can be deduced in the following way: First, from the definition \eqref{defWKB}, one can show that the real part of the integral $\int \varpi/\zeta$ monotonically increases or decreases along the WKB curve.  Second, when $1_{+}$ and $2_{+}$ are both small solutions, $\Re\left( \int \varpi/\zeta\right)$ increases as we {\it move away from} $z_1$ in the vicinity of $z_1$ while it increases as we {\it approach} $z_2$ in the vicinity of $z_2$. From these two observations, one can  conclude that $\Re\left(\int_{\ell_{12}}\varpi/\zeta\right)$ is positive when both of the eigenvectors are small and the punctures are connected by a WKB curve.  Actually, in practice the second condition
above is inessential.  This is because all the punctures are always connected with each other by  WKB curves,  except at discrete values of $\text{Arg}\left(\zeta\right)$,   due to the triangular inequalities, $\Delta_i<\Delta_j+\Delta_k$ (or equivalently $\kappa_i<\kappa_j+\kappa_k$),  which hold  in all  the cases we study in this paper. 

Let us now move on to the study of the $O(V^1)$ contributions. 
When $1_{+}$ and $2_{+}$ are small solutions, the $O(V^1)$ terms in the Born series expansion are given by 
\beq{
\begin{aligned}
&-1_{+}^{(2)}(z_1^{\ast})2_{+}^{(1)}(z_2^{\ast})\int_0 ^1 dt_1 \bra{\text{\bf e}_2}e^{-\int_{t_1}^{1}H_0dt/\zeta}V(t_1)e^{-\int_{0}^{t_1}H_0dt/\zeta}\ket{\text{\bf e}_2}\\
&-1_{+}^{(1)}(z_1^{\ast})2_{+}^{(1)}(z_2^{\ast})\int_0 ^1 dt_1 \bra{\text{\bf e}_2}e^{-\int_{t_1}^{1}H_0dt/\zeta}V(t_1)e^{-\int_{0}^{t_1}H_0dt/\zeta}\ket{\text{\bf e}_1}\\
&+1_{+}^{(2)}(z_1^{\ast})2_{+}^{(2)}(z_2^{\ast})\int_0 ^1 dt_1 \bra{\text{\bf e}_1}e^{-\int_{t_1}^{1}H_0dt/\zeta}V(t_1)e^{-\int_{0}^{t_1}H_0dt/\zeta}\ket{\text{\bf e}_2}\period
\end{aligned}
\label{Ov1}
}
Note that we have omitted the terms of the form, $\bra{\text{\bf e}_1}\ast \ket{\text{\bf e}_1}$, since they are proportional to the factor $\exp\left(\int_{\ell_{12}}\varpi/\zeta \right)$, which, as discussed above,  is exponentially small when $1_{+}$ and $2_{+}$ are small solutions. Since $\ket{\text{\bf e}_1}$ and $\ket{\text{\bf e}_2}$ are the eigenvectors of $H_0$, we can evaluate \eqref{Ov1} as
\beq{
\begin{aligned}
&-1_{+}^{(2)}(z_1^{\ast})2_{+}^{(1)}(z_2^{\ast})e^{\int_{\ell_{12}}\varpi/(2\zeta)} \int_0 ^1 dt_1 \bra{\text{\bf e}_2}V(t_1)\ket{\text{\bf e}_2}\\
&-1_{+}^{(1)}(z_1^{\ast})2_{+}^{(1)}(z_2^{\ast})e^{\int_{\ell_{12}}\varpi/(2\zeta)}\int_0 ^1 dt_1 \bra{\text{\bf e}_2}V(t_1)\ket{\text{\bf e}_1}e^{-\int_{0}^{t_1}\varpi/\zeta}\\
&+1_{+}^{(2)}(z_1^{\ast})2_{+}^{(2)}(z_2^{\ast})e^{\int_{\ell_{12}}\varpi/(2\zeta)}\int_0 ^1 dt_1 \bra{\text{\bf e}_1}V(t_1)\ket{\text{\bf e}_2}e^{-\int_{t_1}^{1}\varpi/\zeta}\period
\end{aligned}
\label{Ov1-2}
}
In the limit $\zeta\to 0$, the integral over $t_1$ in the second term  will be exponentially suppressed by the factor $\exp\left(-2 \int_{0}^{t_1}\varpi/\zeta\right)$,  except when the  interval is short, \ie  $0<t_1<  O(\zeta^1)$. Thus, to $O(\zeta^1)$, one can take $\varpi$ in  $\int_{0}^{t_1}\varpi/\zeta$ to be constant and replace $V(t_1)$ with $V(0)$. We can thus approximate the second term in \eqref{Ov1-2} as
\beq{
-\zeta \,1_{+}^{(1)}(z_1^{\ast})2_{+}^{(1)}(z_2^{\ast})e^{\int_{\ell_{12}}\varpi/(2\zeta)}\bra{\text{\bf e}_2}V(0)\ket{\text{\bf e}_1}\left(\sqrt{T(z_1^{\ast})}\,\dot{z}(t=0)\right)^{-1} \period \label{secondap}
}
Since the factor $1_{+}^{(1)}(z_1^{\ast})$ is of $O(\zeta^1)$, \eqref{secondap} as a whole  is of $O(\zeta^2)$ and thus can be neglected  to the order of our approximation. Similarly, one can also show that the third term of \eqref{Ov1-2}  is of $O(\zeta^2)$. Thus, up to $O(\zeta^1)$, the contribution comes only from the first term proportional to 
\beq{
-e^{\int_{\ell_{12}}\varpi/(2\zeta)} \int_0 ^1 dt_1 \bra{\text{\bf e}_2}V(t_1)\ket{\text{\bf e}_2}\period\label{rborn2}
}

Lastly let us examine the $O(V^2)$ terms. The only term which contributes  at $O(\zeta^1)$ is
\beq{
&1_{+}^{(2)}(z_1^{\ast})2_{+}^{(1)}(z_2^{\ast})\int_{0}^{1}dt_2 \int_{0}^{t_2}dt_1 \bra{\text{\bf e}_2}e^{-\int_{t_2}^{1}H_0dt/\zeta}V(t_2)e^{-\int_{t_1}^{t_2}H_0dt/\zeta}V(t_1)e^{-\int_{0}^{t_1}H_0dt/\zeta}\ket{\text{\bf e}_2}\period\label{dysonexp}
}
Inserting the identity $\id = \ket{\text{\bf e}_1}\bra{\text{\bf e}_1} + \ket{\text{\bf e}_2}\bra{\text{\bf e}_2}$, this quantity can be computed as 
\beq{
1_{+}^{(2)}(z_1^{\ast})2_{+}^{(1)}(z_2^{\ast})e^{\int_{\ell_{12}}\varpi /(2\zeta)}&\left( \frac{1}{2}\left[\int^{1}_{0}dt_1 \bra{\text{\bf e}_2}V(t_1)\ket{\text{\bf e}_2}\right]^2 \right.\nn\\
&\left.+\int_{0}^{1}dt_1 \int_{0}^{t_1}dt_2e^{-\int_{t_2}^{t_1}\varpi/\zeta}\bra{\text{\bf e}_2}V(t_1)\ket{\text{\bf e}_1}\bra{\text{\bf e}_1}V(t_2)\ket{\text{\bf e}_2} \right)\period\label{Ov2}
}
As in the discussion of the $O(V^1)$ terms, we can take $\varpi$ in $\int_{t_2}^{t_1} \varpi/\zeta$ to be constant and replace $V(t_2)$ with $V(t_1)$ in the second term of \eqref{Ov2}, thanks to the suppression factor  $\exp\left(-\int_{t_1}^{t_2}\varpi/\zeta\right)$. Then \eqref{Ov2} can be evaluated as
\beq{
e^{\int_{\ell_{12}}\varpi /(2\zeta)}\left( \frac{1}{2}\left[\int^{1}_{0}dt_1 \bra{\text{\bf e}_2}V(t_1)\ket{\text{\bf e}_2}\right]^2 +\zeta \int_{0}^{1}dt_1\frac{\bra{\text{\bf e}_2}V(t_1)\ket{\text{\bf e}_1}\bra{\text{\bf e}_1}V(t_1)\ket{\text{\bf e}_2}}{\dot{z}\sqrt{T}}\right)\period\label{rborn3}
}
Putting together the expressions \eqref{leadingborn}, \eqref{rborn2} and \eqref{rborn3}, we find that the result can be grouped into an exponential 
in the following way:
\beq{
&\sl{2_{+}\comma 1_{+}}
\sim\exp \left( \frac{1}{2\zeta}\int_{\ell_{12}}\varpi-\int_{0}^{1}dt \bra{\text{\bf e}_2}V(t)\ket{\text{\bf e}_2}+\zeta \int_{0}^{1}dt_1\frac{\bra{\text{\bf e}_2}V(t)\ket{\text{\bf e}_1}\bra{\text{\bf e}_1}V(t)\ket{\text{\bf e}_2}}{\dot{z}\sqrt{T}}\right)\period
}
Thus we have obtained the expansion of $\sl{2_{+}\comma 1_{+}}$ to be 
given by 
\beq{
&\sl{2_{+}\comma 1_{+}}=\exp \left( -\frac{1}{2\zeta}\int_{\ell_{21}}\varpi-\int_{\ell_{21}} \alpha - \frac{\zeta}{2}\int_{\ell_{21}}\eta +O(\zeta^2) \right)\comma
}
where the one-form $\alpha$ is given by
\beq{
\alpha = -\frac{\rho}{\sqrt{T}}\cot 2\gamma dz -\frac{\tilde{\rho}}{\sqrt{\bar{T}}\sin 2\gamma} d\barz\period\label{defofal}
}
The expansion of other Wronskians can be worked out 
 in a similar manner leading to \eqref{++wkb} and \eqref{--wkb}. Furthermore, we can apply the same argument to the expansion around $\zeta=\infty$ and obtain \eqref{++wkb2} and \eqref{--wkb2}, where the one-form $\tilde{\alpha}$ appearing in the $O(\zeta^0)$ term  is given by
\beq{
\tilde{\alpha}=-\frac{\rho}{\sqrt{T}\sin 2\gamma} dz-\frac{\tilde{\rho}}{\sqrt{\bar{T}}}\cot 2\gamma d\barz\period\label{defofalbar}
}
%%%%%%%%%%%%%%%
%%%%%%%%%%%%%%%%%%%%%%%
%end of nullify
%%%%%%%%%%%%%%%%%%%%%%%%%%%
%%%%%%%%%%%%%%%%%%%%%%%%%%%

%%%%%%%%%%%%%%%%%%%%%%%%%%%%%%%%%%%%%%%%%%%%%%%%%%%%%%%%%%%%%%%%
\end{document}